\documentclass[11pt,a4paper]{article}
\usepackage{jheppub}
\usepackage{enumitem}
\usepackage{amsmath,amsthm}
\usepackage{lscape}
\usepackage{dsfont}
\usepackage{tikz}
\usepackage{tkz-graph}
\usepackage{tkz-berge}
\usepackage{pgf}
\usepackage{rotating}
\usepackage{xfrac}
\usepackage{booktabs}
\usepackage{caption}
\def\be{ \begin{equation} }
\def\ee{ \end{equation}}

\def\Co0{{\rm Co}_0}

\def\dim{{\rm dim}}

\renewcommand{\Im}{{\rm Im }}
\def\ker{{\rm ker}}

\def\one{{\hbox{ 1\kern-.8mm l}}}

\def\CD {{\cal D}}

\def\IC{\mathbb{C}}

\def\IZ{{\mathbb{Z}}}

\def\Fi{\tsl{Fi}}

\def\Th{\tsl{Th}}
\def\HN{\tsl{HN}}

\def\V{\mathcal{V}}
\def\U{\mathcal{U}}
\def\W{\mathcal{W}}

\theoremstyle{definition}
\newtheorem{exmp}{Example}[section]

\def\bea{\begin{eqnarray}}
\def\eea{\end{eqnarray}}
\def\beas{\begin{align*}}
\def\eeas{\end{align*}}
\def\be{\begin{equation}}
\def\ee{\end{equation}}
\newcommand{\tsl}[1]{{\textsl{#1}}}
\newcommand{\mc}[1]{{\mathcal{#1}}}

\title{Conformal Field Theories with Sporadic Group Symmetry }

\author[1]{Jin-Beom Bae,}
\author[2]{Jeffrey A. Harvey,}
\author[3]{Kimyeong Lee,}
\author[3]{Sungjay Lee,}
\author[4]{and Brandon C. Rayhaun}

\affiliation[1]{Mathematical Institute, University of Oxford \\
$~~ $Andrew Wiles Building, Radcliffe Observatory Quarter \\
$~~ $Woodstock Road, Oxford, OX2 6GG, U.K.}
\affiliation[2]{Enrico Fermi Institute and Department of Physics \\
$~~ $University of Chicago \\
$~~ $933 East 56th Street, Chicago IL 60637, U.S.A.}
\affiliation[3]{Korea Institute for Advanced Study \\
$~~ $85 Hoegiro, Dongdaemun-Gu, Seoul 02455, Korea}
\affiliation[4]{Stanford Institute for Theoretical Physics \\
$~~ $ Stanford University, Stanford, CA 94305}

\emailAdd{bae@maths.ox.ac.uk}
\emailAdd{j-harvey@uchicago.edu} 
\emailAdd{klee@kias.re.kr} 
\emailAdd{sjlee@kias.re.kr} 
\emailAdd{brayhaun@stanford.edu}

\abstract{
The monster sporadic group is the automorphism group of a central charge $c=24$ vertex operator algebra (VOA) or meromorphic conformal field theory (CFT). In addition to its $c=24$ stress tensor $T(z)$, this theory contains many other conformal vectors of smaller central charge; for example, it admits $48$ commuting $c=\frac12$ conformal vectors whose sum is $T(z)$. Such decompositions of the stress tensor allow one to construct new CFTs from the monster CFT in a manner analogous to the Goddard-Kent-Olive (GKO) coset method for affine Lie algebras. We use this procedure to produce evidence for the existence of a number of CFTs with sporadic symmetry groups and employ a variety of techniques, including Hecke operators, modular linear differential equations, and Rademacher sums, to compute the characters of these CFTs. Our examples include  (extensions of) nine of the sporadic groups appearing as subquotients of the monster, as well as the simple groups ${}^2\tsl{E}_6(2)$ and $\tsl{F}_4(2)$ of Lie type. Many of these examples are naturally associated to McKay's $\widehat{E_8}$ correspondence, and we use the structure of Norton's monstralizer pairs more generally to organize our presentation.

 \vskip 0,1in
}


\arxivnumber{}

\begin{document}
\maketitle

\addtocontents{toc}{\protect\enlargethispage{35mm}}

\section{Introduction and Summary}\label{sec:Intro}

\begin{figure}
\begin{center}
\includegraphics[width=\textwidth]{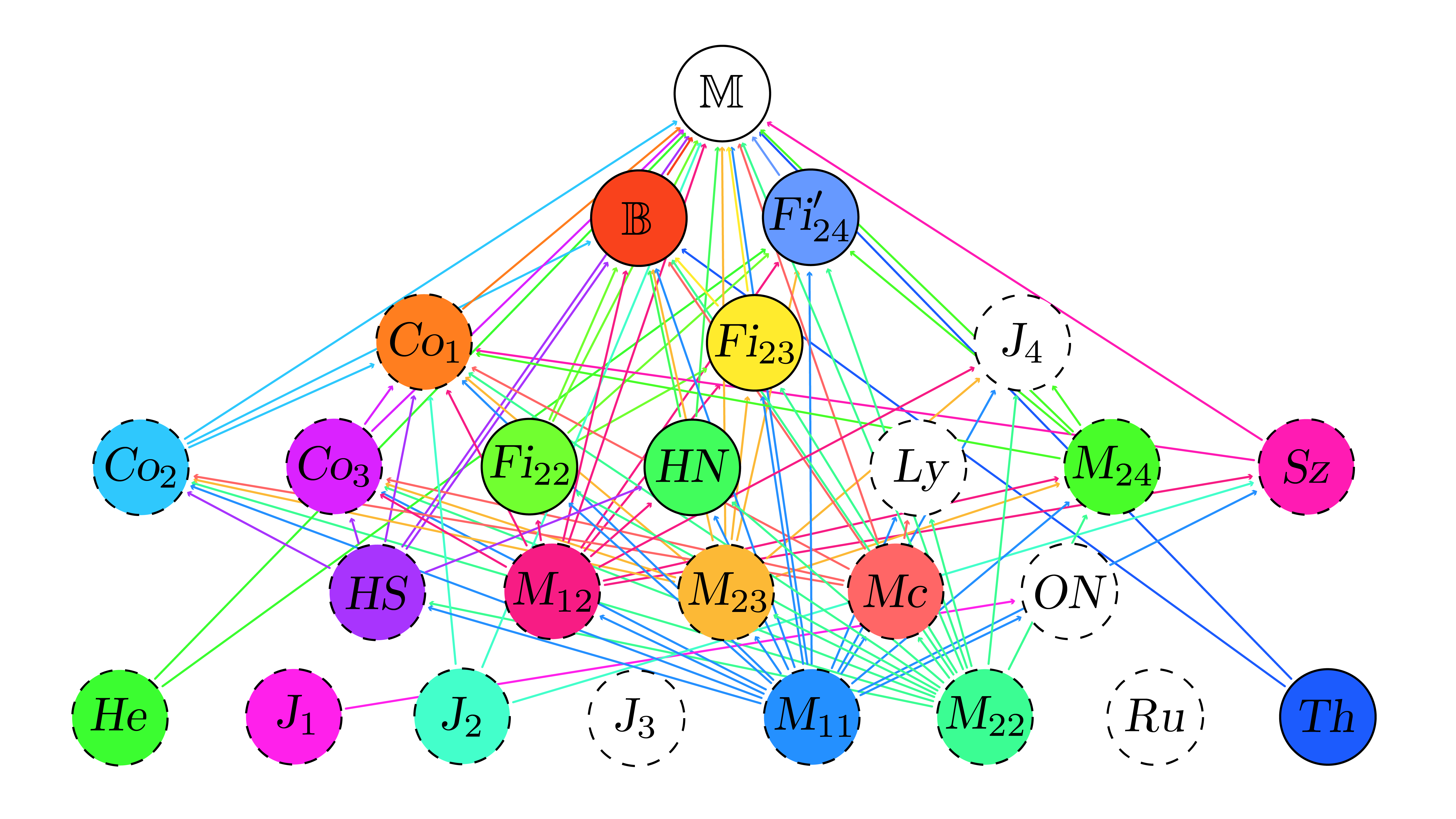}
\caption{A diagram of the simple sporadic groups, based on data taken from the Atlas of Finite Groups \cite{atlas}. An arrow from $H$ to $G$ indicates that $H$ is a subquotient of $G$. The groups $G$ which are encircled by solid lines as opposed to dashed lines are those for which a chiral algebra $\tsl{VG}^\natural$ with inner automorphism group $G$ is known. These chiral algebras embed into one another in the same way their associated groups do as subquotients.}
\label{sporadic diagram}
\end{center}
\end{figure}

The classification of finite simple groups is a remarkable achievement of pure mathematics which occupied the efforts of about 100 mathematicians for a significant part of the 20th century. The main theorem states that every finite simple group is either 
\begin{enumerate}[label=(\alph*)]
    \item a cyclic group of prime order,
    \item an alternating group of degree at least 5,
    \item a group of Lie type, or
    \item one of 26 exceptional groups, called the \emph{sporadic groups}.
\end{enumerate}
Of the sporadic groups, 20 of them are realized inside the largest one --- the monster group $\mathbb{M}$ --- as subquotients\footnote{A subquotient is a quotient of a subgroup.}; more generally these 20, referred to by Griess as the \emph{happy family}, participate in a web of subquotients with one another, as depicted in Figure \ref{sporadic diagram}. Decades after their discovery, the reason for the existence of these sporadic groups remains a mystery.

To get one's hands on a group, it can be fruitful to study it through the objects it acts on by symmetries. Indeed, a concrete representation can reveal properties of the group masked by its abstract presentation, and tie questions about the group to questions about other structures in mathematics. Historically, the sporadic groups have often been implicated as the automorphism groups of a wide variety of auxiliary structures. In the case of the monster group, its existence was proven when Griess demonstrated that it furnishes the symmetries of a 196884 dimensional commutative, non-associative algebra $\mathcal{B}$ called the \emph{Griess algebra} \cite{griess1982friendly}. Witt realized the Mathieu group $\tsl{M}_{24}$ as the automorphism group of the Golay code \cite{witt19375}. The Thompson sporadic group $\Th$ \cite{thompson1976conjugacy} was first constructed as the automorphism group of a 248 dimensional even unimodular lattice in the Lie algebra of $E_8$ \cite{smith1976simple}. And so on and so forth.

It is natural to seek a more unified description of the sporadic groups, perhaps as the automorphisms of a single kind of structure, as opposed to many different kinds of objects. Among the numerous insights it has offered, the program of moonshine has improved this situation by implicating many of the sporadic groups as symmetries of conformal field theories (CFTs) or other CFT-inspired constructions. For example, Frenkel, Lepowsky, and Meurman promoted the Griess algebra $\mathcal{B}$ to a meromorphic CFT/vertex operator algebra (VOA) $V^\natural$ --- the moonshine module --- which is widely thought of as the most natural representation of $\mathbb{M}$ \cite{flm}. In this picture, the operator algebra and stress tensor of $V^\natural$ are preserved by the Monster, and $\mathcal{B}$ arises as an algebra defined on the subspace of dimension 2 operators; furthermore the connection to conformal field theory led to a proof of the main genus zero conjecture of monstrous moonshine \cite{Borcherds}.
The Conway group $\tsl{Co}_1$ plays a similar role in a $c=12$ superconformal field theory \cite{flma, Duncan-super, Duncan-Mack}. More mysteriously, the Mathieu group $\tsl{M}_{24}$ was shown by Eguchi, Ooguri, and Tachikawa \cite{Eguchi:2010ej} to arise when one decomposes the elliptic genus of K3, which counts special supersymmetric states in superconformal sigma models with K3 target, into characters of the $\mathcal{N}=4$ superconformal algebra, though the precise sense in which the Mathieu group is a symmetry in this context remains elusive in spite of detailed studies of Mathieu moonshine \cite{Cheng, Gaberdiel, Eguchi} including a proof of the main conjecture \cite{gannon} and the extension to umbral moonshine \cite{UM,UMNL,DGO}. Even more mysteriously, it was conjectured in \cite{Harvey:2015mca}, and proven in \cite{griffin}, that the Thompson group $\Th$ acts on an infinite-dimensional graded module whose graded characters are modular forms of a particular kind, a tell-tale sign of the existence of some kind of vertex algebraic structure, though no such structure has yet been discovered. The O'Nan group, which is not a member of the happy family, also has a moonshine relation to weight $\frac32$ modular forms \cite{onan}, again suggesting a possible connection to vertex algebras. It thus seems not out of the question that vertex algebras and associated physical structures may eventually provide a setting for a more uniform understanding of the sporadic groups.

While these developments have been interesting in their own right, if one's goal is to understand the relationships between the sporadic groups, the constructions suffer from the shortcoming that, while e.g.\ $\Th$ is a member of the happy family and thus appears inside of the Monster, its corresponding module bears no obvious relation to $V^\natural$. What one would really like is to construct models for these groups that are designed around the lattice of subgroups of $\mathbb{M}$ from the outset. A relevant group theoretic notion, explored by Norton in \cite{anatomy}, is the idea of a \emph{monstralizer pair}: by definition, it consists of two subgroups $G$ and $\widetilde{G}:=\mathrm{Cent}_{\mathbb{M}}(G)$ which mutually centralize each other (c.f.\ Table \ref{monstralizers} for a non-exhaustive list of examples). There are two reasons we might take this as our starting point. First, the idea of a mutually centralizing pair is valid in any group, and has historically lead to very fruitful and natural constructions in representation theory: examples include Schur-Weyl duality \cite{howe1995perspectives} and the general theory of reductive dual pairs \cite{howe1979fi}, of which the Howe-theta correspondence \cite{howe1989remarks} is a well-known application. Second, most of the monster's large perfect\footnote{A group is said to be perfect if it admits no non-trivial abelian quotients. In particular, any non-abelian simple group is a perfect group.} subgroups (or soluble extensions thereof) participate in a monstralizer pair, and so this structure reveals information about relationships between many sporadic groups in the happy family. A natural ambition then is to attempt to give monstralizer pairs a new life in the richer setting of the monster CFT.

In retrospect, first steps in this direction were taken by H{\"o}hn and his collaborators following the discovery of 48 mutually commuting Virasoro subalgebras with central charge $\frac12$ inside the monster VOA \cite{dmz}. Indeed, in his PhD thesis \cite{hoehnbaby}, H{\"o}hn built a rational vertex operator algebra\footnote{In this paper, $\tsl{V}\mathbb{B}^\natural$ will always denote a $\mathbb{Z}$-graded VOA. This differs slightly from H{\"o}hn's $\frac12\mathbb{Z}$-graded vertex operator super algebra, which is constructed by taking a direct sum of $\tsl{V}\mathbb{B}^\natural$ and its unique irreducible module of highest weight $\frac32$.} $\tsl{V}\mathbb{B}^\natural$ of central charge $23\sfrac12$ which admits an action of the baby monster sporadic group $\mathbb{B}$ by automorphisms, and which is realized as a subVOA of $V^\natural$. Later, H{\"o}hn, Lam, and Yamauchi \cite{hohn2012mckaye6} provided a similar construction of a central charge $23\sfrac15$ vertex operator algebra\footnote{The notation $\tsl{VF}^\natural$ was used in \cite{hohn2012mckaye6}, though we use $\tsl{VF}_{24}^\natural$ in anticipation of our construction of similar VOAs $\tsl{VF}_{23}^\natural$ and $\tsl{VF}_{22}^\natural$ associated with the other Fischer groups $\Fi_{23}$ and $\Fi_{22}$.} $\tsl{VF}_{24}^\natural$ whose (inner) automorphism group is the largest Fischer group $\Fi_{24}'$, and which also embeds into $V^\natural$. We will see in a moment that these two examples can be naturally interpreted as VOA uplifts of the monstralizer pairs $(\mathbb{Z}_2,2.\mathbb{B})$ and $(\mathbb{Z}_3,3.\Fi_{24}')$.

The idea behind the construction of $\tsl{V}\mathbb{B}^\natural$ and $\tsl{VF}^\natural_{24}$ is to decompose the stress tensor of the moonshine module into a sum of commuting stress tensors of smaller central charge, $T(z) = t(z)+\widetilde{t}(z)$, a process we will colloquially refer to as deconstruction. One can then consider the VOA which remains once one has subtracted off the subVOA $\W$ of $V^\natural$ associated with $t(z)$; this leaves a subVOA $\widetilde{\W}$ of $V^\natural$ with $\widetilde{t}(z)$ as its stress tensor. This idea is made precise by the notion of a \emph{commutant subalgebra} or a \emph{coset model}, the study of which was first initiated in \cite{Goddard:1984vk,Goddard:1986ee} in the context of affine Lie algebras. 
The commutant $\widetilde{\U}:=\tsl{Com}_{V^\natural}(\U)$ of a subVOA $\U$ in $V^\natural$ is the set of all operators in $V^\natural$ which have regular OPE with every operator in $\U$. In terms of this construction, we can define ``the subVOA associated with $t(z)$'' to mean the commutant of the Virasoro algebra generated by $\widetilde{t}(z)$, i.e.\ $\W :=\tsl{Com}_{V^\natural}(\tsl{Vir}(\widetilde{t}))$. The pair $\W$, $\widetilde{\W}:= \tsl{Com}_{V^\natural}(\W)$ are then each others' commutants, and so form what is called a \emph{commutant pair}.

To any such commutant pair, one can associate two commuting subgroups $\mathbb{M}(\W)$, $\mathbb{M}(\widetilde{\W})$ of the monster which we will refer to as the \emph{subgroups preserved by} $\W$, $\widetilde{\W}$: roughly speaking, $\mathbb{M}(\W)$ is the subgroup of $\mathbb{M}$ which acts trivially on $\widetilde{\W}$ and its modules while preserving $\W$ and its modules, and similarly for $\mathbb{M}(\widetilde{\W})$. In the case that the groups $(\mathbb{M}(\W),\mathbb{M}(\widetilde{\W}))$ furnish a monstralizer pair, we will refer to $(\W,\widetilde{\W})$ as a \emph{monstralizing commutant pair}, or $\mathbb{M}$\emph{-com pair} for short. The intuition behind this definition is that, although there are a rather large number of ways to cut up the monster CFT into commutant pairs, there are far fewer ways to cut it into monstralizing commutant pairs, and our expectation is that such $\mathbb{M}$-com pairs will in general play more nicely with respect to the action of $\mathbb{M}$ on $V^\natural$ than will a generic commutant pair. Within the framework we've laid out, the algebra $\tsl{V}\mathbb{B}^\natural$ arises when one carries out this procedure for $t(z)$ a conformal vector of central charge $\frac12$ in the moonshine module (so that $\W$ is the chiral algebra of the Ising model), and corresponds to the monstralizer pair $(\mathbb{M}(\W),\mathbb{M}(\widetilde{\W})) = (\mathbb{Z}_2,2.\mathbb{B})$. The algebra $\tsl{VF}_{24}^\natural$ corresponds to choosing $t(z)$ to be the stress tensor of a $\mathbb{Z}_3$ parafermion theory \cite{fz}, and has $(\mathbb{M}(\W),\mathbb{M}(\widetilde{\W})) = (\mathbb{Z}_3,3.\Fi'_{24})$. In light of these results, there is a natural question: 
\emph{can this idea be generalized to produce other $\mathbb{M}$-com pairs and, in the process, other subVOAs of the moonshine module which naturally realize sporadic or otherwise exceptional symmetry groups?} 

One of our main results is an answer to this question in the affirmative. Namely, we construct several examples of monstralizing commutant pairs $(\W_{G},\W_{\widetilde{G}})$  corresponding to monstralizer pairs $(G,\widetilde{G})$ which satisfy the following properties\footnote{In particular, our examples are suggestive of the existence of a (unique) functor from the category of monstralizer pairs to the category of monstralizing commutant pairs.}:
\begin{enumerate}[label=(\alph*)]
    \item The subgroups preserved by $\W_G$ and  $\W_{\widetilde{G}}$ are $\mathbb{M}(\W_{G}) = G$ and $\mathbb{M}(\W_{\widetilde{G}})=\widetilde{G}$.
    \item The inner automorphism group of $\W_G$ is given by $G/Z(G)$, where $Z(G)$ is the center of $G$. Therefore, the $\W_G$ realize \emph{subquotients} (rather than just subgroups) of the monster as their inner automorphism groups. The same goes for the $\W_{\widetilde{G}}$.
    \item (A closed subalgebra of) the fusion algebra of both $\W_G$ and $\W_{\widetilde{G}}$ admits an action of $Z(G) = Z(\widetilde{G})$ by automorphisms.
    \item The commutant of $\W_G$ in $V^\natural$ is $\W_{\widetilde{G}}$ and vice versa, i.e.\ $\widetilde{\W}_G = \W_{\widetilde{G}}$ and $\widetilde{\W}_{\widetilde{G}} = \W_G$.
    \item Whenever one monstralizer pair $(H,\widetilde{H})$ includes into another $(G,\widetilde{G})$ in the sense that $\widetilde{H} < \widetilde{G}$ and $G < H$, the associated chiral algebras mirror these inclusions, i.e.\ $\W_{\widetilde{H}}\hookrightarrow \W_{\widetilde{G}}$ and $\W_{G}\hookrightarrow \W_{H}$.
\end{enumerate}  
In each of our examples, we take $G$ to be either cyclic or dihedral, in which case the $\W_{\widetilde{G}}$ furnish chiral algebras with interesting symmetry groups, many of which belong to the happy family (c.f.\ Table \ref{chiral algebra table} for a summary of these results). The cases we consider for which $G = \mathbb{Z}_{k\mathrm{X}}$ (where we use $\mathbb{Z}_{k\mathrm{X}}$ to denote a $\mathbb{Z}_k$ subgroup of $\mathbb{M}$ which is generated by any element in the $k$X conjugacy class) involve taking $\W_{\mathbb{Z}_{k\mathrm{X}}}$ to be a $\mathbb{Z}_k$ parafermion subalgebra \cite{flone}. Such deconstructions yield --- in addition to the algebras $\W_{\widetilde{G}}=V^\natural$, $\tsl{V}\mathbb{B}^\natural$, and $\tsl{VF}_{24}^\natural$ discussed previously, which correspond to $G=\mathbb{Z}_{1\mathrm{A}}$, $\mathbb{Z}_{2\mathrm{A}}$, and $\mathbb{Z}_{3\mathrm{A}}$ --- a new $\mathbb{M}$-com uplift of the monstralizer pair $(\mathbb{Z}_{\mathrm{4A}},4.2^{22}.\tsl{Co}_3)$. The remaining cases for which $G$ is a dihedral group are all new, and are organized through a striking connection to McKay's $\widehat{E_8}$ correspondence. In order to cleanly explain these results, we briefly review this correspondence.

\begin{figure}
\begin{center}
\includegraphics[width=.9\textwidth]{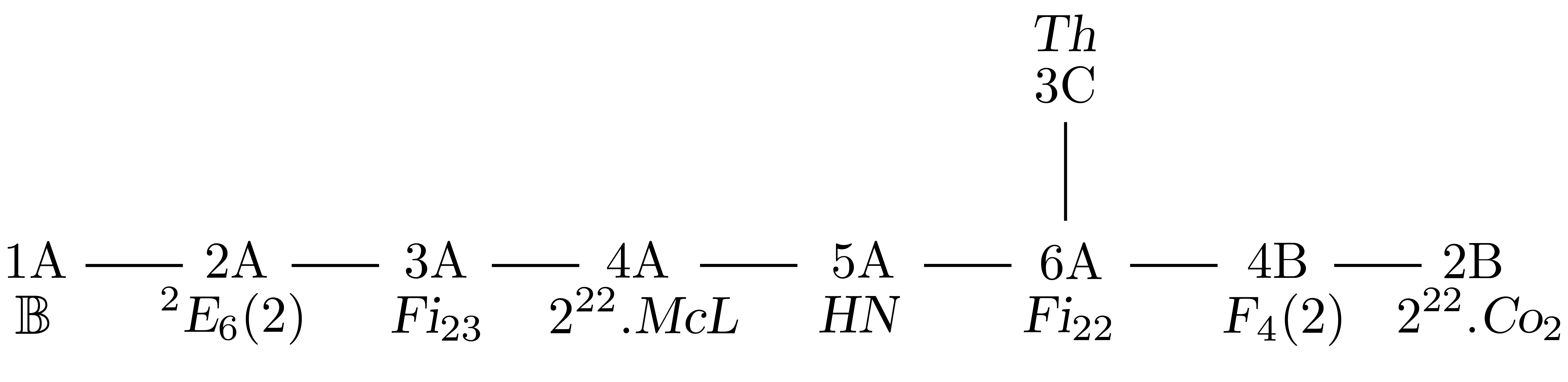}
\caption{Dynkin diagram of $\widehat{E_8}$, decorated by conjugacy classes of the monster sporadic group $\mathbb{M}$. We propose to further decorate each node by the inner automorphism group of the commutant of $\W_{D_{n\mathrm{X}}}$ in $V^\natural$.}
\label{e8}
\end{center}
\end{figure}

The monster has two conjugacy classes of involutions, which are commonly labeled as 2A and 2B, and it is known that the product of any two elements taken from the 2A conjugacy class must lie in one of 1A, 2A, 3A, 4A, 5A, 6A, 2B, 4B, or 3C (c.f.\ \cite{conway1985simple}). It was suggested by McKay that these conjugacy classes can be naturally thought of as nodes of the extended $\widehat{E_8}$ Dynkin diagram, as in Figure \ref{e8}. Crucially for our purposes, any two 2A involutions whose product resides in the $n$X conjugacy class generate a dihedral subgroup\footnote{We use the convention that $D_n$ is the symmetry group of a regular $n$-gon, i.e.\ $|D_n|=2n$.} $D_{n\mathrm{X}}$ of $\mathbb{M}$ \cite{Sakuma,griess2008ee_8} which participates in a monstralizer pair. Now, one would like to associate to $D_{n\mathrm{X}}$ a subalgebra $\W_{D_{n\mathrm{X}}}$ of $V^\natural$ which obeys the desired properties that we have been discussing. In fact, such VOAs have already been constructed in the math literature \cite{miyaconf, sakuma2003vertex, lametal, Sakuma}: the starting point for these constructions is the fact that elements of the 2A conjugacy class are in one-to-one correspondence with central charge $\frac12$ conformal vectors in the moonshine module \cite{conway1985simple,miyamotogriess}. It is natural then to consider the subVOA $\W_{D_{n\mathrm{X}}}$ which is generated by two central charge $\frac12$ conformal vectors whose associated involutions have product lying in the $n\mathrm{X}$ conjugacy class. Each such algebra specifies a deconstruction of the stress tensor of the monster --- i.e.\ a decomposition $T(z) = t(z)+\widetilde{t}(z)$ --- and we may consider ``subtracting off'' $\W_{D_{n\mathrm{X}}}$ and studying the remaining chiral algebra with stress tensor $\widetilde{t}(z)$, i.e.\ the commutant of $\W_{D_{n\mathrm{X}}}$, which we denote by $\widetilde{\W}_{D_{n\mathrm{X}}}$. 

The commutants $\widetilde{\W}_{D_{n\mathrm{X}}}$ beautifully recover chiral algebras whose symmetry groups are either closely related to sporadic groups or are otherwise exceptional. We find that the cases $n\mathrm{X} =$ 1A, 3A, 5A, 6A, and 3C define chiral algebras which we call $\tsl{V}\mathbb{B}^\natural$, $\tsl{VF}_{23}^\natural$, $\tsl{VHN}^\natural$, $\tsl{VF}_{22}^\natural$, and $\tsl{VT}^\natural$ with (inner) automorphism groups\footnote{A few of these algebras inherit an extra order 2 outer automorphism from the monster.} $\mathbb{B}$, $\Fi_{23}$, $\HN$, $\Fi_{22}$, and $\Th$ respectively; each of these groups is precisely a simple sporadic group in the happy family. Moreover, this collection of VOAs reflects the relationships between groups in the happy family in the sense that, for $H$ and $G$ taken from the list 
\begin{align}
    \mathbb{M},\ \mathbb{B}, \ \Fi_{24}',\ \Fi_{23}, \  \Fi_{22},\ \Th, \  \HN,
\end{align} 
whenever $H$ is a subquotient of $G$, then $\tsl{VH}^\natural$ is a subalgebra of $\tsl{VG}^\natural$. These chiral algebras therefore furnish an intriguing mirroring of the group theory of the happy family within the theory of vertex operator algebras, by relating the structure of subquotients in $\mathbb{M}$ to deconstructions of the stress tensor of $V^\natural$. The remaining cases $n\mathrm{X}$ = 2A, 4A, 2B, and 4B recover chiral algebras\footnote{Although the 4A and 2B cases involve the McLaughlin group and Conway's second group, both of which belong to the happy family, we hesitate to give them the names $\tsl{VMcL}^\natural$ and $\tsl{VCo}_2^\natural$ in case it is possible in the future to define chiral algebras with $\tsl{McL}$ and $\tsl{Co}_2$ symmetry \emph{on the nose}, as opposed to our current constructions which realize extensions of these groups.} with inner automorphism groups ${^2}E_6(2)$, $2^{22}.\tsl{McL}$, $2^{22}.\tsl{Co}_2$, and $\tsl{F}_4(2)$ respectively.

We also initiate a study of the representation theory of these models. On general grounds, since $\W_G$ and $\widetilde{\W}_G = \W_{\widetilde{G}}$ are commutant pairs in the moonshine module, one has a Schur-Weyl like decomposition of the form \cite{lin2017mirror,creutzig2019schur}
\begin{align}\label{moonshinedecomp}
\begin{split}
    V^\natural &\cong \bigoplus_{\alpha} \W_G(\alpha)\otimes \W_{\widetilde{G}}(\alpha) 
\end{split}
\end{align}
where the $\W_G(\alpha)$ (respectively $\W_{\widetilde{G}}(\alpha)$) are mutually inequivalent irreducible modules of $\W_G$ (respectively $\W_{\widetilde{G}}$). Such a decomposition furnishes a one-to-one pairing between a subset of irreducible modules of $\W_G$ and a subset of irreducible modules of $\W_{\widetilde{G}}$, and so we henceforth refer to the $\W_{\widetilde{G}}$-module $\W_{\widetilde{G}}(\alpha)$ as the representation \emph{dual} to $\W_G(\alpha)$ in $V^\natural$, and its graded dimension
\begin{align}
    \widetilde{\chi}_{\alpha}(\tau) := \chi_{\W_{\widetilde{G}}(\alpha)}(\tau) = \mathrm{Tr}_{\W_{\widetilde{G}}(\alpha)} q^{\widetilde{l}_0-\frac{c_{\widetilde{t}}}{24}},
\end{align}
as a \emph{dual character}, where $\widetilde{t}(z) = \sum_n \widetilde{l}_n z^{-n-2}$ is the stress tensor of $\W_{\widetilde{G}}$ of central charge $c_{\widetilde{t}}$. To provide evidence for our claims, we construct these dual characters for all of the $\W_{\widetilde{G}}$ and demonstrate that they exhibit the kind of connection to representation theory expected in moonshine: namely, a decomposition of the coefficients of the $\widetilde{\chi}_{\alpha}(\tau)$ into dimensions of irreducible representations of the proposed symmetry group $\widetilde{G}$ which is consistent with how $V^\natural$ decomposes by restriction into representations of $\widetilde{G}$. These dual characters always transform with a unitary and symmetric modular S-matrix so that it is possible in each of our examples to define a full, modular invariant CFT with partition function 
\begin{align}
    \mathcal{Z}_{\widetilde{G}}(\tau,\bar\tau) = \sum_\alpha \overline{{\chi}_{\W_{\widetilde{G}}(\alpha)}(\tau)}{\chi}_{\W_{\widetilde{G}}(\alpha)}(\tau).
\end{align}

At the level of characters, the decomposition of $V^\natural$ in \eqref{moonshinedecomp} presents a highly constrained modular bootstrap problem. We recruit three main technical tools to solve it for the characters of the modules $\W_{\widetilde{G}}(\alpha)$: Rademacher sums, Hecke operators, and modular linear differential equations (MLDEs). Rademacher sums and MLDEs are familiar techniques in the analysis of chiral algebras; however, one relatively new element of our analysis is that, for some choices of $G$, we are able to obtain the $\widetilde{\chi}_\alpha(\tau)$ as the image of the irreducible characters of $\W_G$, 
\begin{align}
    \chi_{\alpha}(\tau) := \mathrm{Tr}_{\W_G(\alpha)}q^{l_0-\frac{c_t}{24}},
\end{align}
under the action of a suitable Hecke operator \cite{HarveyWu}. The simplest application of this technology is to the baby monster algebra $\tsl{V}\mathbb{B}^\natural$, whose characters arise as a Hecke image of the characters of the Ising model. Our constructions thus showcase an intriguing application of Hecke operators, as well as an interpretation of Hecke related RCFTs, which we believe will hold in some generality: namely, that chiral algebras whose characters are related by Hecke operators arise, in favorable circumstances, as commutant pairs embedded inside a larger meromorphic CFT. 

This paper is organized as follows. In \S\ref{subsec:voacft}, we review some basic notions related to chiral algebras/VOAs, and establish notations. In \S\ref{subsec:deconstruction} we exposit, in generality and in a few familiar examples, the technique of ``deconstructing'' the stress tensor of a chiral algebra and using such a deconstruction to obtain new chiral algebras as commutants. In \S\ref{subsec:MLDE} and \S\ref{subsec:Hecke}, we review two technical tools which we will use to study the dual characters of our models: modular linear differential equations and Hecke operators. We then move on to applying these techniques in \S \ref{sec:mainresults}, which contains our main results. After reviewing the constructions of the monster, baby monster, and Fischer VOAs, we use them in \S\ref{subsec:startingmodels}-\S\ref{subsec:otherexamples} to obtain new chiral algebras from the procedure of deconstruction. We propose directions for future research in \S\ref{sec:Conc}. 

\section{Review of Techniques}\label{sec:review}

\subsection{VOA/chiral algebra basics}\label{subsec:voacft}

In order to make this paper readable for both physicists and mathematicians, we begin with a brief, informal review of chiral algebras (also known as vertex operator algebras), and establish a few notations/conventions. For more thorough treatments of these subjects, see any of the following references \cite{frenkel2004vertex,francesco2012conformal,ginsparg1988applied}. 

\subsubsection{Relationship to conformal field theory}

The most familiar starting point for physicists is a two-dimensional (Euclidean) quantum field theory that is invariant under the group $\operatorname{\textsl{SL}}_2(\mathbb{C})$ of global conformal transformations. The Green's functions of such a theory can be analytically continued from functions on $\mathbb{R}^2$ to functions on a larger domain in $\mathbb{C}^2$, so it is natural to work with complex coordinates $(z,\bar{z})$, thought of as independent variables, and only equate $\bar{z}$ with the complex conjugate of $z$ at the end of the problem. The complexification $\mathfrak{sl}(2)\oplus \overline{\mathfrak{sl}(2)}$ of the Lie algebra of $\operatorname{\textsl{SL}}_2(\mathbb{C})$ is then the natural object to consider; its generators are 
\begin{align}
\begin{split}
    & L_{-1} = -\partial_z, \ \ \ \ L_0 = -z\partial_z, \ \ \ \ L_1 = -z^2\partial_z \\
 &\bar{L}_{-1} = -\partial_{\bar{z}}, \ \ \ \ \bar{L}_0 = -\bar{z}\partial_{\bar{z}}, \ \ \ \ \bar{L}_1 = -\bar{z}^2\partial_{\bar{z}}.
\end{split}
\end{align}
The holomorphic and anti-holomorphic generators commute with each other, and each satisfy familiar relations,
\begin{align}
\begin{split}
    [L_1,L_{-1}] = 2L_0 \ \ \ \ \ \  [L_0,L_{\pm 1}] = \mp L_{\pm} \\
    [\bar{L}_1,\bar{L}_{-1}] = 2\bar{L}_0 \ \ \ \ \ \ [\bar{L}_0,\bar{L}_{\pm 1}] = \mp \bar{L}_{\pm}.
    \end{split}
\end{align}
The Hilbert space transforms in a representation of this algebra (which we assume is a direct sum of highest weight representations of $\mathfrak{sl}(2)\oplus \overline{\mathfrak{sl}(2)}$), so it is natural to organize states in the theory in terms of their eigenvalues $(h,\bar{h})$ under $(L_0,\bar{L}_0)$. 

Conformal field theories enjoy a one-to-one correspondence between their states and their local operators, 
\begin{align}
    \varphi(z) \leftrightarrow |\varphi\rangle \equiv \lim_{z,\bar{z}\to 0}\varphi(z,\bar{z})|\Omega\rangle
\end{align}
where $|\Omega\rangle$ is the unique $\operatorname{\textsl{SL}}(2,\mathbb{C})$-invariant vacuum. Equipped with this \emph{state-operator correspondence}, we will often drop the vertical line and angular bracket around $|\varphi\rangle$ and think of $\varphi$ as a state in the Hilbert space, safely distinguishing it from its corresponding operator $\varphi(z,\bar{z})$ by the suppression of its arguments $(z,\bar z)$. 

The power of conformal symmetry in two dimensions is closely related to the power of meromorphy in complex analysis. Indeed, consider any meromorphic operator in the theory, 
\begin{align}
    \partial_{\bar{z}}\varphi(z,\bar{z}) = 0.
\end{align}
From the expressions for the anti-holomorphic generators, it is evident that such operators arise from states with $\bar{h}=0$. Locality then ensures that $h$ is either integral or half-integral, so that the corresponding space of states admits a natural grading, $\mathcal{V}=\bigoplus_{h\in\frac{1}{2}\mathbb{Z}}\mathcal{V}_h$. We will mainly focus on the case where $\V$ is $\mathbb{Z}$-graded. The operator product algebra closes on such operators, so the set of meromorphic operators forms a consistent truncation of the full operator algebra. We will use the following notation for this operator product expansion (OPE) between meromorphic operators,
\begin{align}\label{ope}
   \varphi(z)\varphi'(w) \sim \sum_{0<n\leq h+h'} \frac{(\varphi\varphi')_n(w)}{(z-w)^n} 
\end{align}
where the symbol $\sim$ indicates that we have only retained the singular terms on the right-hand side. This structure is what is known to physicists as a \emph{chiral algebra}, and to mathematicians as a \emph{vertex operator algebra}. From now on, we reserve the symbol $\mathcal{V}$ (resp. $\overline{\mathcal{V}}$) for the holomorphic (resp. anti-holomorphic) chiral algebra.

To see the utility of these meromorphic operators, note that standard contour integral arguments imply that the modes
\begin{align}
    \varphi_n \equiv \oint \frac{dz}{2\pi i}z^{n+h-1}\varphi(z) \hspace{.5in} \left(\varphi(z) = \sum_{n\in\mathbb{Z}} \varphi_n z^{-n-h}\right)
\end{align}
do not depend on the precise choice of contour encircling the origin. In radial quantization, in which the ``time'' slices are concentric circles about the origin, these $\varphi_n$ can therefore be thought of as conserved charges of sorts. We see that the presence of even a single meromorphic operator ensures the existence of an infinite-dimensional symmetry algebra. This large symmetry is a useful jumping-off point for analyzing the full CFT, particularly when the CFT is rational, which by definition means that the Hilbert space decomposes into a direct sum of \emph{finitely} many representations of $\mathcal{V}\otimes\overline{\mathcal{V}}$. 

The chiral algebra of a two-dimensional local CFT is never empty. Indeed, from conservation and tracelessness of the stress tensor, one can prove that %
\begin{align}
    T(z)\equiv T_{zz}(z,\bar{z})=\sum_{n\in\mathbb{Z}}L_nz^{-n-2}
\end{align} 
is a meromorphic dimension 2 operator, and similarly for $\overline{T}(\bar{z}) \equiv T_{\bar{z}\bar{z}}(z,\bar{z})$. The $TT$ OPE is constrained by conformal symmetry to take the form 
\begin{align}\label{TTOPE}
    T(z)T(w) \sim \frac{c/2}{(z-w)^4}+\frac{2T(w)}{(z-w)^2}+\frac{\partial T(w)}{z-w}.
\end{align}
This OPE is equivalent by standard arguments to the commutation relations of the Virasoro algebra,
\begin{align}
    [L_m,L_n] = (m-n)L_{m+n} + \frac{c}{12}m(m^2-1)\delta_{m+n,0}.
\end{align}
We recognize the closed subalgebra furnished by $L_0,L_{\pm 1}$ as the same $\mathfrak{sl}(2)$ described earlier arising from global conformal invariance. The chiral algebra generated by the stress tensor should be thought of as an enhancement of this global conformal algebra to a larger symmetry algebra which implements \emph{local} conformal transformations. In a generic CFT, this Virasoro algebra makes up all of $\V$; in special circumstances, in particular for most rational CFTs, additional operators populate the chiral algebra. We will mainly work with such special theories. 

\subsubsection{Modules and characters}\label{subsubsec:modules}

As we mentioned earlier, the Hilbert space of a CFT decomposes into a direct sum of representations of $\V\otimes\overline{\V}$, 
\begin{align}
\mathcal{H} = \bigoplus_{\alpha,\overline{\alpha}} \Omega_{\alpha,\overline{\alpha}} \V(\alpha)\otimes \overline{\V}(\overline{\alpha})
\end{align}
where the sum over $\alpha$ (resp. $\overline\alpha$) is a sum over the irreducible representations of $\V$ (resp. $\overline{\V}$), and the $\Omega_{\alpha,\overline{\alpha}}$ are non-negative integer coefficients with $\Omega_{0,0}=1$ which give the multiplicities of each representation. We will also find it helpful in this paper to decompose $\V$ and its irreducible modules with respect to representations of some known subalgebra\footnote{A subVOA $\W$ need not have the same stress tensor as $\V$. In the case that it does, we say that $\W$ is a \emph{full} subVOA.} $\W$, e.g.
\begin{align}\label{isotypical component}
\begin{split}
    \V = \bigoplus_\alpha \V_{(\alpha)}
\end{split}
\end{align}
where $\V_{(\alpha)}$ is the subspace of $\V$ generated by all irreducible $\W$-modules isomorphic to $\W(\alpha)$. For example, $\mathcal{W}$ might be the Virasoro subalgebra of a larger chiral algebra $\V$. In the other direction, we may be interested in extending $\W$ to a larger chiral algebra $\V$ by taking a direct sum of $\W$-modules. 

The basic data of a module $M$ is the assignment of an operator $\varphi_M(z)$ acting on $M$ to each $\varphi\in \V$, which is subject to standard axioms (see e.g.\ \S 2 of \cite{Dong:1994wn}). We will use the notation $\mathcal{V}^{(\pm)}$ for the subalgebras spanned by modes $\varphi_n$ with $\pm n>0$, and $\mathcal{V}^{(0)}$ the Lie subalgebra spanned by the zero modes $\varphi_0$. In this paper, we will work with \emph{heighest weight modules} $M$. These are modules built on top of a highest weight vector space $M_{h}$ which carries a representation of $\mathcal{V}^{(0)}$ with $L_0$ acting as $h\mathds{1}$, and which is annihilated by positive modes,
\begin{align}
    U(\mathcal{V}^{(+)})M_h = 0 
\end{align}
where $U(\mathfrak{g})$ is the universal enveloping algebra of a Lie algebra $\mathfrak{g}$. We somewhat loosely refer to states in $M_h$ as $\mathcal{V}$-primaries and any state in $U(\mathcal{V}^{(-)})M_h$ as a descendant. We will always assume that $L_0$ acts semi-simply, and that $M = \bigoplus_{q\in\mathbb{Z}+h}M_q$ is graded by the eigenvalues of $L_0$. The chiral algebra $\mathcal{V}$ itself is always a module, and we refer to it as the \emph{vacuum} module. 

One can consider the OPEs between primaries\footnote{Technically there is a vector space of highest weight states, so e.g. $\varphi^{(\alpha)}$ should be thought of as carrying an extra index which we are suppressing.}, 
\begin{align}
    \varphi^{(\alpha)}(z)\varphi^{(\beta)}(w) = \sum_\gamma C_{\alpha\beta}^\gamma (z-w)^{-h_\alpha -h_\beta + h_\gamma}[\varphi^{(\gamma)}(w) + \cdots ]
\end{align}
where the ellipsis represent universal contributions from descendants, and $h_\alpha$ indicates the eigenvalue of the action of $L_0$ on the highest weight subspace of $\V(\alpha)$. The full operator algebra is in general very complicated, and so it is often useful to pass to coarser information. For example, the fusion coefficients $\mathcal{N}_{\alpha\beta}^\gamma$ are defined as the number of distinct channels along which $\varphi^{(\alpha)}$ and $\varphi^{(\beta)}$ can fuse into $\varphi^{(\gamma)}$. In particular, each non-zero $\mathcal{N}_{\alpha\beta}^\gamma$ thus has to be a positive integer. Note that $\mathcal{N}_{\alpha\beta}^\gamma \neq 0$ if and only if $\varphi^{(\gamma)}$ and its descendants appear in the $\varphi^{(\alpha)}\varphi^{(\beta)}$ OPE. These fusion coefficients serve as the structure constants of an auxiliary \emph{fusion algebra} one can associate to $\mathcal{V}$, which admits as a distinguished basis the irreducible modules of $\V$, and whose product is defined as
\begin{align}
    \mathcal{V}(\alpha)\times \mathcal{V}(\beta) = \sum_\gamma \mathcal{N}_{\alpha\beta}^\gamma \mathcal{V}(\gamma).
\end{align}
There exists an intricate relation between the fusion
algebra and the modular transformation properties of 
characters of  $\V$,
which are complex-valued functions on the upper half-plane $\mathbb{H} = \{\tau \in\mathbb{C}\mid \Im(\tau) >0\}$ defined as
\begin{align}
    \chi_\alpha(\tau) = \mathrm{Tr}_{\mathcal{V}(\alpha)}q^{L_0-\frac{c}{24}} \ \ \ \ \ \ (q=e^{2\pi i \tau}).
\end{align}
On general grounds \cite{zhu1996modular,moore1989classical}, in an RCFT, such characters transform in a finite-dimensional representation of the modular group $\operatorname{\textsl{SL}}_2(\mathbb{Z})$, 
\begin{align}
    \chi_\alpha(-\tfrac{1}{\tau}) = \sum_{\beta}S_{\alpha\beta}\chi_\beta(\tau) \ \ \ \ \ \ \chi_\alpha(\tau+1) = e^{2\pi i (h_\alpha - \frac{c}{24})}\chi_\alpha(\tau)
\end{align}
where $h_\alpha$ is the eigenvalue of $L_0$ on the subspace of highest weight states of $\mathcal{V}(\alpha)$. One can compute 
the fusion coefficients $\mathcal{N}_{\alpha\beta}^\gamma$ 
from the above S-matrix via the Verlinde formula \cite{verlinde1988fusion}, 
\begin{align}\label{Verlinde}
    \mathcal{N}_{\alpha\beta}^\gamma = \sum_{\delta} \frac{S_{\alpha\delta}S_{\beta\delta}(S^{-1})_{\delta\gamma} }{S_{0\delta}},
\end{align}
where the index $0$ labels the identity. Equation \eqref{Verlinde} reflects the fact 
that the fusion matrices $(N_\alpha)_{\beta}^{~\gamma}=\mathcal{N}_{\alpha\beta}^\gamma$ 
can be diagonalized simultaneously with diagonal matrices $D_\alpha$,
\begin{align}
 (D_\alpha)_\delta^{~ \delta} = \frac{S_{\alpha\delta}}{S_{0\delta}}.    
\end{align}

In later sections, we will often construct chiral algebras $\W$ as extensions of tensor products of theories whose irreducible modules are known, including minimal models, parafermion theories, and lattice VOAs. In lieu of a more detailed analysis of the representation theory of such extensions, we offer a strategy for discovering their characters, which is based on the constraint that they should transform into one another under modular transformations. 

Let $\U^{(i)}$ for $i=1,\dots,r$ be chiral algebras whose modules we write as $\U^{(i)}(\alpha_i)$, and whose characters $\chi^{(i)}_{\alpha_i}(\tau)$ transform under a modular S-matrix $\mathcal{S}^{(i)}$. We want to search for an extension of $\U^{(1)}\otimes\cdots \otimes\U^{(r)}$ of the form
\begin{align}
    \W = \bigoplus_{\alpha_1,\dots,\alpha_r}M_{0,(\alpha_1,\dots,\alpha_r)}\U^{(1)}(\alpha_1)\otimes\cdots \otimes \U^{(r)}(\alpha_r)
\end{align}
whose irreducible modules, on general grounds, can be expressed as
\begin{align}
    \W(\alpha) = \bigoplus_{\alpha_1,\dots,\alpha_r}M_{\alpha,(\alpha_1,\dots,\alpha_r)} \U^{(1)}(\alpha_1)\otimes\cdots\otimes \U^{(r)}(\alpha_r)
\end{align}
where $M_{\alpha,(\alpha_1,\dots,\alpha_r)}$ are non-negative integers. Now, the characters of $\U^{(1)}\otimes\cdots\otimes \U^{(r)}$ are 
\begin{align}
     \chi_{(\alpha_1,\dots,\alpha_r)}(\tau) := \chi^{(1)}_{\alpha_1}(\tau)\cdots\chi^{(r)}_{\alpha_r}(\tau)
\end{align}
and transform under the S-matrix $\mathcal{S} = \mathcal{S}^{(1)}\otimes\cdots\otimes\mathcal{S}^{(r)}$. In order for $\W(\alpha)$ to furnish the irreducible modules of a VOA, the vector-valued function 
\begin{align}
    \sum_{\alpha_1,\dots,\alpha_r}M_{\alpha,(\alpha_1,\dots,\alpha_r)}\chi_{\alpha_1}^{(1)}(\tau)\cdots\chi_{\alpha_r}^{(r)}(\tau)  = M\cdot \chi(\tau)
\end{align}
should transform covariantly under a modular S transformation. This means that there should exist a matrix $\mathcal{S}'$ such that $M\cdot \mathcal{S} = \mathcal{S}'\cdot M$, in which case 
\begin{align}
\begin{split}
    M\cdot \chi\left(-\tfrac{1}{\tau}\right) =  M \cdot \mathcal{S} \cdot  \chi(\tau) = \mathcal{S}'\cdot M \cdot \chi(\tau)
    \end{split}
\end{align}
so that $\mathcal{S}'$ is the modular S-matrix of the extended theory with characters $M\cdot\chi(\tau)$. Similar comments apply to the T transformation. This procedure is the same as block diagonalizing the $\tsl{SL}_2(\mathbb{Z})$ representation furnished by $\mathcal{S}$ and $\mathcal{T}$, and in this language $M$ is the projection matrix onto one of the blocks. We will use this method in several places in \S\ref{sec:mainresults}.

\subsubsection{Symmetries}\label{subsubsec:symmetries}

There are several different notions of symmetry which arise in the study of chiral algebras, and they differ slightly from what one would call a symmetry in a full blown CFT, so we define these notions carefully. For more details, see e.g.\ \cite{Dong:1994wn}. An \emph{automorphism} of a chiral algebra $\V$ is an invertible linear map $X:\V\to \V$ which
\begin{enumerate}[label=(\alph*)]
    \item preserves the vacuum, $X|\Omega\rangle = |\Omega\rangle$,
    \item preserves the stress tensor, $XT = T$, and 
    \item respects the state/operator correspondence,
\begin{align}\label{autdef}
    (X\varphi)(z) = X\varphi(z)X^{-1}
\end{align}
for every state $\varphi \in\V$, where $(X\varphi)(z)$ is the operator corresponding to the state $X\varphi$.
\end{enumerate}   
Alternatively, we may replace \eqref{autdef} with the equivalent condition that
\begin{align}\label{altautdef}
    (X\varphi)_n(X\phi) = X(\varphi_n\phi) \text{ for all }n\in \mathbb{Z} \text{ and }\varphi,\phi\in\V.
\end{align}
It is useful to distinguish between inner and outer automorphisms of $\V$. To define these notions, note that if $X$ is an automorphism of $\V$ of finite order, and $M$ is a module, we can define another module $X\circ M$ whose underlying vector space is by definition $M$, and for which the state operator mapping is $\varphi\mapsto \varphi_{X\circ M}(z):= (X\varphi)_M(z)$. If $X\circ M$ is isomorphic to $M$ for every module $M$ of $\V$, then we say that $X$ is an inner automorphism, and outer otherwise. The idea is that outer automorphisms will in general permute the different irreducible modules amongst themselves, while the inner automorphisms preserve them. We will use the notation $\mathrm{Aut}(\V)$ for the automorphism group of $\V$, and $\mathrm{Inn}(\V)$ for its subgroup of inner automorphisms.

We will often look for signatures of such symmetries in the characters of the chiral algebra. For example, because automorphisms preserve the stress tensor, they will also preserve the grading, and so each graded component $\V_h$ transforms in a finite-dimensional representation of $\mathrm{Aut}(\V)$. The graded components of its irreducible modules $\V(\alpha)_h$ will transform in a finite-dimensional representation of the \emph{inner} automorphism group $\mathrm{Inn}(\V)$, though in general they will transform \emph{projectively}, i.e.\ up to a phase \cite{Dong:1994wn}. The projective representations of a finite group can be lifted to ordinary representations of a covering group, and so we will often work with an extension of the inner automorphism group of the form $H.\mathrm{Inn}(\V)$, for $H$ suitably chosen. For example, although $\mathbb{B}$ is the automorphism group of the baby monster VOA $\tsl{V}\mathbb{B}^\natural$, its irreducible modules will in general transform under the group $2.\mathbb{B}$. 
The characters of $\V$ reflect this representation theory through the coefficients in their $q$-expansions: in particular, these coefficients can be written as sums of dimensions of irreducible representations of $H.\mathrm{Inn}(\V)$. All of these observations will provide nontrivial checks on our proposed characters and automorphism groups.

One can also consider a character which is twined by an automorphism $X$,
\begin{align}\label{twined characters}
    \chi_{X,\alpha}(\tau) = \mathrm{Tr}_{\V(\alpha)}Xq^{L_0-\frac{c}{24}}.
\end{align}
On general grounds, this will coincide with a vector-valued modular form of higher level \cite{Dong:1997ea,Dong:2005xj}, i.e. one that transforms covariantly with respect to the action of a discrete subgroup $\Gamma_X$ of $\operatorname{\textsl{SL}}_2(\mathbb{R})$ which includes at least a subgroup $\Gamma_0(n)\subset\Gamma_X$, where
\begin{align}
 \Gamma_0(n) = \left\{\left(\begin{array}{rr} a & b \\ c & d \end{array}\right) \in\operatorname{\textsl{SL}}_2(\mathbb{Z}) \  \bigg\vert\  c\equiv 0~\mathrm{mod}~n\right\}   
\end{align}
and with the value of $n$ being closely related\footnote{In fact, $n$ \emph{is} the order of $X$ if $X$ is non-anomalous; in general, $n$ is some multiple of the order of $X$.} to the order of $X$. 

One can also consider automorphisms of the fusion algebra of $\V$. These are simply invertible linear maps on the fusion algebra which preserve the fusion rules. Of special interest to us will be ``diagonal'' automorphisms which take the form 
\begin{align}\label{diagonalfusion}
    \W(\alpha) \mapsto \zeta_\alpha \W(\alpha) \ \ \ \ \ \ \ \ (\zeta_\alpha \in \mathbb{C}).
\end{align}
Their main utility for us lies in the fact that, if $\W$ is a subalgebra of $\V$, such mappings can often\footnote{We are not aware of a general condition which determines when such a lifting goes through, but we are confident that it does in all the cases that we invoke this structure.} be lifted to ordinary automorphisms of $\V$. Indeed, using the decomposition of $\V$ into $\W$-modules, 
\begin{align}
    \V = \bigoplus_\alpha \V_{(\alpha)}
\end{align}
we can define its lift $\tau_\W$ to be the linear mapping $\V\to\V$ which acts on vectors in the subspace $\V_{(\alpha)}$ as $\varphi \mapsto \zeta_\alpha \varphi$. This map will often be a non-trivial automorphism of $\V$. Furthermore, for any automorphism $X$ of $\V$, the automorphism associated with the subalgebra $X\W$ satisfies 
\begin{align}\label{autconj}
    \tau_{X\W} = X\tau_{\W}X^{-1}.
\end{align}
In particular, if $X$ belongs to $\mathrm{Cent}_{\mathrm{Aut}(\V)}(\tau_{\W})$ then $\tau_\W = \tau_{X\W}$. 
\subsubsection{Conformal vectors}

Generically, the chiral algebra of a CFT involves only the Virasoro algebra, in which case, the only operator with dimension $(2,0)$ is the holomorphic stress tensor. However, in CFTs with an enhanced chiral algebra, there may be additional dimension $(2,0)$ operators. When this happens, it is often possible to find fields distinct from the stress tensor which nonetheless have the canonical $TT$ OPE, albeit for a different value of the central charge. The states corresponding to such operators are known as \emph{conformal vectors}, and they will play an important role in the rest of our paper. To characterize the conformal vectors of a theory, it is useful to define the notion of a \emph{Griess algebra}, to which we turn next.

The Griess algebra \cite{flmbook,beauty} of a VOA $\V$ is an algebraic structure on its space of dimension 2 operators. Two pieces of data define it: a commutative product $\star:\mathcal{V}_2\times \mathcal{V}_2\to \mathcal{V}_2$ and a bilinear form $(\cdot,\cdot):\mathcal{V}_2\times \mathcal{V}_2\to \mathbb{R}$. They are commonly defined in the math literature as
\begin{align}
\begin{split}
\varphi \star \varphi' &= \varphi_0 \varphi'\\
(\varphi,\varphi') &= \langle\Omega|\varphi_2|\varphi'\rangle
\end{split}
\end{align}
where we have used the notation $\varphi = |\varphi\rangle = \varphi(0)|\Omega\rangle$ to denote the state corresponding to the operator $\varphi(z)$. The Griess algebra is meant to encode information related to the OPEs between dimension 2 operators. First, let us see that $\varphi\star\varphi'$ is the state corresponding to the operator which appears in the $1/(z-w)^2$ term of the $\varphi\varphi'$ OPE, i.e. that
\begin{align}
    \varphi\star \varphi' = (\varphi\varphi')_2.
\end{align} 
A quick calculation verifies that this is true,
\begin{align}
(\varphi\varphi')_2  = \oint \frac{dz}{2\pi i} z \varphi(z)\varphi'(0)|\Omega\rangle  = \sum_n \oint  \frac{dz}{2\pi i}z^{-n-1}\varphi_n\varphi' = \varphi_0\varphi' = \varphi\star \varphi'.
\end{align}
Similarly, the bilinear form encodes the coefficient of $1/(z-w)^4$ in the OPE,
$$(\varphi\varphi')_4(z) = (\varphi,\varphi')\mathds{1}.$$
Indeed, the vacuum is the only dimension 0 operator so $(\varphi\varphi')_4 \propto\mathds{1}$, and the constant of proportionality is fixed by the following calculation:
\begin{align}
\begin{split}
\langle \Omega |(\varphi\varphi')_4(0)|\Omega\rangle &= \oint \frac{dz}{2\pi i}z^3 \langle\Omega| \varphi(z)\varphi'(0)|\Omega\rangle  \\
&= \sum_n \oint \frac{dz}{2\pi i} z^{-n+1}\langle\Omega|\varphi_n|\varphi'\rangle \\
&= \langle\Omega|\varphi_2|\varphi'\rangle = (\varphi,\varphi').
\end{split}
\end{align}
By the state-operator correspondence, the Griess algebra is then the same as the two terms $(\varphi\varphi')_{4}$ and $(\varphi\varphi')_2$ in the OPE. As an example, observe that taking $\varphi=\varphi'=T$ to be the stress tensor implies 
\begin{align}
(T,T) = \langle\Omega | L_2 |T\rangle = \langle \Omega| L_2 L_{-2}|\Omega\rangle = \frac{c}{2}
\end{align}
and similarly,
\begin{align}
T\star T = L_0 T = 2T
\end{align}
which recovers the first two non-trivial terms in the $TT$ OPE. 

As the data of a chiral algebra necessarily involves the choice of a conformal vector, they are the first place one should look if one is interested in locating subalgebras. We comment that the collection of conformal vectors of a VOA $\V$ can be characterized using the language of the Griess algebra. Indeed, it is straightforward to check that conformal vectors $t(z)$ of central charge $c_t$ give rise to idempotents of the Griess algebra, $\frac{t}{2}\star\frac{t}{2}=\frac{t}{2}$, with $(t,t)=\frac{c_t}{2}$. If $\V$ does not have currents, (i.e. if $\V_1 = 0$, as is the case for the moonshine module $V^\natural$), then the converse is true as well \cite{miyamotogriess}: every idempotent of the Griess algebra gives rise to a stress tensor with central charge $c_t=2(t,t)$.

\subsubsection{Examples}\label{sec:examples}

In this section, we summarize the salient features of several examples of chiral algebras which will make appearances in the sequel. 

\subsubsection*{Minimal models}

The minimal models are a special class of two-dimensional conformal field theories: those for which the Hilbert space decomposes into a \emph{finite} direct sum of modules of the Virasoro algebra. Such theories only occur for values of the central charge labeled by a pair of coprime integers $p'>p\geq 2$. In terms of these integers, the central charge and the conformal dimensions of the primary operators in the theory are given by
\begin{align}\label{mmcc}
c_{p,p'} = 1-6\frac{(p-p')^2}{p p'},\quad  h^{p,p'}_{r,s} = \frac{(p' s-p r)^2-(p'-p)^2}{4p p'},
\end{align}
where $1 \le r < p', \ 1 \le s < p$, and $sp'<rp$. 

For these values of $c$, we will define $\mathcal{L}(c,0)$ to be the VOA at central charge $c$ after one has taken the quotient by all null vectors. Its highest weight modules are denoted $\mathcal{L}(c,h)$, when $h$ is as in equation (\ref{mmcc}). The characters of these highest weight modules are given by
\begin{align}
\label{character of minimal model}
\begin{split}
&\chi_{r,s}^{p,p'}(\tau) := \mathrm{Tr}_{\mathcal{L}(c_{p,p'},h^{p,p'}_{r,s})}q^{L_0-\frac{c_{p,p'}}{24}} =  K_{r,s}^{p,p'}(q)  - K_{r,-s}^{p,p'}(q), \\
& \ \ \ \ \ \ \ \ \  K_{r,s}^{p,p'}(q) = \frac{1}{\eta(q)} \sum_{n \in \mathbb{Z}} q^{\frac{(2 p p' n + p r- p' s)^2}{4 p p'}}.
\end{split}
\end{align}
The characters \eqref{character of minimal model} form a vector-valued modular form and under the $S$-transformation they transform as
\begin{align}
\chi_{r,s}^{p,p'}(-\tfrac{1}{\tau}) = \sum_{\rho,\sigma} {^\text{mm}}\mathcal{S}^{p,p'}_{rs;\rho \sigma} \chi_{\rho, \sigma}^{p,p'}(\tau).
\end{align}
where
\begin{align}
\label{S-matrix of minimal model}
{^\text{mm}}\mathcal{S}^{p,p'}_{rs;\rho \sigma} = 2\sqrt{\frac{2}{pp'}} (-1)^{1+ s \rho + r \sigma} \mbox{sin}\left( \pi \frac{p}{p'} r \rho \right) \mbox{sin}\left( \pi \frac{p'}{p} s \sigma\right).
\end{align}

Of interest to us will be the \emph{unitary} minimal models. Unitary requires in particular that no representation appearing in the decomposition of the Hilbert space feature states with negative norm; a necessary condition for the absence of negative norm states is that the highest weight states should have non-negative dimension. Examination of equation (\ref{mmcc}) shows that $\min_{r,s} h^{p,p'}_{r,s}<0$ unless $|p-p'| = 1$. It turns out that this primary must be included in the full theory in order to ensure modular invariance of the CFT, which implies that the CFT at central charge $c_{p,p'}$ is non-unitary if $|p-p'|\neq 1$. We therefore parametrize the unitary minimal models by a single integer $m = 2,$ $3, \dots$ by taking $p = m$, $p'=m+1$. We will often replace $p,p'$ with $m$ in the notation when we are working with the unitary theories, e.g. $h^{(m)}_{r,s}:= h^{p,p'}_{r,s}$, $c_m := c_{p,p'}$, and so on. 

These models admit $\mathbb{Z}_2$ automorphisms of their fusion algebras. For the unitary models, these take the form 
\begin{align}
    \mathcal{L}(c_{m},h^{(m)}_{r,s}) \mapsto 
    \begin{cases}
    (-1)^{r+1}\mathcal{L}(c_{m},h^{(m)}_{r,s}) &  \text{if } m\text{ is even} \\
    (-1)^{s+1}\mathcal{L}(c_{m},h^{(m)}_{r,s}) & \text{if }m \text { is odd}.
    \end{cases}
\end{align}
Moreover, these automorphisms are always ``liftable'' in the sense that, if $\V$ admits a minimal model subalgebra $\W\cong \mathcal{L}(c_{m},0)$, then the induced map $\tau_\W$ of \S\ref{subsubsec:symmetries} is guaranteed to be an automorphism of $\V$ \cite{miyamotogriess}. 

A particularly important example of the above is when $\W \cong \mathcal{L}(\tfrac{1}{2},0)$, which corresponds to the Ising model. Then, the fusion algebra automorphism takes the form 
\begin{align}\label{minimal model fusion aut}
\begin{split}
    \mathcal{L}(\tfrac{1}{2},0) &\mapsto \mathcal{L}(\tfrac{1}{2},0) \\
    \mathcal{L}(\tfrac{1}{2},\tfrac{1}{2}) &\mapsto \mathcal{L}(\tfrac{1}{2},\tfrac{1}{2}) \\
    \mathcal{L}(\tfrac{1}{2},\tfrac{1}{16}) &\mapsto - \mathcal{L}(\tfrac{1}{2},\tfrac{1}{16})
\end{split}
\end{align}
Its corresponding lift $\tau_\W$ is referred to as a \emph{Miyamoto involution} \cite{miyamotogriess}. When the decomposition of $\V$ does not have any $\mathcal{L}(\tfrac{1}{2},\tfrac{1}{16})$ modules appearing, 
\begin{align}
    \V = \V_{(0)}\oplus \V_{(\frac12)},
\end{align}
then $\tau_\W$ is trivial, but it is still possible to define another involution $\sigma_\W$ on $\V$ which acts trivially on the subspace $\V_{(0)}$ and sends $\varphi\mapsto -\varphi$ when $\varphi$ belongs to $\V_{(\frac12)}.$ The stress tensor of such an $\mathcal{L}(\tfrac12,0)$ subalgebra is referred to as being of ``$\sigma$-type'' in $\V$.

A lesser known fact is that the chiral algebra $\mathcal{L}(c_{m},0)$ can be extended to a larger (potentially super) chiral algebra by taking a direct sum with one of its irreducible modules \cite{lam2003extension}. Indeed, $h^{(m)}_{1,m+1}=\max_{r,s}h^{(m)}_{r,s}$ is either integral or half-integral depending on whether $m\equiv 1,2~\mathrm{mod}~4$ or $m\equiv 0,3~\mathrm{mod}~4$, and $\mathcal{L}(c_{m},0)\oplus \mathcal{L}(c_{m},h^{(m)}_{1,m+1})$ carries the structure of a chiral algebra in the first case and a super chiral algebra in the second case. We will make use of these simple-current\footnote{A simple-current is an operator $J$ such that the OPE of $J$ with any primary contains only a single term, which is itself a primary.} extensions in later sections.

\subsubsection*{Parafermion theories}

Minimal model CFTs have only $\mathbb{Z}_2$ or $\mathbb{Z}_3$ fusion algebra automorphisms so it will be useful to have rational CFTs with larger automorphism groups available in studying deconstruction of the monster CFT. 
A well studied example are the parafermion CFTs whose Hilbert space decomposes into a finite direct sum of modules of its chiral algebra. Here, we briefly review the properties of two-dimensional $\IZ_k$ parafermion chiral algebras \cite{zam}, following the conventions of \cite{Mercat:2001pu}.

The central charge of the $\IZ_k$ parafermion theory is
\be
c_k = \frac{2(k-1)}{k+2}
\ee
and takes the values $c=\frac12, \frac45, 1,\frac87,\frac54, \dots$ for $k=2,3,4,5,6, \dots$.  There are $\frac{k(k+1)}{2}$ independent primary fields with conformal dimensions
\begin{align}\label{parafermion dimensions}
h^{(k)}_{\ell,m}= \frac{\ell(\ell+2)}{4(k+2)} - \frac{m^2}{4k}
\end{align}
where the independent primary fields $\phi_{\ell,m}$ can be labeled by pairs of integers $(\ell,m)$ in the set
\be
\{ (\ell,m) \mid 0 \le \ell \le k,\ -\ell+2 \le m \le \ell,\  \ell-m \in 2 \IZ \}.
\ee
The operator $\phi_{k,k}$ is the identity operator with dimension $0$. We use the notation $\mathcal{P}(k,[\ell,m])$ for the associated highest weight module, and use the abbreviated notation $\mathcal{P}(k) = \mathcal{P}(k,[k,k])$ for the chiral algebra itself. Its characters are given by
\begin{align}
\label{Parafermion CFT character}
\begin{split}
\psi^{(k)}_{\ell,m}(\tau) &= \mathrm{Tr}_{\mathcal{P}(k,[\ell,m])}q^{L_0-\frac{c_k}{24}} \\
&=\frac{1}{\eta(\tau)^2} \Big\{ \Big(\sum_{i,j \le 0} - \sum_{i,j<0} \Big) (-1)^i q^{\frac{\left(\ell+1+(i+2j)(k+2) \right)^2}{4(k+2)}-\frac{(m+ik)^2}{4k}} \\
                                                & \ \ \ \ \ \ \ \ \ \ \ - \Big(\sum_{i \le 0, j > 0} - \sum_{i<0, j\le0} \Big) (-1)^i q^{\frac{\left(\ell+1-(i+2j)(k+2) \right)^2}{4(k+2)}-\frac{(m+ik)^2}{4k}} \Big\}
\end{split}
\end{align}
where $\eta$ is the Dedekind-eta function, given by 
\begin{align}\label{dedekindeta}
    \eta(\tau) = q^{\frac1{24}}\prod_{n=1}^\infty(1-q^n).
\end{align}
Their behavior under modular transformations is governed by the following modular S-matrix,
\begin{align}
\label{Parafermion S-matrix}
\begin{split}
{^\text{pf}}\mathcal{S}^{(k)}_{\ell m;\ell' m'} = \frac{2}{\sqrt{k(k+2)}}  e^{2 \pi i \frac{m m'}{2k}} \mbox{sin}\left( \pi \frac{(\ell+1)(\ell'+1)}{k+2}  \right)
\end{split}
\end{align}
These theories enjoy a $\IZ_k$ symmetry of their fusion algebra; the generator acts on the highest weight module $\mathcal{P}(k,[\ell,m])$ according to
\be
\label{Zk transformation rule}
\mathcal{P}(k,[\ell,m])\mapsto e^{2 \pi i \frac{m}{k}} \mathcal{P}(k,[\ell,m]) \,.
\ee
In later sections, we will locate parafermionic subalgebras of the moonshine module, and lift (a quotient of) this $\mathbb{Z}_k$ symmetry to an automorphism in the monster group.

In some cases, we are able to extend the parafermion chiral algebras by taking direct sums with their modules with integral highest weight, analogously to the simple current extensions of the previous section. We content ourselves with demonstrating these extensions as needed.

\subsubsection*{Lattice VOAs}

Rational toroidal compactifications of bosonic string theory are described by a $d$-dimensional torus $T^d={\mathbb R}^d/L$ with $L$ an even positive definite lattice. The holomorphic part of the corresponding CFT has the structure of a lattice VOA. Lattice VOAs are described in detail in many places including \cite{dgm,frenkel1992vertex,flmbook} so we will be very brief. The simplest example of lattice VOAs are the $c=1$ examples based on the lattice $L=\sqrt{2N} \mathbb{Z}$. For every $\lambda \in L$ there is a state $| \lambda \rangle$ and a set of oscillators $a_n$, $n \in \mathbb{Z}$ obeying
\be
[a_m,a_n]=m \delta_{m,-n}
\ee
with $a_n^\dagger=a_{-n}$. The state $|\lambda\rangle$ obeys  $a_n |\lambda \rangle=0$ for $n>0$ and $a_0 |\lambda \rangle = \lambda | \lambda \rangle$. Physical states are constructed by acting with the creation oscillators $a_{-n}$ with $n>0$ on the states $| \lambda \rangle$.
Vertex operators $V_\lambda(z)$ that create the states $|\lambda \rangle$ are given by normal ordered exponentials
\be
V_\lambda(z) = :e^{i \lambda X(z)}:
\ee
up to cocycle factors which are discussed in the literature. These vertex operators have conformal weight $\lambda^2/2$. The vacuum state is $|0 \rangle$ and the conformal state (i.e.\ the state corresponding to the stress tensor) is 
\be
\psi_{\tsl{Vir}}= \frac{1}{2} a_{-1} a_{-1} |0  \rangle \, .
\ee
When $N=1$ (which corresponds in the physics picture to a circle compactification at a radius which maps to itself under T-duality), the states $a_{-1} |0 \rangle$, $|\hspace{-.05in} \pm \hspace{-.05in}\sqrt{2} \rangle$ all correspond to vertex operators of conformal weight one and these form the basis of the Segal-Frenkel-Kac construction of affine $A_1$ at level one.

The rank one case has a generalization to lattice VOAs based on an even positive definite lattice $L$ and with central charge $c= {\rm rank}(L)$. The inequivalent simple modules of a lattice VOA can be labeled by elements of the discriminant group $L^*/L$ where $L^*$ is the dual lattice. The characters of these modules are given by
\be
\label{lattice theta}
\chi_{\lambda^*}(\tau)= \frac{\theta_{\lambda^*}(\tau)}{\eta(\tau)^d} \ \ \ \ \ (\lambda^\ast \in L^\ast/L)
\ee
where $\eta$ is the Dedekind eta function in equation \eqref{dedekindeta}, and the theta function is given by
\be
\theta_{\lambda^*}(\tau)= \sum_{\lambda \in L+\lambda^*} q^{\lambda^2/2} \, .
\ee
The S-matrix of this theory can be deduced from the multiplier system of the $\eta$ function, 
\begin{align}
  \eta(-\tfrac1\tau) = \sqrt{-i\tau}\eta(\tau)
\end{align}
as well as the fact that the theta functions transform into one another according to the Weil representation,
\begin{align}
    \theta_{\lambda^\ast}(-\tfrac1\tau) = \frac{i^{\mathrm{rank}(L)}}{\sqrt{L^\ast/L}|}\sum_{\gamma^\ast\in L^\ast/L} e^{-2\pi i(\lambda^\ast,\gamma^\ast) }\theta_{\gamma^\ast}(\tau)
\end{align}
where $(\cdot,\cdot):L^\ast\times L^\ast \to \mathbb{R}$ is the inner-product defined on the dual lattice. 

When $L$ is the root lattice of a simply laced ADE Lie algebra, the lattice VOA has the corresponding affine ADE symmetry. In what follows we will make use of the lattice VOA $\sqrt{2} R$ where $R$ is the root lattice of a simple laced Lie algebra. These lattice VOAs have dimension two operators dual to the states $|\sqrt{2} p \rangle$ with $p \in R$ and $p^2=2$ which play a crucial role in deconstructing VOAs into VOAs with smaller central charge. 

In addition to the lattice VOA $\V_L$, one can also consider the \emph{charge conjugation orbifold} $\V_L^+$, defined as the $\theta$-invariant subspace of $\V_L$, where $\theta$ is the automorphism induced by the canonical lattice involution $v\mapsto-v$. For a physicist, $\V_L^+$ would appear as the chiral algebra of the usual modular-invariant $\mathbb{Z}_2$ orbifold of a sigma model with target $T^d = \mathbb{R}^d/L$ \cite{Dijkgraaf:1989hb}. In the rank 1 case, the charge conjugation orbifolds are all of the form $\V_{\sqrt{2N}\mathbb{Z}}^+$ with $N$ a positive integer. The discriminant group of the associated lattice has representatives 
\begin{align}
    (\tfrac{1}{\sqrt{2N}}\mathbb{Z})/(\sqrt{2N}\mathbb{Z})\cong \{0,\tfrac{1}{\sqrt{2N}},\tfrac{2}{\sqrt{2N}},\dots,\tfrac{2N-1}{\sqrt{2N}}\}
\end{align} 
and, in terms of the characters $\chi_{\frac{k}{\sqrt{2N}}}(\tau)$ of $\V_{\sqrt{2N}\mathbb{Z}}$, and the generalized theta functions
\begin{align}\label{generalizedtheta}
    \Phi_{\alpha,\beta}(\tau) =  \frac{1}{\eta(\tau)} \sum_{m\in\mathbb{Z}} q^{(m+\frac\alpha4)^2}e^{2\pi i m\frac{\beta}{2}},
\end{align}
the characters of $\V_{\sqrt{2N}\mathbb{Z}}^+$ are given by \cite{Dijkgraaf:1989hb,dong1999representations}
\begin{align}
\label{DVVV ch}
\begin{split}
h=0:~\xi^{(N)}_{\mathds{1}}(\tau) &= \frac{1}{2}\left(  \chi_0(\tau) +  \Phi_{0,1}(\tau)\right) \\
h=1:~\xi^{(N)}_j(\tau) &= \frac{1}{2}\left(  \chi_0(\tau) -  \Phi_{0,1}(\tau)\right) \\
h=\frac{N}{4}:~\xi_{N,i}^{(N)}(\tau) &= \frac{1}{2}\chi_{\sqrt{\frac N2}}(\tau) \ \ \ \ (i=1,2) \\
h = \frac{k^2}{4N}:~\xi^{(N)}_k(\tau) &= \chi_{\frac{k}{\sqrt{2N}}}(\tau) \ \ \ (k=1,\dots,N-1) \\
h = \frac{1}{16}:~\xi^{(N)}_{\sigma,i}(\tau) &= \frac{1}{2}\left( \Phi_{1,0}(\tau) +  \Phi_{1,1}(\tau) \right) \ \ \ (i=1,2) \\
h = \frac{9}{16}:~\xi^{(N)}_{\tau,i}(\tau) &= \frac{1}{2}\left( \Phi_{1,0}(\tau) -  \Phi_{1,1}(\tau) \right) \ \ \ (i=1,2)
\end{split}
\end{align}
where we have borrowed from the notation of \S 7.b of \cite{Dijkgraaf:1989hb}. In the case of a general even, $d$-dimensional lattice $L$
\cite{flm,abe2004classification}, the characters are given by
\begin{align}\label{chargeconjorbchars}
\begin{split}
\xi_{\mathds{1}}^{(L)}(\tau) &= \frac{1}{2}\left( \chi_0(\tau) + \Phi_{0,1}(\tau)^d     \right) \\
\xi_{j}^{(L)}(\tau) &= \frac{1}{2}\left( \chi_0(\tau) - \Phi_{0,1}(\tau)^d     \right) \\
\xi^{(L)}_{\lambda^\ast,i}(\tau) &= \frac{1}{2}\chi_{\lambda^\ast}(\tau) \ \ \ (i=1,2, \text{ and } \lambda^\ast \in L^\ast /L \text{ such that } 2\lambda^\ast = 0) \\
\xi^{(L)}_{\lambda^\ast}(\tau) &= \chi_{\lambda^\ast}(\tau) \ \ \ (\lambda^\ast \in (L^\ast/L)/\sim \text{ such that }2\lambda^\ast\neq 0) \\
\xi_{\sigma,i}^{(L)}(\tau) &= \frac{|L/R|^{\frac12}}{2}\left(  \Phi_{1,0}(\tau)^d + \Phi_{1,1}(\tau)^d  \right) \ \ \ (i=1,\dots,|R/2L| )\\
\xi_{\tau,i}^{(L)}(\tau) &= \frac{|L/R|^{\frac12}}{2}\left(  \Phi_{1,0}(\tau)^d + \Phi_{1,1}(\tau)^d  \right) \ \ \ (i=1,\dots,|R/2L| )
\end{split}
\end{align}
where $R = \{\lambda \in L \mid (\lambda,L)\subset 2\mathbb{Z}\}$ and $\sim$ denotes the equivalence relation which identifies $\lambda^\ast$ with $-\lambda^\ast$ (i.e.\ $\xi_{\lambda^\ast}^{(L)}$ and $\xi_{-\lambda^\ast}^{(L)}$ are not regarded as inequivalent characters).

\subsection{Deconstruction generalities}\label{subsec:deconstruction}

It is often useful to build up a complicated theory out of simpler building blocks. An example of this is Gepner's description \cite{Gepner:1987vz} of sigma models with Calabi-Yau target at special points in their moduli space as a suitable (orbifold of a) tensor product of superconformal minimal models. In the other direction, one might begin with a known theory, and discover an alternative way of looking at it in terms of more tractable constituents. For example, after the original construction of the moonshine module by Frenkel, Lepowsky, and Meurman \cite{flmbook} as a $\mathbb{Z}_2$ asymmetric orbifold of the Leech lattice VOA, it was discovered that $V^\natural$ admits an $\mathcal{L}(\tfrac12,0)^{\otimes 48}$ subalgebra \cite{Dong:1997xy} with respect to which its fields can be alternatively organized; this construction is similar in spirit to the decomposition of a theory into representations of its current algebra (an affine Kac-Moody algebra), a procedure which is unavailable in the case of $V^\natural$ due to its lack of dimension 1 operators. One may even find new structures entirely in their efforts to ``deconstruct'' a theory into its pieces, as we will indeed find to be true for us in \S\ref{sec:mainresults}. In this subsection, we will outline a somewhat systematic procedure for discovering such deconstructions, and provide several examples. 

\subsubsection*{Stress tensor decompositions}

In order for a chiral algebra to admit a tensor product subalgebra, it should be at least possible to write its stress tensor with central charge $c$ as the sum of two commuting conformal vectors whose central charges add to $c$. So we will begin our analysis at the level of the stress tensor. Consider first the case where there is a single Virasoro primary field $\varphi(z)$ of dimension $2$. We then have two dimension $2$ fields --- the
stress tensor $T(z)$ and $\varphi(z)$ --- with the following OPEs,
\begin{align}
\begin{split}
T(z) T(w) & \sim  \frac{c/2}{(z-w)^4}+\frac{2 T(w)}{(z-w)^2}+ \frac{\partial T(w)}{z-w} \\
T(z) \varphi(w) & \sim  \frac{2 \varphi(w)}{(z-w)^2} + \frac{\partial \varphi(w)}{z-w}   \\
\varphi(z) \varphi(w) & \sim  \frac{1}{(z-w)^4} + \frac{ 4T(w)/c  + b \varphi(w)}{(z-w)^2} + \frac{ \partial\big(4 T(w)/c+ b \varphi(w)\big)/2}{z-w} .
\end{split}
\end{align}
In the above, the second line follows from the fact that $\varphi$ is primary and has dimension $2$. In the third line, the first term involves
a choice of normalization of $\varphi$, the vanishing of the $1/(z-w)^3$ term follows from Bose symmetry, and the other two terms involve
a single undetermined coefficient $b$ and have a form dictated by associativity of the OPE.

It is now natural to try to construct a conformal vector as a linear combination of $T(z)$ and $\varphi(z)$, that is
to consider
\be
t(z)= \alpha T(z) + \beta \varphi(z)
\ee
and to solve for the constants $\alpha, \beta$ by demanding that $t(z)$ have the OPE of a stress tensor,
\be
t(z)t(w)\sim \frac{c_t/2}{(z-w)^4} + \frac{2 t(w)}{(z-w)^2} + \cdots
\ee
for some $c_t$. Using the OPEs of the two dimension 2 fields, we compute 
\begin{align}
\begin{split}
t(z) t(w) &\sim \frac{1}{(z-w)^4} \left( \alpha^2 c/2+ \beta^2 \right) \\
& \ \ \ \ \ \ + \frac{1}{(z-w)^2} \left( 2 \alpha^2 T(w) + \beta^2(4T(w)/c+ b \varphi(w)) + 4 \alpha \beta \partial \varphi(w) \right) + \cdots
\end{split}
\end{align}
Equating these two expressions, we get the quadratic equations
\begin{align}
\begin{split}
2 \alpha &= 2 \alpha^2 + \frac{4}{c} \beta^2 \\
2 \beta &= b \beta^2 + 4 \alpha \beta 
\end{split}
\end{align}
and 
\be
c_t = \alpha^2 c + 2 \beta^2.
\ee
Provided that $b \ne 0$, one finds two solutions $c_\pm$ for $c_t$ with
\be
c_\pm = \frac{c}{2}\left(1 \pm \left(1+ \frac{32}{b^2 c}\right)^{-\frac{1}{2}} \right),
\ee
and two solutions for $\alpha$ and $\beta$,
\be
\alpha_\pm = \frac{1}{2}\left(1 \pm \left(1+ \frac{32}{b^2 c}\right)^{-\frac{1}{2}} \right), \quad \beta_{\pm} = \mp \frac{2}{b} \left(1+ \frac{32}{b^2 c}\right)^{-\frac{1}{2}}.
\ee
When $b=0$, we have $\alpha=\frac12$ and $\beta = \mp \frac{1}{2} \sqrt{\frac{c}{2}}$. One can check that the OPE between $t^{(+)}$ and $t^{(-)}$ is regular, and further that $c_++c_-=c$. Thus, the stress tensor of any theory with two dimension 2 operators can always be deconstructed into a sum of commuting conformal vectors for two chiral algebras with smaller central charge,
\be
T(z) = t^{(+)}(z)+t^{(-)}(z),
\ee
a fact which we will put to use in later sections. If $c>1$ but $c_\pm <1$, then we can organize the Virasoro primaries of the original theory into a finite set of primary fields with respect to $t^{(+)}$ and $t^{(-)}$.

\begin{exmp}[Bosonization]
Take the starting CFT to be that of a free boson on a circle of radius $R$ with $c=1$. Such a theory arises in many places in physics. For example, it describes the continuum limit of the 2-dimensional statistical mechanical XY model at low temperatures. One may also take the bosonic field $X(z,\bar z)$ as the coordinate of a string propagating on a circle $S^1$
with periodicity $X \sim X+ 2 \pi R$; such a model arises as part of a fuller world-sheet string theory whenever spacetime is compactified on a circle. 

To describe the theory, we split the bosonic field into a left-moving and right-moving part,
\be
X(z, \bar z)= X_L(z) + X_R(\bar z).
\ee
The holomorphic stress tensor is
\be
T= - \frac{1}{2} :\partial X_L \partial X_L:
\ee
where $:\mathcal{O}(z):$ indicates that $\mathcal{O}(z)$ should be normal-ordered. There is a corresponding anti-holomorphic stress tensor $\overline T$ defined analogously. The primary fields include vertex operators of the form
\be
V_{n,m}(z,\bar{z}) = :e^{i p_L X_L(z) + i p_R X_R(\bar{z})}:
\ee
where 
\be\label{vertex_operator_momenta}
(p_L,p_R)= \left( \frac{m}{2R}+nR, \frac{m}{2R}-nR \right).
\ee
They have conformal dimensions $(h,\bar h) = (\tfrac{p_L^2}{2},\tfrac{p_R^2}{2})$ with respect to $T, \overline T$. 

Now, if $R^2$ is irrational, then it is easy to see from inspection of equation (\ref{vertex_operator_momenta}) that there are no holomorphic operators besides the identity and $\partial X_L$; in this case, the chiral algebra is an enhancement of the Virasoro algebra known as an affine $\mathrm{U}(1)$ current algebra (with $\partial X_L$ playing the role of the current). However, if $R^2$ is rational, then the holomorphic chiral algebra enhances further to a VOA known in the math literature as a rank 1 lattice VOA (c.f.\ \S\ref{sec:examples}). To get operators with conformal dimension $(2,0)$, we can go to $R=1$ where we have the purely holomorphic operator
\be
\varphi(z)= \frac{1}{\sqrt{2}} \left( :e^{2 i X_L(z)}:+ :e^{-2 i X_L(z)}: \right).
\ee
The OPE of $\varphi$ with itself trivially gives $b=0$ and so one finds
\be
c_\pm = \frac12
\ee
and that
\be
t^{(\pm)}=\frac{1}{2} \left( - \frac{1}{2} :\partial X_L \partial X_L: \pm \frac{1}{\sqrt{2}}\varphi \right) 
\ee
are commuting conformal vectors with central charge $\frac{1}{2}$, and so give a subalgebra of the chiral algebra which is isomorphic to two copies of the Ising VOA. Thus, we have deconstructed the stress tensor of this $c=1$ model into a sum of conformal vectors with central charge $\frac{1}{2}$. This is closely related to the famous bosonization of fermions in two dimensions.
\end{exmp}

This analysis can be extended to theories which have an arbitrary number $n_2$ of dimension $2$ primary fields $\varphi^i$. The OPE takes the general form
\be
\varphi^i(z) \varphi^j(w) \sim \frac{\delta_{ij}}{(z-w)^4} + \frac{a^{ij \beta} \psi^\beta(w)}{(z-w)^3} + \frac{b^{ijk} \phi^k(w) + \cdots}{(z-w)^2} + \frac{c^{ij \rho} \chi^\rho(w) + \cdots }{z-w}
\ee
where repeated indices are summed, $\psi^\beta$ are the set of dimension $1$ operators, and $\chi^\rho$ the set of dimension $3$ primary operators
of the CFT. The ellipsis indicate contributions from descendants. This now looks rather hopeless, but in fact if one looks at the OPE of
\be
t = \alpha_0 T + \sum_{i=1}^{n_2} \alpha_i \varphi_i
\ee
with itself, the coefficients $a^{ij \beta}$ and $c^{i j \rho}$ cancel out and the equations that have to be solved to deconstruct the stress tensor involve only
the $b^{ijk}$,
\bea \label{deconeqn}
\alpha_0 &=& \alpha_0^2 + \frac{2}{c} \sum_{i=1}^{n_2} \alpha_i^2 \\
\alpha_i &=& 2 \alpha_0 \alpha_i + \frac{1}{2} \sum_{j,k=1}^{n_2} \alpha_j \alpha_k b^{ijk}.
\eea
The decoupling of $a^{ij\beta}$ and $c^{ij\rho}$ should be expected from our earlier claims that conformal vectors can be characterized in terms of the Griess algebra. Indeed, the Griess algebra involves only the data of the $1/(z-w)^4$ and $1/(z-w)^2$ terms of the OPE, and so doesn't witness the coefficients $a^{ij\beta}$ and $c^{ij\rho}$; the $b^{ijk}$ can be thought of as its structure constants.

Assuming there's a solution $t$ with central charge $c_t$, it is known \cite{frenkel1992vertex} that if $(Tt)_3=0$ then $\widetilde{t} = T-t$ is also a conformal vector with central charge $c-c_t$, and in particular one can successfully deconstruct the stress tensor as $T = t+\widetilde{t}$. In a theory without currents, (i.e.\ if $\V_1 = 0$, as is true for the moonshine module), we are guaranteed that $(Tt)_3=0$, and so any conformal vectors in addition to the stress tensor will lead to a deconstruction. 

\begin{exmp}[Toroidal CFT]

We can generalize the previous example by considering chiral algebras which arise in the world sheet CFTs which describe strings propagating on certain special tori \cite{DH}. We will see that the stress tensor can be iteratively deconstructed into multiple conformal vectors with smaller central charge.

As an example, consider the $c=2$ theory of $2$ free bosons $\vec{X}(z,\bar{z}) = (X_1(z,\bar{z}),X_2(z,\bar{z}))$ with stress tensor
\be
T= - \frac{1}{2} :\partial \vec{X}_{L}\cdot  \partial \vec{X}_{L}: 
\ee
We will work at a special point in the moduli space of such theories where the chiral algebra consists of primaries of the form
\begin{align}
    V_{\vec{p}}(z) = :e^{i\vec{p}\cdot \vec{X}_L(z)}:
\end{align}
where $\vec{p}$ belongs to the (rescaled) root lattice $\sqrt{2} \Lambda_{\mathrm{root}}(A_2)$. Such operators have holomorphic dimension $h = \frac{p^2}{2}$, so we can obtain 3 operators with $h=2$ by taking $p_1$, $p_3$ to be the positive simple roots of $A_2$, $p_2=p_1+p_3$, and defining 
\begin{align}
    \varphi^i = \frac{1}{\sqrt{2}}\left(:e^{i\vec{p}_i\cdot \vec{X}_L}: + :e^{-i\vec{p}_i\cdot \vec{X}_L}:\right).
\end{align}
Solving equation \eqref{deconeqn}, one can show that there is a conformal vector $t^{(\frac45)}$ with central charge $\frac45$, given by 
\begin{align}
    t^{(\frac45)}(z) = \frac{2}{5}T(z)-\frac{\sqrt{2}}{5}\sum_i \varphi^i(z)
\end{align}
which deconstructs the stress tensor as
\be
T(z) = t^{(\frac45)}(z)+t^{(\frac65)}(z).
\ee
One can then work in the subspace of dimension 2 operators which have regular OPE with $t^{(\frac45)}$ (which includes $t^{(\frac65)}$), and again search for conformal vectors. If one does this, it is possible to show that the central charge $\frac65$ conformal vector can be further deconstructed as
\be
t^{(\frac65)}(z)= t^{(\frac12)}(z)+t^{(\frac{7}{10})}(z)
\ee
so that all together, the stress tensor of this toroidal CFT can be decomposed as the sum of conformal vectors whose central charges agree with those of the first two non-trivial minimal models and the $\mathbb{Z}_3$ parafermion theory,
\be
T(z)= t^{(\frac12)}(z)+t^{(\frac{7}{10})}(z)+t^{(\frac45)}(z).
\ee
This can be generalized, e.g.\ by working with the lattice VOA associated to $\sqrt{2}\Lambda_{\mathrm{root}}(A_{N-1})$ (which arises as the chiral algebra of a CFT whose target space is an $N-1$ dimensional torus at a special point in its moduli space). It was conjectured in \cite{DH} and proved in \cite{dong1996associative} that the stress tensor of this $c=N-1$ CFT can be deconstructed as
\begin{align}\label{ANpf}
N-1=\frac{2(N-1)}{N+2} + \sum_{m=3}^{N+1} \left( 1- \frac{6}{m(m+1)} \right)
\end{align}
where the first term is the central charge of the $\mathbb{Z}_N$ parafermion theory and the second term is the sum of the central
charges of the first $N-1$ minimal models. 

This construction can be generalized to the VOA associated to any simply laced root system. If $R$ is a simply laced root system with rank $\ell$ and Coxeter number $h$, then it was shown in \cite{dong1996associative}
that the lattice VOA $\V_{\sqrt{2}\Lambda_{\mathrm{root}}(R)}$ contains a conformal vector $\tilde \omega$
with central charge $c_R=2 \ell/(h+2)$, i.e.\ $c_R = \frac{2\ell}{\ell+3}$ if $R=A_\ell$, $c_R=1$ if $R = D_\ell$, $c_R=\frac{6}{7}$ if $R = E_6$, $c_R = \frac{7}{10}$ if $R=E_7$, and $c_R=\frac12$ if $R = E_8$.
\end{exmp}

\subsubsection*{Commutant subalgebras}

So far, we have only described the (not necessarily unique) deconstruction of the stress tensor, 
\begin{align}\label{deconstruct_stress_tensor}
    T(z) = \sum_it^{(i)}(z) 
\end{align}
into commuting conformal vectors $t^{(i)}$ with $c = \sum_i c_i$. Of course, there is more to a chiral algebra than its stress tensor. Equation (\ref{deconstruct_stress_tensor}) should be thought of as the first step towards reaching a more refined statement of the form
\begin{align}\label{deconstruct_chiral_algebra}
    \V \supset \bigotimes_{i}\V^{(i)}.
\end{align}
Here, each $\V^{(i)}$ is a chiral algebra whose stress tensor is $t^{(i)}$, so in particular, equation (\ref{deconstruct_chiral_algebra}) generally contains more information than equation (\ref{deconstruct_stress_tensor}). 

Let us specialize for simplicity to the case where we have deconstructed the stress tensor into just two commuting conformal vectors, $T = t^{(1)}+t^{(2)}$, though it is straightforward to generalize. There is a well-known construction which is useful for describing $\V^{(1)}$ and $\V^{(2)}$, known as the commutant: if $\W$ is a subalgebra of $\V$, the \emph{commutant} of $\W$ in $\V$, which we will denote $\widetilde{\W} := \operatorname{\textsl{Com}}_\V(\W)$, is defined as the set of operators in $\V$ which have regular OPE with every operator in $\W$. If $t(z) = \sum_{n\in\mathbb{Z}}l_n z^{-n-2}$ is the stress tensor of $\W$, then the corresponding space of states consists of those of $\V$ which are vacua with respect to $t$, 
\begin{align}\label{commutant definition}
    \operatorname{\textsl{Com}}_{\V}(\W) = \{ \varphi \in \V \mid l_{-1}\varphi = 0   \}.
\end{align}
This description makes it clear that the commutant of $\W$ in $\V$ depends only on the stress tensor of $\W$, 
\begin{align}
    \operatorname{\textsl{Com}}_\V(\W) = \operatorname{\textsl{Com}}_\V(\operatorname{\textsl{Vir}}(t))
\end{align}
where $\operatorname{\textsl{Vir}}(t)$ is the Virasoro subalgebra of $\V$ generated by $t$. Now, if $(Tt)_3=0$, then $\widetilde{\W} = \operatorname{\textsl{Com}}_\V(\W)$ has the structure of a chiral algebra with $\widetilde{t} = T-t$ as its stress tensor \cite{frenkel1992vertex}. It follows that the maximal subalgebras which can appear in equation (\ref{deconstruct_chiral_algebra}) are simply commutant subalgebras,
\begin{align}
    \V^{(1)} = \operatorname{\textsl{Com}}_\V(\operatorname{\textsl{Vir}}(t^{(2)})),  \ \ \ \ \V^{(2)} = \operatorname{\textsl{Com}}_\V(\operatorname{\textsl{Vir}}(t^{(1)})).
\end{align}
These subalgebras are further each others' commutants, i.e. $\tsl{Com}_\V(\V^{(1)}) = \V^{(2)}$ and $\V^{(1)}=\tsl{Com}_\V(\V^{(2)})$, and so we refer to $(\V^{(1)},\V^{(2)})$ as a \emph{commutant pair}. Our main strategy for building chiral algebras with large, exceptional symmetry groups will be to take $\V = V^\natural$, locate known, simple subalgebras $\W$ (whose stress tensor we denote by $t$), and consider their commutants $\widetilde{\W}$. Since $V^\natural$ has no currents, $(Tt)_3=0$ is trivially satisfied by dimensional analysis, and so $\widetilde{\W}$ is guaranteed to carry the structure of a chiral algebra/VOA. 

We will also occasionally perform iterated deconstructions, as in the toroidal CFT example. For example, we will often find chains of subalgebras of the form
\begin{align}
    \V = \U^{(0)} \supset \U^{(1)} \supset \cdots \supset \U^{(n)} \supset \U^{(n+1)} = 0.
\end{align} 
where each $\U^{(i)}$ is the result of deconstructing $t^{(i)}$ off of $\U^{(i-1)}$, i.e.
\begin{equation}\label{iterated deconstruction}
\begin{aligned}
    \U^{(0)} &\supset \tsl{Vir}(t^{(1)})\otimes \U^{(1)} \ \ & & \U^{(1)}= \tsl{Com}_{\U^{(0)}}(\tsl{Vir}(t^{(1)}))  \\
    &\supset \tsl{Vir}(t^{(1)})\otimes \tsl{Vir}(t^{(2)})\otimes \U^{(2)} \ \ & & \U^{(2)} = \tsl{Com}_{\U^{(1)}}(\tsl{Vir}(t^{(2)})) \\
    & \ ~ \vdots \\
    &\supset \tsl{Vir}(t^{(1)})\otimes \cdots\otimes \tsl{Vir}(t^{(n-1)})\otimes \U^{(n)} \ \ & &\U^{(n)} = \tsl{Com}_{\U^{(n-1)}}(\tsl{Vir}(t^{(n)}))
\end{aligned}
\end{equation}
Of course, since the commutant depends only on the stress tensor, one can alternatively describe the $\U^{(i)}$ as 
\begin{align}\label{alternative iterated deconstruction}
    \U^{(i)} = \tsl{Com}_{\V} \left(  \tsl{Vir}\left( \sum_{j=1}^i t^{(j)}  \right)   \right).
\end{align}

\subsubsection*{Symmetries in deconstruction}
Having control over the properties of $\W$ will allow us to infer various properties of its commutant $\widetilde{\W}$. For example, we will frequently make use of the homomorphism
\begin{align}\label{authom}
\begin{split}
    \Phi_t:\mathrm{Stab}_{\mathrm{Aut}(\V)}(t) = \{X\in\mathrm{Aut}(\V) \mid Xt = t\} &\to \mathrm{Aut}(\operatorname{\textsl{Com}}_\V(\operatorname{\textsl{Vir}}(t))) = \mathrm{Aut}(\widetilde{\W}) \\
    X & \mapsto X\vert_{\widetilde{\W}}
    \end{split}
\end{align}
to infer symmetries of $\widetilde{\W}.$ The reason this map is well-defined is that any $X$ which stabilizes $t$ will preserve $\widetilde{\W} = \operatorname{\textsl{Com}}_\V(\operatorname{\textsl{Vir}}(t))$,
\begin{align}
    l_{-1}(X\varphi) = (Xt)_{-1}(X\varphi) = X(l_{-1}\varphi) = 0
\end{align}
where we have used the definition of an automorphism in \eqref{altautdef}. The map $\Phi_t$ makes it clear why the procedure of deconstruction is well-suited for probing the structure of \emph{subquotients} of $\mathrm{Aut}(\V)$ as opposed to just subgroups: it is because $\mathrm{Stab}_{\mathrm{Aut}(\V)}(t)$ is a subgroup of $\mathrm{Aut}(\V)$, and its image under $\Phi_t$ in $\mathrm{Aut}(\widetilde{\W})$ by the first isomorphism theorem is simply $\mathrm{Stab}_{\mathrm{Aut}(\V)}(t)/\ker(\Phi_t)$. It is this fact which underlies the reason why we are able to obtain precisely simple sporadic groups, as opposed to extensions of them. For example, as we will see in \S\ref{subsubsec:BMe7}, it is $\mathbb{B}$ which is the automorphism group of the baby monster VOA and not $2.\mathbb{B}$, even though $2.\mathbb{B}$ is a subgroup of $\mathbb{M}$, and not $\mathbb{B}$.

We will also be interested in the automorphisms of $\V$ which preserve the decomposition of $\V$ into representations of $\W\otimes \widetilde{\W}$. Let us assume for simplicity that $\V$ is a meromorphic CFT, i.e.\ that it has exactly one irreducible module and its partition function $\mathcal{Z}(\tau)$ is modular invariant, though it is not difficult to generalize. The fact that $(\W,\widetilde{\W})$ furnish a commutant pair inside $\V$ implies (under suitable assumptions, which hold for $\V=V^\natural$) that there is a decomposition of the form \cite{lin2017mirror}
\begin{align}\label{paired decomposition}
\begin{split}
    \V &\cong \bigoplus_{\alpha}\W(\alpha)\otimes \widetilde{\W}(\alpha) 
\end{split}
\end{align}
where the $\W(\alpha)$ (resp.\ $\widetilde{\W}(\alpha)$) are mutually inequivalent irreducibles modules of $\W$ (resp.\ $\widetilde{\W}$). Thus, the realization of a commutant pair $(\W,\widetilde{\W})$ inside of $\V$ establishes a bijective pairing between a subset of irreducible modules of $\W$ and a subset of irreducible modules of $\widetilde{\W}$ in a manner similar to Schur-Weyl duality, and relates\footnote{Their fusion categories are ``braid-reversed equivalent''} their fusion categories \cite{creutzig2019glueing} (e.g.\ $\W$ and $\widetilde{\W}$ will have conjugate modular S-matrices, as we will see in the next subsection). We will say that an automorphism $X$ of $\V$ \emph{preserves} $\W$ if it can be realized in the form 
\begin{align}\label{preserveaut}
    X = \sum_\alpha X_\alpha \otimes\mathds{1}
\end{align}
such that $X_\alpha:\W(\alpha)\to \W(\alpha)$. We denote the collection of such automorphisms $\mathrm{Aut}(\V\vert \W)$, or $\mathbb{M}(\W)$ in the case that $\V = V^\natural$; the analogous notions go through when $\W$ is replaced with $\widetilde{\W}$. It is not hard to see that $\mathrm{Aut}(\V\vert\W)$ commutes with $\mathrm{Aut}(\V\vert\widetilde{\W})$, so that they are contained in each others' centralizers. In the case that they are equal to each others' centralizers, we call the groups $(G,\widetilde{G}) := (\mathrm{Aut}(\V\vert\W),\mathrm{Aut}(\V\vert\widetilde{\W}))$ a \emph{centralizer pair}, and $(\W,\widetilde{\W})$ a \emph{centralizing commutant pair}; in the case that $\V=V^\natural$, we call such a pair a \emph{monstralizing commutant pair}, or $\mathbb{M}$-\emph{com pair} for short. Although we will only work in the monster CFT, we suspect that the notion of a centralizing commutant pair is a natural one in any CFT, as it furnishes decompositions of $\V$ which play well with respect to the action of its automorphism group. 

The centers of a centralizer pair must agree, $Z(G) = Z(\widetilde{G}) = G\cap \widetilde{G}$, as must their normalizers, $N_{\mathrm{Aut}(\V)}(G) = N_{\mathrm{Aut}(\V)}(\widetilde{G})=:N$. The only way the centers can be the same is if e.g.\ the $X_\alpha$ in equation \eqref{preserveaut} are each proportional to the identity, $X_\alpha = \zeta_\alpha\mathds{1}$. Thus, we learn that $G\cap\widetilde{G}$ acts by diagonal fusion algebra automorphisms, as in \eqref{diagonalfusion}, on both the modules of $\W$ and the modules of $\widetilde{\W}$, as one might expect from the fact that their fusion categories are related. We can further define $H := N/\langle G,\widetilde{G}\rangle$, which will be a subgroup of the outer automorphism groups of both $G$ and $\widetilde{G}$. In general, we will report the data of a centralizer pair as $[G\circ \widetilde{G}].H$, or $[G\times\widetilde{G}].H$ in the case that the centers of $G$ and $\widetilde{G}$ are trivial, where $G\circ\widetilde{G}$ denotes the central product of $G$ with $\widetilde{G}$, which will be a subgroup of $\mathrm{Aut}(\V)$. In all our examples, we will find that the outer automorphisms which $\widetilde{\W}$ inherits from $\mathrm{Aut}(\V)$ come from the group $H$. 

How does the group e.g.\ $\widetilde{G}$ compare with $\mathrm{Aut}(\widetilde{\W})$? It simultaneously includes too much and too little. In general, it includes too much because it contains fusion algebra automorphisms coming from its center which act trivially on the vacuum module, and so we must quotient by these to reach what we would properly call the inner automorphism group. It also contains too little because it misses out on outer automorphisms. Therefore, we have the expectation that %
\begin{align}
    \mathrm{Inn}(\widetilde{\W})\sim\widetilde{G}/Z(\widetilde{G}), \ \ \ \ \  \mathrm{Aut}(\widetilde{\W}) \sim (\widetilde{G}/Z(\widetilde{G})).H'
\end{align}
with $H'$ a subgroup of $H$. The group $\widetilde{G}$ is closely related to the projective representation theory of $\mathrm{Inn}(\widetilde{\W})$ because any projective representation of $\mathrm{Inn}(\widetilde{\W})$ that is realized on the modules $\widetilde{\W}(\alpha)$ can be lifted to an honest representation of $\widetilde{G}$. In all of our monster examples, we find that $\widetilde{G}$ is a (usually trivial) quotient of the Schur cover of $\mathrm{Inn}(\widetilde{\W})$.

\subsubsection*{Characters}

Finally, we comment that we are also able to put constraints on the characters and modular S and T matrices of $\widetilde{\W}$ from our knowledge of the analogous data for $\W$ and $\V$. Passing to characters, the decomposition \eqref{paired decomposition} becomes
\begin{align}\label{characterpairing}
\begin{split}
   \mathcal{Z}(\tau) &=   \sum_{\alpha}\chi_{\alpha}(\tau) \widetilde{\chi}_{\alpha}(\tau) 
\end{split}
\end{align}
which can be satisfied if
\begin{align}\label{matrixpairing}
   \mathcal{S}^T \widetilde{\mathcal{S}}  = \mathds{1}, \ \ \ \ \ \  \mathcal{T}^T \widetilde{\mathcal{T}} = \mathds{1}
\end{align}
with $\mathcal{S}$ ($\widetilde{\mathcal{S}}$) the modular S-matrix of $\W$ ($\widetilde{\W}$) and similarly for $\mathcal{T}$ ($\widetilde{\mathcal{T}}$). Because the S and T matrices of a rational CFT are always symmetric and unitary, we in fact have that 
\begin{align}
\widetilde{\mathcal{S}}=\mathcal{S}^\ast, \ \ \ \ \ \ \ \widetilde{\mathcal{T}} = \mathcal{T}^\ast.
\end{align}
In the moonshine module $\V = V^\natural$, which has no currents, the simplest and most natural ansatz for the conformal dimension $\widetilde{h}_\alpha$ of the highest weight subspace of $\widetilde{\W}(\alpha)$ which is consistent with $\widetilde{\mathcal{T}} = \mathcal{T}^\ast$ is $\widetilde{h}_\alpha = 2-h_\alpha$ when $\alpha \neq 0$ (i.e.\ for the non-vacuum modules) and $\widetilde{h}_0 = 0$. We will make this assumption throughout. 

These considerations set up a bootstrap problem of sorts which can be explicitly solved in all the cases we consider in this paper, owing to the relative sparsity of modular forms. Three methods will come to our aid, each producing identical results: Rademacher sums, modular linear differential equations, and Hecke operators. We turn to their treatment next.

\subsection{Character methods}

\subsubsection{Modular linear differential equations}\label{subsec:MLDE}

The characters of a VOA often satisfy a modular linear differential equation (MLDE), a fact which has been successfully exploited in many contexts, see e.g.\ \cite{Mathur:1988na,Mukhi:2019xjy,arakawa2018quasi,gaberdiel2008modular,beem2018vertex}. In the present setting, our knowledge of the conformal dimensions $\widetilde{h}_\alpha$ of the highest weight subspaces of the modules $\widetilde{\W}(\alpha)$ is in many cases sufficient to uniquely determine an MLDE satisfied by the corresponding characters $\widetilde{\chi}_\alpha(\tau)$. We briefly provide some details on the theory of such equations. 

Let $E_4(\tau)$ and $E_6(\tau)$ be the weight four and six Eisenstein series respectively, and
$E_2(\tau)$ the quasi-modular Eisenstein series of weight $2$. The Ramanujan-Serre derivative
${\cal D}_k =\frac{d}{d\tau}- \frac{1}{6} i \pi k E_2 $ maps modular forms of weight $k$ to modular
forms of weight $k+2$ and we can iterate this to define a differential operator
\be \label{mlde}
{\cal D}^n = {\cal D}_{2n-2} {\cal D}_{2n-4} \cdots {\cal D}_2 {\cal D}_0
\ee
that acts on weight zero modular functions to produce a weakly holomorphic modular form
of weight $2n$. A general $n$th order modular linear differential equation has the form \cite{Mathur:1988na}
\be
\left({\cal D}^n + \sum_{k=0}^{n-1} \phi_k {\cal D}^k \right)f=0,
\ee
where the $\phi_k(\tau)$ are modular forms of weight $2(n-k)$. 

If $f_1(\tau)$, $f_2(\tau)$, $\dots$, $f_n(\tau)$ are the $n$ linearly independent solutions of \eqref{mlde} then
we can express the coefficient functions $\phi_k(\tau)$ as
\be
\phi_k = (-1)^{n-k} W_k/W
\ee
where as usual the Wronskians are given by
\be
W_k =  \begin{vmatrix} f_1 & f_2 & \cdots & f_n \\
         {\cal D} f_1 & {\cal D} f_2 & \cdots & {\cal D} f_n \\
         \vdots & \vdots &      & \vdots \\
         {\cal D}^{k-1} f_1 & {\cal D}^{k-1} f_2 & \cdots & {\cal D}^{k-1} f_n \\
        {\cal D}^{k+1} f_1 & {\cal D}^{k+1} f_2 & \cdots  & {\cal D}^{k+1} f_n \\
        \vdots & \vdots &      & \vdots \\
        {\cal D}^{n} f_1 & {\cal D}^{n} f_2 & \cdots  & {\cal D}^{n} f_n
         \end{vmatrix} \, ,
\ee
and we set $W=W_n$. It is useful to classify solutions to MLDEs by the value of the integer
\be
\ell = 3~{\rm ord}_i(W) + 2 ~{\rm ord}_\omega(W) + 6~ \sum'_{p \in \cal F} {{\rm ord}_p( W)}
\ee
where $i=\sqrt{-1}$ and $\omega = e^{2 \pi i/3}$ are the orbifold points of the fundamental domain ${\cal F}$ of the modular group $\tsl{PSL}_2(\IZ)$. For an MLDE of order $n$,
the Wronskian $W$ has weight $n(n-1)$; if the leading behavior of the solutions $f_i$ at the cusp at infinity takes the form $f_i \simeq q^{\alpha_i}$, then it follow from the Riemann-Roch theorem that
\be
\ell = \frac{n(n-1)}{2} - 6 \sum_i \alpha_i \, .
\ee
The solutions $f_i(\tau)$ can be found using the Frobenius method and take the form
\be
f_i(\tau) = q^{\alpha_i} \sum_{n=0}^\infty c_{i,n} q^n
\ee
with $q=e^{2 \pi i \tau}$. The proof that solutions $f_i(\tau)$ of an $n$th order MLDE transform as a vector-valued modular form for $\tsl{SL}_2( \IZ)$ follows from results on the monodromy properties of solutions to complex linear differential equations and the fact that the only singularities of the MLDE occur at the cusp at infinity and at the orbifold points $\tau=i$, $e^{2 \pi i/3}$ of the fundamental domain of $\tsl{PSL}_2(\IZ)$. Note that the function $1728/j(\tau)$ maps the cusp and orbifold points to $(0,1,\infty)$ so one is essentially studying solutions on $P^1(\IC)-\{0,1,\infty \}$. For details see the discussion in \cite{francmason}.

In what follows, we provide MLDEs for the characters of CFTs with exceptional symmetry in \S\ref{subsubsec:BMe7}, \S\ref{subsubsec:Fischere6}, \S\ref{subsubsec:Lietype}, \S\ref{subsubsec:Fi23}, \S\ref{subsubsec:Th}, \S\ref{subsubsec:HN}, and \S\ref{subsubsec:Co2},
where we use MLDEs with ($n,\ell$) values given by $(3,0)$, $(4,0)$, $(4,0)$, $(6,0)$, $(5,0)$, $(6,3)$, and $(6,3)$
 respectively. 
 
\subsubsection{Hecke operators}\label{subsec:Hecke}

One of the tools that we use in later sections to construct the dual characters of commutant chiral algebras is that of Hecke images of RCFT characters, developed in \cite{HarveyWu}. Although not universally applicable, when it works, this technique is particularly powerful since it constructs the dual characters analytically, and determines their modular properties. The reasons behind the effectiveness of this technique are first that the Hecke images have the same number of components as the characters of the original RCFT, and this is a necessary ingredient by virtue of the decomposition in equation \eqref{paired decomposition}. Second, the modular representation of the Hecke image is precisely determined in a way that makes it easy to search for bilinears in the characters and their Hecke images that give the modular $J(\tau)$ function, as is also required. This is explained in more detail below. 

This section has a brief summary of the results in  \cite{HarveyWu} that are relevant to our discussion. The irreducible characters $\chi_\alpha(\tau)$,  $\alpha=0$, $1$, $\dots$, $n-1$, of a rational conformal field theory (RCFT), whose chiral algebra we denote $\W$, are
weakly holomorphic functions which transform under $\operatorname{\textsl{SL}}_2( \IZ)$ according to an $n$-dimensional unitary representation
\be
\rho: \operatorname{\textsl{SL}}_2( \IZ) \rightarrow \operatorname{\textsl{GL}}_n(\mathbb{C}).
\ee
More explicitly, the characters obey
\begin{align}
    \chi_\alpha\left(\frac{a\tau+b}{c\tau+d}\right) = \sum_{\beta} \rho_{\alpha\beta}\begin{pmatrix} a & b \\ c & d\end{pmatrix} \chi_\beta(\tau)
\end{align}
Each of the components $\chi_\alpha$ is separately a weakly holomorphic modular function for the principal congruence subgroup 
\begin{align}
 \Gamma(N) = \left\{\left(\begin{array}{rr} a & b \\ c & d \end{array}\right) \in\operatorname{\textsl{SL}}_2(\mathbb{Z}) \  \bigg\vert\  a, d\equiv 1~\mathrm{mod}~N \text{ and }  b,c\equiv 0~\mathrm{mod}~N\right\}   
\end{align}
where $N$ is the smallest integer such that $\rho(T)^N$ is the identity. Here $T$ is the modular transformation taking $\tau$ to $\tau+1$.  For each integer $q$ with $(q,N)=1$, one can define
a Hecke operator ${\mathsf T}_q$ such that the Hecke images of the RCFT characters $({\mathsf T}_q \chi)_\alpha$ are also individually weakly holomorphic modular functions for $\Gamma(N)$. Moreover, $\mathsf{T}_q\chi$
transforms as a vector-valued modular function according to a representation $\rho^{(q)}$ of the modular group which is defined on the generators $S,T$ as
\be
\rho^{(q)}(T) = \rho( T^{\bar q}), \qquad \rho^{(q)}(S) = \rho(\sigma_q S)
\ee
where $\bar q$ is the multiplicative inverse of $q$ in the group $(\IZ/N \IZ)^\times$ and $\sigma_q$ is the preimage of $\left(\begin{smallmatrix} \bar q & 0 \\ 0 & q \end{smallmatrix}\right)$
under the natural map $\operatorname{\textsl{SL}}_2( \IZ)\to\operatorname{\textsl{SL}}_2( \IZ/N \IZ)$. A short computation using results of \cite{bantay} gives the explicit formula
\be
\rho(\sigma_q) = \rho(T^{\bar q} S^{-1} T^q S T^{\bar q} S) \, .
\ee
The Hecke operators can be described abstractly in terms of a double coset construction, but for practical computations the most useful definition is in terms of the coefficients of the Fourier expansion
of the characters: if
\be
\chi_\alpha(\tau) = \sum_n b_\alpha(n) q^{\frac{n}{N}} \ \ \ \ \ \ \  (q=e^{2\pi i \tau})
\ee
is the Fourier expansion of the character of the irreducible module $\W(\alpha)$, and $p$ is prime, then the Fourier expansions of the components of the Hecke image are given by
\be
\chi^{(p)}_\alpha(\tau) = \sum_\alpha b^{(p)}_\alpha (n) q^{\frac{n}{N}}
\ee
where
\be
b^{(p)}_\alpha(n) = \begin{cases}  p b_\alpha(pn) &  p \nmid n \\ p b_\alpha(pn) + \sum_\beta \rho_{\alpha\beta}(\sigma_p) b_\beta(\frac np) & p|n. \end{cases}
\ee
If $(q,N)=1$ but $q$ is not prime, then the Hecke operator ${\mathsf T}_q$ can be computed in terms of Hecke operators ${\mathsf T}_p$ with $p$ prime using the formulae
\be
{\mathsf T}_{mn} ={\mathsf T}_m {\mathsf T}_n
\ee
for $m,n$ relatively prime and
\be
{\mathsf T}_{p^{m+1}} = {\mathsf T}_p {\mathsf T}_{p^m} - p \sigma_p \circ {\mathsf T}_{p^{m-1}}
\ee
for $p$ prime.

These Hecke operators are well suited for solving the bootstrap problem presented at the end of \S\ref{subsec:deconstruction}. Indeed, it is shown in \cite{HarveyWu} that if $\chi$
is a vector of characters of an RCFT with conductor $N$ transforming in the representation $\rho$ of $\tsl{SL}_2( \IZ)$, and $\chi^{(q)}= 
{\mathsf T}_q \chi$ is the vector of their Hecke images under ${\mathsf T}_q$, then the bilinear
\be
\chi^{(q)}(\tau)^T\cdot  G_\ell \cdot \chi(\tau)
\ee
is modular invariant provided that
\be
\bar q + \ell^2 = 0~{\rm mod~}N \, .
\ee
Here $\bar q$ is the inverse of $q$ mod $N$ and
\be
G_\ell= \rho(T^\ell S^{-1} T^{\bar \ell} S T^\ell S) \,.
\ee
The goal is to locate a linear combination $M = \sum a_\ell G_\ell$ with non-negative, integer entries such that 
\begin{align}
    \chi^{(q)}(\tau)^T\cdot M \cdot \chi(\tau) = \mathcal{Z}(\tau)
\end{align}
where $\mathcal{Z}$ is the modular-invariant partition function of the meromorphic CFT $\V$ in which $\W\otimes \widetilde{\W}$ is realized as a subalgebra. If one is successful, then one could conjecturally identify the components of $M^T\cdot \mathsf{T}_q\chi$ with the dual characters of the modules $\widetilde{\W}(\alpha)$ which arise in the decomposition \eqref{paired decomposition}, i.e.
\begin{align}
    \widetilde{\chi}(\tau)\sim M^T\cdot (\mathsf{T}_q\chi)(\tau)
\end{align}
since these functions would then satisfy equation \eqref{characterpairing}. Because the most singular exponent of $\mathsf{T}_q\chi$ is $q$ times the most singular exponent of $\chi$, such a construction can only go through\footnote{At least for the Hecke operators considered in this paper, though they may generalize.} if $c_{\widetilde{t}}=q c_t$ for $q$ an integer coprime to the conductor $N$ of the characters of $\W$, where $c_t$ ($c_{\widetilde{t}}$) is the central charge of $\W$ $(\widetilde{\W})$.

If we specialize to the case that $\V=V^\natural$ is the moonshine module, then the partition function
\begin{align}
    \mathcal{Z}(\tau) = J(\tau) = q^{-1}+0+196884q + \cdots
\end{align}
is the modular invariant $J$-function. In all the examples we consider, for which $c_t+c_{\widetilde{t}}=24$ (with 24 the central charge of $V^\natural$), this bilinear must yield a function of the form $aJ+b$, with $a$ and $b$ two numbers. Imposing the further physical restriction that the chosen linear combination $M$ have non-negative integral entries, and that the entry of $M$ which pairs the components of $\chi$ and $\mathsf{T}_q\chi$ with the most singular exponents be 1, forces the bilinear to yield $J+b$ with $b$ an integer. It is then usually straightforward to determine whether or not there is physical $M$ for which $b=0$.

\subsubsection{Rademacher sums}\label{subsubsec:Rademacher}

Our last technique for computing the dual characters $\widetilde{\chi}_\alpha(\tau)$ is the theory of Rademacher sums. The method originally goes back to Poincar{\'e} \cite{poincare1911fonctions}, followed by key insights from Rademacher \cite{rademacher1938fourier,rademacher1938partition}, and has since been the subject of extensive study and application by both mathematicians and physicists \cite{niebur1974construction,whalen2014vector,Duncan:2009sq,Cheng:2011ay,dijkgraaf2000black,deBoer:2006vg,Manschot:2007ha,Maloney:2016kee,Alday:2019vdr,nally2019exact,Ferrari:2017kbp} (c.f.\ \cite{cheng2014rademacher} for a review). 

The philosophy behind this approach is that meromorphic vector-valued modular forms typically belong to vector spaces of small dimension, and are therefore determined by only their first few Fourier coefficients along with some information about how they transform under the modular group. In some cases, it is more or less sufficient to specify only the poles, and in such cases Rademacher summation furnishes formulae for the rest of the $q$-expansion in terms of the singular terms. In this sense, Rademacher sums are the spiritual successor of the Cardy formula \cite{Cardy:1986ie}: they determine the entire spectrum of the CFT in terms of finitely many light states, as opposed to only the leading asymptotics of the density of high energy states.

At a technical level, the method of Rademacher summation is the specialization of a very general idea to the theory of automorphic forms: the starting point is the idea that one can obtain an object which is invariant under the action of a group $G$ by summing together the $G$-images of a seed object (or the $G/H$-images of the seed object if it is already invariant with respect to a subgroup $H$ of $G$), 
\begin{align}\label{symmetrize}
``\mathcal{O}_{\mathrm{symmetric}} \equiv \sum_{g\in G/H} g\cdot \mathcal{O}_{\mathrm{seed}}."
\end{align}
The procedure applied to vector-valued modular functions of weight 0 takes the following as its input:
\begin{enumerate}
\item A representation $\rho$ of $\tsl{SL}_2(\mathbb{Z})$ which specifies how the symmetric target function transforms. In the case of the characters of an RCFT, $\rho$ is determined by the modular matrices $\mathcal{S}$ and $\mathcal{T}$.
\item A list $P = \{c_\alpha(\mu)\}$ of the singular terms in the $q$-expansion of the target modular function, which can be determined from the central charge $c$ and the conformal dimensions $h_\alpha$ of the highest weight subspaces of the modules $\W(\alpha)$.
\end{enumerate}
In terms of this data, the seed object is simply the vector-valued function whose $\alpha$-th component consists of the singular terms in the $\alpha$-th component of the target function,
\begin{align}
 \sum_{\mu<0} c_\alpha(\mu) q^\mu. 
\end{align}
Notice that this seed function already transforms correctly with respect to the translation subgroup $\Gamma_\infty = \{ \pm T^r \mid r\in\mathbb{Z}\}$ of $\tsl{SL}_2(\mathbb{Z})$, and we therefore should only sum together its images under $\Gamma_\infty \backslash\tsl{SL}_2(\mathbb{Z})$, as opposed to all of $\tsl{SL}_2(\mathbb{Z})$. Generalizing slightly the constructions in \cite{cheng2014rademacher} to the vector-valued case, the Rademacher sum\footnote{Convergence of this sum has not been proven in general, however we conjecture that in all the cases we consider in this paper, the sum dooes in fact converge.} which then makes equation \eqref{symmetrize} precise is
\begin{align}
R^{P}_{\tsl{SL}_2(\mathbb{Z}),\rho,\alpha}(\tau) = \sum_{\beta}\sum_{\mu<0}c_\beta(\mu) \sum_{\gamma\in\Gamma_\infty\backslash\tsl{SL}_2(\mathbb{Z})} e^{2\pi i \mu\gamma\tau}r^{[\mu]}(\gamma,\tau)\rho(\gamma)^{-1}_{\beta\alpha}
\end{align}
where 
\begin{align}
r^{[\mu]}(\gamma,\tau) =  e^{-2\pi i \mu(\gamma\tau-\gamma\infty)}\sum_{n\geq 0}\frac{[2\pi i \mu(\gamma\tau-\gamma\infty)]^{n+1}}{\Gamma(n+2)}, \ \ \ \ \ \ \ \ \left(\gamma\infty\equiv \lim_{\tau \to i\infty}\gamma\tau\right)
\end{align}
is a regularization term which is needed in order for the sum to converge. A somewhat involved computation of the Fourier expansion of this Rademacher sum, 
\begin{align}
R^{P}_{\tsl{SL}_2(\mathbb{Z}),\rho,\alpha}(\tau) = \sum_{\nu \in h_\alpha - \frac{c}{24} + \mathbb{Z}_{\geq 0}}c_{\alpha}(\nu)q^\nu
\end{align}
yields the following exact expressions\footnote{Similar formulae appear in \cite{pmo} for weight $\frac12$ vector-valued modular forms transforming in the Weil representation.} for the Fourier coefficients,
\begin{align}\label{rad coefficients neq 0}
c_{\alpha}(\nu) = 
\sum_{\beta}\sum_{\mu<0}c_\beta(\mu)\sum_{c=1}^\infty \sum_{\substack{0\leq a < c \\ (a,c) = 1}} e^{2\pi i \mu \frac{a}{c}}e^{2\pi i\nu\frac{d}{c}}\rho\left(\begin{smallmatrix} a & \ast \\ c & \ast\end{smallmatrix}\right)_{\beta\alpha}^\ast\sqrt{-\frac{\mu}{\nu}}\frac{2\pi}{c} I_1\left(\frac{4\pi}{c}\sqrt{-\mu \nu}\right)
\end{align}
 whenever $\nu> 0$, and 
\begin{align}\label{rad coefficients eq 0}
c_{\alpha}(0) = -4\pi^2\sum_\beta\sum_{\mu<0}c_\beta(\mu)\sum_{c=1}^\infty \sum_{\substack{0\leq a < c \\ (a,c) = 1}} \mu \frac{e^{2\pi i\mu\frac{a}{c}}}{c^2}\rho\left(\begin{smallmatrix} a & \ast \\ c & \ast \end{smallmatrix}\right)^\ast_{\beta\alpha}.
\end{align}
In the above, $I_\alpha(z)$ is the modified Bessel function of the first kind, 
\begin{align}
I_\alpha(z) = \sum_{n\geq 0} \frac{1}{\Gamma(n+\alpha+1)n!}\left(\frac{z}{2}\right)^{2n+\alpha}
\end{align}
and $\left(\begin{smallmatrix} a & \ast \\ c & \ast\end{smallmatrix}\right)$ is any element of $\tsl{SL}_2(\mathbb{Z})$ whose first column is given by $\left(\begin{smallmatrix} a \\ c \end{smallmatrix}\right)$. So defined, these Rademacher sums are expected in suitable circumstances (modulo exceptions, which we will mention in a moment) to transform as
\begin{align}\label{expected rademacher}
R^{P}_{\tsl{SL}_2(\mathbb{Z}),\rho}\left(\gamma\tau\right) = \rho(\gamma) R^{P}_{\tsl{SL}_2(\mathbb{Z}),\rho}(\tau) \ \ \ \ \ (\gamma\in\tsl{SL}_2(\mathbb{Z})).
\end{align}
In general, however, there will be obstructions to this procedure recovering the characters of the CFT under consideration. For example, there may be more than one modular function with the desired transformation properties and polar terms, in which case the Rademacher sum is not guaranteed to recover the characters of interest. A more serious issue is that this Rademacher sum may converge to a \emph{mock modular form} rather than a genuine one, in which case equation \eqref{expected rademacher} will not hold. The fact that this problem does not arise for the twined characters of the moonshine module (a.k.a.\ McKay-Thompson series) is equivalent \cite{Duncan:2009sq} to the fact that each transforms under a genus zero subgroup $\Gamma$ of $\tsl{SL}_2(\mathbb{R})$ (c.f.\ \S\ref{monstercft} for the definition of genus zero). This genus zero property (alternatively, the Rademacher summability property) of monstrous moonshine is one of its defining features. Therefore, in addition to its practical utility, our description of the characters of the RCFTs we consider as Rademacher sums can be thought of as the first steps towards investigating whether an analogous genus zero property holds for them as well.

\section{Monster deconstructions}\label{sec:mainresults}

\begin{table}
\begin{small}
\begin{center}
\begin{tabular}{rl}
    $|g|$ &  \hspace{.075in}  $[G\circ \widetilde{G}].H$ \\\midrule
    $71$ & $\star~\mathbb{Z}_{\mathrm{1A}}\times \mathbb{M}$ \\
    $47$ & $\star~\mathbb{Z}_{\mathrm{2A}}\circ 2.\mathbb{B}$ \\
    $31$ & $\star~ D_{\mathrm{3C}}\times\Th$ \\
    $29$ & $\star~ [\mathbb{Z}_{\mathrm{3A}}\circ 3.\Fi_{24}'].2$ \\
    $23$ & \hspace{.075in} $S_4\times 2^{11}.\tsl{M}_{23}$ \\
    & \hspace{.075in} $[A_4\times 2^{11}.\tsl{M}_{24}].2$ \\
    & \hspace{.075in} $D_8\circ 2^{1+22}.\tsl{M}_{23}$ \\
    & $\star~ D_{\mathrm{3A}}\times \tsl{Fi}_{23}$ \\
    & \hspace{.075in}  $[2^2\circ 2^{2+11+22}.\tsl{M}_{24}].S_3$ \\
    & $\star~[D_{\mathrm{2B}}\circ 2^{1+23}.\tsl{Co}_2].2$ \\
    & $\star~[\mathbb{Z}_{\mathrm{4A}}\circ 4.2^{22}.\tsl{Co}_3].2$ \\
    & \hspace{.075in} $\mathbb{Z}_{\mathrm{2B}}\circ 2^{1+24}.\tsl{Co}_1$ \\
    $19$ & \hspace{.075in} $[A_5\times U_3(8).3_1].2$ \\
    & $\star~[D_{\mathrm{5A}}\times \tsl{HN}].2$ \\
    & $\star~[D_{\mathrm{2A}}\circ 2^2.{^2}\tsl{E}_6(2)].S_3$ \\
    $17$ & \hspace{.075in}  $[L_2(7)\times S_4(4).2].2$ \\
    & \hspace{.075in}  $S_4\times S_8(2)$ \\
    & \hspace{.075in}  $[7:3\times \tsl{He}].2$ \\
    & \hspace{.075in}  $[A_4\times O^-_{10}(2)].2$ \\
    & $\star~[D_{\mathrm{4B}}\circ 2.\tsl{F}_4(2)].2$ \\
    13 & \hspace{.075in}  $[13.6\times L_3(3)].2$ \\
    & \hspace{.075in}  $[2.A_4\circ (2\times U_3(4)).2].2$ \\
    & \hspace{.075in}  $[Q_8\circ 2.2^{12}.U_3(4).2].S_3$ \\
    & \hspace{.075in}  $[\mathbb{Z}_{\mathrm{6B}}\circ 6.\tsl{Suz}].2$ \\
    & \hspace{.075in}  $[\mathbb{Z}_{\mathrm{4D}}\circ 4.2^{12}.\tsl{G}_2(4).2].2$ \\
    & \hspace{.075in}  $[\mathbb{Z}_{\mathrm{3B}}\circ 3^{1+12}.2.\tsl{Suz}].2$ \\
     & \hspace{.075in}  $3^2.2.S_4\times L_3(3).2$ \\
    & \hspace{.075in} $[3^{1+2}.2^2\times \tsl{G}_2(3)].2$ \\
    & \hspace{.075in} $[3^2.D_8\times L_4(3).2_2].2$ \\
    & \hspace{.075in} $[2.S_4\circ (2\times {^2}\tsl{F}_4(2)')].2$ \\
    & \hspace{.075in} $[(S_3\times S_3)\times O_7(3)].2$ \\
    & \hspace{.075in} $S_4\times {^3}D_4(2).3$ \\
    & \hspace{.075in} $[3^2.2\times D_4(3)].S_4$ \\
    & $\star~ [D_{\mathrm{6A}}\circ (2.\tsl{Fi}_{22})].2$
\end{tabular}
\quad
\begin{tabular}{cl}
    $|g|$ &  \hspace{.075in}  $[G\circ \widetilde{G}].H$ \\\midrule
    $11$ & \hspace{.075in}  $[\tsl{M}_{12}\times L_2(11)].2$ \\
    & \hspace{.075in}  $S_6.2\times \tsl{M}_{11}$ \\
    & \hspace{.075in}  $[\tsl{L}_2(11)\times \tsl{M}_{12}].2$ \\
    & \hspace{.075in}  $[(2\times S_5)\circ 2.\tsl{M}_{22}].2$ \\
    & \hspace{.075in}  $[3^2.2.\tsl{A}_4\times 3^5.\tsl{L}_2(11)].2$ \\
    & \hspace{.075in}  $[4^2.2.S_3\times 2^{10}L_2(11)].2$ \\
    & \hspace{.075in}  $[4^2.S_3\times 2^{10}.\tsl{M}_{11}].2$ \\
    & \hspace{.075in}  $[3^2.Q_8\times 3^5.\tsl{M}_{11}].S_3$ \\
    & \hspace{.075in}  $[(\tsl{A}_4\times S_3)\times \tsl{U}_5(2)].2$ \\
    & \hspace{.075in}  $[A_5\times A_{12}].2$ \\
    & \hspace{.075in}  $[4^2.3\times 2^{10}.\tsl{M}_{12}.2].D_8$ \\
    & \hspace{.075in}  $[(2\times S_4) \circ 2^{11}.\tsl{M}_{22}].2$ \\
    & \hspace{.075in}  $[(2\times S_4)\circ 2.\tsl{M}_{12}].2$ \\
    & \hspace{.075in}  $[(2\times 5.4)\circ 2.\tsl{HS}].2$ \\
    & \hspace{.075in}  $[(2^{1+4}\circ 2^{1+20}.L_2(11)].(2\times S_3)$ \\
    & \hspace{.075in}  $[8.2^2\circ 2.2^{10}.\tsl{M}_{11}].2$ \\
    & \hspace{.075in}  $[3^{1+2}\circ 3^{1+10}.\tsl{L}_2(11)].D_8$ \\
    & \hspace{.075in}  $[(2^2\times S_3) \circ 2^2.U_6(2)].S_3$ \\
    & \hspace{.075in}  $[(3\times S_3) \times 3^6.\tsl{M}_{11}].2$ \\
    & \hspace{.075in}  $[(2\times D_8) \circ 2^2.2^{10}.\tsl{M}_{12}].(2^2\wr 2)$ \\
    & \hspace{.075in}  $[(2\times D_8)\circ 2^{1+21}.\tsl{M}_{22}].2^2$ \\
    & \hspace{.075in}  $[4.2^2\circ 4.2^{20}.\tsl{M}_{11}].2^2$ \\
    & \hspace{.075in}  $[8.2\circ 2.2^{10}.\tsl{M}_{12}].2^2$ \\
    & \hspace{.075in}  $[3^2\circ 3^{2+5+10}.\tsl{M}_{11}].2.S_4$ \\
    & \hspace{.075in}  $[3^2\circ 3^{1+11}.U_5(2)].(2\times S_3)$ \\
    & \hspace{.075in}  $[2^3\circ 2^3.2^{20}.U_6(2)].S_4$ \\
    & \hspace{.075in}  $[2^3\circ 2^{3+20+10}.\tsl{M}_{22}].S_4$ \\
    & \hspace{.075in}  $[2^3\circ 2^3.2^{20}.\tsl{M}_{12}.2].S_4$ \\
    & \hspace{.075in}  $[(2\times 4)\circ (2\times 4).2^{20}.\tsl{HS}].D_8$ \\
    & \hspace{.075in}  $[(2\times 4)\circ (2\times 4).2^{20}.2.\tsl{M}_{12}].D_8$ \\
    & \hspace{.075in}  $[Q_8\circ 2^{1+22}.3^5.\tsl{M}_{11}].S_3$ \\
    & $\star~[D_{\mathrm{4A}}\circ 2^{1+22}.\tsl{McL}].2$ \\
    & \hspace{.075in}  $S_3\times 3^6.2.\tsl{M}_{12}$ \\
    &
\end{tabular}
\caption{A table taken from \cite{anatomy} of all monstralizer pairs for $|g|\geq 11$, where $|g|$ is the order of the largest prime which divides $\widetilde{G}$. The notation $[G\circ \widetilde{G}].H$ indicates that $G = \mathrm{Cent}_{\mathbb{M}}(\widetilde{G})$ and $\widetilde{G}=\mathrm{Cent}_{\mathbb{M}}(G)$. The centers and normalizers of such pairs agree, $Z(G)=Z(\widetilde{G})$ and $N=N_{\mathbb{M}}(G) = N_{\mathbb{M}}(\widetilde{G})$, and we define $H = N/\langle G,\widetilde{G}\rangle$ which is a subgroup of the outer automorphism groups of both $G$ and $\widetilde{G}$. The group $[G\circ \widetilde{G}].H$ appears as a subgroup of $\mathbb{M}$, where $G\circ \widetilde{G}$ indicates the central product of $G$ with $\widetilde{G}$ (and which is abbreviated to $G\times \widetilde{G}$ when $Z(G)=Z(\widetilde{G})=\{1\}$). The notation $\mathbb{Z}_{n\mathrm{X}}$ denotes a cyclic group which is generated by any element in the $n$X conjugacy class. The notation $D_{n\mathrm{X}}$ denotes a dihedral group which is generated by two 2A involutions whose product lies in the $n$X conjugacy class. A $\star$ next to an entry indicates that it is an example for which $\W_{G}$ and $\W_{\widetilde{G}}$ are known. See \cite{atlas} for further details on notation.}\label{monstralizers}
\end{center}
\end{small}
\end{table}

We now come to our main results. Namely, we construct several examples of commuting subalgebras $(\W_{G},\W_{\widetilde{G}})$ of $V^\natural$ attached to monstralizer pairs $[G\circ \widetilde{G}].H$ (c.f.\ Table \ref{monstralizers} for an explanation of the notation) which satisfy the following properties.
\begin{enumerate}[label=(\alph*)]
    \item The subgroups preserved by $\W_G$ and  $\W_{\widetilde{G}}$ are $\mathbb{M}(\W_{G}) = G$ and $\mathbb{M}(\W_{\widetilde{G}})=\widetilde{G}$.
    \item The inner automorphism group of $\W_G$ is given by $G/Z(G)$, where $Z(G)$ is the center of $G$. Therefore, the $\W_G$ realize \emph{subquotients} (rather than just subgroups) of the monster as their inner automorphism groups. The same goes for the $\W_{\widetilde{G}}$. 
    \item The subalgebra of the full fusion algebra of $\W_G$ (resp.\ $\W_{\widetilde{G}}$) which is spanned by the modules $\W_G(\alpha)$ (resp.\ $\W_{\widetilde{G}}(\alpha)$) appearing in the decomposition \eqref{paired decomposition} admits an action of $Z(G) = Z(\widetilde{G})$ by diagonal fusion algebra automorphisms. 
    \item The full automorphism group of $\W_G$ is given by (at least) $[G/Z(G)].H'$ for some subgroup $H'$ of $H$. The same goes for the $\W_{\widetilde{G}}$.
    \item The commutant of $\W_G$ in $V^\natural$ is $\W_{\widetilde{G}}$ and vice versa, i.e.\ $\widetilde{\W}_G = \W_{\widetilde{G}}$ and $\widetilde{\W}_{\widetilde{G}} = \W_G$.
    \item Whenever one monstralizer pair $(H,\widetilde{H})$ includes into another $(G,\widetilde{G})$ in the sense that $\widetilde{H}\hookrightarrow \widetilde{G}$ and $G\hookrightarrow H$, the associated chiral algebras mirror these inclusions, i.e.\ $\W_{\widetilde{H}}\hookrightarrow \W_{\widetilde{G}}$ and $\W_{G}\hookrightarrow \W_{H}$.
\end{enumerate}  

We will begin by reviewing how known models fit into this framework. In \S\ref{subsec:startingmodels}, we recall various relevant properties of the moonshine module $V^\natural$, the baby monster VOA $\tsl{V}\mathbb{B}^\natural$, and the Fischer VOA $\tsl{VF}_{24}^\natural$, and argue that they participate in $\mathbb{M}$-com pairs with $G:=\mathbb{M}(\W_G)$ a cyclic group. Since the latter two models can be obtained from the first via deconstruction, i.e.\ as commutant subalgebras, they provide an opporunity to exposit the techniques we will be using throughout in the simplest settings. We also describe the relationships of these three theories to the extended Dynkin diagrams of $\widehat{E_8}$, $\widehat{E_7}$, and $\widehat{E_6}$ respectively; in particular, we will describe how the nodes of these diagrams are naturally decorated by conjugacy classes of $\mathbb{M}$, $\mathbb{B}$, and $\Fi_{24}$, and how this leads to distinguished subalgebras $\W_{D_{n\mathrm{X}}}\subset V^\natural$, $\W_{B(m\mathrm{Y})}\subset \tsl{V}\mathbb{B}^\natural$, and $\W_{F(r\mathrm{Z})}\subset \tsl{VF}_{24}^\natural$. We then give our first new example of an $\mathbb{M}$-com pair coming from a parafermion deconstruction, which is associated to the monstralizer $(G,\widetilde{G}) = (\mathbb{Z}_{\mathrm{4A}},4.2^{22}.\tsl{Co}_3)$. 

In \S\ref{subsec:mckaye8}, we study $\mathbb{M}$-coms associated to monstralizer pairs $(G,\widetilde{G})$ with $G=D_{n\mathrm{X}}$ a dihedral group; it will turn out that the VOAs which uplift these dihedral groups are precisely the chiral algebras coming from McKay' $\widehat{E_8}$ correspondence mentioned earlier, $\W_G = \W_{D_{n\mathrm{X}}}$. We compute the symmetry groups and dual characters of the commutants of the algebras $\W_{D_{n\mathrm{X}}}$ in the moonshine module, 
\begin{align}
    \widetilde{\W}_{D_{n\mathrm{X}}} := \tsl{Com}_{V^\natural}(\W_{D_{n\mathrm{X}}}).
\end{align}
and in many cases find that their symmetry groups are closely related to sporadic simple groups in the happy family, i.e.\ sporadic groups which arise as subquotients of the monster. We will also check in many cases that the VOAs embed into one another as subalgebras in the same way their corresponding preserved symmetry groups do as subgroups. We summarize these results in Table \ref{chiral algebra table} and Figure \ref{monstralizers_diagram}. See the appendices for additional data.

\begin{sidewaystable}[ph!]
\begin{center}
\begin{tabular}{ c|c|c|c|c|c|c|c|c|c } 
 \toprule
 $[G\circ \widetilde{G}].H$ & $(\W_G,\W_{\widetilde{G}})$ & $\dim((\W_G)_2,(\W_{\widetilde{G}})_2)$ &  $c_{\widetilde{t}}$ &  $\mathrm{Aut}(\W_{\widetilde{G}})$ & $\mathrm{Inn}(\W_{\widetilde{G}})$ & $\mathrm{Fus}(\W_{\widetilde{G}})$ & H & M & R \\\midrule
 $\mathbb{Z}_{\mathrm{1A}}\times\mathbb{M}$ & $(\mathrm{trivial},V^\natural)$ & $(0,196884)$ & $24$ & $\mathbb{M}$ & $\mathbb{M}$ & $\{1\}$ & & Y & Y \\
 $\mathbb{Z}_{\mathrm{2A}}\circ 2.\mathbb{B}$ & $(\mc{P}(2),\tsl{V}\mathbb{B}^\natural)$ & $(1,96256)$ & $23\sfrac12$ & $\mathbb{B}$ & $\mathbb{B}$ & $\mathbb{Z}_2$ & Y & Y & Y\\
 $[\mathbb{Z}_{\mathrm{3A}}\circ \tsl{Fi}_{24}'].2$ & $(\mc{P}(3),\tsl{VF}_{24}^\natural)$ & $(1,57478)$ & $23\sfrac15$ & $\tsl{Fi}_{24}$ & $\tsl{Fi}'_{24}$ & $\mathbb{Z}_3$ & Y & Y & Y\\
 $[\mathbb{Z}_{\mathrm{4A}}\circ 4.2^{22}.\tsl{Co}_3].2$ & $(\mc{P}(4),\widetilde{\W}_{\mathbb{Z}_{\mathrm{4A}}})$ & $(1,38226)$ & $23$ & $2^{22}.\tsl{Co}_3.2^\ast$ & $2^{22}.\tsl{Co}_3$ & $\mathbb{Z}_4$ & Y & & Y\\
 $[2^2\circ 2^2.{^2}\tsl{E}_6(2)].S_3$ & $(\W_{D_{\mathrm{2A}}},\widetilde{\W}_{D_{\mathrm{2A}}})$ & $(3,48621)$ & $22\sfrac45$ & ${^2}\tsl{E}_6(2).2$ & ${^2}\tsl{E}_6(2)$ & $\mathbb{Z}_2\times\mathbb{Z}_2$ & Y & Y & Y\\
 $D_{\mathrm{3A}}\times\Fi_{23}$ & $(\W_{D_{\mathrm{3A}}},\tsl{VF}_{23}^\natural)$ & $(4,30889)$ & $22\sfrac{12}{35}$ & $\Fi_{23}$ & $\Fi_{23}$ & $\{1\}$ & &Y &Y\\
 $[D_{\mathrm{4A}}\circ 2^{1+22}.\tsl{McL}].2$ & $(\W_{D_{\mathrm{4A}}},\widetilde{\W}_{D_{\mathrm{4A}}})$ & $(5,22528)$ & $22$ & $2^{22}.\tsl{McL}.2^\ast$ & $2^{22}.\tsl{McL}$ & $\mathbb{Z}_2$ &Y & & Y\\
 $[D_{\mathrm{5A}}\times \tsl{HN}].2$ & $(\W_{D_{\mathrm{5A}}},\tsl{VHN}^\natural)$ & $(6,18316)$ & $21\sfrac57$ & $\tsl{HN}.2$ & $\tsl{HN}$ & $\{1\}$ & &Y &Y\\
 $[D_{\mathrm{6A}}\circ 2.\tsl{Fi}_{22}].2$ & $(\W_{D_{\mathrm{6A}}},\tsl{VF}^\natural_{22})$ & $(8,16731)$ & $21\sfrac{9}{20}$ & $\tsl{Fi}_{22}.2$ & $\tsl{Fi}_{22}$ & $\mathbb{Z}_2$ & & &Y\\
 $[D_{\mathrm{4B}}\circ 2.\tsl{F}_4(2)].2$ & $(\W_{D_{\mathrm{4B}}},\widetilde{\W}_{D_{\mathrm{4B}}})$ & $(5,24310)$ & $22\sfrac{1}{10}$ & $\tsl{F}_4(2).2$ & $\tsl{F}_4(2)$ & $\mathbb{Z}_2$ & & & Y \\
 $[D_{\mathrm{2B}}\circ 2^{1+23}.\tsl{Co}_2].2$ & $(\W_{D_{\mathrm{2B}}},\widetilde{\W}_{D_{\mathrm{2B}}})$ & $(2,46851)$ & $23$ & $2^{22}.\tsl{Co}_2.2^\ast$ & $2^{22}.\tsl{Co}_2$ & $\mathbb{Z}_2\times\mathbb{Z}_2$ & Y & Y & Y\\
 $D_{\mathrm{3C}}\times\Th$ & $(\W_{D_{\mathrm{3C}}},\tsl{VT}^\natural)$ & $(3,30876)$ & $22\sfrac{6}{11}$ & $\Th$ & $\Th$ & $\{1\}$ & &Y&Y \\
 \bottomrule
\end{tabular}
\caption{Table of $\mathbb{M}$-com pairs  $(\W_G,\W_{\widetilde{G}})$. The first column indicates the associated monstralizer pair $[G\circ \widetilde{G}].H$ (c.f.\ Table \ref{monstralizers} for an explanation of the notation), from which one can read off the various symmetry groups of $\W_G$ and $\W_{\widetilde{G}}$. Most immediately, the monstralizer pair indicates that the preserved subgroups are $\mathbb{M}(\W_G) = G$ and $\mathbb{M}(\W_{\widetilde{G}}) = \widetilde{G}$; any projective representation which appears in the modules $\W_{\widetilde{G}}(\alpha)$ can be lifted to an honest representation of $\widetilde{G}$. The third column gives the dimensions of the Griess algebras (i.e.\ the space of dimension 2 operators) of $\W_G$ and $\W_{\widetilde{G}}$. The fourth column gives the central charge $c_{\widetilde{t}}$ of $\widetilde{\W}_G = \W_{\widetilde{G}}$. The fifth column indicates the full automorphism group which $\W_{\widetilde{G}}$ inherits from $\mathbb{M}$ in the cases that we have been able to compute it (though we conjecture that it is $\mathrm{Inn}(\W_{\widetilde{G}}).H$ in all of the missing cases). The sixth column specifies the inner automorphism group of $\W_{\widetilde{G}}$ which is obtained as $\widetilde{G}/Z(\widetilde{G})$. The seventh column indicates the fusion algebra automorphisms of $\mathrm{span}_{\mathbb{C}}\{\W_{\widetilde{G}}(\alpha)\}$ which come from the monster, $\mathrm{Fus}(\W_{\widetilde{G}})=Z(\widetilde{G})$. The last three columns indicate whether we were able to obtain the characters from Hecke operators (H), an MLDE (M), or Rademacher summation (R). We have less evidence for the entries with an asterisk $\ast$ next to them because we do not have access to the relevant character tables.}\label{chiral algebra table}
\end{center}
\end{sidewaystable}

\begin{figure}
\begin{center}
\includegraphics[width=.9\textwidth]{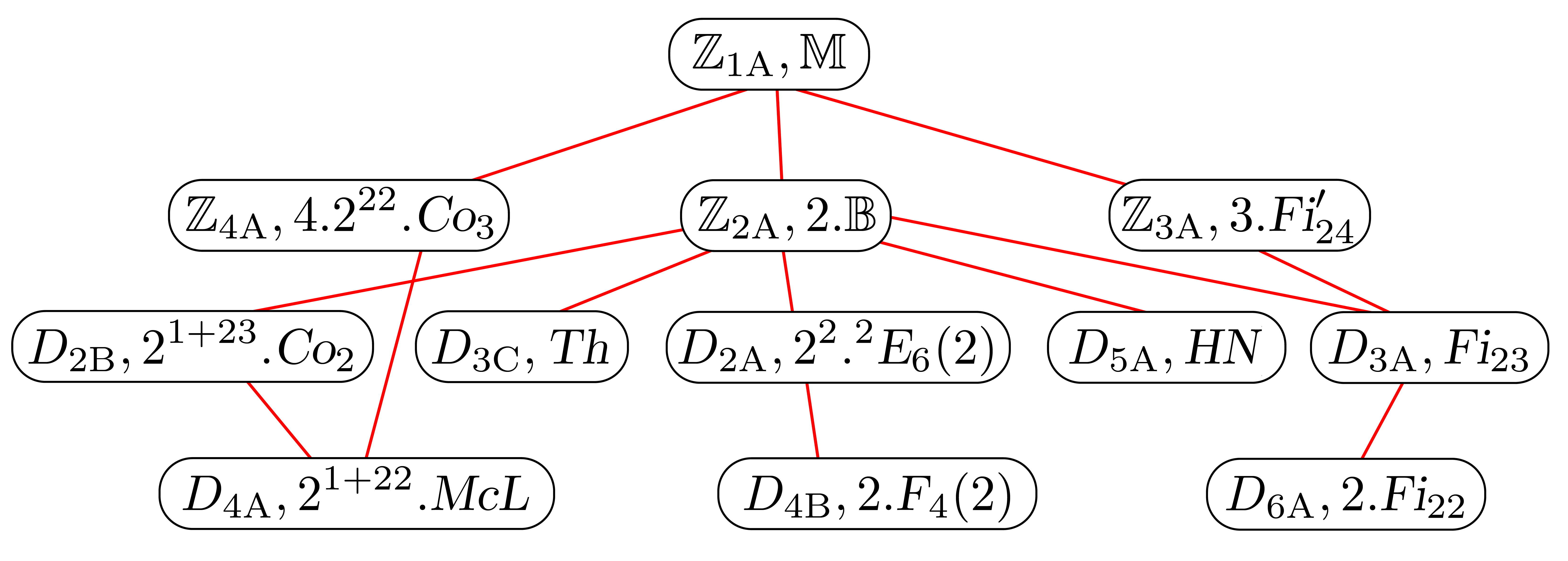}
\caption{The subset of monstralizer pairs $(G,\widetilde{G})$ whose associated $\mathbb{M}$-com pairs $(\W_G,\W_{\widetilde{G}})$ are treated in this paper. A red line from a pair $(H,\widetilde{H})$ up to $(G,\widetilde{G})$ indicates that $\widetilde{H}$ is a subgroup of $\widetilde{G}$ and $G$ is a subgroup of $H$. The associated chiral algebras mirror these inclusions, $\W_{\widetilde{H}}\hookrightarrow \W_{\widetilde{G}}$ and $\W_G\hookrightarrow \W_H$.}
\label{monstralizers_diagram}
\end{center}
\end{figure}

We are able to compute the dual characters of every theory $\W_{\widetilde{G}}$ considered in this paper using at least one of three complementary approaches. First, every model considered in \S\ref{subsec:startingmodels} and \S\ref{subsec:mckaye8} has characters which arise from Rademacher summation, though we will typically only mention this explicitly when the other two methods are ineffective. Second, we will also find that the characters of the $\W_{\widetilde{G}}$ can be obtained as Hecke images of the irreducible characters of $\W_{G}$ whenever $c_{\widetilde{t}} = qc_t$ for some integer $q$, a phenomenon we suspect might provide a useful tool for analyzing commutant subalgebras in more general settings. Finally, in many cases, we are able to exhibit the dual characters as solutions to a suitable MLDE. Whenever more than one method works, we find that they produce identical predictions for the characters, which provides further evidence in support of our conjectures. A nice property that these character-theoretic considerations reveal is that the highest weight spaces of the $\W_{\widetilde{G}}(\alpha)$ always transform under \emph{irreducible} (projective) representations\footnote{We are not aware of any theorem that requires the highest weight subspaces of irreducible modules of a (suitably nice) VOA to transform as irreducible representations of the inner automorphism group, and we suspect that this niceness is related to the fact that $(G,\widetilde{G})$ form a monstralizer pair.} of the inner automorphism group. 

In \S\ref{subsec:otherexamples}, we offer partial results which support the hope that one can derive other $\mathbb{M}$-com pairs, beyond the ones we've studied, which realize chiral algebras with sporadic symmetry. We conclude in \S\ref{subsec:mckaye7} by arguing that many of the algebras which were derived as deconstructions of the monster VOA with respect to the $\W_{D_{n\mathrm{X}}}$ algebras are the same as deconstructions of $\tsl{V}\mathbb{B}^\natural$ and $\tsl{VF}_{24}^\natural$ with respect to the algebras $\W_{B(m\mathrm{Y})}$ and $\W_{F(r\mathrm{Z})}$ coming from McKay's $\widehat{E_7}$ and $\widehat{E_6}$ correspondences.

\subsection{Cyclic monstralizers from parafermion theories}\label{subsec:startingmodels}

We begin by reviewing the properties of the chiral algebras $V^\natural$, $\tsl{V}\mathbb{B}^\natural$, and $\tsl{VF}^\natural_{24}$ which are most relevant for our subsequent analysis, and explain how these models correspond to monstralizers with $G = \mathbb{Z}_{\mathrm{1A}}$, $\mathbb{Z}_{\mathrm{2A}}$, and $\mathbb{Z}_{\mathrm{3A}}$. We then show that these constructions can be straightforwardly generalized to produce a new $\mathbb{M}$-com pair with $G = \mathbb{Z}_{\mathrm{4A}}$.

\subsubsection{$(\mathbb{Z}_{\mathrm{1A}},\mathbb{M})${\normalfont:} The monster and $\widehat{E_8}$}\label{monstercft}

The main model we will deconstruct is the moonshine module $V^\natural$. This chiral algebra was first obtained \cite{flm,flmbook} as a $\mathbb{Z}_2$ asymmetric orbifold of the lattice VOA associated to the Leech lattice, the unique even, unimodular lattice without any vectors of square-length equal to 2. Its automorphism group is the monster, and it trivially constitutes a degenerate $\mathbb{M}$-com pair with $\W_{\mathbb{Z}_{\mathrm{1A}}}$ the trivial CFT. The vacuum module is its unique, irreducible module, and its associated character is famously given up to a constant by the $j$-invariant,
\begin{align}
    \mathrm{Tr}_{V^\natural}q^{L_0-1} = J(\tau) = j(\tau)-744 = q^{-1}+196884 q +21493760q^2+\cdots.
\end{align}
Because it has monster symmetry, we can consider characters \cite{conwaynorton} which are twined by the action of $\mathbb{M}$, known as \emph{McKay-Thompson series},
\begin{align}
J_g(\tau) =  \mathrm{Tr}_{V^\natural}g q^{L_0-1} \ \ \ \ \ \ \ \  \ \  (g\in\mathbb{M}).
\end{align}
These functions are invariant under the action of special discrete subgroups $\Gamma_g$ of  $\operatorname{\textsl{SL}}_2(\mathbb{R})$, 
\begin{align}
    J_g(\tfrac{a\tau+b}{c\tau+d}) = J_g(\tau) \ \ \ \ \ \ \ \ \ \ \   \left(\begin{smallmatrix} a & b \\ c & d\end{smallmatrix}\right)\in \Gamma_g
\end{align}
which have the property that $\Gamma_g\backslash \hat{\mathbb{H}} \cong \hat{\mathbb{C}}$, where $\hat{\mathbb{H}} = \mathbb{H} \cup \mathbb{Q}\cup \{i\infty\}$ is the upper half-plane extended by its cusps, and $\hat{\mathbb{C}} = \mathbb{C}\cup \{i\infty\}$ is the Riemann sphere. Such discrete subgroups of $\operatorname{\textsl{SL}}_2(\mathbb{R})$ are referred to as \emph{genus zero}, and $J_g$ can be described as the unique generator of the field of modular functions for $\Gamma_g$ whose $q$-expansion starts as $q^{-1}+O(q)$; equivalently, and more constructively, all of these McKay-Thompson series can be obtained as Rademacher sums \cite{Duncan:2009sq}. We will refer to conjugacy classes of the monster group using the Atlas notation \cite{atlas}. We will also go back and forth between labeling McKay-Thompson series by group elements versus conjugacy classes, e.g. we allow ourselves to write $J_{\mathrm{2A}}$ in place of $J_g$ whenever $g$ is taken from the 2A conjugacy class of $\mathbb{M}.$

The monster group has an intriguing connection to the extended $\widehat{E_8}$ Dynkin diagram. This is brought about by considering products of elements taken from the 2A conjugacy class. It is known that any pair of 2A involutions $t_1$ and $t_2$ combine to give an element of either the 1A, 2A, 3A, 4A, 5A, 6A, 4B, 2B, or 3C conjugacy class, and generates a corresponding dihedral group  $D_{n\mathrm{X}} = \langle t_1,t_2\rangle$ which participates in a monstralizer pair (c.f.\ Table \ref{monstralizers}). These conjugacy classes can be thought of as corresponding to the nodes of the $\widehat{E_8}$ diagram as follows. If $\alpha_1,$ $\alpha_2,$ $\dots$, $\alpha_8$ are the simple roots of $E_8$ and $\alpha_0$ is the negative of the maximal root, then there are integers $c_i$ known as Coxeter labels which satisfy
\be
\sum_{i=0}^8 c_i \alpha_i=0
\ee
and take the values $1$, $2$, $3$, $4$, $5$, $6$, $4$, $2$, and $3$ for $i=0$, $1$, $\dots$, $8$. It was noted by McKay \cite{mckaye8} and elaborated on by Glauberman-Norton \cite{glaubnorton} that the values of the $c_i$ are precisely the orders of elements in the conjugacy classes appearing in the products of 2A elements. It is natural to try to find signatures of this correspondence in the context of the moonshine module. In doing so, we will see that one is naturally led to a collection of subalgebras $\W_{D_{n\mathrm{X}}}$ --- one for each conjugacy class $n\mathrm{X}$ arising in the $\widehat{E_8}$ correspondence --- with respect to which one may deconstruct $V^\natural$. 

To explain how these subVOAs $\W_{D_{n\mathrm{X}}}$ are defined, we note that the Griess algebra of $V^\natural$ has an interesting structure \cite{conway1985simple,miyamotogriess}. The 2A involutions in the monster group $\mathbb{M}$ are in one-to-one correspondence with conformal vectors of the moonshine module with central charge equal to $\frac{1}{2}$. Equivalently, they are in bijection with idempotents of the Griess algebra $\frac{1}{2}t$ with $(t,t)=\frac{1}{4}$. In one direction, the correspondence is easy to see: each conformal vector of central charge $\frac{1}{2}$ generates a subalgebra $\W\cong \mathcal{L}(\frac{1}{2},0)$ of $V^\natural$, and it turns out that its Miyamoto involution $\tau_\W$ always belongs to the 2A class. In light of the $\widehat{E_8}$ correspondence, it is natural to define the algebra $\W_{D_{n\mathrm{X}}}$ to be the VOA which is generated by any two conformal vectors $e(z)$, $f(z)$ of central charge $\frac12$ whose associated Miyamoto involutions have product $\tau_e\tau_f$ residing in the $n$X conjugacy class of $\mathbb{M}$. It turns out that this algebra depends up to isomorphism only on the conjugacy class $n$X, and so our labeling is consistent. The Miyamoto involutions $\tau_e$ and $\tau_f$ always act by either fusion algebra automorphisms or inner automorphisms, and generate a dihedral subgroup $D_{n\mathrm{X}}$ of $\mathbb{M}$. In fact, the group preserved by $\W_{D_{n\mathrm{X}}}$ (c.f.\ \S\ref{subsec:deconstruction}) is precisely $\mathbb{M}(\W_{D_{n\mathrm{X}}}) = D_{n\mathrm{X}}$.

Our main body of examples will arise by considering the commutants of the $\W_{D_{n\mathrm{X}}}$ in $V^\natural$; we will see that their automorphism groups are often related to sporadic or otherwise exceptional groups.

\subsubsection*{The $\W_{D_{n\mathrm{X}}}$ algebras as subalgebras of $\V_{\sqrt{2}E_8}$}

The abstract description of the algebras $\W_{D_{n\mathrm{X}}}$ above is not explicit, and the connection to $E_8$ is indirect. However, following \cite{lametal}, we now sketch a more concrete description of the $\W_{D_{n\mathrm{X}}}$ by realizing them as subalgebras of the lattice VOA associated to $\sqrt{2}\Lambda_{\mathrm{root}}(E_8)$. 

To describe these subalgebras of $\V_{\sqrt{2}E_8}$, let $i=0,$ $1,$ $\dots$, $8$ label the nodes of the extended $E_8$ Dynkin diagram and let $c_i$ be the Coxeter label of each node. We associate these nodes with the conjugacy classes 1A, 2A, 3A, 4A, 5A, 6A, 4B, 2B, 3C in this order (see also Figure \ref{e8}). Now consider the root lattice $L(i)$ associated to the Dynkin diagram produced by removing the node $i$ in $\widehat{E_8}$. Then $c_i$ is the index of $L(i)$ in $\Lambda_{\mathrm{root}}(E_8)$, and in fact $\Lambda_{\mathrm{root}}(E_8)/L(i) \cong \mathbb{Z}_{c_i}$. The lattices $L(i)$ will in general have several indecomposable components $R^{(i)}_1,$ $\dots$, $R^{(i)}_k$. For each indecomposable component $R^{(i)}_\ell$ of $L(i)$, it is possible \cite{dong1996associative} to write the conformal vector $\omega^{(i)}_\ell$ of $\V_{\sqrt{2}R^{(i)}_\ell}$ as the sum of two commuting conformal vectors of smaller central charge, $\omega^{(i)}_\ell = \tilde{\omega}_\ell^{(i)}+s_\ell^{(i)}$ (c.f.\ \cite{lametal} for the precise definitions, and \S\ref{subsec:deconstruction} for an example), where $\tilde{\omega}_\ell^{(i)}$ has central charge $2n/(n+3)$ if $R_\ell^{(i)} = A_n$, central charge 1 if $R_\ell^{(i)} = D_n$, and central charge $\frac67$, $\frac{7}{10}$, $\frac12$ if $R_\ell^{(i)}=E_6$, $E_7$, $E_8$ respectively. 

With these definitions in place, \cite{lametal} define the subVOA $\U^{(i)}$ of the $\sqrt{2} E_8$ lattice VOA to be the simultaneous commutant of all the $s^{(i)}_\ell$ for $\ell=1,2,$ $\dots$, $k$, which will have $\tilde{\omega}^{(i)}_1+\cdots +\tilde{\omega}^{(i)}_k$ as its stress tensor. For $i=0,$ $1,$ $\cdots$, $8$, the $\U^{(i)}$ are then chiral algebras with central charges $\sfrac12$, $\sfrac65$, $\sfrac{58}{35}$, $2$, $\sfrac{16}{7}$,  $\sfrac{51}{20}$, $\sfrac{19}{10}$, $1$, and $\sfrac{16}{11}$ respectively
which can be
associated to each node of the extended $E_8$ diagram. For example, deleting the node $i=5$ leaves the Dynkin diagram of $A_5 \oplus A_2 \oplus A_1$, and the central charge of the sum of the conformal vectors $\tilde{\omega}^{(5)}_1+\tilde{\omega}^{(5)}_2+\tilde{\omega}^{(5)}_3$ is $\frac54+\frac45+\frac12=\frac{51}{20}$. Furthermore, the VOAs have dihedral groups associated to them which they inherit from the canonical $\mathbb{Z}_2$ symmetry enjoyed by any lattice VOA and the $\mathbb{Z}_{c_i}$ symmetry which is induced from the Abelian group $\sqrt{2}E_8/\sqrt{2}L(i)$. These $\U^{(i)}$ so defined can then be identified (conjecturally in some cases, provably in others) with the algebras $\W_{D_{n\mathrm{X}}}$. 

We end this section by briefly describing a connection to work of Sakuma. In \cite{Sakuma}, Sakuma considers subalgebras $\mathcal{B}_{e,f}$ of the Griess algebra of an arbitrary VOA $\V$ without dimension 1 operators that are generated by a pair of $c=\frac12$ conformal vectors $e,f$. His main theorem is a classification of the possible algebras which can arise in this way: he finds 9 possible algebras $\mathcal{B}_{e,f}$ whose structure depends only on the order of the product of Miyamoto involutions $\tau_e \tau_f$ and the inner product $\langle e,f \rangle$. The resulting algebras $\mathcal{B}_{e,f}$ coincide precisely with the Griess algebras of the $\U^{(i)}$ constructed in \cite{lametal}; in other words, the monster CFT $V^\natural$ realizes all the possible Sakuma algebras. There is some evidence that the VOA generated by $\mathcal{B}_{e,f}$ will always be isomorphic to one of the $\U^{(i)}$, regardless of what VOA $\V$ one is working inside. See \cite{Griess_topics} for a recent discussion.

\subsubsection{$(\mathbb{Z}_{\mathrm{2A}},2.\mathbb{B})${\normalfont:} The baby monster and $\widehat{E_7}$}\label{subsubsec:BMe7}

The simplest example of monster deconstruction was first studied in \cite{hoehnbaby,hohn2010group,yamauchi20052a}, and is obtained by taking $\W_{\mathbb{Z}_{\mathrm{2A}}} \cong \mathcal{L}(\tfrac12,0)$ to be the chiral algebra of the Ising model. As we will see in a moment, the commutant $\widetilde{\W}_{\mathbb{Z}_{\mathrm{2A}}}$, with central charge $c_{\widetilde{t}} = 24-\frac{1}{2} = 23\sfrac{1}{2}$, enjoys an action by the baby monster sporadic simple group $\mathbb{B}$. Just as the moonshine module is denoted $V^\natural$ because it furnishes the ``most natural'' representation of $\mathbb{M}$, the chiral algebra $\widetilde{\W}_{\mathbb{Z}_{\mathrm{2A}}}$ provides the most natural representation of the baby monster, and so is typically denoted $\operatorname{\textsl{V}}\mathbb{B}^\natural$ in the math literature\footnote{In e.g.\ \cite{hoehnbaby}, the notation $\tsl{V}\mathbb{B}^\natural$ is actually used to denote the vertex operator superalgebra obtained by taking a direct sum of $\widetilde{\W}_{\mathbb{Z}_{\mathrm{2A}}}$ with its irreducible module of highest weight $\frac32$. This VOSA has automorphism group $\mathbb{Z}_2 \times \mathbb{B}$, where the extra $\mathbb{Z}_2$ is generated by $(-1)^F$.}; we abide by this convention here. We will show that the pair $(\W_{\mathbb{Z}_{\mathrm{2A}}},\tsl{V}\mathbb{B}^\natural)$ naturally furnishes a VOA uplift of the monstralizer pair $\mathbb{Z}_{\mathrm{2A}}\circ 2.\mathbb{B}$.

It is instructive to sketch the reason that the baby monster acts on $\operatorname{\textsl{V}}\mathbb{B}^\natural$. Let us call $t$ the central charge $\frac{1}{2}$ stress tensor of $\W_{\mathbb{Z}_{\mathrm{2A}}}$. It is known that the Miyamoto involution $\tau_t$ associated to $\W_{\mathbb{Z}_{\mathrm{2A}}}$ always belongs to the 2A conjugacy class of $\mathbb{M}$. Moreover, one can show that, for any $g$ in the monster, the Miyamoto involutions of $t$ and $gt$ are conjugate, $\tau_{gt} = g \tau_t g^{-1}$ (c.f.\ equation \eqref{autconj}). Therefore, if we take $g$ to live in $\mathrm{Cent}_{\mathbb{M}}(\tau_t) \cong \mathrm{Cent}_{\mathbb{M}}(\mathrm{2A}) \cong 2.\mathbb{B}$, then the Miyamoto involutions associated to $t$ and $gt$ are actually the same, $\tau_{gt} = \tau_t$. Because there is a bijection between elements of the 2A conjugacy class of the monster and conformal vectors of central charge $\frac{1}{2}$, it follows that in fact $t=gt$, and so $2.\mathbb{B}$ stabilizes $t$. Indeed, this is made evident if one decomposes the weight 2 subspace $V^\natural_2$ in terms of irreducible representations of $2.\mathbb{B}$. From the well-known fact that $V^\natural_2 = \mathbf{1}\oplus \mathbf{196883}$ as a monster representation, a character theoretic calculation shows that 
\begin{align}
    V_2^\natural\Big\vert_{2.\mathbb{B}} \cong \mathbf{1}\oplus \mathbf{1}\oplus \mathbf{4371}\oplus\mathbf{96255}\oplus\mathbf{96256}
\end{align}
as a $2.\mathbb{B}$ representation. The two-dimensional trivial subspace is spanned by $t$ and $\widetilde{t}$, the stress tensors of $\W_{\mathbb{Z}_{\mathrm{2A}}}$ and $\operatorname{\textsl{V}}\mathbb{B}^\natural$ respectively. 

From the map $\Phi_t$ in \eqref{authom} which relates the stabilizer of $t$ to automorphisms of its commutant, it follows that the image of $2.\mathbb{B}$ under $\Phi_t$ acts via automorphisms on $\tsl{V}\mathbb{B}^\natural$. The central order $2$ element in $2.\mathbb{B}$ is simply the Miyamoto involution $\tau_t$, and the decomposition
\begin{align}\label{babydecomp}
    V^\natural = \mathcal{L}(\tfrac12,0)\otimes \operatorname{\textsl{V}}\mathbb{B}^\natural(0)\oplus \mathcal{L}(\tfrac12,\tfrac12)\otimes \operatorname{\textsl{V}}\mathbb{B}^\natural(1)\oplus \mathcal{L}(\tfrac12,\tfrac{1}{16})\otimes \operatorname{\textsl{V}}\mathbb{B}^\natural(2)
\end{align}
makes it clear that $\tau_t$ can be thought of as a diagonal fusion algebra automorphism of $\tsl{V}\mathbb{B}^\natural$, and in particular acts trivially on $\tsl{V}\mathbb{B}^\natural(0)$. However, the rest of $2.\mathbb{B}/\langle\tau_t\rangle\cong \mathbb{B}$ must act non-trivially on $\tsl{V}\mathbb{B}^\natural(0)$, because the baby monster is a simple group. This establishes that $\mathbb{B}\subset\mathrm{Aut}(\tsl{V}\mathbb{B}^\natural)$; it was later proved \cite{hohn2010group} that the baby monster is actually the entire automorphism group of $\operatorname{\textsl{V}}\mathbb{B}^\natural$. Putting all these observations together, one sees that the Ising chiral algebra and the baby monster VOA furnish an $\mathbb{M}$-com pair corresponding to $G\circ\widetilde{G} = \mathbb{Z}_{\mathrm{2A}}\circ 2.\mathbb{B}$, and that the various symmetry groups agree with those which can be read off from this monstralizer, 
\begin{align}
\mathbb{M}(\W_{\mathbb{Z}_{\mathrm{2A}}}) &= G = \mathbb{Z}_{\mathrm{2A}} & \mathbb{M}(\tsl{V}\mathbb{B}^\natural) &= \widetilde{G}= 2.\mathbb{B} \nonumber\\
\mathrm{Inn}(\W_{\mathbb{Z}_{\mathrm{2A}}}) &= G/Z(G)= \{1\} & \mathrm{Inn}(\tsl{V}\mathbb{B}^\natural) &= \widetilde{G}/Z(\widetilde{G}) = \mathbb{B} \\
\mathrm{Fus}(\W_{\mathbb{Z}_{\mathrm{2A}}}) &= Z(G) = \mathbb{Z}_2 & \mathrm{Fus}(\tsl{V}\mathbb{B}^\natural)&=Z(\widetilde{G}) = \mathbb{Z}_2 \nonumber
\end{align}

It will be useful for us to note that the baby monster VOA enjoys a structure on its Griess algebra which is analogous to the relationship between central charge $\frac12$ conformal vectors in $V^\natural$ and 2A involutions in $\mathbb{M}$ \cite{hohn2012mckaye7}. Namely, there is a one-to-one correspondence between 2A involutions in $\mathbb{B}$ and central charge $\frac{7}{10}$ conformal vectors in $\tsl{V}\mathbb{B}^\natural$ of ``$\sigma$-type''. A central charge $\frac{7}{10}$ conformal vector is said to be of $\sigma$-type in a VOA $\V$ if decomposing $\V$ with respect to the $\W=\mathcal{L}(\tfrac{7}{10},0)$ subalgebra that the conformal vector generates features no modules of highest weight $h=\frac{7}{16}$ or $h=\frac{3}{80}$; alternatively, in the notation of $\eqref{isotypical component}$, we would say that it is of $\sigma$-type if $\V_{(\frac{7}{16})} = 0$ and $\V_{(\frac{3}{80})} = 0$. In this case, the automorphism $\tau_\W$ of $\V$ defined by lifting the $\mathbb{Z}_2$ automorphism in \eqref{minimal model fusion aut} of the fusion algebra of $\W=\mathcal{L}(\tfrac{7}{10},0)$ is trivial, but one can define another automoprhism $\sigma_\W$ which acts as 
\begin{align}\label{sigma type 7/10}
    \sigma_\W(\varphi) = 
    \begin{cases}
    \varphi, & \varphi \in \V_{(0)}\oplus\V_{(\frac35)} \\
    -\varphi, & \varphi \in \V_{(\frac32)}\oplus\V_{(\frac{1}{10})}
    \end{cases}.
\end{align}
In the case that $\V$ is taken to be $\tsl{V}\mathbb{B}^\natural$, this map is a one-to-one correspondence between central charge $\tfrac{7}{10}$ conformal vectors of $\sigma$-type and elements of the 2A conjugacy class of $\mathbb{B}$.

\begin{figure}
\begin{center}
\includegraphics[width=.7\textwidth]{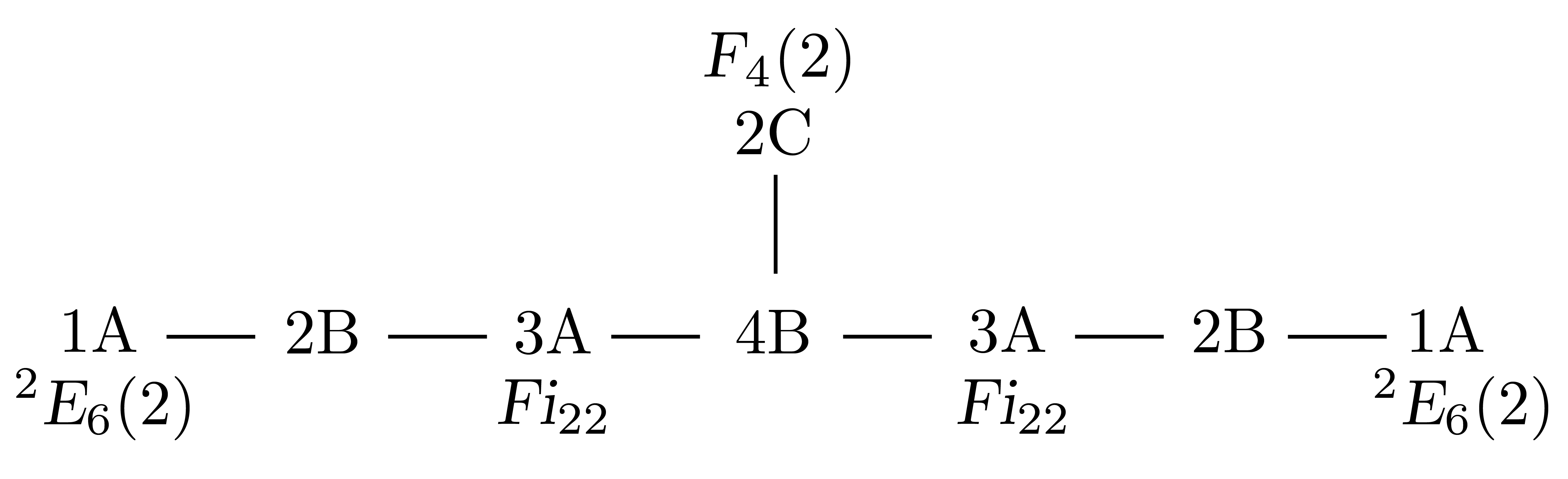}
\caption{Dynkin diagram of $\widehat{E_7}$, decorated by conjugacy classes of the baby monster sporadic group $\mathbb{B}$. We propose to further decorate each node by the inner automorphism group of the commutant of $\W_{B(n\mathrm{X})}$ in $\tsl{V}\mathbb{B}^\natural$ (c.f.\ \S\ref{subsec:mckaye7}). The cases 2B and 4B do not appear to coincide with any of the constructions considered in \S\ref{subsec:mckaye8}, and we leave their study to future work.}
\label{e7}
\end{center}
\end{figure}

The 2A involutions of the baby monster enjoy a relationship to the Dynkin diagram of $\widehat{E_7}$, analogous to the one enjoyed between 2A involutions in the monster and the Dynkin diagram of $\widehat{E_8}$. Namely, the product of any pair of 2A involutions in $\mathbb{B}$ always lies in either 1A, 2B, 3A, 4B or 2C, and it was proposed that these conjugacy classes naturally decorate the nodes of $\widehat{E_7}$, as in Figure \ref{e7}. The correspondence between nodes of $\widehat{E_7}$ and these 5 conjugacy classes is not one-to-one, but only because $\widehat{E_7}$ has a diagram automorphism, and so the conjugacy classes which decorate nodes related by this automorphism should be the same. In \cite{hohn2012mckaye7}, the authors constructed subVOAs $\W_{B(n\mathrm{X})}$ of $\tsl{V}\mathbb{B}^\natural$ for each conjugacy class $n$X arising in the McKay-correspondence. These subVOAs have the property\footnote{Unlike in the case of the $\widehat{E_8}$ correspondence, these VOAs are \emph{not} always generated by their two central charge $\frac{7}{10}$ conformal vectors.} that they contain two central charge $\frac{7}{10}$ conformal vectors whose associated $\sigma$-involutions have product lying in the $n$X class. As in the case of the $\widehat{E_8}$ correspondence, these VOAs come naturally from the lattice VOA based on $\sqrt{2}\Lambda_{\mathrm{root}}(E_7)$, and this helps further justify the relationship between $\mathbb{B}$ and $\widehat{E_7}$. We will also discuss the commutants of these $\W_{B(n\mathrm{X})}$ algebras in $\tsl{V}\mathbb{B}^\natural$ in \S\ref{subsec:mckaye7}; in most cases, we will find that they coincide with commutants of $\W_{m\mathrm{Y}}$ algebras in $V^\natural$.

We now study the characters of the baby monster VOA. The decomposition \eqref{babydecomp} at the level of characters reads 
\begin{align}
    J(\tau) = \chi_{0}(\tau)\chi_{\operatorname{\textsl{V}}\mathbb{B}^\natural(0)}(\tau) + \chi_{1}(\tau) \chi_{\operatorname{\textsl{V}}\mathbb{B}^\natural(1)}(\tau) + \chi_{2}(\tau) \chi_{\operatorname{\textsl{V}}\mathbb{B}^\natural(2)}(\tau)
\end{align}
with $\chi_0(\tau)$, $\chi_{1}(\tau)$, and $\chi_2(\tau)$ the characters of Ising modules $\mathcal{L}(\tfrac12,0)$, $\mathcal{L}(\tfrac12,\tfrac12)$, and $\mathcal{L}(\tfrac12,\tfrac{1}{16})$ respectively, and 
\begin{align}\label{baby characters}
\begin{split}
    \chi_{\operatorname{\textsl{V}}\mathbb{B}^\natural(0)}(\tau) &= q^{-\frac{47}{48}}(1+96256q^2+9646891q^3+366845011q^4+\cdots), \\
    \chi_{\operatorname{\textsl{V}}\mathbb{B}^\natural(1)}(\tau) &= q^{\frac{25}{48}}(4371+1143745q+64680601q^2+1829005611q^3+\cdots), \\
    \chi_{\operatorname{\textsl{V}}\mathbb{B}^\natural(2)}(\tau) &= q^{\frac{23}{24}}(96256+10602496q+420831232q^2+9685952512q^3+\cdots),
\end{split}
\end{align}
the characters of $\operatorname{\textsl{V}}\mathbb{B}^\natural$. The $\chi_{\operatorname{\textsl{V}}\mathbb{B}^\natural(\alpha)}(\tau)$ have been known since $\operatorname{\textsl{V}}\mathbb{B}^\natural$ was first constructed. Later it was pointed out that they are solutions to the following MLDE \cite{Hampapura:2016mmz, Bae:2018qfh}, 
\begin{align}
\begin{split}
\left[\CD^3 + \frac{2315 \pi^2}{576} E_4(\tau) \CD  - i  \frac{27025 \pi^3}{6912} E_6(\tau)\right] \chi_{\tsl{V}\mathbb{B}^\natural(\alpha)}(\tau) = 0.
\end{split}
\end{align}
Even more recently, it was pointed out in \cite{HarveyWu} that they can be described as Hecke images of Ising characters. To understand which Hecke operator can relate these two sets of characters note that $\mathsf{T}_p$ scales the power of the polar term of the vacuum character it acts on by $p$, so in order for $\mathsf{T}_p\chi$ to stand a chance of coinciding with the characters of an RCFT with central charge $\frac{47}{2}$, we must have $p=47$. With this choice, one can indeed show that
\begin{align}
    \chi_{\operatorname{\textsl{V}}\mathbb{B}^\natural(\alpha)}(\tau) = (\mathsf{T}_{47}\chi)_\alpha(\tau).
\end{align}
We refer to loc.\ cit.\ for additional details behind this computation. We will see that this observation generalizes to a number of other examples. 

We conclude by noting that the characters $\chi_{\tsl{V}\mathbb{B}^\natural(\alpha)}(\tau)$ can also be recovered from Rademacher summation, using equations \eqref{rad coefficients neq 0} and \eqref{rad coefficients eq 0}. We have checked this numerically for the first few terms in the $q$-expansion by taking the sum over $c$ from 1 to 900 and checking that the resulting numbers round to the coefficients in \eqref{baby characters}.

\subsubsection{$(\mathbb{Z}_{\mathrm{3A}},3.\normalfont{\tsl{Fi}}_{24}')${\normalfont:} The largest Fischer group and $\widehat{E_6}$}\label{subsubsec:Fischere6}

Another example of monster deconstruction can be obtained by taking $\W_{\mathbb{Z}_{\mathrm{3A}}} \cong \mathcal{L}(\tfrac45,0)\oplus \mathcal{L}(\tfrac45,3)$ to be the chiral algebra of the 3-state Potts model. We will argue that this algebra and its commutant correspond to the monstralizer pair
\begin{align}
    [G\circ \widetilde{G}].H = [\mathbb{Z}_{\mathrm{3A}}\circ 3.\tsl{Fi}_{24}'].2
\end{align} 
To see this, note that the chiral algebra of the 3-state Potts model is the same as the $\mathbb{Z}_3$ parafermion theory, $\W_{\mathbb{Z}_{\mathrm{3A}}}\cong \mathcal{P}(3)$, and thus enjoys a $\mathbb{Z}_3$ automorphism of its fusion algebra which turns out can be lifted to an automorphism $\tau$ of $V^\natural$ belonging to the 3A conjugacy class of $\mathbb{M}$. The theory also has an outer automorphism which acts as $+1$ on $\mc{L}(\tfrac45,0)$ and $-1$ on $\mc{L}(\tfrac45,3)$. These observations are consistent with the symmetry groups for $\W_{\mathbb{Z}_{\mathrm{3A}}}$ that one would read off from the monstralizer,
\begin{align}
\begin{split}
     \mathbb{M}(\W_{\mathbb{Z}_{\mathrm{3A}}}) &= G= \mathbb{Z}_{\mathrm{3A}}, \\
     \mathrm{Inn}(\W_{\mathbb{Z}_{\mathrm{3A}}}) &= G/Z(G) = \{1\}, \\ \mathrm{Aut}(\W_{\mathbb{Z}_{\mathrm{3A}}}) &= (G/Z(G)).H = \mathbb{Z}_2 \\
     \mathrm{Fus}(\W_{\mathbb{Z}_{\mathrm{3A}}}) &= Z(G) = \mathbb{Z}_3.
\end{split}
\end{align}  
Now, it was argued in \cite{hohn2012mckaye6} that the commutant $\widetilde{\W}_{\mathbb{Z}_{\mathrm{3A}}}$ admits an action of $N_{\mathbb{M}}(\langle\tau\rangle)/\langle\tau\rangle \cong \operatorname{\textsl{Fi}}_{24}$ via automorphisms\footnote{We use the notation $N_G(S) = \{ g\in G \mid gS=Sg\} $ to denote the normalizer of the set $S$ in $G$.}, and hence was labeled $\operatorname{\textsl{VF}}^\natural$ there, though we will use the notation $\tsl{VF}_{24}^\natural$ to distinguish it from the chiral algebras we associate to the other Fischer groups, $\Fi_{23}$ and $\Fi_{22}$. The sporadic simple group $\Fi_{24}'$ is a subgroup of $\Fi_{24}$ with index 2, and we will be able to see, once we have expressions for the characters of $\tsl{VF}^\natural_{24}$, that it is the inner automorphism group. Thus, the monstralizer also reproduces the symmetry groups of $\tsl{VF}^\natural_{24}$, 
\begin{align}
\begin{split}
    \mathbb{M}(\tsl{VF}^\natural_{24}) &= \widetilde{G} = 3.\tsl{Fi}_{24}' \\
    \mathrm{Inn}(\tsl{VF}_{24}^\natural) &= \widetilde{G}/Z(\widetilde{G}) = \tsl{Fi}_{24}' \\
    \mathrm{Aut}(\tsl{VF}^\natural_{24}) &= (\widetilde{G}/Z(\widetilde{G})).H = \Fi_{24} \\
    \mathrm{Fus}(\tsl{VF}^\natural_{24}) &= Z(\widetilde{G}) = \mathbb{Z}_3.
\end{split}
\end{align}

\begin{figure}
\begin{center}
\includegraphics[width=.5\textwidth]{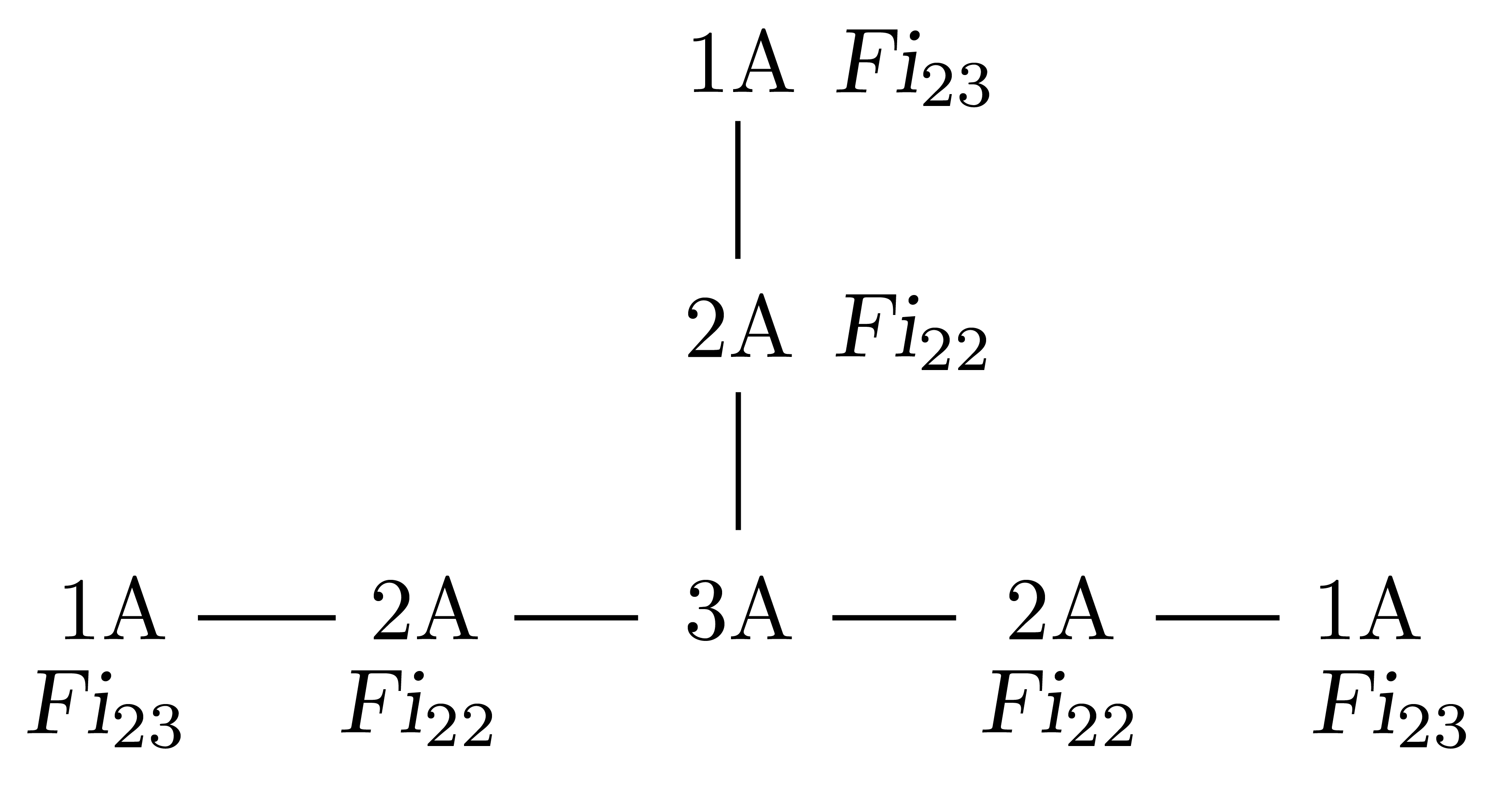}
\caption{Dynkin diagram of $\widehat{E_6}$, decorated by conjugacy classes of the Fischer group $\Fi_{24}$. We propose to further decorate each node by the inner automorphism group of the commutant of $\W_{F(n\mathrm{X})}$ in $\tsl{VF}_{24}^\natural$ (c.f.\ \S\ref{subsec:mckaye7}). The case 3A does not appear to coincide with any of the constructions considered in \S\ref{subsec:mckaye8}, and we leave its study to future work.}
\label{e6}
\end{center}
\end{figure}

As in the case of $\mathbb{M}$ and $\mathbb{B}$, the Fischer group enjoys a relationship to the Dynkin diagram of $\widehat{E_6}$. The product of any two involutions in the 2C conjugacy class of $\Fi_{24}$ always lie in one of 1A, 2A or 3A, and it is natural to label the nodes of $\widehat{E_6}$ with these conjugacy classes, as in Figure \ref{e6}. Again, the correspondence between nodes and conjugacy classes is not one-to-one, but only up to the two non-trivial diagram automorphisms. 

We can again uplift this correspondence from one between $\Fi_{24}$ and $\widehat{E_6}$ to one between $\tsl{VF}_{24}^\natural$ and $\widehat{E_6}$, following \cite{hohn2012mckaye6}. The idea is similar to that of the previous section: the ``derived'' conformal vectors of central charge $\frac67$ in $\operatorname{\textsl{VF}}^\natural_{24}$ are in one-to-one correspondence with involutions of $\operatorname{\textsl{Fi}}_{24}$ in the 2C class (see loc.\ cit.\ for the precise definition of what ``derived'' means). Indeed, it is known that the decomposition of $\operatorname{\textsl{VF}}^\natural_{24}$ with respect to the $\U:=\mathcal{L}(\tfrac67,0)$ subalgebra generated by a derived central charge $\frac67$ conformal vector, 
\begin{align}
    \operatorname{\textsl{VF}}^\natural = \bigoplus_{h\in \Sigma} \tsl{VF}^\natural_{(h)}
\end{align}
features only the modules $\mathcal{L}(\tfrac67,h)$ for $h \in \Sigma= \{0,\frac57,\frac{22}{7},5,\frac{12}{7},\frac17\}$, and so we say that the conformal vector is of $\sigma$-type. These modules fuse among themselves, and admit a $\mathbb{Z}_2$ automorphism $\sigma$ of their fusion algebra, which can be lifted to an automorphisms $\sigma_\U$ of $\operatorname{\textsl{VF}}^\natural$ which acts as 
\begin{align}
    \sigma_\U(\varphi) = \begin{cases}
    \varphi & \varphi \in \tsl{VF}^\natural_{(h)} \text{ for }h = 0,\frac57,\frac{22}{7} \\
    -\varphi & \varphi \in \tsl{VF}^\natural_{(h)} \text{ for }h=5,\frac{12}{7},\frac17
    \end{cases}
\end{align}
and always lies in the 2C class. In light of the $\widehat{E_6}$ correspondence, it is natural to seek subVOAs $\W_{F(n\mathrm{x})}$ of $\tsl{VF}_{24}^\natural$ with the property that they contain two derived central charge $\frac67$ conformal vectors whose associated $\sigma$-involutions have product lying in $n$X, as was done for $V^\natural$ and $\tsl{V}\mathbb{B}^\natural$. These are constructed explicitly in loc.\ cit., again coming naturally from subVOAs of the lattice VOA associated to $\sqrt{2}\Lambda_{\mathrm{root}}(E_6)$. We will consider deconstructing $\tsl{VF}^\natural_{24}$ with respect to such $\W_{F(n\mathrm{X})}$ subVOAs in \S\ref{subsec:mckaye7}, and we will find that in most cases they coincide with VOAs we had already constructed as commutants of $\W_{m\mathrm{Y}}$ in $V^\natural$.

We now study the dual characters of $\tsl{VF}^\natural_{24}$ in $V^\natural$. One way to obtain them is as the solutions of the MLDE
\begin{align}
    \left[\CD^4 + \frac{907 \pi^2}{225} E_4(\tau) \CD^2 -i
    \frac{4289 \pi^3}{675} E_6(\tau) \CD   -\frac{175769 \pi^4}{50625} E_4^2(\tau) \right]\chi_{\tsl{VF}^\natural(\alpha)}(\tau) = 0.
\end{align}
However, it is interesting to note that, as in the case of $\operatorname{\textsl{V}}\mathbb{B}^\natural$, they can also be obtained as Hecke images of the characters $\psi^{(3)}_{\ell,m}$ of $\W_{\mathbb{Z}_{\mathrm{3A}}}=\mc{P}(3)$. First, note that the conductor of the $\IZ_3$ parafermion theory is $N=30$, as one can determine from inspection of the central charge $\frac{4}{5}$ and conformal dimensions of its primary fields: $\frac{1}{15}$, $\frac25$, $\frac{1}{15}$, $\frac23$, $\frac23$, $0$. We will label these primaries with pairs $(\ell,m)=(1,1)$, $(2,0)$, $(2,2)$, $(3,-1)$, $(3,1)$, $(3,3)$ respectively; when we write down matrices, the basis will be in the order we have just written. Note that primaries with dimensions $\frac23$ and $\frac{1}{15}$ appear twice because these operators transform non-trivially under the $\IZ_3$ symmetry, and so one
has the operator and its complex conjugate with the same character. For reference, the first few terms in the $q$-expansions of the characters of $ \mathcal{P}(3)$ are given below,
\begin{align}
\begin{split}
\psi^{(3)}_{1,1}(\tau) &=  q^{\frac{1}{30}}(1+2 q + 2 q^2+4 q^3+5 q^4+8 q^5+11 q^6+ 16 q^7  + \cdots),  \\
\psi^{(3)}_{2,0}(\tau) &=   q^{\frac{11}{30}}(1+q+2 q^2+3 q^3+5 q^4+7 q^5+10 q^6+14 q^7 + \cdots),      \\
\psi^{(3)}_{3,-1}(\tau) &= q^{\frac{19}{30}}(1+q+2 q^2+ 2 q^3+4 q^4+5 q^5 + 8 q^6 + 10 q^7 + \cdots),  \\
\psi^{(3)}_{3,3}(\tau) &= q^{-\frac{1}{30}}(1 + q^2+2 q^3+3 q^4+4 q^5+7 q^6 + 8 q^7 + \cdots ),\\
& \hspace{.5in} \psi^{(3)}_{2,2}(\tau) = \psi^{(3)}_{1,1}(\tau),  \ \ \ \ \ \ \ \ \psi^{(3)}_{3,1}(\tau) = \psi^{(3)}_{3,-1}(\tau).
\end{split}
\end{align}
Calling $\rho$ the $\tsl{SL}_2(\mathbb{Z})$ representation generated by the modular S and T-matrices of the $\mathbb{Z}_3$ parafermion theory (c.f. equations \eqref{parafermion dimensions} and \eqref{Parafermion S-matrix}), the charge conjugation matrix is given by 
\be
\mathcal{C} = \rho(S)\cdot\rho(S)  = \left(
\begin{array}{cccccc}
 0 & 0 & 1 & 0 & 0 & 0 \\
 0 & 1 & 0 & 0 & 0 & 0 \\
 1 & 0 & 0 & 0 & 0 & 0 \\
 0 & 0 & 0 & 0 & 1 & 0 \\
 0 & 0 & 0 & 1 & 0 & 0 \\
 0 & 0 & 0 & 0 & 0 & 1 \\
\end{array}
\right).
\ee
It exchanges the $(1,1)$, $(2,2)$ and $(3,-1)$, $(3,1)$ components, consistent with these being characters of conjugate pairs of primary operators. A straightforward calculation also shows that
\be
\rho(\sigma_{29})= \mathcal{C}.
\ee
These are all the ingredients necessary to define the relevant Hecke operator. Now, if we label the dual modules of $\operatorname{\textsl{VF}}^\natural_{24}$ according to how they appear in the decomposition of $V^\natural$, 
\begin{align}\label{fischer24decomp}
    V^\natural \cong \bigoplus_{(\ell,m)}\mathcal{P}(3,[\ell,m])\otimes \operatorname{\textsl{VF}}_{24}^\natural(\ell,m)
\end{align}
then we claim that their corresponding characters can be realized as 
\begin{align}
    \chi_{\operatorname{\textsl{VF}}_{24}^\natural(\ell,m)}(\tau) = (\mathsf{T}_{29}\psi^{(3)})_{\ell,m}(\tau).
\end{align} 
This leads to the following $q$-expansions,
\begin{align}
\begin{split}
\chi_{\tsl{V}\operatorname{\textsl{F}}^\natural_{24}(1,1)}(\tau) &= q^{\frac{29}{30}}(64584 +6789393 q+261202536 q^2 \\
& \hspace{.5in}+5863550310 q^3+92704262184 q^4+1139097001086 q^5 + \cdots), \\
\chi_{\tsl{V}\operatorname{\textsl{F}}^\natural_{24}(2,0)}(\tau) &= q^{\frac{19}{30}}(8671 +1675504 q+83293626 q^2 \\
& \hspace{.5in}+2175548448 q^3 +38129457201 q^4+505531399264 q^5+\cdots ), \\
\chi_{\tsl{V}\operatorname{\textsl{F}}^\natural_{24}(2,2)}(\tau) &= q^{\frac{29}{30}}(64584 +6789393 q+261202536 q^2 \\
& \hspace{.5in} +5863550310 q^3+92704262184 q^4+1139097001086 q^5 + \cdots), \\
\chi_{\tsl{V}\operatorname{\textsl{F}}^\natural_{24}(3,-1)}(\tau) &=  q^{\frac{11}{30}}(783 +306936 q+19648602 q^2 \\
&\hspace{.5in}+589705488 q^3+11326437954 q^4+160445964456 q^5 + \cdots ), \\
\chi_{\tsl{V}\operatorname{\textsl{F}}^\natural_{24}(3,1)}(\tau) &= q^{\frac{11}{30}}(783 +306936 q+19648602 q^2 \\
&\hspace{.5in}+589705488 q^3+11326437954 q^4+160445964456 q^5+ \cdots), \\
\chi_{\tsl{V}\operatorname{\textsl{F}}^\natural_{24}(3,3)}(\tau) &=  q^{-\frac{29}{30}}(1+57478 q^2+5477520 q^3 \\
&\hspace{.5in}+201424111 q^4+4397752560 q^5+68202269658 q^6+ \cdots). \\
\end{split}
\end{align}
The inner automorphism group in general acts projectively on the non-vacuum modules; the projective representations of $\Fi_{24}'$ correspond to the honest representations of $\widetilde{G}=3.\Fi_{24}'$ in the same way that projective representations of $\tsl{SO}(3)$ correspond to honest representations of $\tsl{SU}(2)$, so we work with this latter triple cover\footnote{This also justifies the consideration of the group $3.\Fi'_{24}$ in  \cite{Bae:2018qfh}.} when analyzing the characters; note that this is consistent with our claims that any projective representation of $\mathrm{Inn}(\W_{\widetilde{G}})$ which is realized on the dual modules $\W_{\widetilde{G}}(\alpha)$ can be lifted to an honest representation of $\widetilde{G}$. Indeed, one can immediately see that the highest weights transform under irreducible representations of $3.\Fi_{24}'$ of dimension $783$, $8671$, and $64584$.
Other low order coefficients also have decompositions into irreducible representations: $57478=57477+1$, $1675504=1666833+8671$, and so on. These identifications are determined by how $V^\natural$ decomposes by restriction into representations of $3.\Fi_{24}'$.

Consistent with \eqref{fischer24decomp}, these dual characters pair diagonally with the characters of the $\mathbb{Z}_3$ parafermion model to yield the partition function of $V^\natural$,
\be
\sum_{(\ell,m)} \psi^{(3)}_{\ell,m}(\tau) \chi_{\tsl{VF}^\natural_{24}(\ell,m)}(\tau) = q^{-1}+196884 q + 21493760 q^2 + \cdots.
\ee
One can also verify at the level of characters that the $\mathbb{Z}_3$ automorphism of the fusion algebra of $\mathcal{P}(3)$ lifts to an automorphism of $V^\natural$ in the 3A conjugacy class of $\mathbb{M}$. Indeed, under this $\mathbb{Z}_3$, the characters transform as $\psi^{(3)}_{\ell,m} \to e^{2\pi i \frac{m}{3}}\psi^{(3)}_{\ell,m}$, and one can straightforwardly compute that
\be
\sum_{(\ell,m)}  e^{2\pi i \frac{m}{3}} \psi^{(3)}_{\ell,m}(\tau)\chi_{\tsl{VF}^\natural_{24}(\ell,m)}(\tau)= q^{-1}+783 q + 8672 q^2 + 65367 q^3+371520 q^4+ \cdots
\ee
which agrees with the $q$-expansion of $J_{\mathrm{3A}}$, the McKay-Thompson series associated to any element belonging to the 3A conjugacy class of $\mathbb{M}$.

The above is consistent with the more general analysis of bilinears in \cite{HarveyWu}. We can construct bilinears from $G_\ell$ with
\be
29 + \ell^2 = 0 ~{\rm mod}~30
\ee
which is solved by $\ell=1,11,19,29$. Explicit computation shows that $G_1$ is the identity matrix giving a diagonal modular invariant.
Since $G_{19}= -G_1$, $\ell=19$ does not lead to an independent solution. Finally, $G_{11}=- G_{29}$ and
\be
G_{29} = \begin{pmatrix} 0 & 0  & 1 & 0 & 0 & 0 \\
					0 & 1 & 0 & 0 & 0 & 0 \\
					1 & 0 & 0 & 0 & 0 & 0 \\
					0 & 0 & 0 & 0 & 1 & 0 \\
					0 & 0 & 0 & 1 & 0 & 0 \\
					0 & 0 & 0 & 0 & 0 & 1 \end{pmatrix}.
\ee
This differs from $G_1$ by the exchange of the first and third characters and the fourth and fifth characters. But since
these characters are identical (there are two characters with the same conformal dimension but which transform differently under $\mathbb{Z}_3$)
this does not give a new modular invariant: there is essentially a single modular invariant bilinear.

\subsubsection{$(\mathbb{Z}_{\mathrm{4A}},4.\normalfont{2^{22}.\tsl{Co}_3})${\normalfont:} The third Conway group}\label{subsubsecCo3}

As we have mentioned in previous sections, the stress tensor of the moonshine module $V^\natural$ can be written as a sum of 48 conformal vectors of central charge $\frac12$. This implies that it admits an $\mc{L}(\tfrac12,0)^{\otimes 48}$ subalgebra, which gives it the structure of a \emph{framed VOA} \cite{Dong:1997xy}. This subalgebra furnishes an alternative description of the moonshine module; indeed, on general grounds, it can be decomposed as \cite{lametal}
\begin{align}
    V^\natural \cong \bigoplus_{h_1,\dots,h_{48}} c_{h_1,\dots,h_{48}} \mc{L}(\tfrac12,h_1)\otimes \cdots \otimes \mc{L}(\tfrac12,h_{48}),
\end{align}
where each $h_i$ runs over the set $\{0,\tfrac12,\tfrac{1}{16}\}$. Parts of this structure are passed on to the baby monster VOA. Indeed, in light of the decomposition \eqref{babydecomp}, one has that 
\begin{align}\label{babyisingdecomp}
    \tsl{V}\mathbb{B}^\natural(\alpha) = \bigoplus_{h_1,\dots,h_{47}}c_{h_1,\dots,h_{47},n_\alpha}\mc{L}(\tfrac12,h_1)\otimes\cdots\otimes\mc{L}(\tfrac12,h_{47}),
\end{align}
where $n_0 = 0$, $n_1 = \tfrac12$, and $n_2 = \tfrac{1}{16}$. The baby monster symmetry ultimately arises from the fact that $2.\mathbb{B}$ centralizes the lift of the $\mathbb{Z}_2$ automorphism of the fusion algebra of any of the Ising factors (c.f. \S\ref{subsubsec:symmetries}). It is natural to wonder whether there are analogs of this kind of structure with the Ising model replaced by some other chiral algebra.

In fact, in \cite{shimakura2002decompositions}, it was shown that the moonshine module admits as a subalgebra a tensor product of 24 copies of the charge conjugation orbifold $\V_{\sqrt{2k}\mathbb{Z}}^+$ (c.f.\ \S\ref{sec:examples} for the definition of $\V_L^+$) for each $k\geq 2$; the special case $k=3$ coincides with the $\mathbb{Z}_4$ parafermion theory, $\V^+_{\sqrt{6}\mathbb{Z}}\cong \mc{P}(4)$. It is possible to construct such a subalgebra because one can always find 24 orthogonal vectors in the Leech lattice of norm-squared $2k$; these generate 24 commuting $\V_{\sqrt{2k}\mathbb{Z}}^+$ subalgebras of $\V_{\Lambda_{\mathrm{Leech}}}^+$, and the latter is in turn a subalgebra of $V^\natural$ by virtue of the original FLM construction \cite{flm} of the monster CFT as a $\mathbb{Z}_2$ orbifold of $\V_{\Lambda_{\mathrm{Leech}}}$. 

Furthermore, the moonshine module was decomposed with respect to these subalgebras, and so in particular one can write
\begin{align}
    V^\natural \cong \bigoplus_{(\ell_1,m_1),\dots,(\ell_{24},m_{24})} c_{(\ell_1,m_1),\dots,(\ell_{24},m_{24})}\mc{P}(4,[\ell_1,m_1])\otimes\cdots\otimes \mc{P}(4,[\ell_{24},m_{24}])
\end{align}
for known multiplicities $c_{(\ell_1,m_1),\dots,(\ell_{24},m_{24})}$.  It is then natural to mimick the construction of the baby monster, but with the role of the Ising model $\mc{L}(\tfrac12,0)$ being played by $\mc{P}(4)$. We will see that doing so produces an $\mathbb{M}$-com pair corresponding to the monstralizer $[G\circ \widetilde{G}].H = [\mathbb{Z}_{\mathrm{4A}}\circ4.2^{22}.\tsl{Co}_3].2$. To carry out this construction, one should choose one of the $\mc{P}(4)$ factors (or alternatively, specify one of the one-dimensional sublattices $L\cong \sqrt{6}\mathbb{Z}$ of $\Lambda_{\mathrm{Leech}}$) and identify it with the algebra $\W_{\mathbb{Z}_{\mathrm{4A}}}$. We may then define, in analogy with equation \eqref{babyisingdecomp}, the modules
\begin{align}\label{Co3decomp}
\begin{split}
    &\widetilde{\W}_{\mathbb{Z}_{\mathrm{4A}}}(\ell,m) := \\
    &\hspace{.3in}\bigoplus_{(\ell_1,m_1),\dots,(\ell_{23},m_{23})}c_{(\ell_1,m_1),\dots,(\ell_{23},m_{23}),(\ell,m)}\mc{P}(4,[\ell_1,m_1])\otimes\cdots\otimes\mc{P}(4,[\ell_{23},m_{23}])
    \end{split}
\end{align}
which leads to a decomposition of the form 
\begin{align}
    V^\natural \cong \bigoplus_{(\ell,m)} \mc{P}(4,[\ell,m])\otimes \widetilde{\W}_{\mathbb{Z}_{\mathrm{4A}}}(\ell,m)
\end{align}
As we will see momentarily, $\widetilde{\W}_{\mathbb{Z}_{\mathrm{4A}}}:= \widetilde{\W}_{\mathbb{Z}_{\mathrm{4A}}}(4,4)$ is the commutant of the chosen $\mc{P}(4)$ subalgebra of $V^\natural$. The other components $\widetilde{\W}_{\mathbb{Z}_{\mathrm{4A}}}(\ell,m)$ will be irreducible modules of $\widetilde{\W}_{\mathbb{Z}_{\mathrm{4A}}}$. 

Another perspective on $\widetilde{\W}_{\mathbb{Z}_{\mathrm{4A}}}$ is that it is the $\mathbb{Z}_2$ orbifold $\V_{\widetilde L}^+$ of the $c=23$ lattice VOA attached to the orthogonal complement $\widetilde{L}:=L^\perp$ of any of the $L\cong \sqrt{6}\mathbb{Z}$ sublattices of  $\Lambda_{\mathrm{Leech}}$ described earlier. We will compute the characters using this explicit description. To proceed, we extract the generators of $\widetilde{L}$ from the data files provided with \cite{hohnmason} and compute the theta function of $\widetilde{L}$ using Magma \cite{MR1484478}. The discriminant group is $\widetilde{L}^\ast/\widetilde{L} \cong \mathbb{Z}_6 := \langle \lambda^\ast\rangle$ and, using the notation $\widetilde{\Theta}_k(\tau) := \theta_{\widetilde{L}+k\lambda^\ast}(\tau)$, we find that the components have $q$-expansions given by
\begin{align}
\begin{split}
\widetilde{\Theta}_0(\tau) &= 1 + 75900 q^2 + 52923000 q^3 + 108706050 q^4 + \cdots \\
\widetilde{\Theta}_1(\tau) &=q^{\frac{23}{12}}( 48600+ 3934656q+ \cdots) \\
\widetilde{\Theta}_2(\tau) &= q^{\frac53}(11178 + 1536975q + \cdots) \\
\widetilde{\Theta}_3(\tau) &= q^{\frac54}(552 + 257600q + \cdots) \\
\widetilde{\Theta}_4(\tau) &= \widetilde{\Theta}_2(\tau), \ \ \ \ \widetilde{\Theta}_5(\tau) = \widetilde{\Theta}_1(\tau).
\end{split}
\end{align}
We can then use the expressions for the characters of charge conjugation orbifolds given in equation \eqref{chargeconjorbchars} to obtain that 
\begin{align}
\begin{split}\label{Co3chargeconjorbchars}
    \xi_{\mathds{1}}^{(\widetilde L)}(\tau) &= \frac12\left( \frac{\widetilde{\Theta}_0(\tau)}{\eta(\tau)^{23}}  + \Phi_{0,1}(\tau)^{23}  \right) \\
    \xi_{j}^{(\widetilde L)}(\tau) &= \frac12\left( \frac{\widetilde{\Theta}_0(\tau)}{\eta(\tau)^{23}}  - \Phi_{0,1}(\tau)^{23}  \right) \\
    \xi^{(\widetilde L)}_{3\lambda^\ast,i}(\tau) &= \frac12 \frac{\widetilde{\Theta}_3(\tau)}{\eta(\tau)^{23}}  \ \ \ (i=1,2)\\
    \xi^{(\widetilde L)}_{k\lambda^\ast}(\tau) &= \frac{\widetilde{\Theta}_k(\tau)}{\eta(\tau)^{23}} \ \ \ (k=1,2) \\
    \xi^{(\widetilde L)}_{\sigma,i}(\tau) &=  2^{10}\left(  \Phi_{1,0}(\tau)^{23} + \Phi_{1,1}(\tau)^{23}  \right) \ \ \ (i=1,2 )\\
\xi_{\tau,i}^{(\widetilde L)}(\tau) &= 2^{10}\left(  \Phi_{1,0}(\tau)^{23} + \Phi_{1,1}(\tau)^{23}  \right) \ \ \ (i=1,2 ).
\end{split}
\end{align}
We will see in a moment that the $q$-expansions of these same characters are reproduced by the method of Hecke operators.

Now, it is known \cite{splag} that the Conway group $\tsl{Co}_3$ is the stabilizer of the rank one lattice $L$, and we will now argue that an extension of $\tsl{Co}_3$ acts naturally on the orbifold VOA associated to its orthogonal complement. To see this, note that the generator of the $\mathbb{Z}_4$ automorphism of the fusion algebra of the level $4$ parafermion theory can be lifted to an automorphism $\sigma$ of $V^\natural$, which a character-theoretic calculation reveals belongs to the 4A conjugacy class. One thus expects that at least a $2^{22}.\tsl{Co}_3\cong \mathrm{Cent}_{\mathbb{M}}(\sigma)/\langle\sigma\rangle$ symmetry group acts by inner automorphisms on $\widetilde{\W}_{\mathbb{Z}_{\mathrm{4A}}}$, where we have quotiented by $\langle\sigma\rangle$ because it acts trivially on $\widetilde{\W}_{\mathbb{Z}_{\mathrm{4A}}}$. We further conjecture that $\widetilde{\W}_{\mathbb{Z}_{\mathrm{4A}}}$ inherits an additional order 2 outer automorphism from the monster, so that the full automorphism group is given by $\mathrm{Aut}(\widetilde{\W}_{\mathbb{Z}_{\mathrm{4A}}}) \cong 2^{22}.\tsl{Co}_3.2$ in accordance with the prediction from the monstralizer. However, without access to the character table of $2^{22}.\tsl{Co}_3.2$, it is difficult to conduct explicit checks of this expectation. In total, 
\begin{align}
    \mathbb{M}(\W_{\mathbb{Z}_{\mathrm{4A}}}) &= G = \mathbb{Z}_{\mathrm{4A}} & \mathbb{M}(\widetilde{\W}_{\mathbb{Z}_{\mathrm{4A}}}) &= \widetilde{G} = 4.2^{22}.\tsl{Co}_3 \nonumber \\
    \mathrm{Inn}(\W_{\mathbb{Z}_{\mathrm{4A}}}) &= G/Z(G) = \{1\} & \mathrm{Inn}(\widetilde{\W}_{\mathbb{Z}_{\mathrm{4A}}}) &= \widetilde{G}/Z(\widetilde{G}) = 2^{22}.\tsl{Co}_3\\
     \mathrm{Aut}(\W_{\mathbb{Z}_{\mathrm{4A}}})&= (G/Z(G)).H = \mathbb{Z}_2 &  \mathrm{Aut}(\widetilde{\W}_{\mathbb{Z}_{\mathrm{4A}}})& = (\widetilde{G}/Z(\widetilde{G})).H = 2^{22}.\tsl{Co}_3.2 \nonumber\\
     \mathrm{Fus}(\W_{\mathbb{Z}_{\mathrm{4A}}}) &= Z(G) = \mathbb{Z}_4 & \mathrm{Fus}(\widetilde{\W}_{\mathbb{Z}_{\mathrm{4A}}}) &= Z(\widetilde{G}) = \mathbb{Z}_4. \nonumber
\end{align}

We will now show that the characters of the $\widetilde{\W}_{\mathbb{Z}_{\mathrm{4A}}}(\ell,m)$ can be computed via Hecke operators. Let us order the characters of $\mc{P}(4)$ as
\begin{align}
\begin{split}\label{z4parafch}
  \chi_{0} &= \psi^{(4)}_{4,4}, \quad ~~ \chi_{1} = \psi^{(4)}_{1,1}, \quad
  \chi_{2} = \psi^{(4)}_{3,3}, \ \quad \chi_{3} = \psi^{(4)}_{2,2}, \quad
  \chi_{4} = \psi^{(4)}_{2,0}, \\             
  \chi_{5} &= \psi^{(4)}_{3,-1}, \quad \chi_{6} = \psi^{(4)}_{3,1}, \quad
  \chi_{7} = \psi^{(4)}_{4,-2}, \quad  \chi_{8} = \psi^{(4)}_{4,2}, \quad
  \chi_{9} = \psi^{(4)}_{4,0},
\end{split}
\end{align}
with highest weights $0$, $\tfrac{1}{16}$, $\tfrac{1}{16}$, $\tfrac{1}{12}$, $\tfrac13$, $\tfrac{9}{16}$, $\tfrac{9}{16}$, $\tfrac34$, $\tfrac34$, and $1$ respectively. The conductor of these characters is $N=48$, and the central charge of $\widetilde{\W}_{\mathbb{Z}_{\mathrm{4A}}}$ is related to the central charge of $\mc{P}(4)$ as $c_{\widetilde{t}}=23=23c_t$. Since $(23,48)=1$, one might guess that the components of $\mathsf{T}_{23}\chi$ can be identified with the graded-dimensions of the $\widetilde{\W}_{\mathbb{Z}_{\mathrm{4A}}}(\ell,m)$. This will indeed turn out to be the case. Using the modular S-matrix of the parafermion theories, one can compute the $q$-expansions of their Hecke images,
\begin{align}
\begin{split}\label{Co3}
  ({\mathsf T}_{23} \chi)_{0}(\tau) &= q^{-\frac{23}{24}}(1
  + 38226 q^2+3519529 q^3+126577280 q^4 + \cdots), \\
  ({\mathsf T}_{23} \chi)_{1}(\tau) &= q^{\frac{23}{48}}(2048
  +565248 q+31700992 q^2 + \cdots), \\
  ({\mathsf T}_{23} \chi)_{3}(\tau) &= q^{\frac{23}{24}}(48600 +5052456 q+192216888 q^2 +
  \cdots), \\
 ({\mathsf T}_{23} \chi)_{4}(\tau) &= q^{\frac{17}{24}}(11178 +1794069 q+82286847 q^2 +
 \cdots), \\
 ({\mathsf T}_{23} \chi)_{5}(\tau) &= q^{\frac{47}{48}}(47104 +4757504 q+178382848 q^2+
 \cdots), \\
 ({\mathsf T}_{23} \chi)_{7}(\tau) &= q^{\frac{7}{24}}(276 +135148 q+9192824 q^2+283316852
 q^3 + \cdots), \\
 ({\mathsf T}_{23} \chi)_{9}(\tau) &= q^{\frac{1}{24}}(23  + 37973 q+3521323 q^2+126567896
 q^3 + \cdots), \\
 ({\mathsf T}_{23} \chi)_{2}(\tau) &= ({\mathsf T}_{23} \chi)_{1}(\tau), \quad ({\mathsf
 T}_{23} \chi)_{6}(\tau) = ({\mathsf T}_{23} \chi)_{5}(\tau), \quad ({\mathsf T}_{23}
 \chi)_{8}(\tau) = ({\mathsf T}_{23} \chi)_{7}(\tau).
\end{split}
\end{align}
If these functions are to be consistently identified with the characters of $\widetilde{\W}_{\mathbb{Z}_{\mathrm{4A}}}$, it should be possible to bilinearly pair them with the $\chi_\alpha$ to produce the $J$ function. To find candidate matrices $G_{\ell}$ which implement this bilinear pairing (c.f.\ \S\ref{subsec:Hecke}), we solve the equation
\begin{align}
23 + \ell^2 = 0 \ \text{mod} \ 24.
\end{align}
One can find solutions for $\ell = 1,5,19,23$.\footnote{The integers $\ell = 7, 11, 13, 17$ also satisfy $23 + \ell^2 = 0 \ \text{mod} \ 24$. However, $G_7,$ $G_{11},$ $G_{13}$, and $G_{17}$ have negative entries and so we do not consider them.} It turns out that only two of them are independent, namely  $G_{1} = G_{5}$ and $G_{19}=G_{23}$. Furthermore, one can check that $G_{1}$ is the identity matrix. On the other hand, $G_{19}$ is the matrix which is zero in each entry, except for the entries $(0,0)$, $(2,1)$, $(1,2)$, $(3,3)$, $(4,4)$, $(5,6)$, $(6,5)$, $(7,8)$, $(8,7)$, and $(9,9)$ where it is 1. Because $\chi_2(\tau) = \chi_1(\tau)$, $\chi_6(\tau) = \chi_5(\tau)$, and $\chi_8(\tau) = \chi_7(\tau)$, $G_{1}$ and $G_{19}$ act identically on $\chi$. This motivates the identification $\widetilde{\chi}_\alpha(\tau) = (\mathsf{T}_{23}\chi)_\alpha(\tau)$ which one can check leads to the bilinear relation
\begin{align}
    J(\tau) = \sum_\alpha \chi_\alpha(\tau)  \widetilde{\chi}_\alpha(\tau) 
\end{align}
consistent with the decomposition in \eqref{Co3decomp}. As an independent check on this result, one can easily see that, to low orders in the $q$-expansion, the functions obtained by Hecke operators agree with the characters presented in \eqref{Co3chargeconjorbchars} from the description of $\widetilde{\W}_{\mathbb{Z}_{\mathrm{4A}}}$ as a charge conjugation orbifold. Explicitly,
\begin{align}
\begin{split}
    &(\mathsf{T}_{23}\chi)_0 = \xi_{\mathds{1}}^{(\widetilde L)}, \quad (\mathsf{T}_{23}\chi)_1 = \xi_{\sigma,1}^{(\widetilde L)}, \quad (\mathsf{T}_{23}\chi)_2 = \xi_{\sigma,2}^{(\widetilde L)}, \quad (\mathsf{T}_{23}\chi)_3 = \xi^{(\widetilde L)}_{\lambda^\ast}, \quad (\mathsf{T}_{23}\chi)_4 = \xi^{(\widetilde L)}_{2\lambda^\ast}, \\
    &(\mathsf{T}_{23}\chi)_5 = \xi^{(\widetilde L)}_{\tau,1}, \quad (\mathsf{T}_{23}\chi)_6 = \xi^{(\widetilde L)}_{\tau,2},\quad (\mathsf{T}_{23}\chi)_7 = \xi^{(\widetilde L)}_{3\lambda^\ast,1}, \  (\mathsf{T}_{23}\chi)_8 = \xi^{(\widetilde L)}_{3\lambda^\ast,2}, \  (\mathsf{T}_{23}\chi)_9 = \xi_j^{(\widetilde L)}.
    \end{split}
\end{align}

As a final consistency check, we may also confirm our earlier claim that the generator of the fusion algebra automorphism of $\mc{P}(4)$ lifts to an element of $\mathbb{M}$ in the 4A class. From the transformation rules in equation \eqref{Zk transformation rule}, the $\mathbb{Z}_4$
twined characters are
\begin{align}
\begin{split}
 \chi_{\omega,0} &= \psi^{(4)}_{4,4}, \quad \ \ \ \ \ \chi_{\omega,1} = i \psi^{(4)}_{1,1}, \quad \chi_{\omega,2} =-i \psi^{(4)}_{3,3},  \quad \chi_{\omega,3} = -\psi^{(4)}_{2,2},  \quad \chi_{\omega,4} = \psi^{(4)}_{2,0}, \\
  \chi_{\omega,5} &= -i \psi^{(4)}_{3,-1}, \quad \chi_{\omega,6} = i \psi^{(4)}_{3,1}, \quad \chi_{\omega,7} = - \psi^{(4)}_{4,-2} ,\quad \chi_{\omega,8} = -\psi^{(4)}_{4,2}, \quad  \chi_{\omega,9} = \psi^{(4)}_{4,0},
\end{split}
\end{align}
and one can check to low orders in the $q$-expansion that
\begin{align}
\begin{split}
  J_{\mathrm{4A}}(\tau)  &= 
  \frac{1}{q} +276 q +2048 q^2 +11202 q^3 +49152 q^4  +184024 q^5+ \cdots 
  \\  &=   \sum_\alpha \widetilde{\chi}_\alpha(\tau)  \chi_{\omega,\alpha}(\tau).
\end{split}
\end{align}
In summary, reverting back to the natural basis for parafermion characters indexed by pairs $(\ell,m)$, we have argued that we can identify the graded dimensions $ \mathrm{Tr}_{\widetilde{\W}_{\mathbb{Z}_{\mathrm{4A}}}(\ell,m)}q^{\widetilde{l}_0-\frac{23}{24}}$ for $(\ell,m) =$ $(4,4)$, $(3,-1)$, $(3,1)$, $(2,2)$, $(2,0)$, $(1,1)$, $(3,3)$, $(4,-2)$, $(4,2)$, $(4,0)$ with the functions $\widetilde{\chi}_\alpha$ for $\alpha = 0,\dots,9$ respectively.

\subsection{Dihedral monstralizers from McKay's $\widehat{E_8}$ correspondence}\label{subsec:mckaye8}

We now turn to monstrous deconstructions which arise from McKay's correspondence. In each subsection, we take as our known subVOA $\W_{D_{n\mathrm{X}}}$, the chiral algebra generated by two conformal vectors $e(z)$ and $f(z)$ of central charge $\frac12$ whose associated Miyamoto involutions $\tau_e$ and $\tau_f$ have product lying in the $n$X conjugacy class, where $n$X is either 1A, 2A, 3A, 4A, 5A, 6A, 2B, 4B, or 3C. These algebras have been studied extensively in e.g. \cite{miyaconf, sakuma2003vertex, lametal, Sakuma}, and we rely heavily on their results. In each case, we study the commutant subalgebra $\widetilde{\W}_{D_{n\mathrm{X}}}$, and in particular establish its symmetries and its dual characters in $V^\natural$. In each case, it will be clear from our character theoretic decomposition of $V^\natural$ that $\W_{D_{n\mathrm{X}}}$ and $\widetilde{\W}_{D_{n\mathrm{X}}}$ are each others' commutants. 

\subsubsection*{Symmetries}
Before treating each example in detail, we provide the sketch of an argument that
\begin{align}\label{dihedralsymmetries}
    \mathbb{M}(\widetilde{\W}_{D_{n\mathrm{X}}}) = \widetilde{G}, \ \ \ \mathrm{Inn}(\widetilde{\W}_{D_{n\mathrm{X}}}) = \widetilde{G}/Z(\widetilde{G}), \ \ \ \mathrm{Fus}(\widetilde{\W}_{D_{n\mathrm{X}}}) = Z(\widetilde{G}) ,
\end{align}
where $\widetilde{G}:=\mathrm{Cent}_{\mathbb{M}}(D_{n\mathrm{X}})$. It is essentially a recapitulation of the arguments given in \S\ref{subsubsec:BMe7} for the baby monster. The starting point is the fact that the subgroup of $\mathbb{M}$ which is preserved by $\W_{D_{n\mathrm{X}}}$ is $\mathbb{M}(\W_{D_{n\mathrm{X}}}) = D_{n\mathrm{X}}$. We then would like to argue that the centralizer of this group acts trivially on $\W_{D_{n\mathrm{X}}}$. To see this, we note that the dihedral group is generated by the Miyamoto involutions of $e$ and $f$, i.e.\ $G:= D_{n\mathrm{X}} = \langle\tau_e,\tau_f\rangle$, and so $\widetilde{G} := \mathrm{Cent}_{\mathbb{M}}(D_{n\mathrm{X}}) = \mathrm{Cent}_{\mathbb{M}}(\tau_e)\cap \mathrm{Cent}_{\mathbb{M}}(\tau_f)$. Because any $X$ in $\mathrm{Cent}_{\mathbb{M}}(\tau_e)\cong 2.\mathbb{B}$ stabilizes $e$ by the arguments presented in \S\ref{subsubsec:BMe7}, and similarly for $f$, any $X$ which lies in the intersection $\mathrm{Cent}_{\mathbb{M}}(\tau_e)\cap \mathrm{Cent}_{\mathbb{M}}(\tau_f)$ will stabilize both $e$ and $f$ simultaneously. Since $\W_{D_{n\mathrm{X}}}$ is generated by $e$ and  $f$ by definition, it follows that $\W_{D_{n\mathrm{X}}}$ is stabilized by $\widetilde{G}$, and thus $\widetilde{G}$ can at most act by diagonal fusion algebra automorphisms. It follows that $\mathbb{M}(\widetilde{\W}_{D_{n\mathrm{X}}}) = \widetilde{G}$. 

Because $(G,\widetilde{G})$ furnish a monstralizer pair, their centers agree and must act by diagonal fusion algebra automorphisms, as discussed in \S\ref{subsec:deconstruction}. Taking the quotient by this center thus leaves the inner automorphism group. This completes the justification of \eqref{dihedralsymmetries}.

\subsubsection*{Griess algebras}
In the following subsections, we will provide constructions of vector-valued modular forms which we tentatively identify with the characters of the VOAs $\widetilde \W_{D_{n\mathrm{X}}}$. One piece of evidence which supports these assignments is the fact that they solve the bootstrap problem presented in equation \eqref{characterpairing}. Another piece of evidence is that the Griess algebras that they lead to satisfy a consistency check which we now describe. 

The putative characters, along with the decomposition of $V^\natural$ by restriction into representations of $D_{n\mathrm{X}}\circ \mathrm{Cent}_{\mathbb{M}}(D_{n\mathrm{X}}) = D_{n\mathrm{X}}\circ\widetilde{D}_{n\mathrm{X}}$, uniquely determines a $\mathrm{Inn}(\widetilde{\W}_{D_{n\mathrm{X}}})\cong \widetilde{D}_{n\mathrm{X}}/Z(\widetilde{D}_{n\mathrm{X}})$ representation $R_{n\mathrm{X}}$ such that the Griess algebra of $\widetilde{\W}_{D_{n\mathrm{X}}}$ decomposes as $(\widetilde{\W}_{D_{n\mathrm{X}}})_2\cong \mathbf{1}\oplus R_{n\mathrm{X}}$; explicitly, these are
\begin{align}
\begin{split}
    ({\rm Inn}(\widetilde{\W}_{n\mathrm{X}}),R_{n\mathrm{X}}) &= (\mathbb{B},\mathbf{96255}), \ ({^2}\tsl{E}_6(2),\mathbf{48620}), \ (\tsl{Fi}_{23}, \mathbf{30888}),\ (\tsl{Th},\mathbf{30875}),\\
    & \hspace{-.5in}  (\tsl{HN},\mathbf{8910}\oplus\mathbf{9405}), \ (\tsl{Fi}_{22},\mathbf{3080}\oplus\mathbf{13650}), \ (\tsl{F}_4(2),\mathbf{1377}\oplus\mathbf{22932}).
\end{split}
\end{align}
Here, irreducible representations are labeled by their dimensions which in each case uniquely specifies the representation, and we have omitted examples where we do not have access to the character tables or are unsure of the characters. In the monster VOA, the existence of an $\mathbb{M}$-invariant inner product and $\mathbb{M}$-invariant algebra on the dimension two operators $V_2^\natural\cong \mathbf{1}\oplus\mathbf{196883}$ relies on the group theoretical fact that the symmetric square of the 196883-dimensional irreducible representation contains both the trivial and the 196883-dimensional representation itself as subrepresentations \cite{MN}. Consistency of the characters and groups we have proposed requires that a similar statement should be true in all our examples, namely it should be the case that ${\rm Sym}^2(R_{n\mathrm{X}})$ contains both the trivial representation as well as $R_{n\mathrm{X}}$, thus allowing an ${\rm Inn}(\widetilde{\W}_{D_{n\mathrm{X}}})$-invariant inner product and algebra. We have used GAP software to check that this is indeed the case in each of the examples listed above.

Let us now look more closely at each example.

\subsubsection{$(D_{\mathrm{1A}},2.\mathbb{B})${\normalfont:} Baby monster}\label{subsubsec:BM}

Two involutions combine to produce the identity element if and only if they are the same involution. Since 2A involutions are in one-to-one correspondence with conformal vectors of central charge $\frac12$, this implies that $\W_{D_{\mathrm{1A}}}$ is a degenerate case of the $\W_{D_{n\mathrm{X}}}$ construction in which the VOA is generated by just one Ising conformal vector, i.e.\ $\W_{D_{\mathrm{1A}}} \cong \mathcal{L}(\tfrac12,0)$. The commutant of $\W_{D_{\mathrm{1A}}}$ in $V^\natural$ is nothing but the baby monster VOA $\tsl{V}\mathbb{B}^\natural$, which was already considered in detail in \S\ref{subsubsec:BMe7}.

\subsubsection{$(D_{\mathrm{2A}},2^2.{^2}\normalfont{\tsl{E}}_6(2))${\normalfont:} Steinberg group}\label{subsubsec:Lietype}

We now consider the subVOA $\W_{D_{\mathrm{2A}}}$ generated by two central charge $\frac12$ conformal vectors whose associated Miyamoto involutions combine to yield an element of the 2A conjugacy class of $\mathbb{M}$. This central charge $c_t = \frac65$ VOA admits an $\mathcal{L}(\tfrac12,0)\otimes \mathcal{L}(\tfrac{7}{10},0)$ subalgebra, in terms of which its 8 irreducible modules decompose as (c.f.\ Theorem 5.2 of \cite{lam2000z2})
\begin{align}
    \begin{split}
    \W_{D_{\mathrm{2A}}}(0)&\cong [0,0]\oplus [\tfrac12,\tfrac32], \\
    \W_{D_{\mathrm{2A}}}(1)&\cong[0,\tfrac35]\oplus[\tfrac12,\tfrac{1}{10}], \\
    \W_{D_{\mathrm{2A}}}(2)&\cong[\tfrac{1}{16},\tfrac{7}{16}] \cong \W_{D_{\mathrm{2A}}}(3), \\
    \W_{D_{\mathrm{2A}}}(4)&\cong [\tfrac12,0]\oplus[0,\tfrac32], \\
    \W_{D_{\mathrm{2A}}}(5)&\cong [\tfrac{1}{16},\tfrac{3}{80}]\cong \W_{D_{\mathrm{2A}}}(6), \\
    \W_{D_{\mathrm{2A}}}(7)&\cong [\tfrac12,\tfrac35]\oplus[0,\tfrac{1}{10}],
    \end{split}
\end{align}
where $[h_1,h_2]:=\mathcal{L}(\tfrac12,h_1)\otimes \mathcal{L}(\tfrac{7}{10},h_2).$ In particular, this gives a prescription for writing its characters $\chi_\alpha(\tau)$ as sums of products of Ising and tricritical Ising characters, 
\begin{align}
\begin{split}
\label{character of c=6/5 CFT}
\chi_{0} &= \chi^{(3)}_{1,1} \chi^{(4)}_{1,1} + \chi^{(3)}_{2,1} \chi^{(4)}_{1,4}, \qquad
\chi_{1} = \chi^{(3)}_{1,1} \chi^{(4)}_{1,3} + \chi^{(3)}_{2,1} \chi^{(4)}_{1,2}, \\
\chi_{2} &= \chi^{(3)}_{1,1} \chi^{(4)}_{1,4} + \chi^{(3)}_{2,1} \chi^{(4)}_{1,1} , \qquad
\chi_3 = \chi_4 = \chi^{(3)}_{1,2} \chi^{(4)}_{2,1} , \\
\chi_5 &= \chi^{(3)}_{1,1} \chi^{(4)}_{1,2} + \chi^{(3)}_{2,1} \chi^{(4)}_{1,3} , \qquad
\chi_6 = \chi_{7} =  \chi^{(3)}_{1,2} \chi^{(4)}_{2,2}.
\end{split}
\end{align}
The resulting modular $S$-matrix is given by
\begin{align}
\label{S-matrix of c=6/5 CFT}
\mathcal{S} =
\begin{pmatrix}
\alpha_- & \alpha_+ & \alpha_- & \alpha_- & \alpha_- & \alpha_+ & \alpha_+ & \alpha_+ \\
\alpha_+ & -\alpha_- & \alpha_+ & \alpha_+ & \alpha_+ & -\alpha_- & -\alpha_- & -\alpha_- \\
\alpha_- & \alpha_+ & \alpha_- & - \alpha_- & -\alpha_- & \alpha_+ & -\alpha_+ & -\alpha_+ \\
\alpha_- & \alpha_+ & -\alpha_- & \alpha_- & -\alpha_- & -\alpha_+ & \alpha_+ & -\alpha_+ \\
\alpha_- & \alpha_+ & -\alpha_- & -\alpha_- & \alpha_- & -\alpha_+ & -\alpha_+ & \alpha_+ \\
\alpha_+ & -\alpha_- & \alpha_+ & -\alpha_+ & -\alpha_+ & -\alpha_- & \alpha_- & \alpha_- \\
\alpha_+ & -\alpha_- & -\alpha_+ & \alpha_+ & -\alpha_+ & \alpha_- & -\alpha_- & \alpha_- \\
\alpha_+ & -\alpha_- & -\alpha_+ & -\alpha_+ & \alpha_+ & \alpha_- & \alpha_- & -\alpha_-
\end{pmatrix}
,
\end{align}
where $\alpha_{\pm}=\frac{1}{2}\sqrt{\frac{5\pm \sqrt{5}}{10}}$, and $T$-matrix read
\begin{align}
\mathcal{T} =\text{diag} \left( e^{-\frac{\pi i}{10}},e^{-\frac{9\pi i}{10}},e^{\frac{9\pi i}{10}},e^{\frac{9\pi i}{10}},e^{\frac{9\pi i}{10}},e^{\frac{\pi i}{10}},e^{\frac{\pi i}{10}},e^{\frac{\pi i}{10}}\right).
\end{align}

Now, we would like to study the commutant $\widetilde{\W}_{D_{\mathrm{2A}}}$. We comment that, in addition to the general argument given at the beginning of \S\ref{subsec:mckaye8}, one can alternatively get a handle on its symmetry group by realizing it as a commutant subalgebra of the baby monster VOA. To do this, we note that $\widetilde{\W}_{D_{\mathrm{2A}}}$ can be obtained via an iterated deconstruction,
\begin{align}
\begin{split}
    V^\natural &\supset \mathcal{L}(\tfrac12,0)\otimes \tsl{V}\mathbb{B}^\natural \\
    &\supset \mathcal{L}(\tfrac12,0)\otimes \mathcal{L}(\tfrac{7}{10},0)\otimes \widetilde{\W}_{D_{\mathrm{2A}}}
    \end{split}
\end{align}
and so in particular, $\widetilde{\W}_{D_{\mathrm{2A}}}$ embeds into $\tsl{V}\mathbb{B}^\natural$. In fact, we can define it as 
\begin{align}\label{2E6(2).2 in B}
\begin{split}
    \widetilde{\W}_{D_{\mathrm{2A}}} = \tsl{Com}_{\tsl{V}\mathbb{B}^\natural}(\mathcal{L}(\tfrac{7}{10},0))
\end{split}
\end{align}
where the $\mathcal{L}(\tfrac{7}{10},0)$ subalgebra is chosen so that its stress tensor is of $\sigma$-type in $\tsl{V}\mathbb{B}^\natural$ (c.f. \S\ref{subsubsec:BMe7}). One can therefore define an involution $\sigma$ as in equation \eqref{sigma type 7/10} which lies in the 2A conjugacy class of $\mathbb{B}$. Using arguments similar to the ones we used in the case of the baby monster, we know that $\mathrm{Cent}_{\mathbb{B}}(\sigma)\cong \mathrm{Cent}_{\mathbb{B}}(\mathrm{2A})\cong 2.{^2}E_6(2).2$ \cite{wilson1999maximal} stabilizes the stress tensor of the $\mathcal{L}(\tfrac{7}{10},0)$ subalgebra. Indeed, the Griess algebra of the baby monster VOA is $\tsl{V}\mathbb{B}^\natural_2\cong \mathbf{1}\oplus \mathbf{96255}$ as a representation of $\mathbb{B}$, and a character-theoretic calculation shows that this decomposes into $2.{^2}E_6(2).2$ representations\footnote{Whenever a group has multiple irreducible representations of dimension $d$, we use the notation $\mathbf{d}_{(i)}$ to denote the $i$th irrep of dimension $d$, according to how they are ordered in Gap.} as
\begin{align}\label{Griess baby E6}
    \tsl{V}\mathbb{B}^\natural_2\Big\vert_{2.{^2}\tsl{E}_6(2).2} \cong \mathbf{1}_{(1)}\oplus \mathbf{1}_{(1)} \oplus \mathbf{1938}_{(1)}\oplus\mathbf{45696}_{(1)}\oplus \mathbf{48620}_{(1)}
\end{align}
where $\mathbf{1}_{(1)}$, $\mathbf{1938}_{(1)}$, $\mathbf{48620}_{(1)}$, and $\mathbf{45696}_{(1)}$ are the 1st, 3rd, 5th, and 192nd irreps of $2.{^2}E_6(2).2$ in the order in which they are recorded in the Gap library \cite{GAP}. The two-dimensional invariant subspace is spanned by the stress tensor of the $\mathcal{L}(\tfrac{7}{10},0)$ subalgebra as well as its commutant in $\tsl{V}\mathbb{B}^\natural$, which implies that the former is stabilized by $2.{^2}E_6(2).2$. As in the case of the baby monster VOA, the central order 2 element of $2.{^2}E_6(2).2$ acts trivially on $\widetilde{\W}_{D_{\mathrm{2A}}}$; this is the same as the statement that the map \eqref{authom} has a $\mathbb{Z}_2$ kernel. Therefore, we must quotient by this $\mathbb{Z}_2$ to get (a subgroup of) the automorphism group of $\widetilde{\W}_{D_{\mathrm{2A}}}$. This leads us to propose that ${^2}E_6(2).2\subset \mathrm{Aut}(\widetilde{\W}_{D_{\mathrm{2A}}})$. We will see momentarily that in fact some of these are outer automorphisms\footnote{The naive prediction from the monstralizer $[D_{\mathrm{2A}}\circ 2^2.{^2}\tsl{E}_6(2)].S_3$ is that the full automorphism group should be ${^2}\tsl{E}_6(2).S_3$ as opposed to ${^2}\tsl{E}_6(2).2$. However, if one decomposes the Griess algebra $V^\natural_2$ into representations of ${^2}\tsl{E}_6(2).S_3$ one does not find any singlets besides the usual stress tensor, which indicates that a deconstruction is not possible. This example is the reason why we say in general that $\mathrm{Aut}(\W_{\widetilde{G}}) = (\widetilde{G}/Z(\widetilde{G})).H'$ for $H'$ a subgroup of the group $H$ which appears in the monstralizer $[G\circ \widetilde{G}].H$. It is the only example we consider for which $H'\neq H$. }, and in particular that ${^2}\tsl{E}_6(2) =\mathrm{Inn}(\widetilde{\W}_{D_{\mathrm{2A}}})$.

Let us try to find evidence for these claims in the dual characters of $\widetilde{\W}_{D_{\mathrm{2A}}}$ in $V^\natural$. We will first describe these as the Hecke image of the characters of $\W_{D_{\mathrm{2A}}}$. The central charge $c_{\widetilde{t}}$ of $\widetilde{\W}_{D_{\mathrm{2A}}}$ is an integer multiple of the central charge $c_t$ of $\W_{D_{\mathrm{2A}}}$, $c_{\widetilde{t}} =24-c_t = 19 c_t$, where $c_t = \frac65$. Moreover, the conductor of the $\chi_\alpha$ is $N=20$, so we may consider applying the Hecke operator $\mathsf{T}_{19}$ to the characters $\chi_\alpha$ of $\W_{D_{\mathrm{2A}}}$, which can be computed using the modular S matrix in \eqref{S-matrix of c=6/5 CFT}. This yields the dual characters $\widetilde{\chi}_{\alpha}$ of $\widetilde{\W}_{D_{\mathrm{2A}}}$ in $V^\natural$, i.e.
\begin{align}
    \widetilde{\chi}_{\alpha}(\tau) = (\mathsf{T}_{19}\chi)_\alpha(\tau).
\end{align}
Explicitly, the $q$-expansions of the Hecke images are
\begin{align}
\begin{split}
\label{characters of c=114/5 CFT}
&\widetilde{\chi}_{0}(\tau) =  q^{-\frac{19}{20}}\left( 1 + 48621 q^2 + 4327402 q^3 + 152207784 q^4 + \cdots \right), \\
&\widetilde{\chi}_{1}(\tau) = q^{\frac{9}{20}}\left(1938 + 556206 q + 31485546 q^2 + 875268630 q^3 + \cdots \right), \\
&\widetilde{\chi}_{2}(\tau) = q^{\frac{11}{20}} \left( 2432 + 539904 q + 27826944 q^2 + 734726656 q^3 + \cdots \right), \\
&\widetilde{\chi}_{5}(\tau) = q^{\frac{19}{20}}\left(45696 + 4713216 q + 177241728 q^2 + 3893072640 q^3  + \cdots \right), \\
&\widetilde{\chi}_{3}(\tau) = \widetilde{\chi}_{4}(\tau) = \widetilde{\chi}_{2}(\tau), \quad \widetilde{\chi}_{6}(\tau) = \widetilde{\chi}_{7}(\tau) = \widetilde{\chi}_{5}(\tau). \\
\end{split}
\end{align}
Another way to obtain these characters is with a modular linear differential equation. Evidently the three irreducible modules of $\widetilde{\W}_{D_{\mathrm{2A}}}$ of highest weight $\frac32$ have identical characters, and the same is true of the three modules with highest weight $\frac{19}{10}$. Therefore, there are only 4 inequivalent characters, and we may therefore try to find the characters of the commutant algebra with a fourth order modular linear differential equation. The following MLDE has the dual characters as solutions,
\begin{align}
\begin{split}
\left[ \mathcal{D}^4 + \frac{1729 \pi ^2}{450} E_4(\tau) \mathcal{D}^2  -\frac{4159 i \pi ^3}{675} E_6(\tau) \mathcal{D} -\frac{35739 \pi ^4}{10000} E_4^2(\tau) \right] \widetilde{\chi}_{\alpha}(\tau) = 0.
\end{split}
\end{align}
As a consistency check we verify that these characters obey a bilinear relation with the partition function of the moonshine module,
\begin{align}
\label{bilinear m3m4}
\begin{split}
J(\tau) = \sum_{\alpha} \chi_\alpha(\tau)\widetilde{\chi}_{\alpha}(\tau)
\end{split}
\end{align}
which is consistent with the desired decomposition of the moonshine module in terms of its $\W_{D_{\mathrm{2A}}}\otimes \widetilde{\W}_{D_{\mathrm{2A}}}$ subalgebra, 
\begin{align}
    V^\natural \cong \bigoplus_\alpha \W_{D_{\mathrm{2A}}}(\alpha)\otimes \widetilde{\W}_{D_{\mathrm{2A}}}(\alpha).
\end{align}
Since $\widetilde{\W}_{D_{\mathrm{2A}}}$ appears as a subalgebra of $\tsl{V}\mathbb{B}^\natural$, it should also be possible on general grounds to decompose the modules/characters of the baby monster VOA into bilinears involving the characters of $\mathcal{L}(\tfrac{7}{10},0)$ and $\widetilde{\W}_{D_{\mathrm{2A}}}$, and indeed one can check that
\begin{align}
\begin{split}
    \chi_{\operatorname{\textsl{V}}\mathbb{B}^\natural(0)}(\tau) &= \chi^{(4)}_{1,1}(\tau) \widetilde{\chi}_{0}(\tau) + \chi^{(4)}_{1,3}(\tau) \widetilde{\chi}_{1}(\tau) + \chi^{(4)}_{1,4}(\tau) \widetilde{\chi}_{4}(\tau) + \chi^{(4)}_{1,2}(\tau) \widetilde{\chi}_{7}(\tau) ,\\
    \chi_{\operatorname{\textsl{V}}\mathbb{B}^\natural(1)}(\tau) &= \chi^{(4)}_{1,4}(\tau) \widetilde{\chi}_{0}(\tau) + \chi^{(4)}_{1,2}(\tau) \widetilde{\chi}_{1}(\tau) + \chi^{(4)}_{1,1}(\tau) \widetilde{\chi}_{4}(\tau) + \chi^{(4)}_{1,3}(\tau) \widetilde{\chi}_{7}(\tau),\\
    \chi_{\operatorname{\textsl{V}}\mathbb{B}^\natural(2)}(\tau) &= \chi^{(4)}_{2,1}(\tau)(\widetilde{\chi}_{2}(\tau)+\widetilde{\chi}_{3}(\tau)) +  \chi^{(4)}_{2,2}(\tau) (\widetilde{\chi}_{5}(\tau)+\widetilde{\chi}_{6}(\tau)),
\end{split}
\end{align}
which suggests that 
\begin{align}\label{baby e6 decomps}
\begin{split}
\tsl{V}\mathbb{B}^\natural(0)&\cong \langle 0,0\rangle \oplus \langle \tfrac35,1\rangle \oplus \langle \tfrac32,4\rangle \oplus \langle \tfrac{1}{10},7\rangle \\
\tsl{V}\mathbb{B}^\natural(1)&\cong \langle \tfrac32,0\rangle \oplus \langle\tfrac{1}{10},1\rangle \oplus \langle 0, 4\rangle \oplus \langle \tfrac35,7\rangle \\
\tsl{V}\mathbb{B}^\natural(2)&\cong \langle \tfrac{7}{16},2\rangle \oplus\langle\tfrac{7}{16},3\rangle \oplus \langle \tfrac{3}{80},5\rangle \oplus \langle \tfrac{3}{80},6\rangle
\end{split}
\end{align}
where we have introduced the notation $\langle h,\alpha\rangle := \mathcal{L}(\tfrac{7}{10},h)\otimes \widetilde{\W}_{D_{\mathrm{2A}}}(\alpha)$.

For concreteness, we illustrate how one can determine how the various coefficients of the characters decompose into irreducible representations of the inner/full automorphism groups. We will see in a moment that the modules $\widetilde{\W}_{D_{\mathrm{2A}}}(\alpha)$ for $\alpha = 0,$ $1,$ $4,$ $7$ are acted upon by the full automorphism group ${^2}\tsl{E}_6(2).2$, while the modules $\widetilde{\W}_{D_{\mathrm{2A}}}(\alpha)$ for $\alpha = 2,$ $3,$ $5,$ $6$ are in general permuted amongst each other by the outer automorphisms, and so are only representations of the inner automorphism group ${^2}\tsl{E}_6(2)$. For uniformity, we will actually work with the groups $2^2.{^2}\tsl{E}_6(2).2$ and $2^2.{^2}\tsl{E}_6(2)$ as opposed to ${^2}\tsl{E}_6(2).2$ and ${^2}\tsl{E}_6(2)$ because the latter pair are in general realized projectively on the modules of $\widetilde{\W}_{D_{\mathrm{2A}}}$, and all the projective representations of relevance to us are honest representations of the covering groups $2^2.{^2}\tsl{E}_6(2).2$ and $2^2.{^2}\tsl{E}_6(2)$. Similarly, we will work with the group $2.\mathbb{B}$ as opposed to $\mathbb{B}$. 

Then, the only possibility that is consistent with the decomposition of $\tsl{V}\mathbb{B}^\natural_2$ into $2.{^2}\tsl{E}_6(2).2$ irreps\footnote{Irreps of $2.{^2}\tsl{E}_6(2).2$ are the same as irreps of $2^2.{^2}\tsl{E}_6(2).2$ in which the central $\mathbb{Z}_2$ which is generated by the 2A conjugacy class acts trivially. In the Gap ordering, the first 320 irreps of $2^2.{^2}\tsl{E}_6(2).2$ map onto the irreps of $2.{^2}\tsl{E}_6(2).2$.} (equation \eqref{Griess baby E6}), the decomposition of $\tsl{V}\mathbb{B}^\natural(0)$ into $\mathcal{L}(\tfrac{7}{10},0)\otimes\widetilde{\W}_{D_{\mathrm{2A}}}$ modules (equation \eqref{baby e6 decomps}), and the characters of the $\widetilde{\W}_{D_{\mathrm{2A}}}(\alpha)$ (equation \eqref{characters of c=114/5 CFT}) is if
\begin{align}
    \widetilde{\W}_{D_{\mathrm{2A}}}(0)_2 \cong \mathbf{1}_{(1)}\oplus \mathbf{48620}_{(1)}, \ \ \ \widetilde{\W}_{D_{\mathrm{2A}}}(1)_{\frac75}\cong\mathbf{1938}_{(1)}, \ \ \ \widetilde{\W}_{D_{\mathrm{2A}}}(7)_{\frac{19}{10}}\cong \mathbf{45696}_{(1)}
\end{align}
as $2^2.{^2}\tsl{E}_6(2).2$ representations. Analogously, the fact that the $2.\mathbb{B}$ representation
\begin{align}
    \tsl{V}\mathbb{B}^\natural_3 \cong \mathbf{1}\oplus \mathbf{96255}\oplus \mathbf{9550635}
\end{align}
decomposes as
\begin{align}
\begin{split}
    \tsl{V}\mathbb{B}^\natural_3\Big\vert_{2^2.{^2}\tsl{E}_6(2).2} &\cong 2\cdot \mathbf{1}_{(1)} \oplus 2\cdot \mathbf{1938}_{(1)} \oplus \mathbf{48620}_{(1)} \oplus \mathbf{554268}_{(1)} \oplus \mathbf{1322685}_{(1)} \\
    & \hspace{-.25in} \oplus \mathbf{2956096}_{(1)} \oplus \mathbf{2432}_{(1)} \oplus 2\cdot \mathbf{45696}_{(1)} \oplus \mathbf{4667520}_{(1)}
\end{split}
\end{align}
determines that
\begin{align}
\begin{split}
   & \widetilde{\W}_{D_{\mathrm{2A}}}(0)_3 \cong \mathbf{1}_{(1)}\oplus\mathbf{48620}_{(1)}\oplus\mathbf{1322685}_{(1)}\oplus \mathbf{2956096}_{(1)}, \ \ \ \widetilde{\W}_{D_{\mathrm{2A}}}(4)_{\frac32} \cong \mathbf{2432}_{(1)},  \\
    & \ \ \ \  \widetilde{\W}_{D_{\mathrm{2A}}}(1)_{\frac{12}{5}}\cong \mathbf{1938}_{(1)} \oplus \mathbf{554268}_{(1)}, \ \ \  \widetilde{\W}_{D_{\mathrm{2A}}}(7)_{\frac{29}{10}}\cong \mathbf{45696}_{(1)}\oplus\mathbf{4667520}_{(1)}. 
\end{split}
\end{align}
One can continue this process to higher orders to obtain constraints on how the $\widetilde{\W}_{D_{\mathrm{2A}}}(\alpha)_h$ transform for $\alpha = 0,$ $1,$ $4$, $7$. 

On the other hand, if we attempt to restrict the $2.\mathbb{B}$ representation $\tsl{V}\mathbb{B}^\natural(2)_{\frac{31}{16}}\cong \mathbf{96256}$ to a $2^2.{^2}\tsl{E}_6(2).2$ representation, we find that it decomposes as 
\begin{align}
    \tsl{V}\mathbb{B}^\natural_{\frac{31}{16}}\Big\vert_{2^2.{^2}\tsl{E}_6(2).2}\cong \mathbf{4864}\oplus \mathbf{91392}.
\end{align}
This seems to contradict the fact that equation \eqref{baby e6 decomps} requires a decomposition of the form $96256 = 2432+2432+45696+45696.$ This can be resolved by further restricting to $2^2.{^2}\tsl{E}_6(2)$, in which case we find that the representations $\mathbf{4864}$ and $\mathbf{91392}$ split in half,
\begin{align}
    \tsl{V}\mathbb{B}^\natural_{\frac{31}{16}}\Big\vert_{2^2.{^2}\tsl{E}_6(2)}\cong \mathbf{2432}_{(2)}\oplus \mathbf{2432}_{(3)}\oplus \mathbf{45696}_{(2)}\oplus \mathbf{45696}_{(3)}.
\end{align}
The interpretation of this is that the extra involution in $2^2.{^2}\tsl{E}_6(2).2$ is acting as an outer automorphism which mixes $\widetilde{\W}_{D_{\mathrm{2A}}}(2)\leftrightarrow \widetilde{\W}_{D_{\mathrm{2A}}}(3)$ and $\widetilde{\W}_{D_{\mathrm{2A}}}(5)\leftrightarrow \widetilde{\W}_{D_{\mathrm{2A}}}(6)$, but the $2^2.{^2}\tsl{E}_6(2)$ subgroup preserves these modules and therefore acts by inner automorphisms, namely as
\begin{align}
\begin{split}
   & \ \ \  \widetilde{\W}_{D_{\mathrm{2A}}}(2)_{\frac32} \cong \mathbf{2432}_{(2)}, \ \ \ \widetilde{\W}_{D_{\mathrm{2A}}}(3)_{\frac32} \cong \mathbf{2432}_{(3)}, \\ &
   \widetilde{\W}_{D_{\mathrm{2A}}}(5)_{\frac{19}{10}} \cong \mathbf{45696}_{(2)}, \ \ \ \widetilde{\W}_{D_{\mathrm{2A}}}(6)_{\frac{19}{10}}\cong \mathbf{45696}_{(2)}.
   \end{split}
\end{align}
Continuing in this manner to one order higher, it is straightforward to see that the decomposition  
\begin{align}
\begin{split}
    \tsl{V}\mathbb{B}^\natural_{\frac{47}{16}}\Big\vert_{2^2.{^2}\tsl{E}_6(2)}&\cong 2\cdot \mathbf{2432}_{(2)}\oplus 2\cdot \mathbf{2432}_{(3)} \oplus 2\cdot \mathbf{45696}_{(2)} \oplus 2\cdot \mathbf{45696}_{(3)} \\
    & \ \ \ \ \ \ \oplus \mathbf{537472}_{(2)} \oplus \mathbf{537472}_{(3)} \oplus \mathbf{4667520}_{(2)}\oplus\mathbf{4667520}_{(3)}.
\end{split}
\end{align} 
 implies that 
\begin{align}
\begin{split}
   & \ \widetilde{\W}_{D_{\mathrm{2A}}}(2)_{\frac52} \cong \mathbf{2432}_{(2)}\oplus \mathbf{537472}_{(2)}, \ \ \ \widetilde{\W}_{D_{\mathrm{2A}}}(3)_{\frac52}\cong \mathbf{45696}_{(3)}\oplus\mathbf{4667520}_{(3)} \\
    & \widetilde{\W}_{D_{\mathrm{2A}}}(5)_{\frac{29}{10}}\cong \mathbf{45696}_{(2)}\oplus \mathbf{4667520}_{(2)}, \ \ \   \widetilde{\W}_{D_{\mathrm{2A}}}(6)_{\frac{29}{10}}\cong \mathbf{45696}_{(3)}\oplus \mathbf{4667520}_{(3)}
    \end{split}
\end{align}
To double check these proposed decompositions, one can check that the twined characters they lead to can be combined bilinearly with the characters of $\W_{D_{\mathrm{2A}}}$ to yield monstrous MT series (at least to low orders in the $q$-expansion),
\begin{align}
    J_g(\tau) = \sum_{\alpha} \chi_\alpha(\tau)\widetilde{\chi}_{g,\alpha}(\tau) \ \ \ \ \ \ \ \ (g\in 2^2.{^2}\tsl{E}_6(2))
\end{align}
where here we are implicitly using the fact that $2^2.{^2}\tsl{E}_6(2)$ is a subgroup of $\mathbb{M}$ on the left hand side of the equation.

\subsubsection{$(D_{\mathrm{3A}},\normalfont{\tsl{Fi}}_{23})${\normalfont:} Second largest Fischer group}\label{subsubsec:Fi23}

For our next example, we consider the commutant of $\W_{D_{\mathrm{3A}}}$, the VOA generated by two Ising conformal vectors whose Miyamoto involutions have product lying in the 3A conjugacy class of $\mathbb{M}$. We will see that the commutant $\widetilde{\W}_{D_{\mathrm{3A}}}$ admits an action by the second largest Fischer group $\operatorname{\textsl{Fi}}_{23}$, and for this reason we denote it with the symbol 
\begin{align}
  \tsl{VF}_{23}^\natural = \tsl{Com}_{V^\natural}(\W_{D_{\mathrm{3A}}}).  
\end{align} 
We will also realize $\tsl{VF}^\natural_{23}$ as a commutant subalgebra of $\operatorname{\textsl{VF}}^\natural_{24}$, which will allow us to leverage the relationship between the corresponding Fischer groups $\Fi_{23}$ and $\Fi_{24}$ and get a complementary perspective on its symmetry groups to the one obtained from the general argument at the beginning of \S\ref{subsec:mckaye8}.

The chiral algebra $\W_{D_{\mathrm{3A}}}$ of central charge $c_t = \frac{58}{35}$ has been studied in \cite{lametal,miyaconf,sakuma2003vertex}, where it was found that it contains a subalgebra of the form, 
\begin{align}\label{m=67 subalgebra}
    \mathcal{N}\otimes\mathcal{P}(3)\cong \mathcal{N}\otimes \left(\mathcal{L}(\tfrac45,0)\oplus \mathcal{L}(\tfrac45,3)\right)
\end{align}
where $\mathcal{N}:= \mathcal{L}(\tfrac67,0)\oplus\mathcal{L}(\tfrac67,5)$. It has 6 irreducible modules whose heighest weights have dimensions $h=(0,\frac17,\frac57,\frac25, \frac{19}{35},\frac{4}{35})$. They decompose into $\mc{N}\otimes\mc{P}(3)$ representations as (c.f.\ Theorem 4.9 of \cite{sakuma2003vertex})
\begin{align}
\begin{split}
    \W_{D_{\mathrm{3A}}}(0) &\cong \mathcal{P}(3)\otimes\mc{N}(0)\oplus\mc{P}(3,[0,2])\otimes\mc{N}(\tfrac43)^+\oplus \mc{P}(3,[3,1])\otimes\mc{N}(\tfrac43)^- \\
    \W_{D_{\mathrm{3A}}}(1) &\cong \mc{P}(3)\otimes\mc{N}(\tfrac17)\oplus \mc{P}(3,[0,2])\otimes\mc{N}(\tfrac{10}{21})^+\oplus \mc{P}(3,[3,1])\otimes\mc{N}(\tfrac{10}{21})^-  \\
    \W_{D_{\mathrm{3A}}}(2) &\cong \mc{P}(3)\otimes\mc{N}(\tfrac57)\oplus P(3,[0,2])\otimes\mc{N}(\tfrac{1}{21})^+\oplus \mc{P}(3,[3,1])\otimes \mc{N}(\tfrac{1}{21})^- \\
    \W_{D_{\mathrm{3A}}}(3) &\cong \mc{P}(3,[2,0])\otimes \mc{N}(0)\oplus \mc{P}(3,[1,1])\otimes\mc{N}(\tfrac43)^+ \oplus \mc{P}(3,[2,2])\otimes\mc{N}(\tfrac43)^-  \\
    \W_{D_{\mathrm{3A}}}(4) &\cong \mc{P}(3,[2,0])\otimes \mc{N}(\tfrac17) \oplus \mc{P}(3,[1,1])\otimes \mc{N}(\tfrac{10}{21})^+\oplus \mc{P}(3,[2,2])\otimes \mc{N}(\tfrac{10}{21})^- \\
    \W_{D_{\mathrm{3A}}}(5) &\cong  \mc{P}(3,[2,0])\otimes\mc{N}(\tfrac57)\oplus \mc{P}(3,[1,1])\otimes\mc{N}(\tfrac{1}{21})^+ \oplus \mc{P}(3,[2,2])\otimes\mc{N}(\tfrac{1}{21})^-\\
\end{split}
\end{align}
where the modules of $\mc{N}$ decompose into $\mc{L}(\tfrac67,0)$ representations as
\begin{align}
\begin{split}
    &\mc{N}(0) \cong \mc{L}(\tfrac67,0)\oplus \mc{L}(\tfrac67,5), \  \mc{N}(\tfrac17)\cong \mc{L}(\tfrac67,\tfrac17)\oplus \mc{L}(\tfrac67,\tfrac{22}{7}), \  \mc{N}(\tfrac57) \cong \mc{L}(\tfrac67,\tfrac57)\oplus \mc{L}(\tfrac67,\tfrac{12}{7}) \\
    & \mc{N}(\tfrac43)^\pm \cong \mc{L}(\tfrac67,\tfrac43)^\pm, \ \ \ \ \ \ \ \  \   \mc{N}(\tfrac{1}{21})^\pm \cong \mc{L}(\tfrac67,\tfrac{1}{21})^\pm, \ \ \ \ \ \ \  \ \ \mc{N}(\tfrac{10}{21})^\pm \cong \mc{L}(\tfrac67,\tfrac{10}{21})^\pm.
    \end{split}
\end{align}
Although $\mc{N}(h)^+$ and $\mc{N}(h)^-$ are isomorphic as $\mc{L}(\tfrac67,0)$ representations, they are distinct as $\mc{N}\cong \mc{L}(\tfrac67,0)\oplus \mc{L}(\tfrac67,5)$ representations. In particular, $\mc{N}^-\cong \sigma \circ \mc{N}^+$ (c.f.\ \S\ref{subsubsec:symmetries}) where $\sigma$ is the order 2 automorphism of $\mc{N}$ which acts trivially on $\mc{L}(\tfrac67,0)$ and as $-1$ on $\mc{L}(\tfrac67,5)$. For convenience, we provide expressions for the characters of $\W_{D_{\mathrm{3A}}}$ in terms of minimal model characters,
\begin{align}
\label{character of 5x6 CFT}
\begin{split}
\chi_0 &= \left( \chi_{1,1}^{(5)} + \chi_{1,5}^{(5)} \right) \left( \chi_{1,1}^{(6)} + \chi_{1,6}^{(6)} \right) + 2\chi_{1,3}^{(5)} \chi_{3,1}^{(6)}, \\
\chi_{1} &= \left( \chi_{1,1}^{(5)} + \chi_{1,5}^{(5)} \right) \left( \chi_{1,2}^{(6)} + \chi_{1,5}^{(6)} \right) + 2\chi_{1,3}^{(5)} \chi_{3,2}^{(6)}, \\
\chi_{2} &= \left( \chi_{1,1}^{(5)} + \chi_{1,5}^{(5)} \right) \left( \chi_{1,3}^{(6)} + \chi_{1,4}^{(6)} \right) + 2\chi_{1,3}^{(5)} \chi_{3,3}^{(6)}, \\
\chi_{3} &= \left( \chi_{2,1}^{(5)} + \chi_{2,5}^{(5)} \right) \left( \chi_{1,1}^{(6)} + \chi_{1,6}^{(6)} \right) + 2\chi_{2,3}^{(5)} \chi_{3,1}^{(6)}, \\
\chi_{4} &= \left( \chi_{2,1}^{(5)} + \chi_{2,5}^{(5)} \right) \left( \chi_{1,2}^{(6)} + \chi_{1,5}^{(6)} \right) + 2\chi_{2,3}^{(5)} \chi_{3,2}^{(6)}, \\
\chi_{5} &= \left( \chi_{2,1}^{(5)} + \chi_{2,5}^{(5)} \right) \left( \chi_{1,3}^{(6)} + \chi_{1,4}^{(6)} \right) + 2\chi_{2,3}^{(5)} \chi_{3,3}^{(6)}.
\end{split}
\end{align}
Now, the fact that $\W_{D_{\mathrm{3A}}}$ (whose stress tensor we write as $t$) has an $\mathcal{L}(\frac45,0)\otimes \mathcal{L}(\frac67,0)$ subalgebra means that there is a decomposition of the stress tensor of $V^\natural$ of the form 
\begin{align}
T(z) = t(z) + \widetilde{t}(z) = t^{(\frac{4}{5})}(z) + t^{(\frac67)}(z) + \widetilde{t}(z).
\end{align}
We can apply an iterated deconstruction to this decomposition, as in \eqref{iterated deconstruction}. The first iteration, $\tsl{Com}_{V^\natural}(\tsl{Vir}(t^{(\frac{4}{5})}))\cong \tsl{Com}_{V^\natural}(\mathcal{L}(\frac45,0)\oplus \mathcal{L}(\frac45,3))$, was computed in \S\ref{subsubsec:Fischere6} and is simply $\tsl{VF}^\natural_{24}$. The second iteration is then $\tsl{Com}_{\tsl{VF}_{24}^\natural}(\tsl{Vir}(t^{(\frac{6}{7})}))\cong \tsl{Com}_{\tsl{VF}_{24}^\natural}(\mathcal{L}(\frac67,0)\oplus \mathcal{L}(\frac67,5))$. Using \eqref{alternative iterated deconstruction}, this implies that 
\begin{align}\label{Fi23 in Fi24}
   \tsl{VF}_{23}^\natural \cong \operatorname{\textsl{Com}}_{V^\natural}\left(\W_{D_{\mathrm{3A}}}\right) \cong \operatorname{\textsl{Com}}_{\operatorname{\textsl{VF}}^\natural_{24}}\left(\mathcal{L}(\tfrac67,0)\oplus \mathcal{L}(\tfrac67,5)\right).
\end{align}
In other words, we have realized $\tsl{VF}_{23}^\natural$ as a commutant subalgebra of $\tsl{VF}^\natural_{24}$. We will argue for the symmetries of $\tsl{VF}_{23}^\natural$ using this description.  We start by noting that the centralizer $\mathrm{Cent}_{\operatorname{\textsl{Fi}}_{24}}(\sigma_\U)\cong \mathbb{Z}_2\times\Fi_{23}$ \cite{linton1991maximal} stabilizes $t^{(\frac67)}$. To see this, note that the Griess algebra $(\tsl{VF}_{24}^\natural)_2\cong \mathbf{1}\oplus \mathbf{57477}$ decomposes into $\mathbb{Z}_2\times \Fi_{23}$ representations as 
\begin{align}
    (\tsl{VF}_{24}^\natural)_2 \Big\vert_{\mathbb{Z}_2\times \tsl{Fi}_{23}}\cong \mathbf{1}_+\oplus \mathbf{1}_+\oplus \mathbf{782}_+\oplus\mathbf{30888}_+\oplus\mathbf{25806}_-
\end{align}
where the $+/-$ indicates how the $\mathbb{Z}_2$ acts. The two-dimensional trivial subspace is spanned by $t^{(\frac67)}$ and $\widetilde{t}$, and so $\mathbb{Z}_2\times\Fi_{23}$ acts by automorphisms on $\tsl{VF}^\natural_{23}$. However, $\sigma_\U$ acts trivially, and so after taking the quotient by this $\mathbb{Z}_2$, we get that the true automorphism group is simply $\Fi_{23}\subset \mathrm{Aut}(\tsl{VF}_{23}^\natural)$.

We move on to deriving the characters of $\tsl{VF}_{23}^\natural$. From the known modular matrices of minimal models,
one can show that the characters $\chi_\alpha$ transform under $S$ and $T$ as
\begin{align}
  \chi_\alpha(-\tfrac1\tau) = \sum_{\beta} \mathcal{S}_{\alpha\beta} \chi_\beta, \qquad
  \chi_\alpha(\tau+1) = \sum_{\beta} \mathcal{T}_{\alpha\beta} \chi_\beta(\tau).
\end{align}
where the S-matrix is
\begin{align}
\label{Fi23S}
    \mathcal{S}=
    \begin{pmatrix}
    \alpha_{-} \mbox{sin}\frac{\pi}{7}& \alpha_{-} \mbox{cos}\frac{3\pi}{14}
    & \alpha_{-} \mbox{cos}\frac{\pi}{14} & \alpha_{+} \mbox{sin}\frac{\pi}{7}
    & \alpha_{+} \mbox{cos}\frac{3\pi}{14} & \alpha_{+} \mbox{cos}\frac{\pi}{14} \\
    \alpha_{-} \mbox{cos}\frac{3\pi}{14} & -\alpha_{-} \mbox{cos}\frac{\pi}{14}
    & \alpha_{-} \mbox{sin}\frac{\pi}{7} & \alpha_{+} \mbox{cos}\frac{3\pi}{14}
    & -\alpha_{+} \mbox{cos}\frac{\pi}{14} & \alpha_{+} \mbox{sin}\frac{\pi}{7} \\
    \alpha_{-} \mbox{cos}\frac{\pi}{14} & \alpha_{-} \mbox{sin}\frac{\pi}{7}
    & - \alpha_{-} \mbox{cos}\frac{3\pi}{14} & \alpha_{+} \mbox{cos}\frac{\pi}{14}
    & \alpha_{+} \mbox{sin}\frac{\pi}{7} & - \alpha_{+} \mbox{cos}\frac{3\pi}{14}  \\
    \alpha_{+} \mbox{sin}\frac{\pi}{7} & \alpha_{+} \mbox{cos}\frac{3\pi}{14}
    & \alpha_{+} \mbox{cos}\frac{\pi}{14} & -\alpha_{-} \mbox{sin}\frac{\pi}{7}
    & -\alpha_{-} \mbox{cos}\frac{3\pi}{14} & -\alpha_{-} \mbox{cos}\frac{\pi}{14} \\
    \alpha_{+} \mbox{cos}\frac{3\pi}{14} & -\alpha_{+} \mbox{cos}\frac{\pi}{14}
    & \alpha_{+} \mbox{sin}\frac{\pi}{7} & - \alpha_{-} \mbox{cos}\frac{3\pi}{14}
    & \alpha_{-} \mbox{cos}\frac{\pi}{14} & -\alpha_{-} \mbox{sin}\frac{\pi}{7}\\
    \alpha_{+} \mbox{cos}\frac{\pi}{14} & \alpha_{+} \mbox{sin}\frac{\pi}{7}
    & -\alpha_{+} \mbox{cos}\frac{3\pi}{14} & -\alpha_{-} \mbox{cos}\frac{\pi}{14}
    & -\alpha_{-} \mbox{sin}\frac{\pi}{7} & \alpha_{-} \mbox{cos}\frac{3\pi}{14},
    \end{pmatrix},
\end{align}
with $\alpha_\pm = \sqrt{\frac{2}{35}(5\pm\sqrt{5})}$,
and the T-matrix reads
\begin{align}
    \mathcal{T} = \mbox{diag} \left( e^{-\frac{29 \pi i}{210}}, e^{\frac{31 \pi i}{210}},
    e^{-\frac{149 \pi i}{210}}, e^{\frac{139 \pi i}{210}}, e^{\frac{199 \pi i}{210}},
    e^{\frac{19 \pi i}{210} }\right).
\end{align}

Labeling the dual modules of $\tsl{VF}_{23}^\natural$ according to how they appear in the decomposition of $V^\natural$ with respect to its $\W_{D_{\mathrm{3A}}}\otimes \tsl{VF}^\natural_{23}$ subalgebra,
\begin{align}
    V^\natural \cong \bigoplus_\alpha \W_{D_{\mathrm{3A}}}(\alpha) \otimes \tsl{VF}_{23}^\natural(\alpha)
\end{align}
we claim that the characters $\chi_{\tsl{VF}^\natural_{23}(\alpha)}(\tau)$ are solutions to the $6$th-order MLDE,
\begin{align}
    \Big[ \mathcal{D}^6 + \mu_1 E_4 \mathcal{D}^4  + i \mu_2 E_6 \mathcal{D}^3
    + \mu_3 E_4^2 \mathcal{D}^2
    + i \mu_4 E_4 E_6 \mathcal{D} + \mu_5 E_4^3
    + \mu_6 E_6^2\Big] \chi_{\tsl{VF}^\natural_{23}(\alpha)} = 0,
\end{align}
with
\begin{align}
\begin{split}
    & \ \ \ \ \ \ \ \ \mu_1 = \frac{24679 \pi ^2}{6300}, \quad \mu_2 = \frac{2531831 \pi ^3}{231525}, \quad
    \mu_3 = -\frac{4706513771 \pi ^4}{277830000},  \\
    &\mu_4 = -\frac{7514276039 \pi ^5}{486202500}, \quad \mu_5 = \frac{2236826255833 \pi ^6}{583443000000},
    \quad \mu_6 = \frac{294432149791 \pi ^6}{107207651250}.
    \end{split}
\end{align}
The Fourier expansions of the six independent solutions to this equation are
\begin{align}
\begin{split}
\label{character of c=782/35}
\chi_{\tsl{VF}^\natural_{23}(0)}(\tau) &=  q^{-\frac{391}{420}}\left( 1 + 30889 q^2 + 2546974 q^3 + 85135558 q^4  + \cdots \right), \\
\chi_{\tsl{VF}^\natural_{23}(1)}(\tau) &= q^{\frac{389}{420}}\left( 25806 + 2622828 q + 96358822 q^2 + 2067752532 q^3 + \cdots \right), \\
\chi_{\tsl{VF}^\natural_{23}(2)}(\tau) &= q^{\frac{149}{420}} \left( 782 + 280347 q + 16687166 q^2 + 470844155 q^3 + \cdots \right), \\
\chi_{\tsl{VF}^\natural_{23}(3)}(\tau) &= q^{\frac{281}{420}}\left( 5083 + 817972 q + 36460359 q^2 + 877212478 q^3 + \cdots \right), \\
\chi_{\tsl{VF}^\natural_{23}(4)}(\tau) &= q^{\frac{221}{420}} \left( 3588 + 792948 q + 39982878 q^2 + 1031142072 q^3 + \cdots \right), \\
\chi_{\tsl{VF}^\natural_{23}(5)}(\tau) &= q^{\frac{401}{420}}\left( 60996 + 5926778 q + 213547709 q^2 + 4527955950 q^3 + \cdots \right).
\end{split}
\end{align}
As a consistency check, one can verify that, to low orders in the $q$-expansion, these characters pair with the characters of $\W_{D_{\mathrm{3A}}}$ to yield the partition function of $V^\natural$,
\begin{align}
\label{bilienar Fi23}
    \begin{split}
    J(\tau)   = \sum_{\alpha} \chi_\alpha(\tau) \chi_{\tsl{VF}^\natural_{23}(\alpha)}(\tau).
\end{split}
\end{align}
Furthermore, we see that the highest weight subspaces have dimensions which coincide with irreducible representations of $\Fi_{23}$, and that higher order terms decompose naturally, as described in Table \ref{decomposition by Fi23}.
\begin{table}
\begin{center}
\begin{tabular}{c|c|c}
\toprule
  $\alpha$   & $h$ & $\tsl{VF}_{23}^\natural(\alpha)_h$  \\\midrule
  $0$ & $0$ & $\mathbf{1}$ \\
      & $2$ & $\mathbf{1}\oplus \mathbf{30888}$ \\
      & $3$ & ${\bf 1} \oplus {\bf 30888} \oplus {\bf 279565}\oplus {\bf 2236520}$ \\
  $1$ & $\sfrac{13}{7}$ & $\mathbf{25806}$ \\
      & $\sfrac{20}{7}$ & ${\bf 25806} \oplus {\bf 274482} \oplus {\bf 2322540}$ \\
  $2$ & $\sfrac97$ & $\mathbf{782}$ \\
      & $\sfrac{16}{7}$ & ${\bf 782} \oplus {\bf 279565}$ \\
  $3$ & $\sfrac85$ & $\mathbf{5083}$ \\
      & $\sfrac{13}{5}$ & ${\bf 5083} \oplus {\bf 812889}$ \\
$4$  & $\sfrac{51}{35}$ & $\mathbf{3588}$ \\
     & $\sfrac{86}{35}$ & ${\bf 3588} \oplus {\bf 789360}$ \\
$5$ & $\sfrac{66}{35}$ & $\mathbf{60996}$ \\
    & $\sfrac{101}{35}$ & ${\bf 60996} \oplus {\bf 1951872}
  \oplus {\bf 3913910}$\\\bottomrule
\end{tabular}
\caption{Decompositions of the graded components of the modules $\tsl{VF}_{23}^\natural(\alpha)$ into irreducible representations of $\Fi_{23}$.}\label{decomposition by Fi23}
\end{center}
\end{table}
These decompositions can be derived by demanding consistency with the restriction of the graded components $V^\natural_n$ to $\Fi_{23}$, in a manner analogous to the previous section.

To summarize, the chiral algebra $\tsl{VF}_{23}^\natural$ has central charge $22{\sfrac{12}{35}}$ and $\Fi_{23}$ as (a subgroup of) its automorphism group, all of which acts by \emph{inner} automorphisms. It can be embedded into $\tsl{VF}^\natural_{24}$, 
\begin{align}
    \big(\mathcal{L}(\tfrac67,0)\oplus \mathcal{L}(\tfrac67,5)\big)\otimes \tsl{VF}^\natural_{23} \subset \tsl{VF}^\natural_{24}
\end{align}
or into the moonshine module as 
\begin{align}
   \W_{D_{\mathrm{3A}}}\otimes\tsl{VF}^\natural_{23}\subset  V^\natural
\end{align}
and so naturally fits into a chain of embeddings, 
\begin{align}
    \tsl{VF}^\natural_{23}\hookrightarrow \tsl{VF}^\natural_{24} \hookrightarrow V^\natural.
\end{align}
The algebra $\W_{D_{\mathrm{3A}}}$ also has a conformal vector of central charge $\frac12$ (c.f. Appendix A of \cite{lametal}), and so $\widetilde{t}(z)$ also fits into a decomposition of the stress tensor of $V^\natural$ of the form 
\begin{align}
    T(z) = t^{(\frac12)}(z) + t^{(\frac{81}{70})}(z) + \widetilde{t}(z).
\end{align}
If one performs an iterated deconstruction with respect to this decomposition, it follows that $\tsl{VF}^\natural_{23}$ also embeds into the baby monster VOA $\tsl{V}\mathbb{B}^\natural$. We will deconstruct this theory one step further to obtain a chiral algebra with $\Fi_{22}$ symmetry in \S\ref{subsubsec:Fi22}.

\subsubsection{$(D_{\mathrm{4A}},\normalfont{2^{1+22}.\tsl{McL}})${\normalfont:} McLaughlin group}\label{subsubsec:Mcl}

The 4A algebra $\W_{D_{\mathrm{4A}}}$ is most readily described as a \emph{charge conjugation orbifold} (c.f.\ \S\ref{sec:examples}) $\V_L^+$, i.e.\ the subspace of a lattice VOA which is fixed by its canonical involution $\theta$. The central charge $c_t=2$ algebra $\W_{D_{\mathrm{4A}}}$ corresponds \cite{lametal} to the specific choice 
\begin{align}\label{4A lattice}
    L = \left\{ \sum_{i=1}^8 a_i\epsilon_i~ \bigg \vert~ \text{all }a_i\in\mathbb{Z} \text{ or all }a_i\in\tfrac12 + \mathbb{Z} \text{ with } \sum_{i=1}^3 a_i = \sum_{i=4}^8a_i = 0     \right\}
\end{align}
where $\epsilon_i$ are any 8 vectors in $\mathbb{R}^8$ with $\langle\epsilon_i,\epsilon_j\rangle = 2\delta_{ij}$; it is generated by two vectors,
\begin{align}
  \vec v_1  = \frac12 \left( \sum_{i=1}^3 \epsilon_i +
  \sum_{j=4}^8 \epsilon_j \right) ,  \quad 
  \vec v_2 = \frac12 \left( - \sum_{i=1}^3 \epsilon_i +
  \sum_{j=4}^8 \epsilon_j \right).
\end{align}
This lattice can be realized as a sublattice of $\Lambda_{\mathrm{Leech}}$ \cite{hohnmason} and, following logic similar to that used in \S\ref{subsubsecCo3}, we claim that the commutant $\widetilde{\W}_{D_{\mathrm{4A}}}$ can be described explicitly as the canonical $\mathbb{Z}_2$ orbifold of the lattice VOA attached to the orthogonal complement $\widetilde{L}:=L^\perp$ of $L$ in $\Lambda_{\mathrm{Leech}}$. This description makes it transparent that $\widetilde{\W}_{D_{\mathrm{4A}}} \hookrightarrow \widetilde{W}_{\mathbb{Z}_{\mathrm{4A}}}$ in accordance with the fact that the associated monstralizers $(D_{\mathrm{4A}},2^{1+22}.\tsl{McL})$ and $(\mathbb{Z}_{\mathrm{4A}},4.2^{22}.\tsl{Co}_3)$ include into one another: indeed, this follows simply because the lattice on which $\widetilde{\W}_{D_{\mathrm{4A}}}$ is based is a sub-lattice of the one on which $\widetilde{\W}_{\mathbb{Z}_{\mathrm{4A}}}$ is based. We will also use this description to perform an explicit check on the characters produced by the Hecke method.

The Miyamoto involutions associated to the two Ising vectors which generate $\W_{D_{\mathrm{4A}}}$ can be lifted to automorphisms of $\W_{D_{\mathrm{4A}}}$ and $V^\natural$ which generate a dihedral subgroup $D_{\mathrm{4A}}$ of the monster $\mathbb{M}$. The centralizer of this subgroup is $\mathrm{Cent}_{\mathbb{M}}(D_{\mathrm{4A}}) \cong 2^{1+22}.\tsl{McL}$ \cite{anatomy}, and we conjecture that this example respects the symmetry properties which are predicted from its associated monstralizer, i.e.
\begin{align}
\begin{split}
    \mathbb{M}(\widetilde{\W}_{D_{\mathrm{4A}}}) &= 2^{1+22}.\tsl{McL} \\
    \mathrm{Inn}(\widetilde{\W}_{D_{\mathrm{4A}}}) &= 2^{22}.\tsl{McL} \\ \mathrm{Aut}(\widetilde{\W}_{D_{\mathrm{4A}}}) &= 2^{22}.\tsl{McL}.2  \\ 
    \mathrm{Fus}(\widetilde{\W}_{D_{\mathrm{4A}}}) &= \mathbb{Z}_2
\end{split}
\end{align}
Further evidence for these symmetry groups comes from the fact that the lattice $L$ is stabilized by a $\tsl{McL}$ subgroup of $\mathrm{Aut}(\Lambda_{\mathrm{Leech}})\cong \tsl{Co}_0$, and so we expect $\V_{\widetilde{L}}^+$ to admit an action by some extension of $\tsl{McL}$ by automorphisms.

As illustrated in \S\ref{sec:examples}, the characters of the modules $\W_{D_{\mathrm{4A}}}(\alpha)$ can be expressed in terms of the vector-valued theta function $\theta_{L+\lambda^\ast}(\tau)$ with $\lambda^\ast$ running over elements of the discriminant group $L^\ast/L$. We can compute this information using the data files published with \cite{hohnmason} and functions implemented in Magma \cite{MR1484478}. We find that the discriminant group is given by $L^\ast/L \cong \mathbb{Z}_{15}\cong \mathbb{Z}_3\times\mathbb{Z}_5$ and, using the notation $\Theta_{a,b}(\tau) := \theta_{L+a\gamma_3^\ast + b\gamma_5^\ast}(\tau)$ with $\gamma_3^\ast=(\vec v_1-\vec v_2)/3$ and $\gamma_5^\ast=(\vec v_1+\vec v_2)/5$ generators of the $\mathbb{Z}_3$ and $\mathbb{Z}_5$ respectively, we recover the $q$-expansions
\begin{align}
\begin{split}
\Theta_{0,0}(\tau) &= 1+4 q^2+2 q^3+2 q^5+8 q^8+6 q^{12}+4 q^{17}+4 q^{18}+\cdots, \\
\Theta_{0,1}(\tau) &= q^{\frac15}(1+2 q+5 q^2+2 q^3+3 q^4 + \cdots), \\
\Theta_{0,2}(\tau) &= q^{\frac45}(3 +q+2 q^2+4 q^3+4 q^4+q^5 + \cdots) ,\\
\Theta_{1,0}(\tau) &= q^{\frac13}(1+3 q+2 q^2+5 q^3+2 q^4 + \cdots), \\
\Theta_{1,1}(\tau) &= q^{\frac{2}{15}}(1+q+3 q^2+q^3+2 q^4+3 q^5+ \cdots), \\
\Theta_{1,2}(\tau) &= q^{\frac{8}{15}}(2 +q+2 q^2+q^3+3 q^4 + \cdots), \\
\Theta_{a,b}(\tau) &= \Theta_{3-a,b}(\tau) =\Theta_{a,5-b}(\tau)=\Theta_{3-a,5-b}(\tau).
\end{split}
\end{align}
The identifications on the last line follow from the fact that the dual lattice $L^\ast$ has a $\mathbb{Z}_2\times\mathbb{Z}_2$ symmetry group which preserves $L$ and induces the automorphisms $(a,b)\mapsto (-a,b)$ and $(a,b)\mapsto (a,-b)$ on the discriminant group $L^\ast/L$.
Now, one can construct the characters $\chi_\alpha(\tau)$ of the modules $\W_{D_{\mathrm{4A}}}(\alpha)$ in terms of these theta functions as below,
\begin{align}
\label{4A characters}
\begin{split}
\chi_0(\tau) &:= \xi_{\mathds{1}}^{(L)}(\tau) = \frac12\left(\frac{\Theta_{0,0}(\tau)}{\eta(\tau)^2} + \Phi_{0,1}(\tau)^2 \right), \ \ \chi_1(\tau) := \xi_j^{(L)}(\tau) =\frac12\left(\frac{\Theta_{0,0}(\tau)}{\eta(\tau)^2} - \Phi_{0,1}(\tau)^2 \right),  \\
\chi_2(\tau) &:= \xi_{\gamma_5^\ast}^{(L)}(\tau)= \frac{\Theta_{0,1}(\tau)}{\eta(\tau)^2}, \ \ \chi_3(\tau) := \xi_{2\gamma_5^\ast}^{(L)}(\tau) = \frac{\Theta_{0,2}(\tau)}{\eta(\tau)^2}, \ \ \chi_4(\tau) := \xi_{\gamma_3^\ast}^{(L)}(\tau) = \frac{\Theta_{1,0}(\tau)}{\eta(\tau)^2}, \\ 
\chi_5(\tau) &:= \xi^{(L)}_{\gamma_3^\ast+\gamma_5^\ast}(\tau) =\frac{\Theta_{1,1}(\tau)}{\eta(\tau)^2}, \quad  \chi_6(\tau) = \xi^{(L)}_{\gamma_3^\ast+4\gamma_5^\ast}(\tau) =\frac{\Theta_{1,4}(\tau)}{\eta(\tau)^2} , \\
\chi_7(\tau) &:=  \xi^{(L)}_{\gamma_3^\ast+2\gamma_5^\ast}(\tau) = \frac{\Theta_{1,2}(\tau)}{\eta(\tau)^2}, \quad \chi_8(\tau) := \xi^{(L)}_{\gamma_3^\ast+3\gamma_5^\ast}(\tau) = \frac{\Theta_{1,3}(\tau)}{\eta(\tau)^2},  \\ 
\chi_9(\tau) &:= \xi^{(L)}_\sigma(\tau) = \Phi_{1,0}(\tau)^2 + \Phi_{1,1}(\tau)^2, \quad \chi_{10}(\tau) := \xi_{\tau}^{(L)}(\tau)= \Phi_{1,0}(\tau)^2 - \Phi_{1,1}(\tau)^2.
\end{split}
\end{align}
Here, $\eta(\tau)$ is the Dedekind $\eta$ function and $\Phi_{\alpha,\beta}$ is a generalized theta function defined in equation \eqref{generalizedtheta}. We note that $\chi_5=\chi_6$ and $\chi_7=\chi_8$, which is a consequence of the fact that $\W_{D_{\mathrm{4A}}}$ inherits a $\mathbb{Z}_2\cong\mathrm{Aut}(L)/\mathbb{Z}_2$ outer automorphism from $L$ which exchanges these modules. 

It is straightforward to see that the behavior of the characters \eqref{4A characters} under modular transformations is described by the $S$-matrix
\begin{align}
\label{N3N5Smatrix}
\mathcal{S} =
    \begin{pmatrix}
    \frac{1}{2\sqrt{15}} & \frac{1}{2\sqrt{15}} & \frac{1}{\sqrt{15}} & \frac{1}{\sqrt{15}} & \frac{1}{\sqrt{15}} & \frac{1}{\sqrt{15}} & \frac{1}{\sqrt{15}} & \frac{1}{\sqrt{15}} & \frac{1}{\sqrt{15}} & \frac{1}{2} & \frac{1}{2} \\
   \frac{1}{2\sqrt{15}} & \frac{1}{2\sqrt{15}} & \frac{1}{\sqrt{15}} & \frac{1}{\sqrt{15}} & \frac{1}{\sqrt{15}} & \frac{1}{\sqrt{15}} & \frac{1}{\sqrt{15}} & \frac{1}{\sqrt{15}} & \frac{1}{\sqrt{15}} & -\frac{1}{2} & -\frac{1}{2} \\
    \frac{1}{\sqrt{15}} & \frac{1}{\sqrt{15}} & -\alpha_+ & \alpha_- & \frac{2}{\sqrt{15}} & \alpha_- & \alpha_- & -\alpha_+ &  -\alpha_+ & 0 & 0 \\
   \frac{1}{\sqrt{15}} & \frac{1}{\sqrt{15}} & \alpha_- & -\alpha_+ & \frac{2}{\sqrt{15}} & -\alpha_+ & -\alpha_+ & \alpha_- & \alpha_- & 0 & 0 \\
   \frac{1}{\sqrt{15}} & \frac{1}{\sqrt{15}} & \frac{2}{\sqrt{15}} & \frac{2}{\sqrt{15}} & -\frac{1}{\sqrt{15}} & -\frac{1}{\sqrt{15}} & -\frac{1}{\sqrt{15}} & -\frac{1}{\sqrt{15}} & -\frac{1}{\sqrt{15}} & 0 & 0 \\
    \frac{1}{\sqrt{15}} & \frac{1}{\sqrt{15}} & \alpha_- & -\alpha_+ & -\frac{1}{\sqrt{15}} & \beta_- & \beta_+ & \gamma_- & -\gamma_+ & 0 & 0 \\
    \frac{1}{\sqrt{15}} & \frac{1}{\sqrt{15}} & \alpha_- & -\alpha_+ & -\frac{1}{\sqrt{15}} & \beta_+ & \beta_- &  -\gamma_+& \gamma_- & 0 & 0 \\
    \frac{1}{\sqrt{15}} & \frac{1}{\sqrt{15}} & -\alpha_+ & \alpha_- &-\frac{1}{\sqrt{15}} & \gamma_- & -\gamma_+ &\beta_+ & \beta_-& 0 & 0 \\
    \frac{1}{\sqrt{15}} & \frac{1}{\sqrt{15}} & -\alpha_+ & \alpha_- &-\frac{1}{\sqrt{15}} & -\gamma_+ & \gamma_- &\beta_- & \beta_+& 0 & 0 \\
    \frac{1}{2} & -\frac{1}{2}  & 0 & 0 & 0 & 0 & 0 & 0 & 0 & \frac{1}{2} & -\frac{1}{2} \\
    \frac{1}{2} & -\frac{1}{2}  & 0 & 0 & 0 & 0 & 0 & 0 & 0 & -\frac{1}{2} & \frac{1}{2}
    \end{pmatrix},
\end{align}
where $\alpha_{\pm} = \sqrt{\frac{3\pm \sqrt{5}}{30}}, \beta_{\pm} =\frac{1+\sqrt{5}\pm\sqrt{30-6\sqrt{5}}}{4\sqrt{15}}$, and $\gamma_{\pm} = \frac{1}{2}\sqrt{\frac{1}{15}\left( 9 + \sqrt{5} \pm \sqrt{30 - 6\sqrt{5}} \right)}$. The $T$-matrix reads 
\begin{align}
\label{4A T matrix}
\mathcal{T} = \mbox{diag} \left( e^{-\frac{i \pi}{6}}, e^{-\frac{i \pi}{6}}, e^{\frac{7i \pi}{30}}, e^{-\frac{17i \pi}{30}}, e^{\frac{i \pi}{2}}, e^{\frac{i \pi}{10}},  e^{\frac{i \pi}{10}}, e^{\frac{9i \pi}{10}}, e^{\frac{9i \pi}{10}}, e^{\frac{i \pi}{12}} , e^{-\frac{11i \pi}{12}}\right).
\end{align}
One can check that these matrices satisfy the relations of $\tsl{PSL}_2(\mathbb{Z})$, i.e.\ $\mathcal{S}^2=1$ and $(\mathcal{ST})^3 = 1$. Moreover, applying the Verlinde formula yields a consistent fusion algebra, whose structure constants we present in appendix \ref{app:FA} (see also \cite{abe2001fusion,abe2005fusion}).

As we said earlier, we expect that the commutant $\widetilde{\W}_{D_{\mathrm{4A}}}$ is described by the $\mathbb{Z}_2$ orbifold $\V_{\widetilde{L}}^+$ of the $c=22$ lattice VOA associated with $\widetilde{L}$. To check this claim at the level of characters, we start by computing the vector-valued theta function of $\widetilde{L}$, using a similar notation $\widetilde{\Theta}_{a,b}(\tau) := \theta_{\widetilde{L}+a\lambda_3^\ast + b\lambda_5^\ast}(\tau)$ to the one used for the theta function of $L$,
\begin{align}
\label{4A dual ch}
\begin{split}
\widetilde{\Theta}_{0,0}(\tau) &= 1+44550 q^2+2525600 q^3+44995500 q^4+418427856 q^5+ \cdots\\
\widetilde{\Theta}_{0,1}(\tau) &= q^{\frac95}(15400 +1269675 q+26908200 q^2+278446300 q^3+ \cdots), \\
\widetilde{\Theta}_{0,2}(\tau) &= q^{\frac65}(275 +113400 q+4833675 q^2+73167600 q^3+ \cdots),\\
\widetilde{\Theta}_{1,0}(\tau) &= q^{\frac53}(7128 +779625 q+18824400 q^2+210097800 q^3+ \cdots), \\
\widetilde{\Theta}_{1,1}(\tau) &= q^{\frac{28}{15}}(22275 +1603800 q+32053725 q^2+319334400 q^3+\cdots),\\
\widetilde{\Theta}_{1,2}(\tau) &= q^{\frac{22}{15}}(2025 +356400 q+10758825 q^2+135432000 q^3+ \cdots),\\
\widetilde{\Theta}_{a,b}(\tau)&= \widetilde{\Theta}_{3-a,b}(\tau) = \widetilde{\Theta}_{a,5-b}(\tau) = \widetilde{\Theta}_{3-a,5-b}(\tau).
\end{split}
\end{align}
One can then find the dual characters $\widetilde{\chi}_\alpha(\tau)$ using these theta functions. Similarly to equation \eqref{4A characters}, the dual characters $\widetilde{\chi}_{\alpha}(\tau)$ are 
\begin{align}
\label{4A dual character}
\begin{split}
\widetilde{\chi}_0(\tau) &:= \xi^{(\widetilde L)}_{\mathds{1}}(\tau) =  \frac12\left(\frac{\widetilde{\Theta}_{0,0}(\tau)}{\eta(\tau)^{22}} +\Phi_{0,1}(\tau)^{22}\right), \\ 
\widetilde{\chi}_1(\tau) &:= \xi^{(\widetilde L)}_{j}(\tau) =  \frac12\left(\frac{\widetilde{\Theta}_{0,0}(\tau)}{\eta(\tau)^{22}} -\Phi_{0,1}(\tau)^{22}\right),  \\
\widetilde{\chi}_2(\tau) &:= \xi^{(\widetilde L)}_{\lambda_5^\ast}(\tau) =  \frac{\widetilde{\Theta}_{0,1}(\tau)}{\eta(\tau)^{22}}, \   \widetilde{\chi}_3(\tau) := \xi^{(\widetilde L)}_{2\lambda_5^\ast}(\tau) = \frac{\widetilde{\Theta}_{0,2}(\tau)}{\eta(\tau)^{22}}, \ 
\widetilde{\chi}_4(\tau) :=\xi^{(\widetilde L)}_{\lambda_3^\ast}(\tau)= \frac{\widetilde{\Theta}_{1,0}(\tau)}{\eta(\tau)^{22}}, \\ \widetilde{\chi}_5(\tau) &:= \xi^{(\widetilde L)}_{\lambda_3^\ast + \lambda_5^\ast}(\tau) = \frac{\widetilde{\Theta}_{1,1}(\tau)}{\eta(\tau)^{22}} , \quad  \widetilde{\chi}_6(\tau) = \xi^{(\widetilde L)}_{\lambda_3^\ast +4 \lambda_5^\ast}(\tau) = \frac{\widetilde{\Theta}_{1,4}(\tau)}{\eta(\tau)^{22}},\\
\widetilde{\chi}_7(\tau) &:=  \xi^{(\widetilde L)}_{\lambda_3^\ast +2 \lambda_5^\ast}(\tau) = \frac{\widetilde{\Theta}_{1,2}(\tau)}{\eta(\tau)^{22}}, \quad  \widetilde{\chi}_8(\tau):=\xi^{(\widetilde L)}_{\lambda_3^\ast +3 \lambda_5^\ast}(\tau) = \frac{\widetilde{\Theta}_{1,3}(\tau)}{\eta(\tau)^{22}}, \\ 
\widetilde{\chi}_9(\tau) &:= \xi^{(\widetilde L)}_{\sigma}(\tau)= 2^{10}(\Phi_{1,0}(\tau)^{22}+\Phi_{1,1}(\tau)^{22}), \\ \widetilde{\chi}_{10}(\tau) &:= \xi^{(\widetilde L)}_\tau(\tau)= 2^{10}(\Phi_{1,0}(\tau)^{22}-\Phi_{1,1}(\tau)^{22}).
\end{split}
\end{align}
With these conventions, the characters \eqref{4A characters} and \eqref{4A dual character} diagonally pair to produce the partition function of the monster CFT,
\begin{align}
\begin{split}
J(\tau) = \sum_{\alpha=0}^{10} \chi_{\alpha}(\tau) \widetilde{\chi}_{\alpha}(\tau)
\end{split}
\end{align}
which is consistent with a decomposition of the form 
\begin{align}
    V^\natural \cong \bigoplus_\alpha \W_{D_{\mathrm{4A}}}(\alpha)\otimes \widetilde{\W}_{D_{\mathrm{4A}}}(\alpha).
\end{align}

We would now like to argue that the characters of $\widetilde{\W}_{D_{\mathrm{4A}}}$ can alternatively be computed using Hecke operators. Let $c_{t}$ and $c_{\widetilde{t}}$ be the central charges of $\W_{D_{\mathrm{4A}}}$ and  $\widetilde{\W}_{D_{\mathrm{4A}}}$, respectively. The fact that $c_{\widetilde{t}}=11c_t$ suggests that we should apply the Hecke operator $\mathsf{T}_{11}$ to the characters $\chi$ in equation \eqref{4A characters}. Doing so recovers\footnote{In most examples, the bilinear which pairs $\chi_\alpha$ with its Hecke images to produce the $J$-function can be obtained as a linear combination of the matrices $G_\ell$ described in \S\ref{subsec:Hecke}. This is one of the few instances for which this is not true, i.e.\ the matrix which relates $\mathsf{T}_{11}\chi$ to $\widetilde{\chi}$ cannot be realized as a linear combination of the matrices $(G_\ell)^T$. This suggests that there are modular invariant ways to combine characters with their Hecke images beyond the bilinears studied in \cite{HarveyWu}, however we leave their study to future work.} the following $q$-expansions,
\begin{align}
\begin{split}
\widetilde{\chi}_0(\tau) &\equiv  ({\mathsf T}_{11} \chi)_{0}(\tau) = q^{-\frac{11}{12}}(1 + 22528 q^2 + 1753334 q^3 + 
 56418362 q^4 + \cdots) \\
\widetilde{\chi}_1(\tau) &\equiv  ({\mathsf T}_{11} \chi)_{1}(\tau) = q^{\frac{1}{12}}(22  + 22297 q + 1754896 q^2 + 56410563 q^3+ \cdots) \\
\widetilde{\chi}_2(\tau) &\equiv  ({\mathsf T}_{11} \chi)_{3}(\tau) = q^{\frac{53}{60}}(15400  + 1608475 q + 59076050 q^2 + \cdots) \\
\widetilde{\chi}_3(\tau) &\equiv  ({\mathsf T}_{11} \chi)_{2}(\tau) = q^{\frac{17}{60}}(275  + 119450 q + 7404100 q^2 + 
 211389200 q^3+ \cdots )\\
\widetilde{\chi}_4(\tau) &\equiv  ({\mathsf T}_{11} \chi)_{4}(\tau) = q^{\frac{3}{4}}(7128  + 936441 q + 37936350 q^2 + 856665315 q^3 + \cdots) \\
\widetilde{\chi}_5(\tau) &\equiv  ({\mathsf T}_{11} \chi)_{7}(\tau) = q^{\frac{19}{20}}(22275  + 2093850 q + 73462950 q^2  + \cdots)\\ 
\widetilde{\chi}_6(\tau) &\equiv  ({\mathsf T}_{11} \chi)_{8}(\tau) = q^{\frac{19}{20}}(22275  + 2093850 q + 73462950 q^2  + \cdots)\\ 
\widetilde{\chi}_7(\tau) &\equiv  ({\mathsf T}_{11} \chi)_{5}(\tau) = q^{\frac{11}{20}}(2025  + 400950 q + 19156500 q^2 + 
 475259400 q^3 + \cdots)\\
\widetilde{\chi}_8(\tau) &\equiv  ({\mathsf T}_{11} \chi)_{6}(\tau) = q^{\frac{11}{20}}(2025  + 400950 q + 19156500 q^2 + 
 475259400 q^3 + \cdots)\\
\widetilde{\chi}_9(\tau) &\equiv  ({\mathsf T}_{11} \chi)_{10}(\tau) = q^{\frac{23}{24}}(45056  + 4190208 q + 146161664 q^2  + \cdots)\\
\widetilde{\chi}_{10}(\tau) &\equiv  ({\mathsf T}_{10} \chi)_{9}(\tau) =  q^{\frac{11}{24}}(2048 + 518144 q + 26898432 q^2 + \cdots) \\ 
\end{split}
\end{align}
These $q$-series perfectly agree with the expressions in \eqref{4A dual ch}. 

Finally, we comment that $\V_L^+$ admits a $\V_{\sqrt{6}\mathbb{Z}}^+\otimes \V_{\sqrt{10}\mathbb{Z}}^+ \cong \mc{P}(4)\otimes \V_{\sqrt{10}\mathbb{Z}}^+$ subalgebra which makes some of its symmetries more manifest. In particular, the characters $\chi_\alpha$ in equation \eqref{4A characters} can be written in terms of $\psi^{(4)}_{\ell,m}$ and the characters of $\V_{\sqrt{10}\mathbb{Z}}$ (c.f.\  equation \eqref{DVVV ch} with $N=5$); this allows one to compute $\mathbb{Z}_4$ twined characters $\chi_{\omega,\alpha}(\tau)$ by dressing $\psi^{(4)}_{\ell,m}$ with the phase $e^{\frac{2\pi i m}{4}}$ wherever it appears in the decomposition of $\chi_\alpha$. It turns out that the twined characters $\chi_{\omega,\alpha}(\tau)$ and $\widetilde\chi_{\alpha}(\tau)$ combine bilinearly to produce the 4A McKay-Thompson series $J_{4\mathrm{A}}$ of $V^\natural$,
\begin{equation}  
\sum_{\alpha=0}^{10} \widetilde{\chi}_{\alpha}(\tau)  \chi_{\omega,\alpha}(\tau) = q^{-1} + 276 q + 2048 q^2 + 11202 q^3 + 49152 q^4 + 184024 q^5 + \cdots
\end{equation}
in harmony with the fact that $\W_{D_{\mathrm{4A}}}$ preserves a dihedral group $\mathbb{M}(\W_{D_{\mathrm{4A}}}) = D_{\mathrm{4A}}$ whose order 4 elements live in the 4A conjugacy class of $\mathbb{M}$.

\subsubsection{$(D_{\mathrm{5A}},\normalfont{\tsl{HN}})${\normalfont:} Harada-Norton group}\label{subsubsec:HN}

In this section, we study the commutant of $\W_{D_{\mathrm{5A}}}$, the subalgebra of $V^\natural$ generated by two conformal vectors $e$ and $f$ whose Miyamoto involutions $\tau_e$ and $\tau_f$ have product residing in the 5A conjugacy class. We claim that the centralizer of this product, $\mathrm{Cent}_{\mathbb{M}}(\tau_e\tau_f) \cong \mathbb{Z}_5\times \tsl{HN}$ \cite{conwaynorton}, stabilizes the conformal vector of $\W_{D_{\mathrm{5A}}}$, and that the $\mathbb{Z}_5\cong \langle\tau_e\tau_f\rangle$ acts trivially on $\widetilde{\W}_{D_{\mathrm{5A}}}$ so that at least $\tsl{HN}$ is the automorphism group. We therefore denote the commutant $\widetilde{\W}_{D_{\mathrm{5A}}}$ by $\tsl{VHN}^\natural$. We will in fact argue momentarily that the slightly larger group $\tsl{HN}.2$ acts by automorphisms, but that the extra $\mathbb{Z}_2$ is outer. 

We will use two different descriptions of $\W_{D_{\mathrm{5A}}}$. In the first, we will characterize $\W_{D_{\mathrm{5A}}}$ in terms of its $\mathcal{P}(5)\otimes \mathcal{P}(5)$ subalgebra. The fusion algebra of $\mathcal{P}(5)\otimes \mathcal{P}(5)$ has a $\mathbb{Z}_5\times \mathbb{Z}_5$ symmetry, and the generator of the diagonal $\mathbb{Z}_5$ lifts to $\tau_e\tau_f$; this is one way to see why the $\mathbb{Z}_5$ acts trivially on $\tsl{VHN}^\natural.$ Another description is in terms of a $\mathcal{L}(\tfrac12,0)\otimes \mathcal{L}(\tfrac{25}{28},0)\otimes \mathcal{L}(\tfrac{25}{28},0)$ subalgebra. This decomposes the stress tensor $t$ of $\W_{D_{\mathrm{5A}}}$ as $t = t^{(\frac12)}+t^{(\frac{25}{14})}$; performing an iterated deconstruction with respect to this decomposition makes it clear that the Harada-Norton VOA embeds into the baby monster VOA,
\begin{align}
\begin{split}
    V^\natural &\supset \tsl{Vir}(t^{(\frac12)})\otimes \tsl{V}\mathbb{B}^\natural \\
    &\supset \tsl{Vir}(t^{(\frac12)})\otimes \tsl{Vir}(t^{(\frac{25}{14})}) \otimes \tsl{VHN}^\natural.
    \end{split}
\end{align}
and that it is in fact a commutant subalgebra of $\tsl{V}\mathbb{B}^\natural$. We can therefore try to determine what automorphisms $\tsl{VHN}^\natural$ inherits from $\mathbb{B}$. The Griess algebra of the baby monster VOA decomposes into $\tsl{HN}$ irreps as 
\begin{align}
    \tsl{V}\mathbb{B}^\natural_2\Big\vert_{\tsl{HN}} \cong \mathbf{1}\oplus \mathbf{1}\oplus\mathbf{1}\oplus (\text{non-trivial reps})
\end{align}
while it decomposes into $\tsl{HN}.2$ (a maximal subgroup of $\mathbb{B}$, \cite{wilson1999maximal}) irreps as 
\begin{align}
    \tsl{V}\mathbb{B}^\natural_2\Big\vert_{\tsl{HN}.2} \cong \mathbf{1}\oplus \mathbf{1} \oplus (\text{non-trivial reps}).
\end{align}
This suggests that $\tsl{HN}$ stabilizes $\widetilde{t}$ and the two central charge $\frac{25}{28}$ conformal vectors individually, while $\tsl{HN}.2$ stabilizes $\widetilde{t}$ and $t^{(\frac{25}{14})}$ but swaps the two central charge $\frac{25}{28}$ conformal vectors. Under this assumption, $\tsl{HN}.2\subset \mathrm{Aut}(\tsl{VHN}^\natural).$ We will see that the extra $\mathbb{Z}_2$ is outer once we have expressions for the characters.

The algebra $\W_{D_{\mathrm{5A}}}$ has 9 irreducible modules, and they decompose into representations of its $\mathcal{L}(\tfrac12,0)\otimes \mathcal{L}(\tfrac{25}{28},0)\otimes \mathcal{L}(\tfrac{25}{28},0)$ subalgebra as (c.f. Theorem 3.19 of \cite{lametal})
\begin{align}
\begin{split}
    M(i,j) &\cong [0,h_{1,i},h_{1,j}]\oplus [0,h_{3,i},h_{5,j}]\oplus [0,h_{5,i},h_{3,j}]\oplus [0,h_{7,i},h_{7,j}] \\
    & \ \ \ \ \oplus [\tfrac12,h_{1,i},h_{7,j}]\oplus [\tfrac12,h_{3,i},h_{3,j}]\oplus [\tfrac12,h_{5,i},h_{5,j}]\oplus [\tfrac12,h_{7,i},h_{1,j}] \\
    & \ \ \ \ \oplus [\tfrac{1}{16},h_{2,i},h_{4,j}]\oplus [\tfrac{1}{16},h_{4,i},h_{2,j}]\oplus [\tfrac{1}{16},h_{6,i},h_{4,j}]\oplus [\tfrac{1}{16},h_{4,i},h_{6,j}]
\end{split}
\end{align}
where $h_{r,s}=h_{r,s}^{(7)}$ and $i,j = 1$, $3$, $5$. We will order these modules as 
\begin{align}
\begin{split}
    &\W_{D_{\mathrm{5A}}}(0) := M(0,0), \ \W_{D_{\mathrm{5A}}}(1) := M(1,5), \ \W_{D_{\mathrm{5A}}}(2) := M(5,1), \\
    &\W_{D_{\mathrm{5A}}}(3) := M(1,3), \ \W_{D_{\mathrm{5A}}}(4) := M(3,1), \ \W_{D_{\mathrm{5A}}}(5) := M(5,5), \\
    & \W_{D_{\mathrm{5A}}}(6) := M(3,5), \ \W_{D_{\mathrm{5A}}}(7) := M(5,3), \ \W_{D_{\mathrm{5A}}}(8) := M(3,3).
    \end{split}
\end{align}
On the other hand, we can also express the characters in terms of products of $\mc{P}(5)$ characters as below:
 \begin{align}
 \label{characterProductZ5}
 \begin{split}
     \chi_{0} & =  \psi^{(5)}_{5,5} \psi^{(5)}_{5,5}
     + \psi^{(5)}_{5,-3} \psi^{(5)}_{5,-1} + \psi^{(5)}_{5,3}  \psi^{(5)}_{5,1}+ \psi^{(5)}_{5,-1} \psi^{(5)}_{5,3}+ \psi^{(5)}_{5,1} \psi^{(5)}_{5,-3},
      \\
     \chi_{1}  & =
     \psi^{(5)}_{5,5} \psi^{(5)}_{4,0} + \psi^{(5)}_{4,-2} \psi^{(5)}_{5,-1} +\psi^{(5)}_{4,2} \psi^{(5)}_{5,1} + \psi^{(5)}_{5,-3} \psi^{(5)}_{1,1} +  \psi^{(5)}_{5,3} \psi^{(5)}_{4,4},
      \\
     \chi_{1}  & =
     \psi^{(5)}_{5,5} \psi^{(5)}_{4,0} + \psi^{(5)}_{5,1} \psi^{(5)}_{4,-2} + \psi^{(5)}_{5,-1} \psi^{(5)}_{4,2} + \psi^{(5)}_{4,4} \psi^{(5)}_{5,-3} + \psi^{(5)}_{1,1} \psi^{(5)}_{5,3},
      \\
     \chi_{3} &= \psi^{(5)}_{5,5} \psi^{(5)}_{2,0}
     + \psi^{(5)}_{5,-1} \psi^{(5)}_{2,2}+ \psi^{(5)}_{5,1} \psi^{(5)}_{3,3}  + \psi^{(5)}_{5,-3} \psi^{(5)}_{3,-1} + \psi^{(5)}_{5,3} \psi^{(5)}_{3,1} ,
      \\
     \chi_{4} &= \psi^{(5)}_{5,5} \psi^{(5)}_{2,0}+ \psi^{(5)}_{2,2} \psi^{(5)}_{5,1}+ \psi^{(5)}_{3,3} \psi^{(5)}_{5,-1} +  \psi^{(5)}_{3,1} \psi^{(5)}_{5,-3} + \psi^{(5)}_{3,-1}  \psi^{(5)}_{5,3} ,
      \\
     \chi_{5} &= \psi^{(5)}_{4,0} \psi^{(5)}_{4,0}
     + \psi^{(5)}_{1,1} \psi^{(5)}_{4,-2} + \psi^{(5)}_{4,4} \psi^{(5)}_{4,2} + \psi^{(5)}_{4,2} \psi^{(5)}_{1,1} + \psi^{(5)}_{4,-2} \psi^{(5)}_{4,4}  ,   \\
     \chi_{6} &= \psi^{(5)}_{2,0} \psi^{(5)}_{4,0}
     + \psi^{(5)}_{4,-2} \psi^{(5)}_{3,-1} + \psi^{(5)}_{4,2} \psi^{(5)}_{3,1} + \psi^{(5)}_{2,2} \psi^{(5)}_{1,1} +  \psi^{(5)}_{3,3} \psi^{(5)}_{4,4} ,
      \\
     \chi_{7} &=  \psi^{(5)}_{2,0} \psi^{(5)}_{4,0}
      + \psi^{(5)}_{3,-1} \psi^{(5)}_{4,2}  + \psi^{(5)}_{3,1} \psi^{(5)}_{4,-2} +\psi^{(5)}_{1,1}\psi^{(5)}_{3,3} + \psi^{(5)}_{4,4} \psi^{(5)}_{2,2} ,
      \\
     \chi_{8} &= \psi^{(5)}_{2,0} \psi^{(5)}_{2,0}
     + \psi^{(5)}_{2,2} \psi^{(5)}_{3,-1} + \psi^{(5)}_{3,3} \psi^{(5)}_{3,1}  + \psi^{(5)}_{3,-1} \psi^{(5)}_{3,3}  +\psi^{(5)}_{3,1} \psi^{(5)}_{2,2}.
\end{split}
\end{align}
One can then show from the known modular matrices of the minimal models that the characters of this theory transform according to
\begin{align}
\label{HN S-matrix}
\mathcal{S} = \frac{4}{7}
    \begin{pmatrix}
    s_1^2 & s_1 c_2 & s_1 c_2 & s_1 c_1 & s_1 c_1 & c_2^2 & c_1 c_2 & c_1 c_2 & c_1^2 \\
    s_1 c_2 & -s_1 c_1 & c_2^2 & s_1^2 & c_1 c_2 & -c_1 c_2 & s_1 c_2 & -c_1^2  & s_1 c_1 \\
    s_1 c_2 & c_2^2 & -s_1 c_1 & c_1 c_2 & s_1^2 & -c_1 c_2 & -c_1^2 & s_1 c_2 & s_1 c_1 \\
    s_1 c_1 & s_1^2 & c_1 c_2 & -s_1 c_2 & c_1^2 & s_1 c_2 & -c_2^2 & s_1 c_1 & -c_1 c_2 \\
    s_1 c_1 & c_1 c_2 & s_1^2 & c_1^2 & -s_1 c_2 & s_1 c_2 & s_1 c_1 & -c_2^2 & -c_1 c_2 \\
    c_2^2 & -c_1 c_2 & -c_1 c_2 & s_1 c_2 & s_1 c_2 & c_1^2 & -s_1 c_1 & -s_1 c_1 & s_1^2 \\
    c_1 c_2 & s_1 c_2 & -c_1^2 & -c_2^2 & s_1 c_1 & -s_1 c_1 & c_1 c_2 & s_1^2 & -s_1 c_2 \\
    c_1 c_2 & -c_1^2 & s_1 c_2 & s_1 c_1 & -c_2^2 & -s_1 c_1 & s_1^2 & c_1 c_2 & -s_1 c_2 \\
    c_1^2 & s_1 c_1 & s_1 c_1 & -c_1 c_2 & -c_1 c_2 & s_1^2 & -s_1 c_2 & -s_1 c_2 & c_2^2
    \end{pmatrix},
\end{align}
and
\begin{align}
    \mathcal{T} =  \mbox{diag}\Big(e^{-\frac{4\pi i}{21}},e^{-\frac{10\pi i}{21}},e^{-\frac{10\pi i}{21}},
    e^{\frac{8\pi i}{21}},e^{\frac{8\pi i}{21}},e^{-\frac{16\pi i}{21}},
    e^{\frac{2\pi i}{21}},e^{\frac{2\pi i}{21}},e^{\frac{20\pi i}{21}}\Big),
\end{align}
where $s_1 = \mbox{sin}\left( \frac{\pi}{7} \right)$,
$c_1 = \mbox{cos}\left( \frac{\pi}{14} \right)$ and
$c_2 = \mbox{cos}\left( \frac{3\pi}{14} \right)$.

The nine characters in \eqref{characterProductZ5}
obey the bilinear relation
\begin{align}
\begin{split}
J(\tau)  &= \sum_{\alpha} \chi_\alpha(\tau)\chi_{\tsl{VHN}^\natural(\alpha)}(\tau),
\end{split}
\end{align}
where we claim that the dual characters of $\tsl{VHN}^\natural:= \widetilde{\W}_{D_{\mathrm{5A}}}$ in $V^\natural$ are solutions to the MLDE below,
\begin{align}
\label{MLDE HN}
\begin{split}
\Big[ E_6(\tau) \mathcal{D}^6 &+ \mu_1 E_4(\tau) E_6(\tau) \mathcal{D}^4  + i \mu_2 E_6^2(\tau) \mathcal{D}^3 + \mu_3 E_4^2(\tau) E_6(\tau) \mathcal{D}^2 \\
&+ i \mu_4 E_4(\tau) E_6^2(\tau) \mathcal{D} + \mu_5 E_4^3(\tau) E_6(\tau) + \mu_6 E_6^3(\tau) \\
&+ i \mu_7 E_4^2(\tau) \mathcal{D}^5 + i \mu_8 E_4^3(\tau) \mathcal{D}^3 + i \mu_9 E_4^4(\tau) \mathcal{D} \Big]\chi_{\tsl{VHN}^\natural(\alpha)}(\tau) = 0, \\
\mu_1 =& \frac{241 \pi ^2}{63}, \ \mu_2 = -\frac{13891 \pi ^3}{1323}, \ \mu_3 = -\frac{215315 \pi ^4}{27783}, \ \mu_4 = -\frac{7902740 \pi ^5}{583443}, \\
\mu_5 =& 0, \ \mu_6 = \frac{181913600 \pi ^6}{85766121}, \ \mu_7 = \pi, \ \mu_8 = \frac{241 \pi^3}{63}, \ \mu_9 = - \frac{215315 \pi^5}{27783}.
\end{split}
\end{align}
This suggests a diagonal decomposition of the moonshine module of the form 
\begin{align}
    V^\natural \cong \bigoplus_\alpha \W_{D_{\mathrm{5A}}}(\alpha)\otimes \tsl{VHN}^\natural(\alpha).
\end{align}
The $q$-expansions of the characters of $\tsl{VHN}^\natural$ are given by
\begin{align}
\label{HN characters}
\begin{split}
&\chi_{\tsl{VHN}^\natural(0)}(\tau) = q^{-\frac{19}{21}}(1 + 18316 q^{2}+1360096 q^3+42393826 q^4 + \cdots),\\
&\chi_{\tsl{VHN}^\natural(1)}(\tau) = q^{\frac{5}{21}}(133 +65968 q+4172476 q^2+119360584 q^3 + \cdots), \\
&\chi_{\tsl{VHN}^\natural(2)}(\tau) = q^{\frac{5}{21}}(133 +65968 q+4172476 q^2+119360584 q^3 + \cdots), \\
&\chi_{\tsl{VHN}^\natural(3)}(\tau) = q^{\frac{17}{21}}(8778 +1003408 q+37866696 q^2 + \cdots ), \\
&\chi_{\tsl{VHN}^\natural(4)}(\tau) = q^{\frac{17}{21}}(8778 +1003408 q+37866696 q^2 + \cdots ), \\
&\chi_{\tsl{VHN}^\natural(5)}(\tau) = q^{\frac{8}{21}}(760 +231705 q+12595936 q^2+333082540 q^3 + \cdots), \\
&\chi_{\tsl{VHN}^\natural(6)}(\tau) = q^{\frac{20}{21}}(35112 +3184818 q+108781232 q^2 + \cdots), \\
&\chi_{\tsl{VHN}^\natural(7)}(\tau) = q^{\frac{20}{21}}(35112 +3184818 q+108781232 q^2 + \cdots), \\
&\chi_{\tsl{VHN}^\natural(8)}(\tau) = q^{\frac{11}{21}}(3344 +680504 q+32364068 q^2+795272512 q^3 + \cdots).
\end{split}
\end{align}
The lowest-order coefficients of (\ref{HN characters}),
$133,$ $760,$ $3344,$ $8778$, and $35112$,
are indeed dimensions of irreducible
representations of the Harada-Norton group $\tsl{HN}$. However, 133 is for example not the dimension of an irreducible representation of $\tsl{HN}.2$. This implies that the extra involution in $\tsl{HN}.2$ mixes the modules of $\tsl{VHN}^\natural$ and is therefore an outer automorphism. Higher order
coefficients of the characters also have decompositions into
irreducible representations of $\tsl{HN}$, as in Table \ref{decomposition by HN}; one can check that they are consistent with the decomposition of the graded components $V_n^\natural$ into irreducible representations of $\tsl{HN}$. 
\begin{table}
\begin{center}
\begin{tabular}{c|c|c}
\toprule
  $\alpha$   & $h$ & $\tsl{VHN}^\natural(\alpha)_h$  \\\midrule
  $0$ & $0$ & $\mathbf{1}$ \\
      & $2$ & $\mathbf{1}\oplus \mathbf{8910}\oplus\mathbf{9405}$ \\
  $1$ & $\sfrac{8}{7}$ & $\mathbf{133}_{(1)}$ \\
      & $\sfrac{15}{7}$ & $\mathbf{133}_{(1)}\oplus \mathbf{65835}_{(1)}$ \\
  $2$ & $\sfrac{8}{7}$ & $\mathbf{133}_{(2)}$ \\
      & $\sfrac{15}{7}$ & $\mathbf{133}_{(2)}\oplus \mathbf{65835}_{(2)}$ \\
   $3$ & $\sfrac{12}{7}$ & $\mathbf{8778}_{(1)}$ \\
   & $\sfrac{19}{7}$ & $\mathbf{8778}_{(1)}\oplus\mathbf{8910}\oplus \mathbf{267520}\oplus\mathbf{718200}_{(1)}$ \\
   $4$ & $\sfrac{12}{7}$ & $\mathbf{8778}_{(2)}$ \\
   & $\sfrac{19}{7}$ & $\mathbf{8778}_{(2)}\oplus\mathbf{8910}\oplus \mathbf{267520}\oplus\mathbf{718200}_{(2)}$ \\
   $5$ & $\sfrac97$ & $\mathbf{760}$ \\
   & $\sfrac{16}{7}$ & $\mathbf{760}\oplus \mathbf{16929}\oplus\mathbf{214016}$ \\
   $6$ & $\sfrac{13}{7}$ & $\mathbf{35112}_{(1)}$ \\
   $7$ & $\sfrac{13}{7}$ & $\mathbf{35112}_{(2)}$ \\
   $8$ & $\sfrac{10}{7}$ & $\mathbf{3344}$ \\
   & $\sfrac{17}{7}$ & $\mathbf{3344}\oplus \mathbf{270864}\oplus\mathbf{406296}$\\
  \bottomrule
\end{tabular}
\caption{Decompositions of the graded components of the modules $\tsl{VHN}^\natural(\alpha)$ into irreducible representations of $\HN$.}\label{decomposition by HN}
\end{center}
\end{table}

To illustrate an example of a twined bilinear, one can verify our earlier claim that the lift of the diagonal of the $\mathbb{Z}_5\times \mathbb{Z}_5$ fusion algebra automorphism of $\mc{P}(5)\times\mc{P}(5)$ lifts to an element in the 5A conjugacy class of $\mathbb{M}$. Indeed, if we denote the generator of this diagonal $\mathbb{Z}_5$ by $\omega$, then one can compute that the twined characters of $\W_{D_{\mathrm{5A}}}$ are
\begin{align}
\label{HN twined Z5}
\begin{split}
    \chi_{\omega,0}(\tau) &= q^{-\frac{2}{21}}(1 + q^{2}
     +2 q^{3}+3 q^{4}+4 q^{5}+6 q^{6}+ \cdots), \\
    \chi_{\omega,1}(\tau) &= q^{\frac{16}{21}}(1+q+2 q^2
    +q^3+3 q^4+3 q^5 + \cdots), \\
    \chi_{\omega,2}(\tau) &= q^{\frac{16}{21}}(1+q+2 q^2
    +q^3+3 q^4+3 q^5 + \cdots), \\
    \chi_{\omega,3}(\tau) &= q^{\frac{4}{21}}(2 +2 q+3 q^2+5 q^3
     +8 q^4+13 q^5 + \cdots), \\
    \chi_{\omega,4}(\tau) &= q^{\frac{4}{21}}(2 +2 q+3 q^2+5 q^3
     +8 q^4+13 q^5 + \cdots), \\
     \chi_{\omega,5}(\tau) &= q^{\frac{13}{21}}(-1-2 q-3 q^2
    -5 q^3-8 q^4-11 q^5 + \cdots),  \\
     \chi_{\omega,6}(\tau) &= q^{\frac{1}{21}}(-1-q-2 q^2
    -3 q^3-5 q^4-7 q^5+ \cdots), \\
     \chi_{\omega,7}(\tau) &= q^{\frac{1}{21}}(-1-q-2 q^2
    -3 q^3-5 q^4-7 q^5+ \cdots), \\
    \chi_{\omega,8}(\tau) &= q^{\frac{31}{21}}(1+q+3 q^2+3 q^3 +6 q^4+8 q^5 + \cdots),
    \end{split}
    \end{align}
and one can show by direct calculation that
\begin{align}
     \sum_{\alpha} \chi_{\omega,\alpha}(\tau) \chi_{\tsl{VHN}^\natural(\alpha)} (\tau)
    = \frac{1}{q} + 134q +  760 q^2 +3345 q^3 + 12256 q^4 +39350 q^5 + \cdots 
\end{align}
agrees with the Mckay-Thompson series of class 5A of the monster group to low orders in its $q$-expansion. The inner automorphisms of $\tsl{VHN}^\natural$ will also lift to corresponding automorphisms of $\mathbb{M}$, 
\begin{align}
\label{HN bilinear}
    J_{g}(\tau) = \sum_\alpha \chi_{\alpha}(\tau)\chi_{g,\tsl{VHN}^\natural(\alpha)}(\tau), \ \ \ \ \ \ \ \  (g\in\tsl{HN})
\end{align}
where for $J_g(\tau)$, we are implicitly using the fact $\tsl{HN}$ is a subgroup of $\mathbb{M}$. One can construct a bilinear \eqref{HN bilinear} for each of the 54 conjugacy classes in $\tsl{HN}$ using Table \ref{HN Fusion table}.

To summarize, $\tsl{VHN}^\natural$ is a central charge $21\sfrac57$ chiral algebra with $\tsl{HN}$ as its inner automorphism group, and $\tsl{HN}.2$ as its full automorphism group. It embeds into both the moonshine module and the baby monster VOA, 
\begin{align}
    \tsl{VHN}^\natural\hookrightarrow \tsl{V}\mathbb{B}^\natural\hookrightarrow V^\natural.
\end{align}

\subsubsection{$(D_{\mathrm{6A}},2.\normalfont{\tsl{Fi}}_{22})${\normalfont:} Third largest Fischer group}\label{subsubsec:Fi22}

We now study the commutant of $\W_{D_{\mathrm{6A}}}$ \cite{lametal,Dong:2019ihr}, the algebra generated by two conformal vectors of central charge $\frac12$ whose associated Miyamoto involutions have product residing in the 6A conjugacy class of $\mathbb{M}$. The commutant $\widetilde{\W}_{\mathrm{6A}}$ we will argue has an automorphism group which at least contains $\Fi_{22}.2$, and so we denote it with the symbol $\tsl{VF}_{22}^\natural$.  

In addition to using the argument provided at the beginning of \S\ref{subsec:mckaye8}, one can derive (part of) the symmetry group by realizing $\tsl{VF}_{22}^\natural$ as a commutant subalgebra of $\tsl{VF}_{23}^\natural$ and using the fact that $2.\Fi_{22} \cong \mathrm{Cent}_{\Fi_{23}}(\mathrm{2A})$. The idea is that the stress tensor $t$ of $\W_{D_{\mathrm{6A}}}$ can be decomposed as 
\begin{align}
    t(z) = t^{(\frac45)}(z) + t^{(\frac67)}(z) + t^{(\frac{25}{28})}(z).
\end{align}
If one iteratively deconstructs these conformal vectors as in \eqref{iterated deconstruction}, one gets 
\begin{align}\label{Fi22 iterated}
\begin{split}
    V^\natural &\supset \tsl{Vir}(t^{(\frac45)}) \otimes \tsl{VF}^\natural_{24} \\ 
    &\supset \tsl{Vir}(t^{(\frac45)}) \otimes \tsl{Vir}(t^{(\frac67)}) \otimes \tsl{VF}_{23}^\natural  \\
    &\supset \tsl{Vir}(t^{(\frac45)})\otimes \tsl{Vir}(t^{(\frac67)})\otimes \tsl{Vir}(t^{(\frac{25}{28})})\otimes \tsl{VF}^\natural_{22}
    \end{split}
\end{align}
so that $\mathcal{L}(\tfrac{25}{28},0)\otimes \tsl{VF}^\natural_{22}\subset \tsl{VF}^\natural_{23}.$ The $\mathcal{L}(\tfrac{25}{28},0)$ chiral algebra has a $\mathbb{Z}_2$ automorphism of its fusion algebra which lifts to an element in the 2A conjugacy class of $\Fi_{23}$. The centralizer of this element is $2.\Fi_{22}$ and fixes $t^{(\frac67)}$, with the central $\mathbb{Z}_2$ acting trivially on $\tsl{VF}^\natural_{22}$. After taking the quotient by this $\mathbb{Z}_2$, one is left with $\Fi_{22}$. When we have the $q$-expansions of the dual characters of $\tsl{VF}_{22}^\natural$ in $V^\natural$, we will see that this $\Fi_{22}$ is in fact the \emph{inner} automorphism group. We turn to this now.

Although $\W_{D_{\mathrm{6A}}}$ and its irreducible modules can be decomposed into representations of its $\mc{L}(\tfrac45,0)\otimes \mc{L}(\tfrac67,0)\otimes\mc{L}(\tfrac{25}{28},0)$ subalgebra \cite{Dong:2019ihr}, we will instead describe it --- using the block-diagonalization method presented at the end of \S\ref{subsubsec:modules} --- in terms of a $\mc{P}(2)\otimes \mc{P}(3)\otimes\mc{P}(6)$ subalgebra which makes some of its symmetries more manifest. From the known modular S-matrices of the parafermion theories, one can check that the modules specified by the 14 characters $\chi_{\alpha}(
\tau)$ presented in equation \eqref{m236 characters} form a consistent extension of $\mc{P}(2)\otimes\mc{P}(3)\otimes\mc{P}(6).$ As in previous subsections, our goal is to find 14 functions to serve as candidates for the dual characters, i.e.\ 14 functions $\chi_{\tsl{VF}^\natural_{22}(\alpha)}(\tau)$ which satisfy
\begin{align}
\label{Bilinear k236}
\begin{split}
J(\tau) &= \sum_{\alpha} \chi_{\alpha}(\tau) \chi_{\tsl{VF}_{22}^\natural(\alpha)}(\tau).
\end{split}
\end{align}
Unfortunately, neither MLDEs nor Hecke operators are effective at recovering these functions: MLDEs are unwieldy because the required differential equation would be high order, and the coefficient functions belong to very large dimensional vector spaces of modular forms. On the other hand, the Hecke operator method does not work for this example because $c_{\widetilde{t}}/c_t$ is not an integer. Nevertheless, we claim\footnote{See Appendix \ref{app:C} for an alternative derivation of these characters.} that the functions can be obtained from the Rademacher sum (c.f.\ \S\ref{subsubsec:Rademacher}) attached to the representation $\rho$ generated by the complex conjugates of the modular matrices presented in Appendix \ref{app:C} with singular part of the form $R^P_{\tsl{SL}_2(\mathbb{Z}),\rho,\alpha}(\tau) = \delta_{\alpha,0}q^{-\frac{143}{160}}+O(q^0)$. After some computation, we arrive at the following $q$-series for $\chi_{\tsl{VF}_{22}^\natural(\alpha)}(\tau)$,
\begin{align}
\label{Fi22 dual characters}
\begin{split}
\chi_{\tsl{VF}_{22}^\natural(0)}(\tau) &= q^{-\frac{143}{160}}(1 +16731 q^2+1188616 q^3+35978085 q^4 + \cdots), \\
\chi_{\tsl{VF}_{22}^\natural(1)}(\tau) &= q^{\frac{137}{160}}(10725 +1097304 q+39095342 q^2 + 808423473 q^3 +  \cdots), \\
{\chi}_{\tsl{VF}_{22}^\natural(2)}(\tau) &= q^{\frac{57}{160}}(429 +131560 q+7147074 q^2 + 187368324 q^3 + \cdots), \\
{\chi}_{\tsl{VF}_{22}^\natural(3)}(\tau) &= q^{\frac{97}{160}}(1001 +163449 q+7054047 q^2 + 163126535 q^3 + \cdots), \\
{\chi}_{\tsl{VF}_{22}^\natural(4)}(\tau) &= q^{\frac{153}{160}}(30030 +2625337 q+87378642 q^2 + 1733413968 q^3 + \cdots), \\
{\chi}_{\tsl{VF}_{22}^\natural(5)}(\tau) &= q^{\frac{73}{160}}(1430 +333762 q+16444857 q^2 + 409283732 q^3 + \cdots), \\
{\chi}_{\tsl{VF}_{22}^\natural(6)}(\tau) &= q^{\frac{33}{160}}(78 +43758 q+2787213 q^2 + 79431495 q^3 + \cdots), \\
{\chi}_{\tsl{VF}_{22}^\natural(7)}(\tau) &=q^{\frac{113}{160}} (3003 +401478 q+15980965 q^2 + 352878240 q^3 +  \cdots), \\
{\chi}_{\tsl{VF}_{22}^\natural(8)}(\tau) &= q^{\frac{19}{20}}(13728 +1212288 q+40511328 q^2 + 805723776 q^3 +  \cdots), \\
{\chi}_{\tsl{VF}_{22}^\natural(9)}(\tau) &= q^{\frac{19}{20}}(13728 +1212288 q+40511328 q^2 + 805723776 q^3 + \cdots), \\
{\chi}_{\tsl{VF}_{22}^\natural(10)}(\tau) &= q^{\frac{13}{40}}(352 +123552 q+6918912 q^2 + 184499744 q^3 +  \cdots), \\
{\chi}_{\tsl{VF}_{22}^\natural(11)}(\tau) &= q^{\frac{37}{40}}(27456 +2517216 q+85543392 q^2 + 1718827968 q^3 + \cdots), \\
{\chi}_{\tsl{VF}_{22}^\natural(12)}(\tau) &= q^{\frac{11}{20}}(2080 +384384 q+17407104 q^2 + 413523968 q^3 + \cdots), \\
{\chi}_{\tsl{VF}_{22}^\natural(13)}(\tau) &= q^{\frac{11}{20}}(2080 +384384 q+17407104 q^2 + 413523968 q^3 + \cdots).
\end{split}
\end{align}
As a consistency check, we note that all the leading coefficients of the dual characters $\chi_{\tsl{VF}^\natural_{22}(\alpha)}$ are dimensions of \emph{irreducible} representations of $2.\tsl{Fi}_{22}$, and that these functions satisfy equation \eqref{Bilinear k236}.

We end this section by analyzing twined characters and their associated bilinears. The fact that $\W_{D_{\mathrm{6A}}}$ admits a $\mc{P}(2)\otimes\mc{P}(3)\otimes\mc{P}(6)$ subalgebra gives the false impression that the theory has some kind of $\mathbb{Z}_2\times\mathbb{Z}_3\times\mathbb{Z}_6$ symmetry, in apparent contradiction with the assertion that $\mathbb{M}(\W_{D_{\mathrm{6A}}})=D_{\mathrm{6A}}$. We will show, however, that a $\mathbb{Z}_6$ subgroup of this group conspires to act trivially. 

Twined characters are dressed by three phases $\omega_k^{m_k}$ where $\omega_k \equiv e^{\frac{2 \pi i}{k}}$ and $k=2,3,6$. Therefore, there are a priori 36 ways to twine the characters $\chi_\alpha(\tau)$ using the transformation rules from \eqref{Zk transformation rule}. We will denote twined characters as $\chi_{\alpha;\omega_2^{m_2},\omega_3^{m_3},\omega_6^{m_6}}(\tau)$. For instance, $\chi_{\alpha;1,1,\omega_6^2}(\tau)$ means we dress the $\mc{P}(6)$ characters with their $\mathbb{Z}_6$ phase $\omega_6^2$ while leaving the $\mc{P}(2)$ and $\mc{P}(3)$ characters unmodified. Combined with the dual characters \eqref{Fi22 dual characters}, each twined character generates a Mckay-Thompson series corresponding to a certain conjugacy class in the monster. Table \ref{8 bilinears of Fi22} reveals that the $\mathbb{Z}_6$ subgroup generated by $\omega_2;\omega_3;\omega_6$ acts trivially, which reconciles the Miyamoto lifts of the fusion algebra automorphisms of the parafermion theories $\mc{P}(2)$, $\mc{P}(3)$, $\mc{P}(6)$ with the observed dihedral group $\mathbb{M}(\W_{D_{\mathrm{6A}}}) = D_{\mathrm{6A}}$.

\begin{table}[h]
\centering
\scalebox{1.0}{
\begin{tabular}{ c c c c c c c c c c c}
\hline
    $1;\omega_3;1$ & $1;\omega_3^2;1$ & $\omega_2;1;1$ & $\omega_2;\omega_3;1$ & $\omega_2;\omega_3^2;1$ \\
\hline
  $\mathrm{3A}$ & $\mathrm{3A}$ & $\mathrm{2A}$ & $\mathrm{6A}$ & $\mathrm{6A}$ &  \\
\hline
    $1;1;\omega_6$ & $1;1;\omega_6^2$ & $1;1;\omega_6^3$ & $1;1;\omega_6^4$ & $1;1;\omega_6^5$ & \\
\hline
  $\mathrm{6A}$ & $\mathrm{3A}$ & $\mathrm{2A}$ & $\mathrm{3A}$ & $\mathrm{6A}$ &  \\
\hline
   $\omega_2;1;\omega_6$ & $\omega_2;1;\omega_6^2$ & $\omega_2;1;\omega_6^3$ & $\omega_2;1;\omega_6^4$ & $\omega_2;1;\omega_6^5$ & \\
\hline
 $\mathrm{3A}$ & $\mathrm{6A}$ & $\mathrm{1A}$ & $\mathrm{6A}$ & $\mathrm{3A}$ \\
\hline
    $1;\omega_3;\omega_6$ & $1;\omega_3;\omega_6^2$ & $1;\omega_3;\omega_6^3$ & $1;\omega_3;\omega_6^4$ & $1;\omega_3;\omega_6^5$ \\
\hline
 $\mathrm{2A}$ & $\mathrm{3A}$ & $\mathrm{6A}$ & $\mathrm{1A}$ & $\mathrm{6A}$  \\
\hline
    $1;\omega_3^2;\omega_6$ & $1;\omega_3^2;\omega_6^2$ & $1;\omega_3^2;\omega_6^3$ & $1;\omega_3^2;\omega_6^4$ & $1;\omega_3^2;\omega_6^5$ \\
\hline
 $\mathrm{6A}$ & $\mathrm{1A}$ & $\mathrm{6A}$ & $\mathrm{3A}$ & $\mathrm{2A}$  \\
\hline
    $\omega_2;\omega_3;\omega_6$ & $\omega_2;\omega_3;\omega_6^2$ & $\omega_2;\omega_3;\omega_6^3$ & $\omega_2;\omega_3;\omega_6^4$ & $\omega_2;\omega_3;\omega_6^5$ \\
\hline
 $\mathrm{1A}$ & $\mathrm{6A}$ & $\mathrm{3A}$ & $\mathrm{2A}$ & $\mathrm{3A}$   \\
\hline
    $\omega_2;\omega_3^2;\omega_6$ & $\omega_2;\omega_3^2;\omega_6^2$ & $\omega_2;\omega_3^2;\omega_6^3$ & $\omega_2;\omega_3^2;\omega_6^4$ & $\omega_2;\omega_3^2;\omega_6^5$ \\
\hline
 $\mathrm{3A}$ & $\mathrm{2A}$ & $\mathrm{3A}$ & $\mathrm{6A}$ & $\mathrm{1A}$  \\
\hline
\end{tabular}
}
\caption{In this table, $\omega_2^{a};\omega_3^{b};\omega_6^{c}$ refers to the corresponding fusion algebra automorphism of the $\mc{P}(2)\otimes\mc{P}(3)\otimes\mc{P}(6)$ subalgebra. This lifts to an element of the monster, whose conjugacy class is determined by a character theory calculation. This table shows that the $\mathbb{Z}_6$ group generated by $\omega_2;\omega_3;\omega_6$ acts trivially on $V^\natural$. \label{8 bilinears of Fi22}}
\end{table}

\subsubsection{$(D_{\mathrm{4B}},2.\normalfont{\tsl{F}}_4(2))${\normalfont:} Chevalley group}

The chiral algebra $\W_{4\mathrm{B}}$ is most easily described as an extension of its $\mathcal{L}(\tfrac12,0)\otimes \mathcal{L}(\tfrac{7}{10},0)\otimes \mathcal{L}(\tfrac{7}{10},0)$ subalgebra \cite{lametal}, in terms of which it and its irreducible modules decompose as 
\begin{align}\label{D4B characters}
\begin{split}
    \W_{D_{\mathrm{4B}}}(0) &\cong [0,0,0]\oplus [\tfrac12,\tfrac32,0]\oplus[\tfrac12,0,\tfrac32]\oplus [0,\tfrac32,\tfrac32]\\
    \W_{D_{\mathrm{4B}}}(1) &\cong [0,0,\tfrac32]\oplus[0,\tfrac32,0]\oplus[\tfrac12,0,0]\oplus[\tfrac12,\tfrac32,\tfrac32] \\
    \W_{D_{\mathrm{4B}}}(2) &\cong [0,0,\tfrac35]\oplus[0,\tfrac32,\tfrac{1}{10}]\oplus[\tfrac12,\tfrac32,\tfrac35]\oplus[\tfrac12,0,\tfrac{1}{10}]\\
    \W_{D_{\mathrm{4B}}}(3) &\cong [0,\tfrac35,0]\oplus[0,\tfrac{1}{10},\tfrac32]\oplus[\tfrac12,\tfrac35,\tfrac32]\oplus[\tfrac12,\tfrac{1}{10},0]\\
    \W_{D_{\mathrm{4B}}}(4) &\cong [0,\tfrac35,\tfrac35]\oplus[0,\tfrac{1}{10},\tfrac{1}{10}]\oplus[\tfrac12,\tfrac{1}{10},\tfrac35]\oplus[\tfrac12,\tfrac35,\tfrac{1}{10}]\\
    \W_{D_{\mathrm{4B}}}(5) &\cong [0,\tfrac35,\tfrac{1}{10}]\oplus[0,\tfrac{1}{10},\tfrac35]\oplus[\tfrac12,\tfrac{1}{10},\tfrac{1}{10}]\oplus[\tfrac12,\tfrac35,\tfrac35]\\
    \W_{D_{\mathrm{4B}}}(6) &\cong [0,0,\tfrac{1}{10}]\oplus[0,\tfrac32,\tfrac35]\oplus[\tfrac12,\tfrac32,\tfrac{1}{10}]\oplus[\tfrac12,0,\tfrac35]  \\
    \W_{D_{\mathrm{4B}}}(7) &\cong [0,\tfrac{1}{10},0]\oplus[0,\tfrac35,\tfrac32]\oplus[\tfrac12,\tfrac35,0]\oplus[\tfrac12,\tfrac{1}{10},\tfrac32] \\
    \W_{D_{\mathrm{4B}}}(8) &\cong [\tfrac{1}{16},\tfrac{7}{16},\tfrac{7}{16}]\otimes Q, \ \ \ \W_{D_{\mathrm{4B}}}(9) \cong [\tfrac{1}{16},\tfrac{7}{16},\tfrac{3}{80}]\otimes Q \\
    \W_{D_{\mathrm{4B}}}(10) &\cong [\tfrac{1}{16},\tfrac{3}{80},\tfrac{7}{16}]\otimes Q, \ \ \ \W_{D_{\mathrm{4B}}}(11) \cong [\tfrac{1}{16},\tfrac{3}{80},\tfrac{3}{80}]\otimes Q
    \end{split}
\end{align}
where $[h_1,h_2,h_3] := \mathcal{L}(\tfrac12,h_1)\otimes \mathcal{L}(\tfrac{7}{10},h_2)\otimes \mathcal{L}(\tfrac{7}{10},h_3)$, and $Q$ is the 2-dimensional representation of the quaternion group of order 8. The resulting modular S matrix of this theory is then 
\begin{align}\label{D4BSmatrix}
\mathcal{S}=\left(
\begin{array}{cccccccccccc}
 \gamma_- & \gamma_- & \xi & \xi & \gamma_+ & \gamma_+ & \xi & \xi & -\omega & \sigma & \sigma & \lambda \\
 \gamma_- & \gamma_- & \xi & \xi & \gamma_+ & \gamma_+ & \xi & \xi & \omega & -\sigma & -\sigma & -\lambda \\
 \xi & \xi & -\gamma_- & \gamma_+ & -\xi & -\xi &
   -\gamma_- & \gamma_+ & \sigma & \omega & \lambda & -\sigma \\
 \xi & \xi & \gamma_+ & -\gamma_- & -\xi & -\xi &
   \gamma_+ & -\gamma_- & \sigma & \lambda & \omega & -\sigma \\
 \gamma_+ & \gamma_+ & -\xi & -\xi & \gamma_- & \gamma_- & -\xi & -\xi & \lambda & -\sigma & -\sigma &
   -\omega \\
 \gamma_+ & \gamma_+ & -\xi & -\xi & \gamma_- & \gamma_- & -\xi & -\xi & -\lambda & \sigma & \sigma &
   \omega \\
 \xi & \xi & -\gamma_- & \gamma_+ & -\xi & -\xi &
   -\gamma_- & \gamma_+ & -\sigma & -\omega & -\lambda & \sigma \\
 \xi & \xi & \gamma_+ & -\gamma_- & -\xi & -\xi &
   \gamma_+ & -\gamma_- & -\sigma & -\lambda & -\omega & \sigma \\
 -\omega & \omega & \sigma & \sigma & \lambda & -\lambda & -\sigma & -\sigma & 0 & 0 & 0 & 0 \\
 \sigma & -\sigma & \omega & \lambda & -\sigma & \sigma &
   -\omega & -\lambda & 0 & 0 & 0 & 0 \\
 \sigma & -\sigma & \lambda & \omega & -\sigma & \sigma &
   -\lambda & -\omega & 0 & 0 & 0 & 0 \\
 \lambda & -\lambda & -\sigma & -\sigma & -\omega & \omega & \sigma & \sigma & 0 & 0 & 0 & 0 \\
\end{array}
\right)
\end{align}
where we have defined $\gamma_\pm = \frac{1}{20}(5 \pm \sqrt5)$, $\omega = \frac{\sqrt{5}-5}{10\sqrt2}$, $\lambda = \frac12 \sqrt{\frac35 + \frac{1}{\sqrt5}}$, $\xi = \frac{1}{2\sqrt{5}}$, and $\sigma = \frac{1}{\sqrt{10}}$. The $\mc{L}(\tfrac12,0)\otimes\mc{L}(\tfrac{7}{10},0)\otimes \mc{L}(\tfrac{7}{10},0)$ subalgebra of $\W_{4\mathrm{B}}$ defines a deconstruction of the monster stress tensor as 
\begin{align}
    T(z) = t^{(\frac12)}(z) + t_1^{(\frac{7}{10})}(z) + t_2^{(\frac{7}{10})}(z) + \widetilde{t}(z).
\end{align}
This decomposition makes it clear that this chiral algebra can be obtained from an iterated deconstruction whose intermediate steps involve the baby monster VOA and the VOA $\widetilde{\W}_{D_{\mathrm{2A}}}$ with ${^2}\tsl{E}_6(2).2$ symmetry. Indeed, 
\begin{align}\label{F4(2).2 iterated}
\begin{split}
    V^\natural &\supset \tsl{Vir}(t^{(\frac12)})\otimes \tsl{V}\mathbb{B}^\natural \\
    &\supset \tsl{Vir}(t^{(\frac12)})\otimes \tsl{Vir}(t_1^{(\frac{7}{10})}) \otimes \widetilde{\W}_{D_{\mathrm{2A}}} \\
    &\supset \tsl{Vir}(t^{(\frac12)})\otimes \tsl{Vir}(t_1^{(\frac{7}{10})})\otimes \tsl{Vir}(t_2^{(\frac{7}{10})}) \otimes \widetilde{\W}_{D_{\mathrm{4B}}}
\end{split}
\end{align}
so that in particular, $\widetilde{\W}_{D_{\mathrm{4B}}}\hookrightarrow \widetilde{\W}_{D_{\mathrm{2A}}} \hookrightarrow \tsl{V}\mathbb{B}^\natural \hookrightarrow V^\natural$. 

The monstralizer $[D_{\mathrm{4B}}\circ 2.\tsl{F}_4(2)].2$ predicts that the inner automorphism group should be $\tsl{F}_4(2)$. One can see this from the embedding chain above as follows. The algebra $\widetilde{\W}_{D_{\mathrm{4B}}}$ arises from deconstructing a central charge $\frac{7}{10}$ conformal vector $t_2^{(\frac{7}{10})}$ off of $\widetilde{\W}_{D_{\mathrm{2A}}}$, and the $\mathbb{Z}_2$ fusion algebra automorphism of the $\mc{L}(\frac{7}{10},0)$ VOA it generates lifts to an element of the 2D conjugacy class of ${^2}\tsl{E}_6(2).2\subset \mathrm{Aut}(\widetilde{\W}_{D_{\mathrm{2A}}})$. We can thus consider $\mathrm{Cent}_{{^2}\tsl{E}_6(2).2}(\mathrm{2D})\cong \mathbb{Z}_2\times \tsl{F}_4(2)$, which will stabilize $t_2^{(\frac{7}{10})}$ as well as the stress tensor of $\widetilde{\W}_{D_{\mathrm{4B}}}$, and after taking the quotient by the $\mathbb{Z}_2$ which acts trivially on $\widetilde{\W}_{D_{\mathrm{4B}}}$, we are left with $\tsl{F}_4(2)$ acting by inner automorphisms on $\widetilde{\W}_{D_{\mathrm{4B}}}$. A character theoretic calculation in the monster shows that the full automorphism group contains the slightly larger group $\tsl{F}_4(2).2$

Neither Hecke operators nor MLDEs are effective at recovering the dual characters $\widetilde{\chi}_\alpha(\tau)$, for the same reasons as in the previous section. We can nonetheless consider the Rademacher sum attached to the $\tsl{SL}_2(\mathbb{Z})$ representation $\rho$ generated by $\mathcal{S}^\ast$ and $\mathcal{T}^\ast$ and whose singular part takes the shape $\widetilde{\chi}_\alpha(\tau) \equiv R^P_{\tsl{SL}_2(\mathbb{Z}),\rho,\alpha}(\tau) = \delta_{\alpha,0}q^{-\frac{221}{240}}+O(q^0)$. If we compute the $q$-expansion of this Rademacher sum to low orders, we find 
\begin{align}
\begin{split}
\widetilde{\chi}_0(\tau) &= q^{-\frac{221}{240}} \left( 1 + 24310 q^2 + 1923805 q^3 + 62596703q^4 + 1240094427q^5 + \cdots \right), \\
\widetilde{\chi}_1(\tau) &= q^{\frac{139}{240}} \left( 1326 + 252603 q + 11865711q^2+ 292523335q^3+\cdots \right), \\
\widetilde{\chi}_2(\tau) &= q^{\frac{23}{48}} \left( 1105 + 262990 q + 13533325 q^2 + 350645230q^3 + 6008564575q^4+\cdots\right), \\
\widetilde{\chi}_3(\tau) &= q^{\frac{23}{48}} \left( 1105 + 262990 q + 13533325 q^2 + 350645230q^3 + 6008564575q^4+\cdots \right), \\
\widetilde{\chi}_4(\tau) &= q^{\frac{211}{240}} \left( 21658 + 2303925 q + 85629544q^2+ 1841109010q^3+ \cdots \right), \\
\widetilde{\chi}_5(\tau) &= q^{\frac{91}{240}}\left( 833 + 270725 q + 15330770 q^2 + 418641184q^3+ 7423572325q^4+\cdots\right) , \\
\widetilde{\chi}_6(\tau) &= q^{\frac{47}{48}}\left( 23205 + 2114970 q + 73396140q^2+ 1513535075q^3+\cdots \right) , \\
\widetilde{\chi}_7(\tau) &= q^{\frac{47}{48}}\left( 23205 + 2114970 q + 73396140q^2+ 1513535075q^3+\cdots\right) , \\
\widetilde{\chi}_8(\tau) &= q^{\frac{17}{120}}\left( 52 + 46852 q + 3452020 q^2 + 107984680q^3+ 2088333312q^4+\cdots \right) , \\
\widetilde{\chi}_9(\tau) &= q^{\frac{13}{24}}\left( 2380 + 490620 q + 23832640q^2+ 598421980q^3+\cdots\right) , \\
\widetilde{\chi}_{10}(\tau) &= q^{\frac{13}{24}}\left( 2380 + 490620 q + 23832640q^2+ 598421980q^3+\cdots\right) , \\
\widetilde{\chi}_{11}(\tau) &= q^{\frac{113}{120}}\left( 43316 + 4176900 q +148652556q^2+ 3113186336q^3 \cdots\right).
\end{split}
\end{align}
One can check that this set of characters pairs bilinearly with the characters $\chi_\alpha(\tau)$ of $\W_{D_{\mathrm{4B}}}$, which can be deduced from the decompositions in \eqref{D4B characters}, to produce the partition function of $V^\natural$, 
\begin{align}
J(\tau) = \sum_\alpha \chi_\alpha(\tau)\widetilde{\chi}_\alpha(\tau)
\end{align}
which is consistent with a diagonal decomposition of the moonshine module of the form 
\begin{align}
V^\natural \cong \bigoplus_\alpha \W_{D_{\mathrm{4B}}}(\alpha) \otimes \widetilde{\W}_{D_{\mathrm{4B}}}(\alpha).
\end{align}
Furthermore, we observe for example that the leading coefficient of each component is always the dimension of an \emph{irreducible} representation of $2.\tsl{F}_4(2)$ (i.e.\ a projective irreducible representation of the predicted inner automorphism group, $\tsl{F}_4(2)$). Furthermore, for each character which appears with multiplicity 2, twice the leading coefficient coincides with the dimension of an irreducible representation of $2.\tsl{F}_4(2).2$, which is consistent with our prediction that this theory inherits an order 2 outer automorphism from the monster; this outer automorphism evidently exchanges pairs of modules that come with the same character. 

We conclude this section by describing the algebra $\W_{D_{\mathrm{4B}}}$ in terms of an $\mc{P}(2)\otimes \mc{P}(8)$ subalgebra, which makes some of its symmetries more manifest. We claim that the characters which arise from \eqref{D4B characters} can be alternatively expressed in terms of parafermion characters as 
\begin{align}
\label{4B character by PF}
\begin{split}
\chi_0 &= \psi^{(2)}_{2,2} \psi^{(8)}_{8,8} + \psi^{(2)}_{2,2} \psi^{(8)}_{8,0} + \psi^{(2)}_{2,0} \psi^{(8)}_{8,4} + \psi^{(2)}_{2,0} \psi^{(8)}_{8,-4}, \\
\chi_1 &= \psi^{(2)}_{2,0} \psi^{(8)}_{8,8} + \psi^{(2)}_{2,2} \psi^{(8)}_{8,-4} + \psi^{(2)}_{2,2} \psi^{(8)}_{8,4} + \psi^{(2)}_{2,0} \psi^{(8)}_{8,0} , \\
\chi_2 &= \psi^{(2)}_{2,2} \psi^{(8)}_{4,0} + \psi^{(2)}_{2,0} \psi^{(8)}_{4,4} , \quad 
\chi_3 = \psi^{(2)}_{2,2} \psi^{(8)}_{4,0} + \psi^{(2)}_{2,0} \psi^{(8)}_{4,4} , \\
\chi_4 &= \psi^{(2)}_{2,2} \psi^{(8)}_{2,0} + \psi^{(2)}_{2,2} \psi^{(8)}_{6,0} + \psi^{(2)}_{2,0} \psi^{(8)}_{6,4} + \psi^{(2)}_{2,4} \psi^{(8)}_{6,-4}, \\
\chi_5 &= \psi^{(2)}_{2,2} \psi^{(8)}_{6,4} + \psi^{(2)}_{2,2} \psi^{(8)}_{6,-4} + \psi^{(2)}_{2,0} \psi^{(8)}_{2,0} + \psi^{(2)}_{2,0} \psi^{(8)}_{6,0}, \\
\chi_6 &= \psi^{(2)}_{2,2} \psi^{(8)}_{4,4} + \psi^{(2)}_{2,0} \psi^{(8)}_{4,0} , \quad \chi_7 = \psi^{(2)}_{2,2} \psi^{(8)}_{4,4} + \psi^{(2)}_{2,0} \psi^{(8)}_{4,0} , \\
\chi_8 &= \psi^{(2)}_{1,1} \psi^{(8)}_{8,6} + \psi^{(2)}_{1,1} \psi^{(8)}_{8,-6} + \psi^{(2)}_{1,1} \psi^{(8)}_{8,2} + \psi^{(2)}_{1,1} \psi^{(8)}_{8,-2}, \\
\chi_9 &= \psi^{(2)}_{1,1} \psi^{(8)}_{4,2} + \psi^{(2)}_{1,1} \psi^{(8)}_{4,-2} , \quad
\chi_{10} = \psi^{(2)}_{1,1} \psi^{(8)}_{4,2} + \psi^{(2)}_{1,1} \psi^{(8)}_{4,-2}, \\
\chi_{11} &= \psi^{(2)}_{1,1} \psi^{(8)}_{2,2} + \psi^{(2)}_{1,1} \psi^{(8)}_{6,6} + \psi^{(2)}_{1,1} \psi^{(8)}_{6,2} + \psi^{(2)}_{1,1} \psi^{(8)}_{6,-2}.
\end{split}
\end{align}
Indeed, from these expressions, one can check that fusion algebra automorphisms of $\mc{P}(8)$ lift to $\mathbb{Z}_{4\mathrm{B}}$ in the monster.

\subsubsection{$(D_{\mathrm{2B}},2.2^{1+22}.\normalfont{\tsl{Co}}_2)${\normalfont:} Second Conway group}\label{subsubsec:Co2}

We now consider the commutant of $\W_{D_{\mathrm{2B}}}$. It turns out that if two 2A involutions have product residing in $2\mathrm{B}$, then their corresponding conformal vectors of central charge $\frac12$ commute. This means that the $\W_{D_{\mathrm{2B}}}$ subalgebra is a decoupled pair of Ising models, $\W_{D_{\mathrm{2B}}}\cong \mathcal{L}(\tfrac12,0) \otimes \mathcal{L}(\tfrac12,0)$, a case which has already been considered in \cite{Bae:2018qfh}. We are therefore brief and refer readers to (loc. cit.) for additional details.

To compute the automorphisms that its commutant inherits from the monster, we note that the group $\mathrm{Cent}_{\mathbb{M}}(\mathbb{Z}_2\times\mathbb{Z}_2)\cong 2.2^{1+22}.\tsl{Co}_2$ stabilizes the stress tensor of $\W_{D_{\mathrm{2B}}}$, and therefore its image under the map in equation \eqref{authom} acts by automorphisms on the commutant $\widetilde{\W}_{D_{\mathrm{2B}}}.$ The $\mathbb{Z}_2\times\mathbb{Z}_2$ being centralized is the automorphism group of the fusion algebra of $\W_{D_{\mathrm{2B}}}$, and it is not difficult to see that it acts trivially on $\widetilde{\W}_{D_{\mathrm{2B}}}$. Therefore, after taking the quotient \footnote{We use the fact that $2^{1+22}.\tsl{Co}_2$ has a $\mathbb{Z}_2$ normal subgroup which leaves $2^{22}.\tsl{Co}_2$ after taking the quotient \cite{pahlings2007character}.} by this group we are left with $2^{22}.\tsl{Co}_2$ as the inner automorphism group of $\widetilde{\W}_{D_{\mathrm{2B}}}$. We conjecture that it inherits a further order 2 outer automorphism from the monster so that $\mathrm{Aut}(\widetilde{\W}_{D_{\mathrm{2B}}})=2^{22}.\tsl{Co}_2.2$, consistent with the monstralizer in which it participates. 

We note also that this VOA embeds into the baby monster VOA. Indeed, it arises from an iterated deconstruction of two Ising models, 
\begin{align}
\begin{split}
V^\natural&\supset \mathcal{L}(\tfrac12,0)\otimes \tsl{V}\mathbb{B}^\natural \\
&\supset \mathcal{L}(\tfrac12,0)\otimes \mathcal{L}(\tfrac12,0)\otimes \widetilde{\W}_{D_{\mathrm{2B}}}
\end{split}
\end{align}
and so in particular, it can be described as the commutant of a particular Ising subalgebra of $\tsl{V}\mathbb{B}^\natural$. This means in particular that the characters of $\widetilde{\W}_{D_{\mathrm{2B}}}$ can be bilinearly combined with the characters of the Ising model to produce the characters of $\tsl{V}\mathbb{B}^\natural$. 

The VOA $\W_{D_{\mathrm{2B}}}$ has 9 irreducible modules $\W_{D_{\mathrm{2B}}}(\alpha)$ for $\alpha = 0, 1, \dots, 8$. One can express their characters as
\begin{align}\label{D2B chars}
\begin{split}
\chi_0(\tau) &= \chi^{(3)}_{1,1}(\tau) \chi^{(3)}_{1,1}(\tau) , \quad \chi_1(\tau) = \chi^{(3)}_{2,1}(\tau) \chi^{(3)}_{1,1}(\tau) , \quad \chi_2(\tau) = \chi^{(3)}_{1,1}(\tau) \chi^{(3)}_{2,1} (\tau), \\
\chi_3(\tau) &= \chi^{(3)}_{1,2}(\tau) \chi^{(3)}_{1,1}(\tau) , \quad \chi_4(\tau) = \chi^{(3)}_{1,1}(\tau) \chi^{(3)}_{1,2}(\tau) , \quad \chi_5(\tau) = \chi^{(3)}_{1,2}(\tau) \chi^{(3)}_{1,2}(\tau) ,
\\
\chi_6(\tau) &= \chi^{(3)}_{2,1}(\tau) \chi^{(3)}_{1,2}(\tau) , \quad \chi_7(\tau) = \chi^{(3)}_{1,2}(\tau) \chi^{(3)}_{2,1}(\tau) , \quad \chi_8(\tau) = \chi^{(3)}_{2,1}(\tau) \chi^{(3)}_{2,1}(\tau).
\end{split}
\end{align}
Its commutant has 9 dual modules, $\widetilde{\W}_{D_{\mathrm{2B}}}(\alpha)$ for $\alpha = 0, 1, \dots, 8$. The moonshine module decomposes simply in terms of its $\W_{D_{\mathrm{2B}}}\otimes \widetilde{\W}_{D_{\mathrm{2B}}}$ subalgebra as 
\begin{align}
    V^\natural \cong \bigoplus_{\alpha=0}^{8} \W_{D_{\mathrm{2B}}}(\alpha)\otimes \widetilde{\W}_{D_{\mathrm{2B}}}(\alpha).
\end{align}
In (loc.\ cit.), the characters $\widetilde{\chi}_{\alpha}(\tau)$ of the 9 dual modules 
were obtained as solutions to an MLDE, 
\begin{align}
\begin{split}
&\Big[\mathcal{D}^6+\mu_1E_4(\tau)\mathcal{D}^4 +  \mu_2 E_6(\tau) \mathcal{D}^3 + \mu_3E_4^2(\tau)\mathcal{D}^2 + \mu_4E_4(\tau)E_6(\tau)\mathcal{D}+ \mu_5 E_4^3(\tau) \\
&\hspace{.8in}  + \mu_6 E_6^2(\tau) + \mu_7 \frac{E_4^2(\tau)}{E_6(\tau)}\mathcal{D}^5 + \mu_8 \frac{E_4^3(\tau)}{E_6(\tau)}\mathcal{D}^3 + \mu_9 \frac{E_4^4(\tau)}{E_6(\tau)}\mathcal{D}\Big]\widetilde{\chi}_{\alpha}(\tau) = 0,
\end{split}
\end{align}
with the coefficients $\mu_i$ given by 
\begin{align}
\begin{split}
   &\mu_1 = \frac{2647\pi^2}{576}, \quad \mu_2 = -i\frac{84271\pi^3}{6912},\quad \mu_3 = -\frac{598979\pi^4}{82944}, \ \  \mu_4 =  i\frac{10884545\pi^5}{995328},\\
  &\mu_5 = 0, \quad \mu_6 = \frac{3555409\pi^6}{5971968},\quad \mu_7=i\pi,\quad\mu_8 = i\frac{2647\pi^3}{576}, \quad \mu_9 = -i\frac{598979\pi^5}{82944}.
\end{split}
\end{align}
Here we comment that they can alternatively be expressed as Hecke images of the characters $\chi_{\alpha}(\tau)$ of $\W_{D_{\mathrm{2B}}}$. Specifically, 
\begin{align}\label{D2B dual chars}
\begin{split}
    &\widetilde{\chi}_{0}(\tau) = (\mathsf{T}_{23}\chi)_{0}(\tau), ~ \widetilde{\chi}_{1}(\tau) = \widetilde{\chi}_{2}(\tau) = (\mathsf{T}_{23}\chi)_{1}(\tau), \\ &\widetilde{\chi}_{3}(\tau) = \widetilde{\chi}_{4}(\tau) = (\mathsf{T}_{23}\chi)_{3}(\tau),   ~ \widetilde{\chi}_{5}(\tau) = (\mathsf{T}_{23}\chi)_{5}(\tau), \\ 
    &\widetilde{\chi}_{6}(\tau) = \widetilde{\chi}_{7}(\tau )=  (\mathsf{T}_{23}\chi)_{6}(\tau),  ~ \widetilde{\chi}_{8}(\tau) = (\mathsf{T}_{23}\chi)_{8}(\tau).
\end{split}
\end{align}
To construct modular invariant bilinears, one can consider the matrices $G_{\ell}$ with $\ell^2 + 23 = 0 ~ \text{mod} ~ 24$. There are eight solutions $\ell = 1,5,7,11,13,17,19,23$, and their corresponding matrices are identical. We choose $G_5$ whose entries are 0 except for the entries
\begin{align}
(0,0), ~(1,2), ~(2,1), ~(3,4), ~(4,3), ~(5,5), ~(6,7), ~(7,6), ~(8,8),
\end{align}
which are equal to 1. One can then show that the characters $\chi(\tau)$ and their Hecke images pair to produce $J(\tau)$,
\begin{align}
J(\tau) = (\mathsf{T}_{23}\chi)^T(\tau) \cdot G_5 \cdot \chi(\tau)
\end{align}
so that we identify $\widetilde{\chi}(\tau) = G_5^T\cdot (\mathsf{T}_{23}\chi)(\tau).$

Finally, we note that analogous remarks apply to this deconstruction as the ones made in \S\ref{subsubsecCo3} and \S\ref{subsubsec:Mcl} about $\widetilde{\W}_{\mathbb{Z}_{\mathrm{4A}}}$ and $\widetilde{\W}_{D_{\mathrm{4A}}}$. In particular, because $\W_{D_{\mathrm{2B}}}\cong \mc{L}(\frac12,0)\otimes\mc{L}(\frac12,0)\cong \V^+_{L}$ with $L\cong 2\mathbb{Z}$ a rank one sublattice of $\Lambda_{\mathrm{Leech}}$ generated by a vector of norm-squared 4 \cite{Ginsparg:1987eb}, we may realize $\widetilde{\W}_{D_{\mathrm{2B}}}$ as a $\mathbb{Z}_2$ orbifold $\V_{\widetilde{L}}^+$ of the $c=23$ lattice VOA attached to its orthogonal complement $\widetilde{L}:=L^\perp$ in $\Lambda_{\mathrm{Leech}}$. From this perspective, the fact that the inner automorphism group of $\widetilde{\W}_{\mathrm{2B}}$ is an extension of $\tsl{Co}_2$ is related to the fact that $\tsl{Co}_2$ stabilizes $L$ \cite{splag}. One can indeed check that the $q$-expansions of the $\widetilde{\chi}_\alpha$ agree with the characters $\xi_{\mathds{1}}^{(\widetilde L)}$, $\xi_{j}^{(\widetilde L)}$, $\xi_{\sigma,i}^{(\widetilde L)}$, $\xi_{\tau,i}^{(\widetilde L)}$, $\xi_{\lambda^\ast}^{(\widetilde L)}$, $\xi_{2\lambda^\ast,i}^{(\widetilde L)}$ where $i=1,2$ and $\lambda^\ast$ is a generator of the discriminant group $\widetilde{L}^\ast/\widetilde{L} \cong \mathbb{Z}_4$. This description also establishes that $\widetilde{\W}_{D_{\mathrm{2B}}}$ contains $\widetilde{\W}_{D_{\mathrm{4A}}}$ as a subalgebra: indeed, this follows from the fact that $\widetilde{L}$ contains the lattice \eqref{4A lattice} on which $\widetilde{\W}_{D_{\mathrm{4A}}}$ is based.

\subsubsection{$(D_{\mathrm{3C}},\normalfont{\tsl{Th}})${\normalfont:} Thompson group}\label{subsubsec:Th}

Finally, we consider two conformal vectors $e$ and $f$ of central charge $\frac{1}{2}$ whose associated involutions $\tau_e$ and $\tau_f$ have product residing in the 3C conjugacy class. We will take our subalgebra of $V^\natural$ to be the VOA $ \W_{D_{\mathrm{3C}}}$ generated by these two conformal vectors, and argue that its commutant $\widetilde{\W}_{D_{\mathrm{3C}}}$, as well as its modules, enjoy an action by the Thompson sporadic group, $\Th$. We therefore refer to this chiral algebra as $\tsl{VT}^\natural$.

First, we discuss some of the properties of $\W_{D_{\mathrm{3C}}}$. It admits several descriptions in terms of simpler models. For example, it is known that its $c = \frac{16}{11}$ stress tensor can be deconstructed into two conformal vectors with central charges $\frac{1}{2}$ and $\frac{21}{22}$, 
\begin{align}
    t(z) = t^{(\frac{1}{2})}(z)+t^{(\frac{21}{22})}(z).
\end{align}
Therefore, $\W_{D_{\mathrm{3C}}}$ admits a subalgebra of the form $\mathcal{L}(\frac{1}{2},0)\otimes \mathcal{L}(\frac{21}{22},0)$, with respect to which it decomposes as 
\begin{align}
    \W_{D_{\mathrm{3C}}} \cong [0,0]\oplus [0,8]\oplus [\tfrac12,\tfrac72]\oplus [\tfrac12,\tfrac{45}{2}]\oplus [\tfrac{1}{16},\tfrac{31}{16}]\oplus [\tfrac{1}{16},\tfrac{175}{16}]
\end{align}
where $[h_1,h_2]:= \mathcal{L}(\frac12,h_1)\otimes \mathcal{L}(\frac{21}{22},h_2)$. Its irreducible modules are also known: there are exactly five with highest weights $(0,\frac{2}{11},\frac{6}{11},\frac{1}{11},\frac{9}{11})$, and the non-vacuum modules decompose as 
\begin{align}
\begin{split}
    \W_{D_{\mathrm{3C}}}(1) &\cong [0,\tfrac{13}{11}]\oplus [0,\tfrac{35}{11}] \ \oplus [\tfrac12,\tfrac{15}{22}] \oplus [\tfrac12,\tfrac{301}{22}] \oplus [\tfrac{1}{16},\tfrac{21}{176}] \oplus [\tfrac{1}{16},\tfrac{901}{176}], \\
    \W_{D_{\mathrm{3C}}}(2) &\cong [0,\tfrac{6}{11}]\oplus [0,\tfrac{50}{11}] \ \oplus [\tfrac12, \tfrac{1}{22}]\oplus [\tfrac12,\tfrac{155}{22}]\oplus [\tfrac{1}{16},\tfrac{85}{176}]\oplus [\tfrac{1}{16},\tfrac{261}{176}], \\
    \W_{D_{\mathrm{3C}}}(3) &\cong [0,\tfrac{1}{11}]\oplus [0,\tfrac{111}{11}]\oplus[\tfrac12,\tfrac{35}{22}]\oplus [\tfrac12,\tfrac{57}{22}] \ \oplus[\tfrac{1}{16},\tfrac{5}{176}]\oplus [\tfrac{1}{16},\tfrac{533}{176}], \\
    \W_{D_{\mathrm{3C}}}(4) &\cong [0,\tfrac{20}{11}]\oplus [0,\tfrac{196}{11}]\oplus[\tfrac12,\tfrac{7}{22}]\oplus [\tfrac12,\tfrac{117}{22}]\oplus [\tfrac{1}{16},\tfrac{133}{176}]\oplus [\tfrac{1}{16},\tfrac{1365}{176}].
\end{split}
\end{align}
Alternatively, there is a parafermionic description which makes some of its symmetries more manifest. Indeed, $\W_{D_{\mathrm{3C}}}$ is an extension of the level 9 parafermion theory $\mathcal{P}(9)$ by its two irreducible modules with integral highest weight. Abbreviating $\langle \ell,m\rangle :=\mathcal{P}(9,[\ell,m])$, we have 
\begin{align}
    \begin{split}
        \W_{D_{\mathrm{3C}}}(0) &\cong \langle9,9 \rangle\oplus \langle 9,3\rangle \oplus \langle 9,-3\rangle, \\
        \W_{D_{\mathrm{3C}}}(1) &= \langle 2,0\rangle \oplus \langle 7,3 \rangle \oplus \langle 7,-3\rangle, \\
        \W_{D_{\mathrm{3C}}}(2) &= \langle 4,0\rangle \oplus \langle 5,3\rangle \oplus \langle 5,-3\rangle, \\
        \W_{D_{\mathrm{3C}}}(3) &= \langle 6,0\rangle \oplus \langle 3,3\rangle \oplus \langle 3,-3\rangle, \\
        \W_{D_{\mathrm{3C}}}(4) &= \langle 8,0\rangle \oplus \langle 8,6\rangle \oplus \langle 8,-6\rangle.
    \end{split}
\end{align}
It follows that the characters,
\begin{align}
    \chi_\alpha(\tau) = \mathrm{Tr}_{\W_{D_{\mathrm{3C}}}(\alpha)}q^{L_0 - \frac{2}{33}}
\end{align}
can be expressed either as sums of products of minimal model characters, or sums of parafermion characters. For example, in terms of parafermion characters, one has 
\begin{align}
\label{Th characters}
\begin{split}
\chi_0(\tau) &=\psi^{(9)}_{9,9}(\tau)+\psi^{(9)}_{9,3}(\tau)+\psi^{(9)}_{9,-3} (\tau), \\
\chi_1(\tau) &= \psi^{(9)}_{2,0}(\tau)+\psi^{(9)}_{7,3}(\tau) + \psi^{(9)}_{7,-3}(\tau),   \\
\chi_2(\tau) &=  \psi^{(9)}_{4,0}(\tau)+ \psi^{(9)}_{5,3}(\tau) + \psi^{(9)}_{5,-3}(\tau),   \\
\chi_3(\tau) &=  \psi^{(9)}_{6,0}(\tau)+\psi^{(9)}_{3,3}(\tau)+ \psi^{(9)}_{3,-3}(\tau),  \\
\chi_4(\tau) &=    \psi^{(9)}_{8,0}(\tau)+\psi^{(9)}_{8,6}(\tau) + \psi^{(9)}_{8,-6}(\tau).
\end{split}
\end{align}
The five characters \eqref{Th characters} transform under $\mathcal{S}$ and $\mathcal{T}$ as 
\begin{equation}
\chi_{\alpha}(-\tfrac{1}{\tau}) = \sum_{\beta} \mathcal{S}_{\alpha \beta} \chi_{\beta} (\tau), \quad \chi_{\alpha}(\tau + 1) =  \sum_{\beta} \mathcal{T}_{\alpha \beta} \chi_{\beta} (\tau)
\end{equation}
where the S-matrix is $\mathcal{S}$ given by
\begin{align}
\label{k=9 Smatrix}
\mathcal{S} = \frac{2}{\sqrt{11}}
\begin{pmatrix}
    \mbox{sin}\frac{\pi}{11}& \mbox{cos}\frac{5\pi}{22} & \mbox{cos}\frac{\pi}{22} &  \mbox{cos}\frac{3\pi}{22}& \mbox{sin}\frac{2\pi}{11}  \\
    \mbox{cos}\frac{5\pi}{22} & \mbox{sin}\frac{2\pi}{11} & - \mbox{cos}\frac{3\pi}{22} & -\mbox{sin}\frac{\pi}{11} & \mbox{cos}\frac{\pi}{22}  \\
    \mbox{cos}\frac{\pi}{22} & -\mbox{cos}\frac{3\pi}{22} &  \mbox{cos}\frac{5\pi}{22} & -\mbox{sin}\frac{2\pi}{11} & \mbox{sin}\frac{\pi}{11}   \\
    \mbox{cos}\frac{3\pi}{22} & -\mbox{sin}\frac{\pi}{11} & -\mbox{sin}\frac{2\pi}{11} &  \mbox{cos}\frac{\pi}{22} &  -\mbox{cos}\frac{5\pi}{22}  \\
    \mbox{sin}\frac{2\pi}{11} & \mbox{cos}\frac{\pi}{22} & \mbox{sin}\frac{\pi}{11} & - \mbox{cos}\frac{5\pi}{22}& -\mbox{cos}\frac{3\pi}{22}
\end{pmatrix}
\end{align}
and the T-matrix $\mathcal{T}$ reads
\begin{align}
\label{k=9 Smatrix}
\mathcal{T} = \mbox{diag} \left( e^{-\frac{4 \pi i}{33}}, e^{\frac{8 \pi i}{33}}, e^{\frac{32 \pi i}{33}}, e^{\frac{2 \pi i}{33}}, e^{\frac{50 \pi i}{33}}\right).
\end{align}

Now, the parafermion theory enjoys a $\mathbb{Z}_9$ automorphism of its fusion algebra, which acts on the characters as
\be
\psi^{(9)}_{\ell,m}(\tau) \rightarrow \zeta^m \psi^{(9)}_{\ell,m}(\tau) \, 
\ee
where $\zeta = e^{\frac{2\pi i}{9}}$. This symmetry can be lifted to an automorphism of $\W_{D_{\mathrm{3C}}}$, and even further to an automorphism of $V^\natural$. However, since the modules $\langle \ell,m\rangle$ of $\mathcal{P}(9)$ which appear in the decomposition of $\W_{D_{\mathrm{3C}}}$ all have $m\equiv 0 ~{\rm mod}~3$, only a $\mathbb{Z}_3$ acts non-trivially. The generator $\omega$ of this $\mathbb{Z}_3$, thought of as an element of the monster group, lives in the 3C conjugacy class, and is equal to the product $\tau_e\tau_f$ of Miyamoto involutions associated to the two $c=\frac{1}{2}$ conformal vectors $e$ and $f$ which generate $\W_{D_{\mathrm{3C}}}$. 

Now, the centralizer of $D_{\mathrm{3C}}$ in the monster is given by  $\mathrm{Cent}_{\mathbb{M}}(D_{\mathrm{3C}}) =  \Th$. If we decompose the dimension 2 subspace $V^\natural_2$ with respect to the action of $\Th$, we find that 
\begin{align}
    V_2^\natural\Big\vert_{\Th} = \mathbf{1}\oplus \mathbf{1}\oplus\mathbf{1} \oplus 3\cdot \mathbf{4123} \oplus \mathbf{30628}\oplus \mathbf{30875}  \oplus  2\cdot \mathbf{61256}. 
\end{align}
We claim that the 3-dimensional subspace on which $\Th$ acts trivially is spanned by $\mathbb{C}(t^{(\frac{1}{2})})\oplus \mathbb{C}(t^{(\frac{21}{22})}) \oplus \mathbb{C}(\widetilde{t})$ with $\widetilde{t}$ the stress tensor of the commutant, $\tsl{VT}^\natural$. This in particular implies that $\Th$ is a subgroup of the stabilizer group of $t = t^{(\frac{1}{2})}+t^{(\frac{11}{12})}$, and so from the homomorphism $\mathrm{Stab}_{\mathrm{Aut}(V^\natural)}(t) \to \mathrm{Aut}(\operatorname{\textsl{Com}}_{V^\natural}(\operatorname{\textsl{Vir}}(t)))= \mathrm{Aut}(\tsl{VT}^\natural)$ in equation \eqref{authom}, $\tsl{VT}^\natural$ admits an action of the image of $\Th$ under this map by automorphisms. Because $\Th$ is simple, it follows that the map restricted to $\Th$ has trivial kernel, and therefore all of $\Th$ acts. Its simplicity also implies that $\Th$ acts by inner automorphisms, and so all the modules of $\tsl{VT}^\natural$ will also be $\Th$ symmetric. Furthermore, because $D_{\mathrm{3C}}\times\tsl{Th}$ is a maximal subgroup of $\mathbb{M}$, and $D_{\mathrm{3C}}$ acts trivially on $\tsl{VT}^\natural$, it follows that $\tsl{VT}^\natural$ can inherit no additional automorphisms from $V^\natural$.

Now, $\Th$ is a maximal subgroup of $\mathbb{B}$, and so it is natural to wonder whether $\tsl{VT}^\natural$ can be obtained by deconstructing the baby monster VOA, $\tsl{V}\mathbb{B}^\natural$. In fact, it is straightforward to see that we can obtain $\tsl{VT}^\natural$ via an iterated deconstruction, in which we first strip off the central charge $\frac{1}{2}$ conformal vector to obtain $\tsl{V}\mathbb{B}^\natural$, and then strip off the central charge $\frac{21}{22}$ conformal vector to obtain $\tsl{VT}^\natural$. This demonstrates that the Thompson VOA embeds into the baby monster VOA, $\mathcal{L}(\tfrac{21}{22},0)\otimes\tsl{VT}^\natural \hookrightarrow \tsl{V}\mathbb{B}^\natural$.

We will now produce evidence for this symmetry by considering the dual characters $\chi_{\tsl{VT}^\natural(\alpha)}(\tau)$ of $\tsl{VT}^\natural$ in $V^\natural$. We claim that they are solutions to the fifth order differential equation,
\begin{align}
\begin{split}
&\left[ \mathcal{D}^5 + \mu_1 E_4(\tau) \mathcal{D}^3  + \mu_2 E_6(\tau) \mathcal{D}^2 + \mu_3 E_4^2(\tau) \mathcal{D} + \mu_4  E_4(\tau)  E_6(\tau)\right] \chi_{\tsl{VT}^\natural(\alpha)}(\tau) = 0, \\
&\ \ \ \ \ \ \mu_1 = \frac{413 \pi^2}{99}, \ \mu_2 = -i \frac{845 \pi^3}{99}, \ \mu_3 = -\frac{861871 \pi^4}{107811},\ \mu_4 = i \frac{125198336 \pi^5}{39135393},
\end{split}
\end{align}
whose $q$-expansions are given by
\begin{align}
\label{Ch of c=248/11}
\begin{split}
\chi_{\tsl{VT}^\natural(0)}(\tau) &=q^{-\frac{31}{33}} (1 + 30876 q^2 + 2634256 q^3 + 90061882 q^4 +
   1855967520 q^5 \\
   & \hspace{.4in} + 27409643240 q^6 + 317985320008 q^7 + 3064708854915 q^8  + \cdots ),  \\
\chi_{\tsl{VT}^\natural(1)}(\tau) &= q^{\frac{29}{33}} (30628 + 3438240 q + 132944368 q^2 + 2954702008 q^3 +
   45976123126 q^4    \\
   & \hspace{.4in} + 554583175040 q^5 + 5510740058664 q^6 +
   46939446922208 q^7  + \cdots ),    \\
\chi_{\tsl{VT}^\natural(2)}(\tau) &= q^{\frac{17}{33}} (4123 + 961248 q + 49925748 q^2 + 1315392496 q^3 +
   22953663126 q^4    \\
   & \hspace{.4in} + 301143085728 q^5 + 3193490344856 q^6 +
   28662439021248 q^7  + \cdots ),    \\
\chi_{\tsl{VT}^\natural(3)}(\tau) &= q^{\frac{32}{33}} (61256 + 5955131 q + 216162752 q^2 + 4622827508 q^3 +
   70051197488 q^4    \\
   & \hspace{.4in} + 828481014062 q^5 + 8106388952544 q^6 +
   68191291976248 q^7 + \cdots ),    \\
\chi_{\tsl{VT}^\natural(4)}(\tau) &= q^{\frac{8}{33}} (248 + 147498 q + 10107488 q^2 + 308975512 q^3 +
   5936748000 q^4   \\
   & \hspace{.4in} + 83455971224 q^5 + 932866634976 q^6 +
   8730997273664 q^7 + \cdots ).
\end{split}
\end{align}
One can check that 
\begin{align}\label{th_bilinear}
    J(\tau) = \sum_\alpha \chi_\alpha(\tau)\chi_{\tsl{VT}^\natural(\alpha)}(\tau)
\end{align}
which suggests that the moonshine module decomposes as 
\begin{align}
    V^\natural \cong \bigoplus_\alpha \W_{D_{\mathrm{3C}}}(\alpha)\otimes \tsl{VT}^\natural(\alpha).
\end{align}
The decompositions of the low-lying coefficients of the $\chi_{\tsl{VT}^\natural(\alpha)}$ into dimensions of irreducible representations of $\Th$ are fixed by the bilinear relation equation (\ref{th_bilinear}) and the known decompositions of the coefficients of $J$ into dimensions of irreducible representations of $\mathbb{M}$. For instance, the first few leading coefficients of \eqref{Ch of c=248/11} are decomposed as in Table \ref{decomposition by Th}.
\begin{table}
\begin{center}
\begin{tabular}{c|c|c}
\toprule
  $\alpha$   & $h$ & $\tsl{VT}^\natural(\alpha)_h$  \\\midrule
  $0$ & $0$ & $\mathbf{1}$ \\
      & $2$ & $\mathbf{1}\oplus\mathbf{30875}$ \\
      & $3$ & $\mathbf{1}\oplus\mathbf{30628}\oplus\mathbf{30875}\oplus\mathbf{2572752}$ \\
$1$   & $\sfrac{20}{11}$ & $\mathbf{30628}$ \\
     & $\sfrac{31}{11}$ & $\mathbf{30628}\oplus\mathbf{30875}\oplus\mathbf{3376737}$ \\
     $2$ & $\sfrac{16}{11}$ & $\mathbf{4123}$ \\
     & $\sfrac{27}{11}$ & $\mathbf{4123}\oplus\mathbf{957125}$\\
$3$ & $\sfrac{21}{11}$ & $\mathbf{61256}$ \\
   & $\sfrac{32}{11}$ & $\mathbf{61256}\oplus \mathbf{957125}\oplus\mathbf{4936750}$ \\
   $4$ & $\sfrac{13}{11}$ & $\mathbf{248}$ \\
   & $\sfrac{24}{11}$ & $\mathbf{248}\oplus\mathbf{147250}$ \\
  \bottomrule
\end{tabular}
\caption{Decompositions of the graded components of the modules $\tsl{VT}^\natural(\alpha)$ into irreducible representations of $\Th$.}\label{decomposition by Th}
\end{center}
\end{table}

As a consistency check, one can twine the characters of $\W_{D_{\mathrm{3C}}}$ by their $D_{\mathrm{3C}}$ automorphism, 
\begin{align}
    \chi_{g,\alpha}(\tau) = \mathrm{Tr}_{\W_{D_{\mathrm{3C}}}(\alpha)}gq^{l_0 - \frac{2}{33}} \ \ \ \ \ \ \ \ (g\in D_{\mathrm{3C}})
\end{align}
and/or the characters of $\tsl{VT}^\natural$ by elements of $\Th$ using Table \ref{decomposition by Th}, 
\begin{align}
    \chi_{\tsl{VT}^\natural(\alpha),h}(\tau) = \mathrm{Tr}_{\tsl{VT}^\natural(\alpha)}hq^{\widetilde{l}_0-\frac{31}{33}} \ \ \ \ \ \ \ \ \ (h\in\Th)
\end{align}
and observe that they satisfy e.g.\ (at least to low order in the $q$-expansion)
\begin{align}
\label{Thompson Gen bil}
    J_{\mathrm{3C}}(\tau) = \sum_\alpha \chi_{\omega,\alpha}(\tau)\chi_{\tsl{VT}^\natural(\alpha)}(\tau), \ \ \ \ J_h(\tau) = \sum_\alpha \chi_\alpha(\tau) \chi_{\tsl{VT}^\natural(\alpha),h}(\tau)
\end{align}
where $\omega$ is one of the elements of order 3 in $D_{\mathrm{3C}}$, and more generally, 
\begin{align}
\label{Thompson Gen bilinear}
    J_{gh}(\tau) = \sum_\alpha \chi_{g,\alpha}(\tau) \chi_{\tsl{VT}^\natural(\alpha),h}(\tau).
\end{align}
In the above, we are thinking of $gh\in D_{\mathrm{3C}}\times\Th \hookrightarrow \mathbb{M}.$ As an illustrative example, we present the twined character for $h$ taken from the 2A conjugacy class of $\tsl{Th}$, which can be computed using the data in Table \ref{decomposition by Th} and Tables \ref{thchartabone}, \ref{thchartabtwo}, \ref{thchartabthree}, and \ref{thchartabfour} in appendix \ref{app:A}.
\begin{align}
\begin{split}
\chi_{\tsl{VT}^\natural(0),\mathrm{2A}}(\tau) &=  q^{-\frac{31}{33}}\left( 1 + 156 q^2 - 1008  q^3 + \cdots \right), \\
\chi_{\tsl{VT}^\natural(1),\mathrm{2A}}(\tau) &= q^{\frac{29}{33}}\left( -92 + 672 q + \cdots \right), \quad \chi_{\tsl{VT}^\natural(2),\mathrm{2A}}(\tau) = q^{\frac{17}{33}} \left( 27 -288 q + \cdots \right), \\
\chi_{\tsl{VT}^\natural(3),\mathrm{2A}}(\tau) &= q^{\frac{32}{33}}\left( 72  - 453 q +  \cdots \right), \quad \chi_{\tsl{VT}^\natural(4),\mathrm{2A}}(\tau) = q^{\frac{8}{33}} \left( -8 + 42 q  + \cdots \right).
\end{split}
\end{align}
It is straightforward to see that
\begin{align}
\begin{split}
\sum_{\alpha} \chi_\alpha(\tau) \chi_{\tsl{VT}^\natural(\alpha),\mathrm{2A}}(\tau) = \frac{1}{q} + 276 q -2048 q^2 +  \cdots ,
\end{split}
\end{align}
where the right-hand side corresponds to the Mckay-Thompson series of the 2B class in $\mathbb{M}$. Generalized bilinear relations for arbitrary conjugacy classes of $\mathbb{Z}_3\times\Th$ can be determined by how they fuse into conjugacy classes of the monster group. These fusion rules can be computed with Gap \cite{GAP}, and we present this data in Table \ref{Th Fusion table}.

Moonshine for the Thompson sporadic group has been studied in another context \cite{Harvey:2015mca}, but to our knowledge, the chiral algebra $\tsl{VT}^\natural$ does not have any obvious relationship to the automorphic forms which arise there.

\subsection{Other monstralizers}\label{subsec:otherexamples}

It is worthwhile to ask whether or not there are $\mathbb{M}$-com pairs beyond the ones we've discussed. One immediate example arises by considering the monstralizer pair $G\circ \widetilde{G}:=\mathbb{Z}_{2\mathrm{B}}\circ 2^{1+24}.\tsl{Co}_1$ which is intimately related to the original construction of $V^\natural$ by Frenkel, Lepowsky, and Meurman \cite{flm,flma,flmbook} as a $\mathbb{Z}_2$ orbifold of the Leech lattice VOA. Calling $\Lambda := \Lambda_{\mathrm{Leech}}$, the orbifold construction implies that the monster CFT can be decomposed as 
\begin{align}
    V^\natural = \V_\Lambda^+ \oplus \V_{\Lambda}^+(\sigma)
\end{align}
where $\V_\Lambda^+(\sigma)$, in physics language \cite{beauty}, is the space of $\mathbb{Z}_2$-invariant states in the twisted sector of the orbifold. In this picture, the generator of $\mathbb{Z}_{2\mathrm{B}}$ is the automorphism which acts as $+1$ on $\V_\Lambda^+$ and as $-1$ on the twisted states in $\V_\Lambda^+(\sigma)$. The group $2^{1+24}.\tsl{Co}_1$ is then interpreted as the collection of automorphisms in the monster which maps untwisted states to untwisted states, and twisted states to twisted states, i.e.\ which does not mix $\V_\Lambda^+$ and $\V_\Lambda^+(\sigma)$. In the language that we've been using in this paper, we would then say that $\mathrm{Aut}(\V_\Lambda^+)=\mathrm{Inn}(\V_\Lambda^+) \cong 2^{24}.\tsl{Co}_1$ (which is related to the fact that $2.\tsl{Co}_1$ is the automorphism group of the Leech lattice) and that the fusion algebra spanned by $\{\V_\lambda^+,\V_\Lambda^+(\sigma)\}$ has a $\mathbb{Z}_2$ automorphism. This suggests defining $\W_{\widetilde{G}}= \V_\Lambda^+$, which will then have most of the nice properties required of an $\mathbb{M}$-com pair. However, this would force us to define $\W_G$ to be the trivial CFT, in which case $(\W_G,\W_{\widetilde{G}})$ are not each others' commutants. This example is therefore a somewhat degenerate case of the $\mathbb{M}$-com pairs we have been considering.

Despite this shortcoming, the VOA $\V_{\Lambda}^+$ does seem to respect the structure of inclusions of monstralizer pairs. For example, the monstralizers $(\mathbb{Z}_{4\mathrm{A}},4.2^{22}.\tsl{Co}_3)$, $(D_{\mathrm{4A}},2^{1+22}.\tsl{McL})$, and $(D_{\mathrm{2B}},2.2^{1+22}.\tsl{Co}_2)$ all include into $(\mathbb{Z}_{2\mathrm{B}},2^{1+24}.\tsl{Co}_1)$, and the chiral algebras $\widetilde{\W}_{\mathbb{Z}_{\mathrm{4A}}}$, $\widetilde{\W}_{D_{\mathrm{4A}}}$, $\widetilde{\W}_{D_{\mathrm{2B}}}$ correspondingly each embed into $\V_\Lambda^+$ because each is a charge conjugation orbifold $\V_{\widetilde L}^+$ with $\widetilde{L}\subset\Lambda$. 
One can see this explicitly at the level of characters. The graded-dimensions of $\V_\Lambda^+$ and $\V_{\Lambda}^+(\sigma)$ are given by $J_+(\tau)$ and $J_-(\tau)$ respectively,
\begin{align}\label{2Beven}
J_+(\tau) &= \frac12\left(\frac{\Theta_\Lambda(\tau)}{\eta(\tau)^{24}}+\Phi_{0,1}(\tau)^{24} \right) =\frac12\left(\frac{\Theta_\Lambda(\tau)}{\eta(\tau)^{24}}+\frac{\eta(\tau)^{24}}{\eta(2\tau)^{24} } \right) , \nonumber \\
J_-(\tau) &=  2^{11}\left( \Phi_{1,0}(\tau)^{24} - \Phi_{1,1}(\tau)^{24}\right) = 2^{11}\left( \frac{\eta(\tau)^{24}}{\eta(\tau/2)^{24}} - \frac{\eta(2\tau)^{24}\eta(\tau/2)^{24}}{\eta(\tau)^{48}} \right), 
\end{align}
where $\Theta_\Lambda(\tau)$ is the theta-function of the Leech lattice, which satisfies $\Theta_\Lambda(\tau)/\eta(\tau)^{24}= J(\tau)+24$. These characters can be decomposed into $\W_{\mathbb{Z}_{\mathrm{4A}}}\otimes\widetilde{\W}_{\mathbb{Z}_{\mathrm{4A}}}$ characters as
\begin{align} 
\begin{split}
J_+(\tau) &= \sum_{\alpha=0,3,4,7,8,9}\chi_\alpha(\tau)\widetilde\chi_\alpha(\tau)\\
J_-(\tau) &=  \sum_{\alpha=1,2,5,6}\chi_\alpha(\tau)\widetilde\chi_\alpha(\tau)   
\end{split}
\end{align}
where $\chi_\alpha(\tau)$ and $\widetilde{\chi}_\alpha(\tau)$ are given in equations \eqref{z4parafch} and \eqref{Co3}. Into $\W_{D_{\mathrm{4A}}}\otimes\widetilde{\W}_{D_{\mathrm{4A}}}$, they decompose as 
\begin{align}
J_+(\tau) &= \sum_{\alpha = 0}^8 \chi_\alpha(\tau)\widetilde{\chi}_\alpha(\tau) \\
J_-(\tau) &= \sum_{\alpha=9,10} \chi_\alpha(\tau) \widetilde{\chi}_\alpha(\tau)
\end{align}
where $\chi_\alpha(\tau)$ and $\widetilde{\chi}_\alpha(\tau)$ are given in equations \eqref{4A characters} and \eqref{4A dual ch}. Into $\W_{D_{\mathrm{2B}}}\otimes \widetilde{\W}_{D_{\mathrm{2B}}}$ characters, 
 \begin{align} 
J_+(\tau) &= \sum_{\alpha=0,1,2,5,8}\chi_\alpha(\tau)\widetilde\chi_\alpha(\tau)   ,\nonumber \\
J_-(\tau) &=    \sum_{\alpha=3,4,6,7} \chi_\alpha(\tau) \widetilde\chi_\alpha(\tau)  .
\end{align}
where $\chi_\alpha(\tau)$ and $\widetilde{\chi}_\alpha(\tau)$ are given in equations \eqref{D2B chars} and \eqref{D2B dual chars}.

It is more interesting to ask if the idea of $\mathbb{M}$-com pairs has a chance at recovering chiral algebras whose inner automorphism groups are precisely sporadic groups. For example, there are at least three other monstralizers 
\begin{align}\label{othermonstralizers}
    [G\circ \widetilde{G}].H = [7:3\times \tsl{He}].2, \ [\mathbb{Z}_{\mathrm{6B}}\circ 6.\tsl{Suz}].2, \ [(2\times 5:4)\circ 2.\tsl{HS}].2
\end{align}
with the property that $\widetilde{G}/Z(\widetilde{G})$ is exactly a sporadic group, so any monstralizing commutant pair which uplifts these would give rise to chiral algebras one might justifiably call $\tsl{VHe}^\natural$, $\tsl{VSz}^\natural$, and $\tsl{VHS}^\natural$. Do such $\mathbb{M}$-com pairs exist? 

For concreteness, we will present an argument for the existence of $\tsl{VHe}^\natural$, and then make some comments about more general monstralizer pairs. \\ \\
\emph{\textbf{Claim:} There exists a chiral algebra $\tsl{VHe}^\natural$ with $\tsl{He}.2$ automorphism group (and conjecturally $\tsl{He}$ inner automorphism group) which is a commutant subalgebra of both $\tsl{VF}_{24}^\natural$ and $V^\natural$. Thus, the VOA participates in an embedding chain 
\begin{align}
    \tsl{VHe}^\natural\hookrightarrow \tsl{VF}_{24}^\natural \hookrightarrow V^\natural
\end{align}
 which mirrors the inclusions of the corresponding monstralizers, 
 \begin{align}
     \tsl{He}\hookrightarrow 3.\tsl{Fi}_{24}' \hookrightarrow \mathbb{M}.
 \end{align}
 }
 
The sketch of a proof is as follows. Our goal is to realize $\tsl{VHe}^\natural$ as a commutant subalgebra of $\tsl{VF}_{24}^\natural$, the Fischer VOA. In order to do this, we must find a decomposition of the stress tensor $T(z)$ of $\tsl{VF}_{24}^\natural$ into a sum of two conformal vectors of smaller central charge, $T(z) = t(z)+\widetilde{t}(z)$. To find a decomposition which respects the Held symmetry, we decompose the Griess algebra of the Fischer VOA into $\tsl{He}$ representations via a character theoretic calculation, 
\begin{align}
    (\tsl{VF}^\natural_{24})_2\Big\vert_{\tsl{He}}\cong \mathbf{1}\oplus \mathbf{1} \oplus (\text{non-trivial})
\end{align}
which reveals a 2-dimensional subspace fixed by $\tsl{He}$. The same is true if one decomposes $(\tsl{VF}^\natural_{24})_2$ with respect to $\tsl{He}.2$. This means that the subVOA $(\tsl{VF}^\natural_{24})^{\tsl{He}}$, consisting of states fixed by $\tsl{He}$, has a 2-dimensional space of dimension 2 operators, i.e.\ a 2-dimensional Griess algebra. On general grounds, it must be spanned by the stress tensor $T(z)$ and a single Virasoro primary. This was precisely the situation considered in \S\ref{subsec:deconstruction}, where it was found that such a theory always admits a unique deconstruction of its stress tensor $T(z) = t(z) + \widetilde{t}(z)$ into two commuting conformal vectors $t$ and $\widetilde{t}$ of smaller central charge. By virtue of the map \eqref{authom} and the fact that $\tsl{He}$ is a simple group, it must act entirely on either $\tsl{Com}_{\tsl{VF}_{24}^\natural}(\tsl{Vir}(t))$ or $\tsl{Com}_{\tsl{VF}_{24}^\natural}(\tsl{Vir}(\widetilde{t}))$ as the only other possibility is that it act by an abelian group of diagonal fusion algebra automorphisms; we assume without loss of generality that it acts on the former and define 
\begin{align}
    \tsl{VHe}^\natural = \tsl{Com}_{\tsl{VF}_{24}^\natural}(\tsl{Vir}(t))
\end{align}
which is a theory with at least $\tsl{He}$ symmetry. Further, because $\tsl{He}.2$ is maximal in $\Fi_{24}$ \cite{linton1991maximal}, $\tsl{VHe}^\natural$ can inherit no more automorphisms from $\Fi_{24}$ than just $\tsl{He}.2$. We suspect that it does inherit the full $\tsl{He}.2$ with the extra order two element acting as an outer automorphism, and that the theory $\tsl{VHe}^\natural$ and its commutant in $V^\natural$ furnish an $\mathbb{M}$-com pair corresponding to the monstralizer $[7:3\times \tsl{He}].2$. \qedsymbol{}\\ \\
There are two important general takeaways from this argument. The first is that, if an $\mathbb{M}$-com uplift of a monstralizer $[G\circ \widetilde{G}].H$ exists, and if the Griess algebra of $V^\natural$ decomposes into $[G\circ \widetilde{G}].H'$ representations with two singlets, i.e. 
\begin{align}
    V_2^\natural\Big\vert_{[G\circ \widetilde{G}].H'} \cong \mathbf{1}\oplus\mathbf{1}\oplus (\text{non-trivial irreducibles})
\end{align}
then the pair $(\W_G,\W_{\widetilde{G}})$ is uniquely determined. This is the case, for example, for $[G\circ \widetilde{G}].H = [\mathbb{Z}_{\mathrm{6B}}\circ 6.\tsl{Suz}].2\cong 6.\tsl{Suz}.2$. The second takeaway is that, in the cases that more than two singlets appear in the decomposition of $V_2^\natural$ with respect to the monstralizer, one may be able to find an intermediate $\mathbb{M}$-com pair $(\W_K,\W_{\widetilde{K}})$ with  $\widetilde{G}\hookrightarrow \widetilde{K}$ and try to realize the putative VOA $\W_{\widetilde{G}}$ as a commutant subalgebra of $\W_{\widetilde{K}}$. We suspect that such arguments could play a useful role in iteratively defining the full suite of $\mathbb{M}$-com pairs $(\W_G,\W_{\widetilde{G}})$, albeit non-constructively.

Unfortunately, such arguments do not immediately reveal the central charge of e.g.\ $\W_{7:3}$, which is necessary for a detailed understanding of $\tsl{VHe}^\natural = \widetilde{\W}_{7:3}$, including the computation of its dual characters. However, there are various properties we can expect. First, one can read off from the monstralizer that $\mathrm{Inn}(\W_{7:3})\cong 7:3$. Furthermore, in order for $\tsl{VHe}^\natural$ to embed into the Fischer VOA $\tsl{VF}_{24}^\natural$, $\W_{7:3}$ should contain a $\mc{P}(3)$ subalgebra. Actually, since the Miyamoto lift of the $\mathbb{Z}_3$ fusion algebra automorphism of this $\mc{P}(3)$ subalgebra will give rise to an order 3 element in $7:3$, there should be 7 $\mc{P}(3)$ subalgebras which are permuted amongst each other by the generator of the $\mathbb{Z}_7$ normal subgroup. We leave a more detailed study of these issues to future work, and for now provide a character theoretic calculation related to an algebra $\widetilde{\U}$ which may sit in between $\tsl{VHe}^\natural$ and $V^\natural$, i.e.\ $\tsl{VHe}^\natural\hookrightarrow \widetilde{\U} \hookrightarrow V^\natural$ (though we emphasize that $\widetilde{\U}$ does \emph{not} participate in an $\mathbb{M}$-com pair).  

Our starting point is the observation that the Held group appears in connection with the centralizer/normalizer of the cyclic subgroup generated by any element from the 7A conjugacy class of the monster, 
\begin{align}
    \mathrm{Cent}_{\mathbb{M}}(\mathrm{7A})\cong \mathbb{Z}_7\times\tsl{He}, \ \ \ \ \ \ N_{\mathbb{M}}(\mathrm{7A}) \cong (7:3\times \tsl{He}).2
\end{align}
with $(7:3\times\tsl{He}).2$ a maximal subgroup of $\mathbb{M}$. A natural way to try to realize order $k$ cyclic subgroups of the monster is by finding a $\mc{P}(k)$ parafermion subalgebra and lifting the order $k$ automorphism of its fusion algebra to an automorphism of $V^\natural$. In this case, we might try to locate a $\mc{P}(7)$ subalgebra of $V^\natural$, and study its commutant. 

Before proceeding, we should establish that such a subalgebra exists. First, it is known (c.f.\ the discussion around equation \eqref{ANpf}) that the lattice VOA based on $\sqrt{2}\Lambda_{\mathrm{root}}(A_{k-1})$ contains a $\mc{P}(k)$ subalgebra, which in fact survives in $\V_{\sqrt{2}\Lambda_{\mathrm{root}}(A_{k-1})}^+$ as well \cite{dong1996associative}. Second, it is also known that $\sqrt{2}N$ is a sublattice of the Leech lattice $\Lambda_{\mathrm{Leech}}$ for each Niemeier lattice\footnote{The Niemeier lattices are the even, positive-definite, unimodular lattices of rank 24.} $N$. Furthermore, there is a Niemeier lattice $N(A_6^4)$ based on the root system $A_6^4$, which implies the chain of embeddings 
\begin{align}
    \mc{P}(7)\hookrightarrow \V^+_{\sqrt{2}\Lambda_{\mathrm{root}}(A_{6})} \hookrightarrow \V^+_{\sqrt{2}N(A_6^4)}\hookrightarrow \V^+_{\Lambda_{\mathrm{Leech}}} \hookrightarrow V^\natural
\end{align}
where in the last step, we used the fact that $V^\natural$ is a $\mathbb{Z}_2$ orbifold of the lattice VOA based on $\Lambda_{\mathrm{Leech}}$. This establishes the existence of a $\mc{P}(7)$ subalgebra;  we call its commutant $\widetilde{\U}$, which we expect admits an action of $\tsl{He}$. There will then be a decomposition 
\begin{align}
    V^\natural \cong \bigoplus_{(\ell,m)} \mc{P}(7,[\ell,m])\otimes \widetilde{\U}(\ell,m)
\end{align}
with $\widetilde{\U}(7,7) \cong \widetilde{\U}$, and the $\widetilde{\U}(\ell,m)$ furnishing representations of $\widetilde{\U}$. We will give evidence momentarily that the $\mathbb{Z}_7$ fusion algebra automorphism of $\mc{P}(7)$ lifts to an element of the 7A conjugacy class of $\mathbb{M}$.

Once again, we employ Hecke operators to get our hands on the graded-dimensions of the $\widetilde{\U}(\ell,m)$. The $\IZ_7$ parafermion theory has $28$ different primary fields. We order the characters $\chi_\alpha$ of the corresponding highest weight modules as 
\begin{align}
\chi_{0} &= \psi^{(7)}_{7,7}, \quad  \chi_{1} = \psi^{(7)}_{1,1}, \quad \chi_{2} = \psi^{(7)}_{6,6}, \quad \chi_{3} = \psi^{(7)}_{2,2}, \quad \chi_{4} = \psi^{(7)}_{5,5}, \nonumber \\
\chi_{5} &= \psi^{(7)}_{3,3}, \quad \chi_{6} = \psi^{(7)}_{4,4}, \quad \chi_{7} = \psi^{(7)}_{2,0}, \quad \chi_{8} = \psi^{(7)}_{3,-1}, \quad \chi_{9} = \psi^{(7)}_{3,1}, \nonumber \\
\chi_{10} &= \psi^{(7)}_{4,-2}, \quad \chi_{11} = \psi^{(7)}_{4,2}, \quad \chi_{12} = \psi^{(7)}_{5,-3}, \quad \chi_{13} = \psi^{(7)}_{5,3}, \quad \chi_{14} = \psi^{(7)}_{4,0}, \\
\chi_{15} &= \psi^{(7)}_{6,-4}, \quad \chi_{16} = \psi^{(7)}_{6,4}, \quad \chi_{17} = \psi^{(7)}_{7,-5}, \quad \chi_{18} = \psi^{(7)}_{7,5}, \quad \chi_{19} = \psi^{(7)}_{5,-1}, \nonumber \\
 \chi_{20} &= \psi^{(7)}_{5,1}, \quad \chi_{21} = \psi^{(7)}_{6,-2}, \quad \chi_{22} = \psi^{(7)}_{6,2}, \quad \chi_{23} = \psi^{(7)}_{6,0}, \quad \chi_{24} = \psi^{(7)}_{7,-3}, \nonumber\\
 \chi_{25} &= \psi^{(7)}_{7,3}, \quad \chi_{26} = \psi^{(7)}_{7,-1}, \quad \chi_{27} = \psi^{(7)}_{7,1}, \nonumber
\end{align}
which have conformal weights 
\begin{align}
\begin{split}
h &= \big(0,\tfrac{1}{21},\tfrac{1}{21},\tfrac{5}{63},\tfrac{5}{63},\tfrac{2}{21},\tfrac{2}{21},\tfrac{2}{9},
\tfrac{8}{21},\tfrac{8}{21},\tfrac{11}{21},\tfrac{11}{21},\tfrac{41}{63},\tfrac{41}{63}, \\
& \quad\quad\quad\quad\quad \tfrac{2}{3},\tfrac{16}{21},\tfrac{16}{21},\tfrac{6}{7},\tfrac{6}{7},\tfrac{59}{63},\tfrac{59}{63}, \tfrac{25}{21},\tfrac{25}{21},\tfrac{4}{3},\tfrac{10}{7},\tfrac{10}{7},\tfrac{12}{7},\tfrac{12}{7}\big).
\end{split}
\end{align}
The conductor is given by $N=126$. 

The central charge of $\mc{P}(7)$ is $c_t=\frac43$, and so the central charge $c_{\widetilde{t}}$ of $\widetilde{\U}$ is $c_{\widetilde{t}} =24-c_t= 17c_t$. Since $(126,17) = 1$, we may consider applying the Hecke operator $\mathsf{T}_{17}$ to $\chi$. Using the modular S matrix of the characters $\chi$ provided in equation \eqref{Parafermion CFT character}, one readily computes the $q$-expansions of their Hecke images,
\begin{align*}
({\mathsf T}_{17} \chi)_{0}(\tau) &= q^{ -\frac{17}{18}} \left(  1 + 15810 q^2+ 1375232 q^3 + 47653839 q^4    +\cdots   \right), \\
({\mathsf T}_{17} \chi)_{1}(\tau) &= q^{\frac{37}{126} } \left( 204 + 97563 q + 6392374 q^2+191594522 q^3 + 3643081640 q^4 + \cdots  \right), \\
({\mathsf T}_{17} \chi)_{3}(\tau) &= q^{\frac{5}{42} } \left(  51 + 55284 q + 4454595 q^2  +148169790 q^3 + 3009822273 q^4 +\cdots   \right), \\
({\mathsf T}_{17} \chi)_{5}(\tau)&= q^{\frac{67}{126}} \left( 1955+ 448902 q + 23218498 q^2 + 612398140 q^3      +\cdots  \right), \\
({\mathsf T}_{17} \chi)_{7}(\tau) &= q^{\frac56 } \left( 11679 + 1432080 q+ 57855030 q^2  +1323345654 q^3 + \cdots   \right), \\
({\mathsf T}_{17} \chi)_{8}(\tau) &= q^{\frac{121}{126}} \left( 27234 + 2721020 q + 100447832 q^2 + 2175623416 q^3   +\cdots  \right), \\
({\mathsf T}_{17} \chi)_{10}(\tau) &= q^{\frac{85}{126} } \left( 5084 + 836961 q + 38286006 q^2 + 941558481 q^3      +\cdots  \right), \\
({\mathsf T}_{17} \chi)_{12}(\tau) &= q^{\frac{41}{42} } \left(  26112 + 2548623 q+ 93075204 q^2 + 2002645050 q^3     +\cdots  \right), \\
({\mathsf T}_{17} \chi)_{14}(\tau)  &= q^{\frac{7}{18} } \left(  680 + 234226 q + 13898520 q^2 + 395054092 q^3     +\cdots  \right), \\
({\mathsf T}_{17} \chi)_{15}(\tau) &= q^{-\frac{17}{126} } \left( 1 + 10404 q + 1210230 q^2 + 47772074 q^3 + 1077610433 q^4    +\cdots  \right), \\
({\mathsf T}_{17} \chi)_{17}(\tau) &= q^{ \frac{43}{126}} \left(  153 + 65212 q + 4054534 q^2 + 118328160 q^3   +\cdots  \right), \\
({\mathsf T}_{17} \chi)_{19}(\tau) &= q^{\frac{17}{42} } \left(  681 + 221646 q + 12948390 q^2 + 364895820 q^3     +\cdots  \right), \\
({\mathsf T}_{17} \chi)_{21}(\tau) &= q^{\frac{127}{126} } \left( 22984 + 2144856 q + 76669456 q^2       +1628223858 q^3+\cdots \right), \\
({\mathsf T}_{17} \chi)_{23}(\tau) &= q^{\frac{13}{18} } \left( 4454 + 668848 q + 29441076 q^2 + 708076656 q^3     +\cdots  \right), \\
({\mathsf T}_{17} \chi)_{24}(\tau) &= q^{\frac{25}{126} } \left( 51 + 32504 q + 2381394 q^2 + 75441121 q^3    +\cdots  \right), \\
({\mathsf T}_{17} \chi)_{26}(\tau) &= q^{\frac{79}{126} } \left(   1275  + 236980 q + 11269249 q^2 + 283574263 q^3    +\cdots \right), 
\end{align*}
and
\begin{align*}
({\mathsf T}_{17} \chi)_{2}(\tau) &= ({\mathsf T}_{17} \chi)_{1}(\tau), \quad ({\mathsf T}_{17} \chi)_{4}(\tau) = ({\mathsf T}_{17} \chi)_{3}(\tau),  \quad ({\mathsf T}_{17} \chi)_{6}(\tau) = ({\mathsf T}_{17} \chi)_{5}(\tau), \\
({\mathsf T}_{17} \chi)_{9}(\tau) &= ({\mathsf T}_{17} \chi)_{8}(\tau), \quad ({\mathsf T}_{17} \chi)_{11}(\tau) = ({\mathsf T}_{17} \chi)_{10}(\tau), \quad ({\mathsf T}_{17} \chi)_{13}(\tau) = ({\mathsf T}_{17} \chi)_{12}(\tau),   \\
({\mathsf T}_{17} \chi)_{16}(\tau) &= ({\mathsf T}_{17} \chi)_{15}(\tau), \quad ({\mathsf T}_{17} \chi)_{18}(\tau) = ({\mathsf T}_{17} \chi)_{17}(\tau), \quad  ({\mathsf T}_{17} \chi)_{22}(\tau) = ({\mathsf T}_{17} \chi)_{21}(\tau),   \\
({\mathsf T}_{17} \chi)_{25}(\tau) &= ({\mathsf T}_{17} \chi)_{24}(\tau), \quad ({\mathsf T}_{17} \chi)_{27}(\tau) = ({\mathsf T}_{17} \chi)_{26}(\tau).
\end{align*}
If we are to identify these functions with the graded dimensions of the $\widetilde{\U}(\ell,m)$, they should fit into a bilinear with the characters $\chi$ which produces $J(\tau)$. From the general theory of such bilinears, we must solve the equation
\be
89 + \ell^2 = 0 \ \mbox{mod} \ 126.
\ee
for $\ell$, and consider the matrices $G_\ell$ (c.f.\ \S\ref{subsec:Hecke}). It is straightforward to see $\ell = 17,$ $53,$ $73,$ $109$ are the available solutions, and the corresponding matrices are identical up to sign. Namely,
\be
G_{17} = G_{109} = -G_{53} = -G_{73}.
\ee
So, without loss of generality, we can work with $G_{17}$, which is 0 in each entry, except for the entries
\begin{align}
\begin{split}
 &(0, 0), \ (1, 15),\ (2, 16),\ (3, 19),\ (4, 20),\ (5, 11),\ (6, 10),\ (7, 7),\ (8, 6),\ (9, 5),\\ 
 &(10, 9), (11, 8),\ (12, 4),\ (13, 3),\ (14, 14),\ (15, 22),\ (16, 21),\ (17, 26),\ (18, 27),\\ 
 &(19, 12),\ (20, 13),\ (21, 1),\ (22, 2),\ (23, 23),\ (24, 18),\ (25, 17),\ (26, 24),\ (27, 25),
 \end{split}
\end{align}
in which it is 1. One can then show that this matrix furnishes a bilinear with the right properties,
\begin{align}
\begin{split}
J(\tau) =  ({\mathsf T}_{17} \chi)^T(\tau) \cdot G_{17} \cdot \chi(\tau).
\end{split}
\end{align}
As a further check, we can show that the same bilinear is consistent with the $\mathbb{Z}_7$ symmetry of the parafermion theory lifting to an element of the 7A conjugacy class of $\mathbb{M}$. Indeed, if one defines the vector-valued function $\chi_\omega$ by making the replacements $\psi^{(7)}_{\ell,m}\to e^{\frac{2\pi i m}{7}}\psi_{\ell,m}^{(7)}$ in the components of $\chi$, then one finds by direct computation that, to low orders in the $q$-expansion,
\begin{align}
\begin{split}
 ({\mathsf T}_{17} \chi)^T(\tau)  \cdot G_{17} \cdot \chi_{\omega}(\tau) = \frac{1}{q} +51 q +204 q^2+681 q^3+1956 q^4+ \cdots,
\end{split}
\end{align}
which agrees with the Mckay-Thompson series of the 7A conjugacy class in $\mathbb{M}$, $J_{\mathrm{7A}}(\tau)$. One can thus identify $\widetilde{\chi}_{\ell,m}(\tau):=\mathrm{Tr}_{\widetilde{\U}(\ell,m)}q^{\widetilde{l}_0-\frac{17}{18}}$ with the component of $G_{17}^T\cdot (\mathsf{T}_{17}\chi)$.

Finally, we mention that low order terms in the $q$-expansion of the $({\mathsf T}_{17} \chi)_{\alpha}(\tau)$ involve coefficients that are consistent with decompositions into small numbers of irreducible representation of the Held group. For example, $51$, $153$ and $680$ are dimensions of Held irreps, $204=153+51$, $1955=1275+680$ and so on.

\subsection{Baby monster and Fischer deconstructions from McKay's correspondence}\label{subsec:mckaye7}

In previous sections, we constructed several examples of VOAs as commutant subalgebras of $V^\natural$, one for each conjugacy class arising in McKay's $\widehat{E_8}$ correspondence. As described in \S\ref{subsubsec:BMe7} and \S\ref{subsubsec:Fischere6}, the baby monster $\mathbb{B}$ and Fischer's group $\Fi_{24}$ enjoy similar relationships with the Dynkin diagrams of $\widehat{E_7}$ and $\widehat{E_6}$ respectively (c.f.\ Figures \ref{e7} and \ref{e6}), and so we may consider repeating the same kind of analysis for $\tsl{V}\mathbb{B}^\natural$ and $\tsl{VF}^\natural_{24}$. We we will see that this naturally leads to centralizing commutant pairs in $\tsl{V}\mathbb{B}^\natural$ and $\tsl{VF}^\natural_{24}$. 

\subsubsection*{$\mathbb{B}$-com pairs}

We start by studying the commutants of the chiral algebras $\W_{B(n\mathrm{X})}$ considered in \cite{hohn2012mckaye7}, whose defining property is that they have two central charge $\frac{7}{10}$ conformal vectors whose associated $\sigma$-type Miyamoto involutions (c.f.\ equation \eqref{sigma type 7/10}) have product lying in the $n$X conjugacy class of $\mathbb{B}$; we refer to (loc.\ cit.) for detailed descriptions of these algebras. In this section, we will label the dihedral subgroup of $\mathbb{B}$ generated by these $\sigma$-type Miyamoto involutions as $D_{n\mathrm{X}}$. We do not attempt to solve this problem completely; we content ourselves here with observing that the cases\footnote{The algebra e.g.\ $\widetilde{\W}_{B(\mathrm{1A})}$ is defined to be the commutant of $\W_{B(\mathrm{1A})}$ in $\tsl{V}\mathbb{B}^\natural$, while the algebra $\widetilde{\W}_{D_{\mathrm{2A}}}$ is defined to be the commutant of $\W_{D_{\mathrm{2A}}}$ in $V^\natural$; we hope that this notation will not cause confusion.} $\widetilde{\W}_{B(\mathrm{2A})}$, $\widetilde{\W}_{B(\mathrm{3A})}$, and $\widetilde{\W}_{B(\mathrm{2C})}$ are identical to the algebras $\widetilde{\W}_{D_{\mathrm{2A}}}$, $\widetilde{\W}_{D_{\mathrm{3A}}}\cong \tsl{VF}_{23}^\natural$, and $\widetilde{\W}_{D_{\mathrm{4B}}}$ respectively. This will follow immediately from the fact that e.g.\ $\W_{B(\mathrm{2A})}\subset \W_{D_{\mathrm{1A}}}$, and so on.  \\ \\
\textbf{1A case:}  $[D_{\mathrm{1A}}\circ 2.{^2}\tsl{E}_6(2)].2$

In analogy with the algebra $\W_{D_{\mathrm{1A}}}$ considered in \S\ref{subsubsec:BM}, the algebra $\W_{B(\mathrm{1A})}$ by definition has two conformal vectors $e$ and $f$ of central charge $\frac{7}{10}$ and $\sigma$-type whose associated Miyamoto involutions $\sigma_e$ and $\sigma_f$ multiply to produce the identity element of $\mathbb{B}$. This implies that their involutions are actually the same, $\sigma_e=\sigma_f$, and since elements of the 2A conjugacy class of $\mathbb{B}$ are in one-to-one correspondence with central charge $\frac{7}{10}$ conformal vectors of $\sigma$-type in $\tsl{V}\mathbb{B}^\natural$, this means that $e=f$. Thus, we simply have that $\W_{B(\mathrm{1A})} \cong \mathcal{L}(\tfrac{7}{10},0)$; further, this algebra sits inside of the 2A case of McKay's $\widehat{E_8}$ correspondence, i.e.\ $\W_{B(\mathrm{1A})}\subset\W_{D_{\mathrm{2A}}}$. The commutant of such a subalgebra of $\tsl{V}\mathbb{B}^\natural$ was computed earlier (c.f. equation \eqref{2E6(2).2 in B}), where it was shown to produce $\widetilde{\W}_{D_{\mathrm{2A}}}$, the chiral algebra with ${^2}E_6(2).2$ symmetry. We thus decorate the 1A node of $\widetilde{E_7}$ with this chiral algebra. \\ \\
\textbf{2C case:} $[D_{\mathrm{2C}}\circ 2^2. \tsl{F}_4(2)].2$

The 2C case can be described as a subalgebra of $\tsl{V}\mathbb{B}^\natural$ isomorphic to %
\begin{align}
    \W_{B(\mathrm{2C})} \cong \mathcal{L}(\tfrac{7}{10},0)\otimes \mathcal{L}(\tfrac{7}{10},0) \oplus \mathcal{L}(\tfrac{7}{10},\tfrac32) \otimes \mathcal{L}(\tfrac{7}{10},\tfrac32)
\end{align}
This is a subalgebra of $\W_{D_{\mathrm{4B}}}$. It follows from the iterated deconstruction in equation \eqref{F4(2).2 iterated} that the commutant of this algebra in $\tsl{V}\mathbb{B}^\natural$ is given by $\widetilde{\W}_{D_{\mathrm{4B}}}$, the chiral algebra with $\tsl{F}_4(2).2$ symmetry. We thus label the 2C node of $\widehat{E_7}$ with this group. \\ \\
\textbf{3A case:} $[D_{\mathrm{3A}}\times \tsl{Fi}_{22}].2$

The 3A algebra $\W_{B(\mathrm{3A})}$ is an extension of $\mathcal{P}(3)\otimes \mathcal{P}(6)$. Using the fact that $\W_{D_{\mathrm{6A}}}$ has a $\mathcal{L}(\tfrac12,0)\otimes \mathcal{P}(3)\otimes \mathcal{P}(6)$ subalgebra, if one performs an iterated deconstruction of $V^\natural$ with respect to the decomposition 
\begin{align}
    T(z) = t^{(\frac12)}(z)+t^{(\frac{41}{20})}(z)+\widetilde{t}(z),
\end{align}
where $t^{(\frac{41}{20})}$ is the stress of $\mathcal{P}(3)\otimes \mathcal{P}(6)$, then it follows immediately that one can realize $\widetilde{\W}_{\mathrm{6A}}$ as the commutant of $\W_{B(\mathrm{3A})}$ in $\tsl{V}\mathbb{B}^\natural$, i.e. $\widetilde{\W}_{B(\mathrm{3A})}\cong \widetilde{\W}_{\mathrm{6A}}$, the algebra with $\Fi_{22}.2$ symmetry. We thus label the 3A node of $\widehat{E_7}$ with this group.

\subsubsection*{$\normalfont{\tsl{Fi}}_{24}$-com pairs}

We may repeat this analysis once more for the Fischer group, which enjoys a relationship to the Dynkin diagram of $\widehat{E_6}$, as described in \S\ref{subsubsec:Fischere6}.  We briefly show that the commutants of $\W_{F(\mathrm{1A})}$ and $\W_{F(\mathrm{2A})}$ in $\tsl{VF}_{24}^\natural$ are isomorphic to $\tsl{VF}_{23}^\natural$ and $\tsl{VF}_{22}^\natural$ respectively. Again, we label the dihedral subgroup of $\tsl{VF}^\natural_{24}$ generated by the Miyamoto involutions as $D_{n\mathrm{X}}$.\\ \\
\textbf{1A case:} $D_{\mathrm{1A}}\times\tsl{Fi}_{23}$

McKay's $\widehat{E_6}$ correspondence concerns the ``derived'' conformal vectors of central charge $\frac67$ in $\tsl{VF}_{24}^\natural$. Thus, just as $\W_{D_{\mathrm{1A}}}\cong \mathcal{L}(\tfrac12,0)$ and $\W_{B(\mathrm{1A})}\cong \mathcal{L}(\tfrac{7}{10},0)$, we have that\footnote{ The appearance of the module $\mathcal{L}(\tfrac67,5)$ is related to the fact that the conformal vector is ``derived.''} $\W_{F(\mathrm{1A})}\cong \mathcal{L}(\tfrac67,0)\oplus \mathcal{L}(\tfrac67,5)$; further, $\W_{F(\mathrm{1A})}$ sits inside $\W_{D_{\mathrm{3A}}}$. It follows from the fact that $\W_{D_{\mathrm{3A}}}\cong\tsl{VF}^\natural_{23}$ admits a $\mathcal{L}(\tfrac45,0)\otimes \mathcal{L}(\tfrac67,0)$ subalgebra, and the iterated deconstruction performed with respect to the resulting decomposition 
\begin{align}
    T(z) = t^{(\frac45)}(z)+t^{(\frac67)}(z) + \widetilde{t}(z)
\end{align}
in e.g.\ \eqref{Fi23 in Fi24} that $\widetilde{\W}_{F(\mathrm{1A})}\cong \widetilde{\W}_{D_{\mathrm{3A}}}\cong \tsl{VF}_{23}^\natural$.\\ \\
\textbf{2A case:} $[D_{\mathrm{2A}}\circ 2^2.\tsl{Fi}_{22}].2$

Finally, the 2A case admits a subalgebra of the form $\mathcal{L}(\tfrac67,0)\otimes \mathcal{L}(\tfrac{25}{28},0)\subset \W_{F(\mathrm{2A})}$ and sits inside $\W_{D_{\mathrm{6A}}}$. It thus follows from the iterated deconstruction considered in \eqref{Fi22 iterated} that %
\begin{align}
    \widetilde{\W}_{F(\mathrm{2A})}\cong \widetilde{\W}_{\mathrm{6A}}\cong\tsl{VF}_{22}^\natural.
\end{align}

\section{Conclusions} \label{sec:Conc}

In this paper, we have studied various chiral algebras that occur as commutant subalgebras of the monster CFT, and which have interesting, usually sporadic symmetry groups. In many cases, we were able to obtain the characters of these theories as Hecke images, as solutions to MLDEs, or both.
Our results are naturally organized via a connection to monstralizer pairs, a concept which we have extended to the chiral algebra setting by defining the notion of a monstralizing commutant pair. Although the characters we have found were not derived directly from the definition of the commutatant subalgebras, they pass a number of nontrivial consistency checks, and in particular are compatible with the existence of a number of non-trivial Griess algebras with sporadic automorphism groups. 

A number of open questions remain.

\begin{enumerate}
    \item Is it possible to find an $\mathbb{M}$-com uplift of every pair of mutually centralizing groups in the monster? Can one iteratively look for centralizing commutant pairs inside of the chiral algebras so produced? Will this lead to candidate algebras $\tsl{VG}^\natural$ for every $G$ in the happy family, i.e.\ for each simple sporadic group $G$ which arises as a subquotient of $\mathbb{M}$? What distinguishes such subalgebras?
    \item What are the general properties of centralizing commutant pairs? Can they be put to good use in more general settings besides the monster CFT?
    \item Our use of Hecke operators to obtain the dual characters of $\widetilde{\W}$ in $V^\natural$ from the characters of $\W$ required that the central charge of $\widetilde{\W}$ be an integer multiple of the central charge of $\W$, $c_{\widetilde{t}}=qc_t$, and also that the conductor $N$ of the characters of $\W$ be coprime to this integer, $(N,q)=1$. Is there a generalization of Hecke operators which works in the more general case?
    \item The baby monster VOA was studied recently in \cite{Lin:2019hks}, where it was used to probe the category of topological lines in $V^\natural$. It was further anticipated in (loc.\ cit.) that analogous decompositions of $V^\natural$ could also shed light on its topological lines. What, if anything, do our deconstructions of $V^\natural$ reveal about the structure of defects in the moonshine module?
    \item Do the chiral algebras we constructed inherit any aspect of the genus zero property from $V^\natural$? What characterizes their McKay-Thompson series? Are they Rademacher summable?
    \item In recent work, \cite{Johnson-Freyd:2019wgb} classified all $\mathcal{N}=1$ SVOAs whose even part is a simply connected WZW algebra (other than $E_{7,2}$, $E_{7,1}^2$, and $E_{8,2}$). The exceptional cases in the classification have automorphism groups related to a chain of exceptional subgroups of $\tsl{Co}_1$ called the \emph{Suzuki chain}; the SVOAs arise as subalgebras in the Conway SVOA $V^{f\natural}$ \cite{Duncan-super,Duncan-Mack} in a similar manner to how our chiral algebras arise in $V^\natural$. Namely, they arise by what one might call supersymmetric deconstruction. Do our chiral algebras arise as exceptional entries in a classification of rational VOAs of some kind?
\end{enumerate}

\section*{Acknowledgements}

We thank G. Mason for very helpful correspondence and understand that he and C. Franc have also been looking at commutants of subVOAs of $V^\natural$. We also thank L.~F.~Alday, H.~Choi, S.~Harrison, T.~Johnson-Freyd, B.~Julia, Y.~Lin, G.~Moore, J.~Sempliner, S.~Shao, and Y. Wu for discussions. JH  would like to thank
L.~Dixon for collaboration on deconstructing CFTs roughly thirty years ago.
JH and SL gratefully acknowledge
the hospitality of the Aspen Center for Physics  (under
NSF Grant No. PHY-1066293) for providing an excellent atmosphere for collaboration.
We have made use of Gap \cite{GAP}, Sage \cite{sage}, and Magma \cite{MR1484478} for various computational aspects of our analysis. JH acknowledges support from the NSF\footnote{Any opinions, findings, and conclusions or recommendations expressed in this material are those of the author(s) and do not necessarily reflect the views of the National Science Foundation.} under grant PHY 1520748. BR gratefully acknowledges support from the NSF under grant PHY 1720397. KL and SL are supported in part by KIAS Individual Grants PG006904 and PG056502, and by the
National Research Foundation of Korea Grants NRF2017R1D1A1B06034369 and NRF2017R1C1B1011440. The work of JB is supported by the European Research Council (ERC) under the European Union's Horizon 2020 research and innovation programme (Grant No. 787185).

\newpage

\appendix

\section{Fusion rules for the $\widetilde{\W}_{D_{n\mathrm{X}}}$ algebras}\label{app:D}

In this appendix, we present the fusion rules for some of the theories discussed in the main text. The structure constants $\mathcal{N}_{\alpha\beta}^{\gamma}$ are computed using the S-matrix of each theory and the Verlinde formula, equation \eqref{Verlinde}. It turns out that each theory has $\mathcal{N}_{\alpha\beta}^{\gamma} = 0 \ \text{or} \ 1$ for all $\alpha,\beta,\gamma$. In the cases that we used the block-diagonalization method to determine the characters of our models (c.f.\ \S\ref{subsubsec:modules}), the consistency of the fusion rules provides a non-trivial check on our results. Because fusion algebra is associative, $\mathcal{N}_{\alpha\beta}^{\gamma} = 1$ imposes that $\mathcal{N}_{\beta\alpha}^{\gamma} = 1$. Below, we present all the non-vanishing fusion algebra coefficients of the five theories in \S \ref{subsubsec:Lietype}, \S\ref{subsubsec:Fi23}, \S\ref{subsubsec:Th}, \S\ref{subsubsec:HN} and \S\ref{subsubsec:Fi22}.

\bigskip

\subsubsection*{List of non-vanishing $\mathcal{N}_{\alpha\beta}^{\gamma}$ for $\widetilde{\W}_{D_{\mathrm{2A}}}$:}
\begin{align}
\begin{split}
&\mathcal{N}_{00}^{0} ,\ \mathcal{N}_{01}^{1},\ \mathcal{N}_{02}^{2},\ \mathcal{N}_{03}^{3} , \ \mathcal{N}_{04}^{4},\ \mathcal{N}_{05}^{5},\ \mathcal{N}_{06}^{6},\ \mathcal{N}_{07}^{7} , \ \mathcal{N}_{11}^{0},\ \mathcal{N}_{11}^{1},\ \mathcal{N}_{12}^{5},\ \mathcal{N}_{13}^{6},\ \mathcal{N}_{14}^{7}, \\ 
& \mathcal{N}_{15}^{2},\ \mathcal{N}_{15}^{5}, \ \mathcal{N}_{16}^{3}, \ \mathcal{N}_{16}^{6},\ \mathcal{N}_{17}^{4},\ \mathcal{N}_{17}^{7},\ \mathcal{N}_{22}^{0},\ \mathcal{N}_{23}^{4},\ \mathcal{N}_{24}^{3},\ \mathcal{N}_{25}^{1},\ \mathcal{N}_{26}^{7}, \ \mathcal{N}_{27}^{6},\ \mathcal{N}_{33}^{0}, \\ 
&\mathcal{N}_{34}^{2},\ \mathcal{N}_{35}^{7},\ \mathcal{N}_{36}^{1},\ \mathcal{N}_{37}^{5},\ \mathcal{N}_{44}^{0},\ \mathcal{N}_{45}^{6}, \ \mathcal{N}_{46}^{5},\ \mathcal{N}_{47}^{1},\ \mathcal{N}_{55}^{0},\ \mathcal{N}_{55}^{1},\ \mathcal{N}_{56}^{4},\ \mathcal{N}_{56}^{7},\ \mathcal{N}_{57}^{3}, \\ 
&\mathcal{N}_{57}^{6}, \ \mathcal{N}_{66}^{0},\ \mathcal{N}_{66}^{1},\ \mathcal{N}_{67}^{2},\ \mathcal{N}_{67}^{5},\ \mathcal{N}_{77}^{0},\ \mathcal{N}_{77}^{1}.
\end{split}
\end{align}

\subsubsection*{List of non-vanishing $\mathcal{N}_{\alpha\beta}^{\gamma}$ for $\normalfont{\tsl{VT}}^\natural$:}\label{app:FA}
\begin{align}
\begin{split}
&\mathcal{N}_{00}^{0} , \ \mathcal{N}_{01}^{1} , \ \mathcal{N}_{02}^{2} , \ \mathcal{N}_{03}^{3} , \ \mathcal{N}_{04}^{4} , \  \mathcal{N}_{11}^{0} ,\ \mathcal{N}_{11}^{1} , \  \mathcal{N}_{11}^{2} , \ \mathcal{N}_{12}^{1} , \ \mathcal{N}_{12}^{2} , \ \mathcal{N}_{12}^{3} , \ \mathcal{N}_{13}^{2} , \ \mathcal{N}_{13}^{3} , \\
&\mathcal{N}_{13}^{4} ,\ \mathcal{N}_{14}^{3} , \  \mathcal{N}_{14}^{4} , \ \mathcal{N}_{22}^{0} , \ \mathcal{N}_{22}^{1} , \ \mathcal{N}_{22}^{2} , \ \mathcal{N}_{22}^{3} , \ \mathcal{N}_{22}^{4} , \  \mathcal{N}_{23}^{1} ,\ \mathcal{N}_{23}^{2} , \  \mathcal{N}_{23}^{3} , \ \mathcal{N}_{23}^{4} , \ \mathcal{N}_{24}^{2} , \\
&  \mathcal{N}_{24}^{3} , \ \mathcal{N}_{33}^{0} , \ \mathcal{N}_{33}^{1} , \  \mathcal{N}_{33}^{2} ,\ \mathcal{N}_{33}^{3} , \  \mathcal{N}_{34}^{1} , \ \mathcal{N}_{34}^{2} , \ \mathcal{N}_{44}^{0} , \ \mathcal{N}_{44}^{1} .
\end{split}
\end{align}

\bigskip

\subsubsection*{List of non-vanishing $\mathcal{N}_{\alpha\beta}^{\gamma}$ for $\normalfont{\tsl{VHN}}^\natural$:}
\begin{align}
\begin{split}
&\mathcal{N}_{00}^{0} , \ \mathcal{N}_{01}^{1} , \ \mathcal{N}_{02}^{2} , \ \mathcal{N}_{03}^{3} , \ \mathcal{N}_{04}^{4} , \  \mathcal{N}_{05}^{5} ,\ \mathcal{N}_{06}^{6} , \  \mathcal{N}_{07}^{7} , \ \mathcal{N}_{08}^{8} , \ \mathcal{N}_{11}^{0} , \ \mathcal{N}_{11}^{3} , \ \mathcal{N}_{12}^{5} , \ \mathcal{N}_{13}^{1} ,  \\
&\mathcal{N}_{13}^{3} ,\ \mathcal{N}_{14}^{7} , \  \mathcal{N}_{15}^{2} , \ \mathcal{N}_{15}^{6} , \ \mathcal{N}_{16}^{5} , \ \mathcal{N}_{16}^{6} , \ \mathcal{N}_{17}^{4} , \ \mathcal{N}_{17}^{8} , \  \mathcal{N}_{18}^{7} ,\ \mathcal{N}_{18}^{8} , \  \mathcal{N}_{22}^{0} , \ \mathcal{N}_{22}^{4} , \ \mathcal{N}_{23}^{6} , \\
& \mathcal{N}_{24}^{2} , \ \mathcal{N}_{24}^{4} , \ \mathcal{N}_{25}^{1} , \  \mathcal{N}_{25}^{7} ,\ \mathcal{N}_{26}^{3} , \  \mathcal{N}_{26}^{8} ,\ \mathcal{N}_{27}^{5} , \ \mathcal{N}_{27}^{7} , \ \mathcal{N}_{28}^{6} , \ \mathcal{N}_{28}^{8} , \ \mathcal{N}_{33}^{0} , \  \mathcal{N}_{33}^{1} ,\ \mathcal{N}_{33}^{3} ,  \\
&\mathcal{N}_{34}^{8} ,\ \mathcal{N}_{35}^{5} , \ \mathcal{N}_{35}^{6} , \ \mathcal{N}_{36}^{2} , \ \mathcal{N}_{36}^{5} , \ \mathcal{N}_{36}^{6} , \  \mathcal{N}_{37}^{7} ,\ \mathcal{N}_{37}^{8} , \  \mathcal{N}_{38}^{4} , \ \mathcal{N}_{38}^{7} , \ \mathcal{N}_{38}^{8} , \ \mathcal{N}_{44}^{0} , \ \mathcal{N}_{44}^{2} , \\
&  \mathcal{N}_{44}^{4} , \  \mathcal{N}_{45}^{5} ,\ \mathcal{N}_{45}^{7} , \  \mathcal{N}_{46}^{6} , \ \mathcal{N}_{46}^{8} , \ \mathcal{N}_{47}^{1} , \ \mathcal{N}_{47}^{5} , \ \mathcal{N}_{47}^{7} , \ \mathcal{N}_{48}^{3} , \  \mathcal{N}_{48}^{6} ,\ \mathcal{N}_{48}^{8} , \  \mathcal{N}_{55}^{0} , \ \mathcal{N}_{55}^{3} , \\
&  \mathcal{N}_{55}^{4} , \ \mathcal{N}_{55}^{8} , \ \mathcal{N}_{56}^{1} , \ \mathcal{N}_{56}^{3} , \  \mathcal{N}_{56}^{7} ,\ \mathcal{N}_{56}^{8} , \  \mathcal{N}_{57}^{2} , \ \mathcal{N}_{57}^{4} , \ \mathcal{N}_{57}^{6} , \ \mathcal{N}_{57}^{8} , \ \mathcal{N}_{58}^{5} , \ \mathcal{N}_{58}^{6} , \  \mathcal{N}_{58}^{7} , \\
& \mathcal{N}_{58}^{8} , \  \mathcal{N}_{66}^{0} ,\ \mathcal{N}_{66}^{1} , \ \mathcal{N}_{66}^{3} , \ \mathcal{N}_{66}^{4} , \ \mathcal{N}_{66}^{7} , \ \mathcal{N}_{66}^{8} , \  \mathcal{N}_{67}^{5} ,\ \mathcal{N}_{67}^{6} , \  \mathcal{N}_{67}^{7} , \ \mathcal{N}_{67}^{8} , \ \mathcal{N}_{68}^{2} , \ \mathcal{N}_{68}^{4} , \\
& \mathcal{N}_{68}^{5} , \ \mathcal{N}_{68}^{6} , \  \mathcal{N}_{68}^{7} ,\ \mathcal{N}_{68}^{8} , \  \mathcal{N}_{77}^{0} , \ \mathcal{N}_{77}^{2} , \ \mathcal{N}_{77}^{3} , \ \mathcal{N}_{77}^{4} , \ \mathcal{N}_{77}^{6} , \ \mathcal{N}_{77}^{8} , \  \mathcal{N}_{78}^{1} ,\ \mathcal{N}_{78}^{3} , \  \mathcal{N}_{78}^{5} , \\
&\mathcal{N}_{78}^{6} , \ \mathcal{N}_{78}^{7} , \ \mathcal{N}_{78}^{8} , \ \mathcal{N}_{88}^{0} , \ \mathcal{N}_{88}^{1} , \  \mathcal{N}_{88}^{2} ,\ \mathcal{N}_{88}^{3} , \  \mathcal{N}_{88}^{4} , \ \mathcal{N}_{88}^{5} , \ \mathcal{N}_{88}^{6} , \ \mathcal{N}_{88}^{7} , \ \mathcal{N}_{88}^{8} .
\end{split}
\end{align}

\subsubsection*{List of non-vanishing $\mathcal{N}_{\alpha\beta}^{\gamma}$ for $\normalfont{\tsl{VF}}_{23}^\natural$:}

\begin{align}
\begin{split}
     &\mathcal{N}_{00}^{0} , \ \mathcal{N}_{01}^{1} , \ \mathcal{N}_{02}^{2} , \ \mathcal{N}_{03}^{3} , \ \mathcal{N}_{04}^{4} , \
     \mathcal{N}_{05}^{5} , \ \mathcal{N}_{11}^{0} , \ \mathcal{N}_{11}^{2} , \ \mathcal{N}_{12}^{1} , \ \mathcal{N}_{12}^{2} , \ \mathcal{N}_{13}^{4} , \ \mathcal{N}_{14}^{3} , \ \mathcal{N}_{14}^{5} ,  \\
     & \mathcal{N}_{15}^{4} , \ \mathcal{N}_{15}^{5} , \ \mathcal{N}_{22}^{0} , \ \mathcal{N}_{22}^{1} , \ \mathcal{N}_{22}^{2} , \ \mathcal{N}_{23}^{5} , \ \mathcal{N}_{24}^{4} , \ \mathcal{N}_{24}^{5} , \ \mathcal{N}_{25}^{3} , \ \mathcal{N}_{25}^{4} , \ \mathcal{N}_{25}^{5} , \ \mathcal{N}_{33}^{0} , \ \mathcal{N}_{33}^{3} ,  \\
     & \mathcal{N}_{34}^{1} , \ \mathcal{N}_{34}^{4} , \ \mathcal{N}_{35}^{2} , \
     \mathcal{N}_{35}^{5} , \ \mathcal{N}_{44}^{0} , \ \mathcal{N}_{44}^{2} , \ \mathcal{N}_{44}^{3} , \ \mathcal{N}_{44}^{5} , \ \mathcal{N}_{45}^{1} , \ \mathcal{N}_{45}^{2} , \ \mathcal{N}_{45}^{4} , \
     \mathcal{N}_{45}^{5} , \ \mathcal{N}_{55}^{0} , \\
     & \mathcal{N}_{55}^{1} , \ \mathcal{N}_{55}^{2} , \ \mathcal{N}_{55}^{3} , \ \mathcal{N}_{55}^{4} , \ \mathcal{N}_{55}^{5} .
\end{split}
\end{align}

\subsubsection*{List of non-vanishing $\mathcal{N}_{\alpha\beta}^{\gamma}$ for $\normalfont{\tsl{VF}}^\natural_{22}$:}

\begin{align}
\begin{split}
&\mathcal{N}_{00}^{0} , \ \mathcal{N}_{01}^{1} , \ \mathcal{N}_{02}^{2} , \ \mathcal{N}_{03}^{3} , \ \mathcal{N}_{04}^{4} , \  \mathcal{N}_{05}^{5} ,\ \mathcal{N}_{06}^{6} , \  \mathcal{N}_{07}^{7} , \ \mathcal{N}_{08}^{8} , \ \mathcal{N}_{09}^{9} , \ \mathcal{N}_{0,10}^{10} , \ \mathcal{N}_{0,11}^{11} , \ \mathcal{N}_{0,12}^{12} ,  \\
&\mathcal{N}_{0,13}^{13} ,\ \mathcal{N}_{11}^{0} , \  \mathcal{N}_{11}^{1} , \ \mathcal{N}_{11}^{2} , \ \mathcal{N}_{12}^{1} , \ \mathcal{N}_{12}^{2} , \ \mathcal{N}_{12}^{3} , \ \mathcal{N}_{13}^{2} , \  \mathcal{N}_{14}^{4} ,\ \mathcal{N}_{14}^{5} , \  \mathcal{N}_{14}^{6} , \ \mathcal{N}_{15}^{4} , \ \mathcal{N}_{15}^{5} , \\
& \mathcal{N}_{15}^{7} , \ \mathcal{N}_{16}^{4} , \ \mathcal{N}_{17}^{5} , \  \mathcal{N}_{18}^{9} ,\ \mathcal{N}_{18}^{10} , \  \mathcal{N}_{19}^{8} ,\ \mathcal{N}_{19}^{10} , \ \mathcal{N}_{1,10}^{8} , \ \mathcal{N}_{1,10}^{9} , \ \mathcal{N}_{1,10}^{10} , \ \mathcal{N}_{1,11}^{11} , \  \mathcal{N}_{1,11}^{12} ,\ \mathcal{N}_{1,11}^{13} ,  \\
&\mathcal{N}_{1,12}^{11} ,\ \mathcal{N}_{1,12}^{13} , \ \mathcal{N}_{1,13}^{11} , \ \mathcal{N}_{1,13}^{12} , \ \mathcal{N}_{22}^{0} , \ \mathcal{N}_{22}^{1} , \  \mathcal{N}_{22}^{2} ,\ \mathcal{N}_{23}^{1} , \  \mathcal{N}_{24}^{4} , \ \mathcal{N}_{24}^{5} , \ \mathcal{N}_{24}^{7} , \ \mathcal{N}_{25}^{4} , \ \mathcal{N}_{25}^{5} , \\
&  \mathcal{N}_{25}^{6} , \  \mathcal{N}_{26}^{5} ,\ \mathcal{N}_{27}^{4} , \  \mathcal{N}_{28}^{8} , \ \mathcal{N}_{28}^{10} , \ \mathcal{N}_{29}^{9} , \ \mathcal{N}_{29}^{10} , \ \mathcal{N}_{2,10}^{8} , \ \mathcal{N}_{2,10}^{9} , \  \mathcal{N}_{2,10}^{10} ,\ \mathcal{N}_{2,11}^{11} , \  \mathcal{N}_{2,11}^{12} , \ \mathcal{N}_{2,11}^{13} , \\
&  \mathcal{N}_{2,12}^{11} , \ \mathcal{N}_{2,12}^{12} , \ \mathcal{N}_{2,13}^{11} , \ \mathcal{N}_{2,13}^{13} , \  \mathcal{N}_{33}^{0} ,\ \mathcal{N}_{34}^{5} , \  \mathcal{N}_{35}^{4} , \ \mathcal{N}_{36}^{7} , \ \mathcal{N}_{37}^{6} , \ \mathcal{N}_{38}^{9} , \ \mathcal{N}_{39}^{8} , \ \mathcal{N}_{3,10}^{10} , \  \mathcal{N}_{3,11}^{11} , \\
& \mathcal{N}_{3,12}^{13} , \  \mathcal{N}_{3,13}^{12} ,\ \mathcal{N}_{44}^{0} , \ \mathcal{N}_{44}^{1} , \ \mathcal{N}_{44}^{2} , \ \mathcal{N}_{44}^{4} , \ \mathcal{N}_{44}^{5} , \  \mathcal{N}_{44}^{7} ,\ \mathcal{N}_{45}^{1} , \  \mathcal{N}_{45}^{2} , \ \mathcal{N}_{45}^{3} , \ \mathcal{N}_{45}^{4} , \ \mathcal{N}_{45}^{5} , \\
& \mathcal{N}_{45}^{6} , \ \mathcal{N}_{46}^{1} , \  \mathcal{N}_{46}^{5} ,\ \mathcal{N}_{47}^{2} , \  \mathcal{N}_{47}^{4} , \ \mathcal{N}_{48}^{11} , \ \mathcal{N}_{48}^{13} , \ \mathcal{N}_{49}^{11} , \ \mathcal{N}_{49}^{12} , \ \mathcal{N}_{4,10}^{11} , \  \mathcal{N}_{4,10}^{12} ,\ \mathcal{N}_{4,10}^{13} , \  \mathcal{N}_{4,11}^{8} , \\
&\mathcal{N}_{4,11}^{9} ,\ \mathcal{N}_{4,11}^{10} , \ \mathcal{N}_{4,11}^{11} , \ \mathcal{N}_{4,11}^{12} , \ \mathcal{N}_{4,11}^{13} , \ \mathcal{N}_{4,12}^{9} , \  \mathcal{N}_{4,12}^{10} ,\ \mathcal{N}_{4,12}^{11} , \  \mathcal{N}_{4,12}^{12} , \ \mathcal{N}_{4,13}^{8} , \ \mathcal{N}_{4,13}^{10} , \ \mathcal{N}_{4,13}^{11} , \\
&  \mathcal{N}_{4,13}^{13} , \ \mathcal{N}_{55}^{0} , \  \mathcal{N}_{55}^{1} ,\ \mathcal{N}_{55}^{2} , \  \mathcal{N}_{55}^{4} , \ \mathcal{N}_{55}^{5} , \ \mathcal{N}_{55}^{7} , \ \mathcal{N}_{56}^{2} , \ \mathcal{N}_{56}^{4} , \ \mathcal{N}_{57}^{1} , \  \mathcal{N}_{57}^{5} ,\ \mathcal{N}_{58}^{11} , \  \mathcal{N}_{58}^{12} , \ \mathcal{N}_{59}^{11} , \\
&  \mathcal{N}_{59}^{13} , \ \mathcal{N}_{5,10}^{11} , \ \mathcal{N}_{5,10}^{12} , \ \mathcal{N}_{5,10}^{13} , \  \mathcal{N}_{5,11}^{8} ,\ \mathcal{N}_{5,11}^{9} , \  \mathcal{N}_{5,11}^{10} , \ \mathcal{N}_{5,11}^{11} , \ \mathcal{N}_{5,11}^{12} , \ \mathcal{N}_{5,11}^{13} , \ \mathcal{N}_{5,12}^{8} , \ \mathcal{N}_{5,12}^{10}  , \\
& \mathcal{N}_{5,12}^{11}, \ \mathcal{N}_{5,12}^{13} , \  \mathcal{N}_{5,13}^{9} ,\ \mathcal{N}_{5,13}^{10} , \ \mathcal{N}_{5,13}^{11} , \ \mathcal{N}_{5,13}^{12} , \ \mathcal{N}_{66}^{0} , \ \mathcal{N}_{66}^{7} , \  \mathcal{N}_{67}^{3} ,\ \mathcal{N}_{67}^{6} , \  \mathcal{N}_{68}^{12} , \ \mathcal{N}_{69}^{13} , \ \mathcal{N}_{6,10}^{11} , \\
&  \mathcal{N}_{6,11}^{10} , \ \mathcal{N}_{6,12}^{8} , \ \mathcal{N}_{6,12}^{13} , \  \mathcal{N}_{6,13}^{9} ,\ \mathcal{N}_{6,13}^{12} , \  \mathcal{N}_{77}^{0} , \ \mathcal{N}_{77}^{7} , \ \mathcal{N}_{78}^{13} , \ \mathcal{N}_{79}^{12} , \ \mathcal{N}_{7,10}^{11} , \ \mathcal{N}_{7,11}^{10} , \  \mathcal{N}_{7,11}^{11} ,\\
&  \mathcal{N}_{7,12}^{9} , \ \mathcal{N}_{7,12}^{12} , \ \mathcal{N}_{7,13}^{8}, \ \mathcal{N}_{7,13}^{13} , \  \mathcal{N}_{88}^{0} ,\ \mathcal{N}_{88}^{2} , \ \mathcal{N}_{89}^{1} , \ \mathcal{N}_{89}^{3} , \ \mathcal{N}_{8,10}^{1} , \ \mathcal{N}_{8,10}^{2} , \  \mathcal{N}_{8,11}^{4} ,\ \mathcal{N}_{8,11}^{5} ,  \\
& \mathcal{N}_{8,12}^{5} ,  \ \mathcal{N}_{8,12}^{6} , \ \mathcal{N}_{8,13}^{4} , \ \mathcal{N}_{8,13}^{7} , \  \mathcal{N}_{99}^{0} ,\ \mathcal{N}_{99}^{2} , \  \mathcal{N}_{9,10}^{1} , \ \mathcal{N}_{9,10}^{2} , \ \mathcal{N}_{9,11}^{4} , \ \mathcal{N}_{9,11}^{5} , \ \mathcal{N}_{9,12}^{4} , \ \mathcal{N}_{9,12}^{7} ,   \\
& \mathcal{N}_{9,13}^{5} , \ \mathcal{N}_{9,13}^{6} , \ \mathcal{N}_{10,10}^{0}, \ \mathcal{N}_{10,10}^{1} , \  \mathcal{N}_{10,10}^{2} ,\ \mathcal{N}_{10,10}^{3} , \ \mathcal{N}_{10,11}^{4} , \ \mathcal{N}_{10,11}^{5} , \ \mathcal{N}_{10,11}^{6} , \ \mathcal{N}_{10,11}^{7} ,  \\
& \mathcal{N}_{10,12}^{4} , \ \mathcal{N}_{10,12}^{5} , \ \mathcal{N}_{10,13}^{4} , \ \mathcal{N}_{10,13}^{5} , \  \mathcal{N}_{11,11}^{0} ,\ \mathcal{N}_{11,11}^{1} , \  \mathcal{N}_{11,11}^{2} , \ \mathcal{N}_{11,11}^{3} , \ \mathcal{N}_{11,11}^{4} , \ \mathcal{N}_{11,11}^{5} ,    \\
& \mathcal{N}_{11,11}^{6},  \ \mathcal{N}_{11,11}^{7} , \ \mathcal{N}_{11,12}^{1} , \ \mathcal{N}_{11,12}^{2} , \ \mathcal{N}_{11,12}^{4} , \  \mathcal{N}_{11,12}^{5} ,\ \mathcal{N}_{11,13}^{1} , \  \mathcal{N}_{11,13}^{2} , \ \mathcal{N}_{11,13}^{4} , \ \mathcal{N}_{11,13}^{5} ,  \\
& \mathcal{N}_{12,12}^{0} , \ \mathcal{N}_{12,12}^{2} , \  \mathcal{N}_{12,12}^{4} ,\ \mathcal{N}_{12,12}^{7} , \  \mathcal{N}_{12,13}^{1} , \ \mathcal{N}_{12,13}^{3} , \ \mathcal{N}_{12,13}^{5} , \ \mathcal{N}_{12,13}^{6} , \ \mathcal{N}_{13,13}^{0} , \ \mathcal{N}_{13,13}^{2} ,  \\
& \mathcal{N}_{13,13}^{4} , \ \mathcal{N}_{13,13}^{7}.
\end{split}
\end{align}

\subsubsection*{List of non-vanishing $\mathcal{N}_{\alpha\beta}^{\gamma}$ for $\widetilde{\W}_{D_{\mathrm{4A}}}$:}
\begin{align}
\begin{split}
&\mathcal{N}_{00}^{0} , \ \mathcal{N}_{01}^{1} , \ \mathcal{N}_{02}^{2} , \ \mathcal{N}_{03}^{3} , \ \mathcal{N}_{04}^{4} , \  \mathcal{N}_{05}^{5} ,\ \mathcal{N}_{06}^{6} , \  \mathcal{N}_{07}^{7} , \ \mathcal{N}_{08}^{8} , \ \mathcal{N}_{09}^{9} , \ \mathcal{N}_{0,10}^{10}, \ \mathcal{N}_{11}^{0}, \ \mathcal{N}_{12}^{2} \\
&\mathcal{N}_{13}^{3} ,\ \mathcal{N}_{14}^{4} , \  \mathcal{N}_{15}^{5} , \ \mathcal{N}_{16}^{6} , \ \mathcal{N}_{17}^{7} , \ \mathcal{N}_{18}^{8} , \ \mathcal{N}_{19}^{10} , \ \mathcal{N}_{1,10}^{9} , \  \mathcal{N}_{22}^{0} ,\ \mathcal{N}_{22}^{1} , \  \mathcal{N}_{22}^{3} , \ \mathcal{N}_{23}^{2} , \ \mathcal{N}_{23}^{3} , \\
& \mathcal{N}_{24}^{7} , \ \mathcal{N}_{24}^{8} , \ \mathcal{N}_{25}^{6} , \  \mathcal{N}_{25}^{8} ,\ \mathcal{N}_{26}^{5} , \  \mathcal{N}_{26}^{7} ,\ \mathcal{N}_{27}^{4} , \ \mathcal{N}_{27}^{6} , \ \mathcal{N}_{28}^{4} , \ \mathcal{N}_{28}^{5} , \ \mathcal{N}_{29}^{9} , \  \mathcal{N}_{29}^{10} ,\ \mathcal{N}_{2,10}^{9} ,  \\
&\mathcal{N}_{2,10}^{10} ,\ \mathcal{N}_{33}^{0} , \ \mathcal{N}_{33}^{1} , \ \mathcal{N}_{33}^{2} , \ \mathcal{N}_{34}^{5} , \ \mathcal{N}_{34}^{6} , \  \mathcal{N}_{35}^{4} ,\ \mathcal{N}_{35}^{7} , \  \mathcal{N}_{36}^{4} , \ \mathcal{N}_{36}^{8} , \ \mathcal{N}_{37}^{5} , \ \mathcal{N}_{37}^{8} , \ \mathcal{N}_{38}^{6} , \\
&  \mathcal{N}_{38}^{7} , \  \mathcal{N}_{39}^{9} ,\ \mathcal{N}_{39}^{10} , \  \mathcal{N}_{3,10}^{9} , \ \mathcal{N}_{3,10}^{10} , \ \mathcal{N}_{44}^{0} , \ \mathcal{N}_{44}^{1} , \ \mathcal{N}_{44}^{4} , \ \mathcal{N}_{45}^{3} , \  \mathcal{N}_{45}^{6} ,\ \mathcal{N}_{46}^{3} , \  \mathcal{N}_{46}^{5} , \ \mathcal{N}_{47}^{2} , \\
&  \mathcal{N}_{47}^{8} , \ \mathcal{N}_{48}^{2} , \ \mathcal{N}_{48}^{7} , \ \mathcal{N}_{49}^{9} , \  \mathcal{N}_{49}^{10} ,\ \mathcal{N}_{4,10}^{9} , \  \mathcal{N}_{4,10}^{10} , \ \mathcal{N}_{55}^{0} , \ \mathcal{N}_{55}^{1} , \ \mathcal{N}_{55}^{8} , \ \mathcal{N}_{56}^{2} , \ \mathcal{N}_{56}^{4} , \  \mathcal{N}_{57}^{3} , \\
& \mathcal{N}_{57}^{7} , \  \mathcal{N}_{58}^{2} ,\ \mathcal{N}_{58}^{5} , \ \mathcal{N}_{59}^{9} , \ \mathcal{N}_{59}^{10} , \ \mathcal{N}_{5,10}^{9} , \ \mathcal{N}_{5,10}^{10} , \  \mathcal{N}_{66}^{0} ,\ \mathcal{N}_{66}^{1} , \  \mathcal{N}_{66}^{7} , \ \mathcal{N}_{67}^{2} , \ \mathcal{N}_{67}^{6} , \ \mathcal{N}_{68}^{3} , \\
& \mathcal{N}_{68}^{8} , \ \mathcal{N}_{69}^{9} , \  \mathcal{N}_{69}^{10} ,\ \mathcal{N}_{6,10}^{9} , \  \mathcal{N}_{6,10}^{10} , \ \mathcal{N}_{77}^{0} , \ \mathcal{N}_{77}^{1} , \ \mathcal{N}_{77}^{5} , \ \mathcal{N}_{78}^{4} , \ \mathcal{N}_{78}^{4} , \  \mathcal{N}_{79}^{9} ,\ \mathcal{N}_{79}^{10} , \  \mathcal{N}_{7,10}^{9} , \\
&\mathcal{N}_{7,10}^{10} , \ \mathcal{N}_{88}^{0} , \ \mathcal{N}_{88}^{1} , \ \mathcal{N}_{88}^{6} , \ \mathcal{N}_{89}^{9} , \  \mathcal{N}_{89}^{10} ,\ \mathcal{N}_{8,10}^{9} , \  \mathcal{N}_{8,10}^{10} , \ \mathcal{N}_{99}^{0} , \ \mathcal{N}_{99}^{2} , \ \mathcal{N}_{99}^{3} , \ \mathcal{N}_{99}^{4} ,\ \mathcal{N}_{99}^{5} ,\\
& \mathcal{N}_{99}^{6} , \ \mathcal{N}_{99}^{7} , \ \mathcal{N}_{99}^{8} , \ \mathcal{N}_{9,10}^{1} , \ \mathcal{N}_{9,10}^{2} , \  \mathcal{N}_{9,10}^{3} ,\ \mathcal{N}_{9,10}^{4} , \  \mathcal{N}_{9,10}^{5} , \ \mathcal{N}_{9,10}^{6} , \ \mathcal{N}_{9,10}^{7} , \ \mathcal{N}_{9,10}^{8} , \\
&\mathcal{N}_{10,10}^{0} , \ \mathcal{N}_{10,10}^{2} , \ \mathcal{N}_{10,10}^{3} , \ \mathcal{N}_{10,10}^{4} , \ \mathcal{N}_{10,10}^{5} , \  \mathcal{N}_{10,10}^{6} ,\ \mathcal{N}_{10,10}^{7} , \  \mathcal{N}_{10,10}^{8}
\end{split}
\end{align}

\subsubsection*{List of non-vanishing $\mathcal{N}_{\alpha\beta}^{\gamma}$ for $\widetilde{\W}_{D_{\mathrm{4B}}}$:}
\begin{align}
\begin{split}
&\mathcal{N}_{00}^{0} , \ \mathcal{N}_{01}^{1} , \ \mathcal{N}_{02}^{2} , \ \mathcal{N}_{03}^{3} , \ \mathcal{N}_{04}^{4} , \  \mathcal{N}_{05}^{5} ,\ \mathcal{N}_{06}^{6} , \  \mathcal{N}_{07}^{7} , \ \mathcal{N}_{08}^{8} , \ \mathcal{N}_{09}^{9} , \ \mathcal{N}_{0,10}^{10}, \ \mathcal{N}_{0,11}^{11}, \ \mathcal{N}_{11}^{0} \\
&\mathcal{N}_{12}^{6} ,\ \mathcal{N}_{13}^{7} , \  \mathcal{N}_{14}^{5} , \ \mathcal{N}_{15}^{4} , \ \mathcal{N}_{16}^{2} , \ \mathcal{N}_{17}^{3} , \ \mathcal{N}_{18}^{8} , \ \mathcal{N}_{19}^{9} , \  \mathcal{N}_{1,10}^{10} ,\ \mathcal{N}_{1,11}^{11} , \  \mathcal{N}_{22}^{0} , \ \mathcal{N}_{22}^{2} , \ \mathcal{N}_{23}^{4} , \\
& \mathcal{N}_{24}^{3} , \ \mathcal{N}_{24}^{4} , \ \mathcal{N}_{25}^{5} , \  \mathcal{N}_{25}^{7} ,\ \mathcal{N}_{26}^{1} , \  \mathcal{N}_{26}^{6} ,\ \mathcal{N}_{27}^{5} , \ \mathcal{N}_{28}^{9} , \ \mathcal{N}_{29}^{8} , \ \mathcal{N}_{29}^{9} , \ \mathcal{N}_{2,10}^{11} , \  \mathcal{N}_{2,11}^{10} ,\ \mathcal{N}_{2,11}^{11} ,  \\
&\mathcal{N}_{33}^{0} ,\ \mathcal{N}_{33}^{3} , \ \mathcal{N}_{34}^{2} , \ \mathcal{N}_{34}^{4} , \ \mathcal{N}_{35}^{5} , \ \mathcal{N}_{35}^{6} , \  \mathcal{N}_{36}^{5} ,\ \mathcal{N}_{37}^{1} , \  \mathcal{N}_{37}^{7} , \ \mathcal{N}_{38}^{10} , \ \mathcal{N}_{39}^{11} , \ \mathcal{N}_{3,10}^{8} , \ \mathcal{N}_{3,10}^{10} , \\
&  \mathcal{N}_{3,11}^{9} , \  \mathcal{N}_{3,11}^{11} ,\ \mathcal{N}_{44}^{0} , \  \mathcal{N}_{44}^{2} , \ \mathcal{N}_{44}^{3} , \ \mathcal{N}_{44}^{4} , \ \mathcal{N}_{45}^{1} , \ \mathcal{N}_{45}^{5} , \ \mathcal{N}_{45}^{6} , \  \mathcal{N}_{45}^{7} ,\ \mathcal{N}_{46}^{5} , \  \mathcal{N}_{46}^{7} , \ \mathcal{N}_{47}^{5} , \\
&  \mathcal{N}_{47}^{6} , \ \mathcal{N}_{48}^{11} , \ \mathcal{N}_{49}^{10} , \ \mathcal{N}_{49}^{11} , \  \mathcal{N}_{4,10}^{9} ,\ \mathcal{N}_{4,10}^{11} , \  \mathcal{N}_{4,11}^{8} , \ \mathcal{N}_{4,11}^{9} , \ \mathcal{N}_{4,11}^{10} , \ \mathcal{N}_{4,11}^{11} , \ \mathcal{N}_{55}^{0} , \ \mathcal{N}_{55}^{2} , \  \mathcal{N}_{55}^{3} , \\
& \mathcal{N}_{55}^{4} , \  \mathcal{N}_{56}^{3} ,\ \mathcal{N}_{56}^{4} , \ \mathcal{N}_{57}^{2} , \ \mathcal{N}_{57}^{4} , \ \mathcal{N}_{58}^{11} , \ \mathcal{N}_{59}^{10} , \  \mathcal{N}_{59}^{11} ,\ \mathcal{N}_{5,10}^{9} , \  \mathcal{N}_{5,10}^{11} , \ \mathcal{N}_{5,11}^{8} , \ \mathcal{N}_{5,11}^{9} , \ \mathcal{N}_{5,11}^{10} , \\
& \mathcal{N}_{5,11}^{11} , \ \mathcal{N}_{66}^{0} , \  \mathcal{N}_{66}^{2} ,\ \mathcal{N}_{67}^{4} , \  \mathcal{N}_{68}^{9} , \ \mathcal{N}_{69}^{8} , \ \mathcal{N}_{69}^{9} , \ \mathcal{N}_{6,10}^{11} , \ \mathcal{N}_{6,11}^{10} , \ \mathcal{N}_{6,11}^{11} , \  \mathcal{N}_{77}^{0} ,\ \mathcal{N}_{77}^{3} , \  \mathcal{N}_{78}^{10} , \\
&\mathcal{N}_{79}^{11} , \ \mathcal{N}_{7,10}^{8} , \ \mathcal{N}_{7,10}^{10} , \ \mathcal{N}_{7,11}^{9} , \ \mathcal{N}_{7,11}^{11} , \  \mathcal{N}_{88}^{0} ,\ \mathcal{N}_{88}^{1} , \  \mathcal{N}_{89}^{2} , \ \mathcal{N}_{89}^{6} , \ \mathcal{N}_{8,10}^{3} , \ \mathcal{N}_{8,10}^{7} , \ \mathcal{N}_{8,11}^{4} ,\ \mathcal{N}_{8,11}^{5} ,\\
& \mathcal{N}_{99}^{0} , \ \mathcal{N}_{99}^{1} , \ \mathcal{N}_{99}^{2} , \ \mathcal{N}_{99}^{6} , \ \mathcal{N}_{9,10}^{4} , \  \mathcal{N}_{9,10}^{5} ,\ \mathcal{N}_{9,11}^{3} , \  \mathcal{N}_{9,11}^{4} , \ \mathcal{N}_{9,11}^{5} , \ \mathcal{N}_{9,11}^{7} , \ \mathcal{N}_{10,10}^{0} , \\
&\mathcal{N}_{10,10}^{1} , \ \mathcal{N}_{10,10}^{3} , \ \mathcal{N}_{10,10}^{7} , \ \mathcal{N}_{10,11}^{2} , \ \mathcal{N}_{10,11}^{4} , \  \mathcal{N}_{10,11}^{5} ,\ \mathcal{N}_{10,11}^{6} , \  \mathcal{N}_{11,11}^{0} ,\ \mathcal{N}_{11,11}^{1} , \  \mathcal{N}_{11,11}^{2},\\
& \mathcal{N}_{11,11}^{3} , \ \mathcal{N}_{11,11}^{4} , \ \mathcal{N}_{11,11}^{5} , \ \mathcal{N}_{11,11}^{6} , \ \mathcal{N}_{11,11}^{7} 
\end{split}
\end{align}

\clearpage

\section{Group theory data}\label{app:A}

In this appendix we provide, for a few examples, the group theoretic data necessary for analyzing twined bilinears. 

Let us start with a somewhat general discussion of how characters of $\W$ and $\widetilde{\W}$, twined by inner automorphisms, can be bilinearly combined to produce twined characters of the VOA $\V$ in which they sit as commutant pairs. We assume for simplicity that $\V$ is a meromorphic CFT with partition function $\mathcal{Z}$, that $\mathrm{Inn}(\W)\times\mathrm{Inn}(\widetilde{\W})\subset\mathrm{Aut}(\V)$, and that the inner automorphism groups are realized honestly on the modules of $\W$ and $\widetilde{\W}$ which appear in the decomposition of $\V$ (as opposed to projectively). Under these assumptions, there will be generalized bilinear relations of the form 
\begin{align}
\mathcal{Z}_{gh}(\tau) = \sum_\alpha \chi_{g,\alpha}(\tau)\widetilde{\chi}_{h,\alpha}(\tau)
\end{align}
which arise by taking the graded trace of both sides of the decomposition
\begin{align}
    \V = \bigoplus_{\alpha}\W(\alpha)\otimes \widetilde{\W}(\alpha).
\end{align}
Because the graded characters are class functions of the associated groups, one only needs to know how the conjugacy classes of $\mathrm{Inn}(\W)\times\mathrm{Inn}(\widetilde{\W})$ fuse into the conjugacy classes of $\mathrm{Aut}(\V)$. 
For illustrative purposes, we take $\V=V^\natural$ and provide the necessary data for the cases $(\W,\widetilde{\W}) = (\W_{5A},\tsl{VHN}^\natural)$ and  $(\W_{D_{\mathrm{3C}}},\tsl{VT}^\natural)$ in Tables \ref{HN Fusion table} and \ref{Th Fusion table}: namely, information about how conjugacy classes of $\tsl{HN}\cong \mathrm{Inn}(\tsl{VHN}^\natural)$ and $\mathbb{Z}_3\times\Th\subset \mathrm{Inn}(\W_{D_{\mathrm{3C}}})\times\mathrm{Inn}(\tsl{VT}^\natural)$ fuse into conjugacy classes of $\mathbb{M}$. One can use this data to conduct checks on our proposals regarding the implementation of the symmetry groups in these two examples. For example, a prediction of Table \ref{Th Fusion table} is that 
\begin{align}
    J_{\mathrm{6F}}(\tau) = \sum_\alpha \chi_{\omega,\alpha}(\tau) \chi_{\mathrm{2A},\tsl{VT}^\natural(\alpha)}(\tau)
\end{align}
where $\chi_{\omega,\alpha}$ are the characters of $\W_{D_{\mathrm{3C}}}$ twined by the generator of its $\mathbb{Z}_3$ automorphism, $\chi_{\mathrm{2A},\tsl{VT}^\natural(\alpha)}$ are the characters of $\tsl{VT}^\natural$ twined by an element of the 2A conjugacy class of $\Th$, and $J_{\mathrm{6F}}$ is the McKay-Thompson series of the 6A conjugacy class in $\mathbb{M}$ (c.f.\ \S\ref{subsubsec:Th} for more details). To compute $\chi_{\mathrm{2A},\tsl{VT}^\natural(\alpha)}$ to low order in its $q$-expansion, one can use the character table of $\Th$, Tables \ref{thchartabone}-\ref{thchartabfour}, as well as the decompositions of the graded-components $\tsl{VT}^\natural(\alpha)_h$ into $\Th$ representations, Table \ref{decomposition by Th}.

\begin{landscape}
\begin{table}
\begin{center}
\smallskip
\begin{small}
\begin{tabular}{c@{ }r@{ }r@{ }r@{ }r@{ }r@{ }r@{ }r@{ }r@{ }r@{ }r@{ }r@{ }r@{ }r@{ }r@{ }r@{ }r@{ }r@{ }r@{ }r@{ }r@{ }r@{ }r@{ }r@{ }r} \toprule
$[g]$ & 1A  & 2A  & 3A  & 3B  & 3C  & 4A  & 4B  & 5A  & 6A  & 6B  & 6C  & 7A  & 8A  & 8B  & 9A  & 9B  & 9C  & 10A  & 12A  & 12B  & 12C  & 12D  & 13A  & 14A  \\ 
	 \midrule$\chi_{1}$ & $1$  & $1$  & $1$  & $1$  & $1$  & $1$  & $1$  & $1$  & $1$  & $1$  & $1$  & $1$  & $1$  & $1$  & $1$  & $1$  & $1$  & $1$  & $1$  & $1$  & $1$  & $1$  & $1$  & $1$ \\
		 $\chi_{2}$ & $248$  & -$8$  & $14$  & $5$  & -$4$  & $8$  & $0$  & -$2$  & $4$  & -$2$  & $1$  & $3$  & $0$  & $0$  & $5$  & -$4$  & $2$  & $2$  & $2$  & $2$  & -$1$  & $0$  & $1$  & -$1$ \\
		 $\chi_{3}$ & $4123$  & $27$  & $64$  & -$8$  & $1$  & $27$  & -$5$  & -$2$  & $9$  & $0$  & $0$  & $7$  & $3$  & -$1$  & -$8$  & $1$  & $4$  & $2$  & $0$  & $0$  & $0$  & $1$  & $2$  & -$1$ \\
		 $\chi_{4}$ & $27000$  & $120$  & -$27$  & $27$  & $0$  & $8$  & $0$  & $0$  & $0$  & -$3$  & $3$  & $1$  & $0$  & $0$  & $0$  & $0$  & $0$  & $0$  & $A$  & $\overline{A}$ & -$1$  & $0$  & -$1$  & $1$ \\
		 $\chi_{5}$ & $27000$  & $120$  & -$27$  & $27$  & $0$  & $8$  & $0$  & $0$  & $0$  & -$3$  & $3$  & $1$  & $0$  & $0$  & $0$  & $0$  & $0$  & $0$  & $\overline{A}$  & $A$  & -$1$  & $0$  & -$1$  & $1$ \\
		 $\chi_{6}$ & $30628$  & -$92$  & $91$  & $10$  & $10$  & $36$  & $4$  & $3$  & $10$  & -$5$  & -$2$  & $3$  & -$4$  & $0$  & $10$  & $10$  & $1$  & $3$  & $3$  & $3$  & $0$  & -$2$  & $0$  & -$1$ \\
		 $\chi_{7}$ & $30875$  & $155$  & $104$  & $14$  & $5$  & $27$  & -$5$  & $0$  & $5$  & $8$  & $2$  & $5$  & $3$  & -$1$  & $14$  & $5$  & $2$  & $0$  & $0$  & $0$  & $0$  & $1$  & $0$  & $1$ \\
		 $\chi_{8}$ & $61256$  & $72$  & $182$  & $20$  & $20$  & $56$  & $0$  & $6$  & $12$  & $6$  & $0$  & $6$  & $0$  & $0$  & -$7$  & -$7$  & $2$  & $2$  & $2$  & $2$  & $2$  & $0$  & $0$  & $2$ \\
		 $\chi_{9}$ & $85995$  & -$21$  & $0$  & -$27$  & $27$  & -$21$  & $11$  & -$5$  & $3$  & $0$  & -$3$  & $0$  & $3$  & -$1$  & $0$  & $0$  & $0$  & -$1$  & $0$  & $0$  & -$3$  & -$1$  & $0$  & $0$ \\
		 $\chi_{10}$ & $85995$  & -$21$  & $0$  & -$27$  & $27$  & -$21$  & $11$  & -$5$  & $3$  & $0$  & -$3$  & $0$  & $3$  & -$1$  & $0$  & $0$  & $0$  & -$1$  & $0$  & $0$  & -$3$  & -$1$  & $0$  & $0$ \\
		 $\chi_{11}$ & $147250$  & $50$  & $181$  & -$8$  & -$35$  & $34$  & $10$  & $0$  & $5$  & $5$  & -$4$  & $5$  & $2$  & -$2$  & $19$  & -$8$  & $1$  & $0$  & $1$  & $1$  & -$2$  & $1$  & -$1$  & $1$ \\
		 $\chi_{12}$ & $767637$  & $405$  & $0$  & $0$  & $0$  & -$27$  & -$3$  & $12$  & $0$  & $0$  & $0$  & $3$  & -$3$  & -$3$  & $0$  & $0$  & $0$  & $0$  & $0$  & $0$  & $0$  & $0$  & $0$  & -$1$ \\
		 $\chi_{13}$ & $767637$  & $405$  & $0$  & $0$  & $0$  & -$27$  & -$3$  & $12$  & $0$  & $0$  & $0$  & $3$  & -$3$  & -$3$  & $0$  & $0$  & $0$  & $0$  & $0$  & $0$  & $0$  & $0$  & $0$  & -$1$ \\
		 $\chi_{14}$ & $779247$  & -$273$  & -$189$  & -$54$  & $0$  & $63$  & -$9$  & -$3$  & $0$  & $3$  & $6$  & $0$  & -$1$  & $3$  & $0$  & $0$  & $0$  & -$3$  & $3$  & $3$  & $0$  & $0$  & $1$  & $0$ \\
		 $\chi_{15}$ & $779247$  & -$273$  & -$189$  & -$54$  & $0$  & $63$  & -$9$  & -$3$  & $0$  & $3$  & $6$  & $0$  & -$1$  & $3$  & $0$  & $0$  & $0$  & -$3$  & $3$  & $3$  & $0$  & $0$  & $1$  & $0$ \\
		 $\chi_{16}$ & $957125$  & -$315$  & $650$  & -$52$  & -$25$  & $133$  & $5$  & $0$  & $15$  & -$6$  & $0$  & $8$  & -$3$  & $1$  & -$25$  & $2$  & $2$  & $0$  & -$2$  & -$2$  & -$2$  & -$1$  & $0$  & $0$ \\
		 $\chi_{17}$ & $1707264$  & -$768$  & $0$  & -$54$  & $54$  & $0$  & $0$  & $14$  & -$6$  & $0$  & $6$  & $6$  & $0$  & $0$  & $0$  & $0$  & $0$  & $2$  & $0$  & $0$  & $0$  & $0$  & $0$  & $2$ \\
		 $\chi_{18}$ & $1707264$  & -$768$  & $0$  & -$54$  & $54$  & $0$  & $0$  & $14$  & -$6$  & $0$  & $6$  & $6$  & $0$  & $0$  & $0$  & $0$  & $0$  & $2$  & $0$  & $0$  & $0$  & $0$  & $0$  & $2$ \\
		 $\chi_{19}$ & $2450240$  & $832$  & $260$  & $71$  & $44$  & $64$  & $0$  & -$10$  & $4$  & $4$  & -$5$  & -$5$  & $0$  & $0$  & $17$  & -$10$  & -$1$  & $2$  & $4$  & $4$  & $1$  & $0$  & $0$  & -$1$ \\
		 $\chi_{20}$ & $2572752$  & -$1072$  & $624$  & $111$  & $84$  & $48$  & $0$  & $2$  & -$4$  & -$16$  & -$1$  & $7$  & $0$  & $0$  & $30$  & $3$  & $3$  & -$2$  & $0$  & $0$  & $3$  & $0$  & $0$  & -$1$ \\
		 $\chi_{21}$ & $3376737$  & $609$  & $819$  & $9$  & $9$  & $161$  & $1$  & -$13$  & $9$  & $3$  & -$3$  & $0$  & $1$  & $1$  & $9$  & $9$  & $0$  & -$1$  & -$1$  & -$1$  & -$1$  & $1$  & $0$  & $0$ \\
		 $\chi_{22}$ & $4096000$  & $0$  & $64$  & -$8$  & -$80$  & $0$  & $0$  & $0$  & $0$  & $0$  & $0$  & -$8$  & $0$  & $0$  & -$8$  & $1$  & $4$  & $0$  & $0$  & $0$  & $0$  & $0$  & -$1$  & $0$ \\
		 $\chi_{23}$ & $4096000$  & $0$  & $64$  & -$8$  & -$80$  & $0$  & $0$  & $0$  & $0$  & $0$  & $0$  & -$8$  & $0$  & $0$  & -$8$  & $1$  & $4$  & $0$  & $0$  & $0$  & $0$  & $0$  & -$1$  & $0$ \\
		 $\chi_{24}$ & $4123000$  & $120$  & $118$  & $19$  & -$80$  & $8$  & $0$  & $0$  & $0$  & $6$  & $3$  & -$7$  & $0$  & $0$  & $19$  & $1$  & $4$  & $0$  & $2$  & $2$  & -$1$  & $0$  & -$2$  & $1$ \\
		 \bottomrule
	\end{tabular}
\captionsetup{width=1.1\textwidth}
\caption{The character table of the Thompson group, part I. The notation follows that of the ATLAS \cite{atlas} and the GAP software package \cite{GAP} with $b_N=(-1+ i\sqrt{N})/2$ for $N \equiv -1 ~{\rm mod}~4$. $A=1+4b_3$, $B=2+8 b_3$, $C=b_{15}$, $D=-i\sqrt{3}$,$E=i\sqrt{6}$, $F= 1+3 b_3$, $G=b_{31}$, $H=2b_3$ and $I=b_{39}$. An overline indicates complex conjugation.}\label{thchartabone}
	\end{small}
	\end{center}
	\end{table}
		\end{landscape}
	
\begin{landscape}
\begin{table}
\begin{center}
\smallskip
\begin{small}
\begin{tabular}{c@{ }r@{ }r@{ }r@{ }r@{ }r@{ }r@{ }r@{ }r@{ }r@{ }r@{ }r@{ }r@{ }r@{ }r@{ }r@{ }r@{ }r@{ }r@{ }r@{ }r@{ }r@{ }r@{ }r@{ }r} \toprule
$[g]$ & 15A  & 15B  & 18A  & 18B  & 19A  & 20A  & 21A  & 24A  & 24B  & 24C  & 24D  & 27A  & 27B  & 27C  & 28A  & 30A  & 30B  & 31A  & 31B  & 36A  & 36B  & 36C  & 39A  & 39B  \\ 
	 \midrule$\chi_{1}$ & $1$  & $1$  & $1$  & $1$  & $1$  & $1$  & $1$  & $1$  & $1$  & $1$  & $1$  & $1$  & $1$  & $1$  & $1$  & $1$  & $1$  & $1$  & $1$  & $1$  & $1$  & $1$  & $1$  & $1$ \\
		 $\chi_{2}$ & $1$  & $1$  & $1$  & -$2$  & $1$  & $0$  & $0$  & $0$  & $0$  & $0$  & $0$  & $2$  & -$1$  & -$1$  & $1$  & -$1$  & -$1$  & $0$  & $0$  & -$1$  & -$1$  & -$1$  & $1$  & $1$ \\
		 $\chi_{3}$ & $1$  & $1$  & $0$  & $0$  & $0$  & $0$  & $1$  & $0$  & $0$  & -$1$  & -$1$  & -$2$  & $1$  & $1$  & -$1$  & -$1$  & -$1$  & $0$  & $0$  & $0$  & $0$  & $0$  & -$1$  & -$1$ \\
		 $\chi_{4}$ & $0$  & $0$  & $0$  & $0$  & $1$  & $0$  & $1$  & -$D$  & $\overline{D}$  & $0$  & $0$  & $0$  & $0$  & $0$  & $1$  & $0$  & $0$  & -$1$  & -$1$  & $2$  & $H$  & $\overline{H}$  & -$1$  & -$1$ \\
		 $\chi_{5}$ & $0$  & $0$  & $0$  & $0$  & $1$  & $0$  & $1$  & $\overline{D}$  & $D$  & $0$  & $0$  & $0$  & $0$  & $0$  & $1$  & $0$  & $0$  & -$1$  & -$1$  & $2$  & $\overline{H}$  & $H$  & -$1$  & -$1$ \\
		 $\chi_{6}$ & $0$  & $0$  & -$2$  & $1$  & $0$  & -$1$  & $0$  & -$1$  & -$1$  & $0$  & $0$  & $1$  & $1$  & $1$  & $1$  & $0$  & $0$  & $0$  & $0$  & $0$  & $0$  & $0$  & $0$  & $0$ \\
		 $\chi_{7}$ & $0$  & $0$  & $2$  & $2$  & $0$  & $0$  & -$1$  & $0$  & $0$  & -$1$  & -$1$  & $2$  & -$1$  & -$1$  & -$1$  & $0$  & $0$  & -$1$  & -$1$  & $0$  & $0$  & $0$  & $0$  & $0$ \\
		 $\chi_{8}$ & $0$  & $0$  & -$3$  & $0$  & $0$  & $0$  & $0$  & $0$  & $0$  & $0$  & $0$  & -$1$  & -$1$  & -$1$  & $0$  & $2$  & $2$  & $0$  & $0$  & -$1$  & -$1$  & -$1$  & $0$  & $0$ \\
		 $\chi_{9}$ & $C$  & $\overline{C}$ & $0$  & $0$  & $1$  & $1$  & $0$  & $0$  & $0$  & -$1$  & -$1$  & $0$  & $0$  & $0$  & $0$  & -$C$  & -$ \overline{C}$  & $1$  & $1$  & $0$  & $0$  & $0$  & $0$  & $0$ \\
		 $\chi_{10}$ & $\overline{C}$  & $C$  & $0$  & $0$  & $1$  & $1$  & $0$  & $0$  & $0$  & -$1$  & -$1$  & $0$  & $0$  & $0$  & $0$  & -$\overline{C}$  & -$C$  & $1$  & $1$  & $0$  & $0$  & $0$  & $0$  & $0$ \\
		 $\chi_{11}$ & $0$  & $0$  & -$1$  & -$1$  & $0$  & $0$  & -$1$  & -$1$  & -$1$  & $1$  & $1$  & $1$  & $1$  & $1$  & -$1$  & $0$  & $0$  & $0$  & $0$  & $1$  & $1$  & $1$  & -$1$  & -$1$ \\
		 $\chi_{12}$ & $0$  & $0$  & $0$  & $0$  & -$1$  & $2$  & $0$  & $0$  & $0$  & $0$  & $0$  & $0$  & $0$  & $0$  & $1$  & $0$  & $0$  & $G$  & $\overline{G}$  & $0$  & $0$  & $0$  & $0$  & $0$ \\
		 $\chi_{13}$ & $0$  & $0$  & $0$  & $0$  & -$1$  & $2$  & $0$  & $0$  & $0$  & $0$  & $0$  & $0$  & $0$  & $0$  & $1$  & $0$  & $0$  & $\overline{G}$  & $G$  & $0$  & $0$  & $0$  & $0$  & $0$ \\
		 $\chi_{14}$ & $0$  & $0$  & $0$  & $0$  & $0$  & $1$  & $0$  & -$1$  & -$1$  & $0$  & $0$  & $0$  & $0$  & $0$  & $0$  & $0$  & $0$  & $0$  & $0$  & $0$  & $0$  & $0$  & $I$  & $\overline{I}$ \\
		 $\chi_{15}$ & $0$  & $0$  & $0$  & $0$  & $0$  & $1$  & $0$  & -$1$  & -$1$  & $0$  & $0$  & $0$  & $0$  & $0$  & $0$  & $0$  & $0$  & $0$  & $0$  & $0$  & $0$  & $0$  & $\overline{I}$  & $I$ \\
		 $\chi_{16}$ & $0$  & $0$  & $3$  & $0$  & $0$  & $0$  & -$1$  & $0$  & $0$  & $1$  & $1$  & -$1$  & -$1$  & -$1$  & $0$  & $0$  & $0$  & $0$  & $0$  & $1$  & $1$  & $1$  & $0$  & $0$ \\
		 $\chi_{17}$ & -$1$  & -$1$  & $0$  & $0$  & $0$  & $0$  & $0$  & $0$  & $0$  & $E$  & $\overline{E}$  & $0$  & $0$  & $0$  & $0$  & -$1$  & -$1$  & $1$  & $1$  & $0$  & $0$  & $0$  & $0$  & $0$ \\
		 $\chi_{18}$ & -$1$  & -$1$  & $0$  & $0$  & $0$  & $0$  & $0$  & $0$  & $0$  & $\overline{E}$  & $E$  & $0$  & $0$  & $0$  & $0$  & -$1$  & -$1$  & $1$  & $1$  & $0$  & $0$  & $0$  & $0$  & $0$ \\
		 $\chi_{19}$ & -$1$  & -$1$  & $1$  & $1$  & $0$  & $0$  & $1$  & $0$  & $0$  & $0$  & $0$  & -$1$  & -$1$  & -$1$  & $1$  & -$1$  & -$1$  & $0$  & $0$  & $1$  & $1$  & $1$  & $0$  & $0$ \\
		 $\chi_{20}$ & -$1$  & -$1$  & $2$  & -$1$  & $0$  & $0$  & $1$  & $0$  & $0$  & $0$  & $0$  & $0$  & $0$  & $0$  & -$1$  & $1$  & $1$  & $0$  & $0$  & $0$  & $0$  & $0$  & $0$  & $0$ \\
		 $\chi_{21}$ & -$1$  & -$1$  & -$3$  & $0$  & $0$  & $1$  & $0$  & $1$  & $1$  & $1$  & $1$  & $0$  & $0$  & $0$  & $0$  & -$1$  & -$1$  & $0$  & $0$  & -$1$  & -$1$  & -$1$  & $0$  & $0$ \\
		 $\chi_{22}$ & $0$  & $0$  & $0$  & $0$  & -$1$  & $0$  & $1$  & $0$  & $0$  & $0$  & $0$  & $1$  & $F$  & $\overline{F}$  & $0$  & $0$  & $0$  & $1$  & $1$  & $0$  & $0$  & $0$  & -$1$  & -$1$ \\
		 $\chi_{23}$ & $0$  & $0$  & $0$  & $0$  & -$1$  & $0$  & $1$  & $0$  & $0$  & $0$  & $0$  & $1$  & $\overline{F}$  & $F$  & $0$  & $0$  & $0$  & $1$  & $1$  & $0$  & $0$  & $0$  & -$1$  & -$1$ \\
		 $\chi_{24}$ & $0$  & $0$  & $3$  & $0$  & $0$  & $0$  & -$1$  & $0$  & $0$  & $0$  & $0$  & -$2$  & $1$  & $1$  & $1$  & $0$  & $0$  & $0$  & $0$  & -$1$  & -$1$  & -$1$  & $1$  & $1$ \\
		 \bottomrule
	\end{tabular}
	\end{small}
    \caption{The character table of the Thompson group, part II.}\label{thchartabtwo}
	\end{center}
	\end{table}
	\end{landscape}
	
\begin{landscape}
\begin{table}
\begin{center} 
\smallskip
\begin{small}
\begin{tabular}{c@{ }r@{ }r@{ }r@{ }r@{ }r@{ }r@{ }r@{ }r@{ }r@{ }r@{ }r@{ }r@{ }r@{ }r@{ }r@{ }r@{ }r@{ }r@{ }r@{ }r@{ }r@{ }r@{ }r@{ }r} \toprule
$[g]$ & 1A  & 2A  & 3A  & 3B  & 3C  & 4A  & 4B  & 5A  & 6A  & 6B  & 6C  & 7A  & 8A  & 8B  & 9A  & 9B  & 9C  & 10A  & 12A  & 12B  & 12C  & 12D  & 13A  & 14A  \\ 
	 \midrule$\chi_{25}$ & $4881384$  & $1512$  & $729$  & $0$  & $0$  & $72$  & $24$  & $9$  & $0$  & $9$  & $0$  & $4$  & $8$  & $0$  & $0$  & $0$  & $0$  & -$3$  & -$3$  & -$3$  & $0$  & $0$  & $1$  & $0$ \\
		 $\chi_{26}$ & $4936750$  & -$210$  & $637$  & -$38$  & -$65$  & $126$  & -$10$  & $0$  & $15$  & -$3$  & $6$  & $0$  & -$2$  & $2$  & $16$  & -$11$  & -$2$  & $0$  & -$3$  & -$3$  & $0$  & -$1$  & $0$  & $0$ \\
		 $\chi_{27}$ & $6669000$  & -$1080$  & -$351$  & $108$  & $0$  & $56$  & $0$  & $0$  & $0$  & $9$  & $0$  & $2$  & $0$  & $0$  & $0$  & $0$  & $0$  & $0$  & $A$  & $\overline{A}$  & $2$  & $0$  & $0$  & -$2$ \\
		 $\chi_{28}$ & $6669000$  & -$1080$  & -$351$  & $108$  & $0$  & $56$  & $0$  & $0$  & $0$  & $9$  & $0$  & $2$  & $0$  & $0$  & $0$  & $0$  & $0$  & $0$  & $\overline{A}$  & $A$  & $2$  & $0$  & $0$  & -$2$ \\
		 $\chi_{29}$ & $6696000$  & -$960$  & -$378$  & $135$  & $0$  & $64$  & $0$  & $0$  & $0$  & $6$  & $3$  & $3$  & $0$  & $0$  & $0$  & $0$  & $0$  & $0$  & $B$  & $\overline{B}$  & $1$  & $0$  & -$1$  & -$1$ \\
		 $\chi_{30}$ & $6696000$  & -$960$  & -$378$  & $135$  & $0$  & $64$  & $0$  & $0$  & $0$  & $6$  & $3$  & $3$  & $0$  & $0$  & $0$  & $0$  & $0$  & $0$  & $\overline{B}$  & $B$  & $1$  & $0$  & -$1$  & -$1$ \\
		 $\chi_{31}$ & $10822875$  & -$805$  & $924$  & $141$  & -$75$  & $91$  & -$5$  & $0$  & $5$  & -$4$  & $5$  & $0$  & $3$  & -$1$  & -$21$  & $6$  & -$3$  & $0$  & $4$  & $4$  & $1$  & $1$  & -$2$  & $0$ \\
		 $\chi_{32}$ & $11577384$  & $552$  & $351$  & $135$  & $0$  & -$120$  & $24$  & $9$  & $0$  & $15$  & $3$  & $7$  & -$8$  & $0$  & $0$  & $0$  & $0$  & -$3$  & $3$  & $3$  & -$3$  & $0$  & $0$  & -$1$ \\
		 $\chi_{33}$ & $16539120$  & $2544$  & $0$  & $297$  & -$54$  & $48$  & $16$  & -$5$  & -$6$  & $0$  & -$3$  & $3$  & $0$  & $0$  & $0$  & $0$  & $0$  & -$1$  & $0$  & $0$  & $3$  & -$2$  & $0$  & $3$ \\
		 $\chi_{34}$ & $18154500$  & $1540$  & -$273$  & $213$  & -$30$  & -$28$  & $20$  & $0$  & $10$  & -$17$  & $1$  & $0$  & -$4$  & $0$  & -$3$  & -$3$  & -$3$  & $0$  & -$1$  & -$1$  & -$1$  & $2$  & $0$  & $0$ \\
		 $\chi_{35}$ & $21326760$  & $168$  & $0$  & -$135$  & -$108$  & -$168$  & $0$  & $10$  & $12$  & $0$  & -$3$  & $0$  & $0$  & $0$  & $0$  & $0$  & $0$  & -$2$  & $0$  & $0$  & $3$  & $0$  & $0$  & $0$ \\
		 $\chi_{36}$ & $21326760$  & $168$  & $0$  & -$135$  & -$108$  & -$168$  & $0$  & $10$  & $12$  & $0$  & -$3$  & $0$  & $0$  & $0$  & $0$  & $0$  & $0$  & -$2$  & $0$  & $0$  & $3$  & $0$  & $0$  & $0$ \\
		 $\chi_{37}$ & $28861000$  & $840$  & $1078$  & -$110$  & $160$  & $56$  & $0$  & $0$  & $0$  & $6$  & -$6$  & $0$  & $0$  & $0$  & -$29$  & -$2$  & -$2$  & $0$  & $2$  & $2$  & $2$  & $0$  & -$1$  & $0$ \\
		 $\chi_{38}$ & $30507008$  & $0$  & $896$  & -$184$  & $32$  & $0$  & $0$  & $8$  & $0$  & $0$  & $0$  & $0$  & $0$  & $0$  & $32$  & $5$  & -$4$  & $0$  & $0$  & $0$  & $0$  & $0$  & -$1$  & $0$ \\
		 $\chi_{39}$ & $40199250$  & $3410$  & -$78$  & $3$  & $165$  & -$62$  & $10$  & $0$  & $5$  & $2$  & -$1$  & -$7$  & -$6$  & $2$  & $3$  & $3$  & $3$  & $0$  & -$2$  & -$2$  & $1$  & $1$  & $0$  & $1$ \\
		 $\chi_{40}$ & $44330496$  & $3584$  & $168$  & $6$  & -$156$  & $0$  & $0$  & -$4$  & -$4$  & $8$  & $2$  & $0$  & $0$  & $0$  & $6$  & $6$  & -$3$  & $4$  & $0$  & $0$  & $0$  & $0$  & $2$  & $0$ \\
		 $\chi_{41}$ & $51684750$  & $2190$  & $0$  & $108$  & $135$  & -$162$  & -$10$  & $0$  & $15$  & $0$  & $12$  & -$9$  & $6$  & -$2$  & $0$  & $0$  & $0$  & $0$  & $0$  & $0$  & $0$  & -$1$  & $0$  & -$1$ \\
		 $\chi_{42}$ & $72925515$  & -$2997$  & $0$  & $0$  & $0$  & $27$  & $51$  & $15$  & $0$  & $0$  & $0$  & -$9$  & $3$  & $3$  & $0$  & $0$  & $0$  & $3$  & $0$  & $0$  & $0$  & $0$  & $0$  & -$1$ \\
		 $\chi_{43}$ & $76271625$  & -$2295$  & $729$  & $0$  & $0$  & $153$  & -$15$  & $0$  & $0$  & $9$  & $0$  & -$11$  & -$7$  & -$3$  & $0$  & $0$  & $0$  & $0$  & -$3$  & -$3$  & $0$  & $0$  & $1$  & $1$ \\
		 $\chi_{44}$ & $77376000$  & $2560$  & $1560$  & -$60$  & -$60$  & $0$  & $0$  & $0$  & -$20$  & -$8$  & $4$  & $2$  & $0$  & $0$  & -$6$  & -$6$  & $3$  & $0$  & $0$  & $0$  & $0$  & $0$  & $0$  & -$2$ \\
		 $\chi_{45}$ & $81153009$  & -$783$  & -$729$  & $0$  & $0$  & $225$  & $9$  & $9$  & $0$  & -$9$  & $0$  & -$7$  & $1$  & -$3$  & $0$  & $0$  & $0$  & -$3$  & $3$  & $3$  & $0$  & $0$  & $2$  & $1$ \\
		 $\chi_{46}$ & $91171899$  & $315$  & $0$  & $243$  & $0$  & -$21$  & -$45$  & $24$  & $0$  & $0$  & -$9$  & $0$  & $3$  & $3$  & $0$  & $0$  & $0$  & $0$  & $0$  & $0$  & -$3$  & $0$  & $0$  & $0$ \\
		 $\chi_{47}$ & $111321000$  & $3240$  & -$1728$  & -$216$  & $0$  & $216$  & $0$  & $0$  & $0$  & $0$  & $0$  & $7$  & $0$  & $0$  & $0$  & $0$  & $0$  & $0$  & $0$  & $0$  & $0$  & $0$  & -$2$  & -$1$ \\
		 $\chi_{48}$ & $190373976$  & -$3240$  & $0$  & $0$  & $0$  & -$216$  & $0$  & -$24$  & $0$  & $0$  & $0$  & $9$  & $0$  & $0$  & $0$  & $0$  & $0$  & $0$  & $0$  & $0$  & $0$  & $0$  & $0$  & $1$ \\
		 \bottomrule
	\end{tabular}
	\end{small}
	\end{center}
    \caption{The character table of the Thompson group, part III. }\label{thchartabthree}
	\end{table}
	\end{landscape}
	
\begin{landscape}
\begin{table}[h!]
\begin{center}
\smallskip
\begin{small}
\begin{tabular}{c@{ }r@{ }r@{ }r@{ }r@{ }r@{ }r@{ }r@{ }r@{ }r@{ }r@{ }r@{ }r@{ }r@{ }r@{ }r@{ }r@{ }r@{ }r@{ }r@{ }r@{ }r@{ }r@{ }r@{ }r} \toprule
$[g]$ & 15A  & 15B  & 18A  & 18B  & 19A  & 20A  & 21A  & 24A  & 24B  & 24C  & 24D  & 27A  & 27B  & 27C  & 28A  & 30A  & 30B  & 31A  & 31B  & 36A  & 36B  & 36C  & 39A  & 39B  \\ 
	 \midrule$\chi_{25}$ & $0$  & $0$  & $0$  & $0$  & -$1$  & -$1$  & $1$  & -$1$  & -$1$  & $0$  & $0$  & $0$  & $0$  & $0$  & $2$  & $0$  & $0$  & $0$  & $0$  & $0$  & $0$  & $0$  & $1$  & $1$ \\
		 $\chi_{26}$ & $0$  & $0$  & $0$  & $0$  & -$1$  & $0$  & $0$  & $1$  & $1$  & -$1$  & -$1$  & $1$  & $1$  & $1$  & $0$  & $0$  & $0$  & $0$  & $0$  & $0$  & $0$  & $0$  & $0$  & $0$ \\
		 $\chi_{27}$ & $0$  & $0$  & $0$  & $0$  & $0$  & $0$  & -$1$  & $\overline{D}$  & $D$  & $0$  & $0$  & $0$  & $0$  & $0$  & $0$  & $0$  & $0$  & $1$  & $1$  & $2$  & $H$  & $\overline{H}$  & $0$  & $0$ \\
		 $\chi_{28}$ & $0$  & $0$  & $0$  & $0$  & $0$  & $0$  & -$1$  & $D$  & $\overline{D}$  & $0$  & $0$  & $0$  & $0$  & $0$  & $0$  & $0$  & $0$  & $1$  & $1$  & $2$  & $\overline{H}$  & $H$  & $0$  & $0$ \\
		 $\chi_{29}$ & $0$  & $0$  & $0$  & $0$  & $1$  & $0$  & $0$  & $0$  & $0$  & $0$  & $0$  & $0$  & $0$  & $0$  & $1$  & $0$  & $0$  & $0$  & $0$  & -$2$  & -$H$  & -$\overline{H}$  & -$1$  & -$1$ \\
		 $\chi_{30}$ & $0$  & $0$  & $0$  & $0$  & $1$  & $0$  & $0$  & $0$  & $0$  & $0$  & $0$  & $0$  & $0$  & $0$  & $1$  & $0$  & $0$  & $0$  & $0$  & -$2$  & -$\overline{H}$  & -$H$  & -$1$  & -$1$ \\
		 $\chi_{31}$ & $0$  & $0$  & -$1$  & -$1$  & $0$  & $0$  & $0$  & $0$  & $0$  & -$1$  & -$1$  & $0$  & $0$  & $0$  & $0$  & $0$  & $0$  & $0$  & $0$  & $1$  & $1$  & $1$  & $1$  & $1$ \\
		 $\chi_{32}$ & $0$  & $0$  & $0$  & $0$  & $0$  & -$1$  & $1$  & $1$  & $1$  & $0$  & $0$  & $0$  & $0$  & $0$  & -$1$  & $0$  & $0$  & $0$  & $0$  & $0$  & $0$  & $0$  & $0$  & $0$ \\
		 $\chi_{33}$ & $1$  & $1$  & $0$  & $0$  & $0$  & $1$  & $0$  & $0$  & $0$  & $0$  & $0$  & $0$  & $0$  & $0$  & -$1$  & -$1$  & -$1$  & $0$  & $0$  & $0$  & $0$  & $0$  & $0$  & $0$ \\
		 $\chi_{34}$ & $0$  & $0$  & $1$  & $1$  & $0$  & $0$  & $0$  & -$1$  & -$1$  & $0$  & $0$  & $0$  & $0$  & $0$  & $0$  & $0$  & $0$  & $1$  & $1$  & -$1$  & -$1$  & -$1$  & $0$  & $0$ \\
		 $\chi_{35}$ & $C$  & $\overline{C}$  & $0$  & $0$  & $1$  & $0$  & $0$  & $0$  & $0$  & $0$  & $0$  & $0$  & $0$  & $0$  & $0$  & $C$  & $\overline{C}$  & $0$  & $0$  & $0$  & $0$  & $0$  & $0$  & $0$ \\
		 $\chi_{36}$ & $\overline{C}$  & $C$  & $0$  & $0$  & $1$  & $0$  & $0$  & $0$  & $0$  & $0$  & $0$  & $0$  & $0$  & $0$  & $0$  & $\overline{C}$  & $C$  & $0$  & $0$  & $0$  & $0$  & $0$  & $0$  & $0$ \\
		 $\chi_{37}$ & $0$  & $0$  & $3$  & $0$  & $0$  & $0$  & $0$  & $0$  & $0$  & $0$  & $0$  & $1$  & $1$  & $1$  & $0$  & $0$  & $0$  & $0$  & $0$  & -$1$  & -$1$  & -$1$  & -$1$  & -$1$ \\
		 $\chi_{38}$ & $2$  & $2$  & $0$  & $0$  & $0$  & $0$  & $0$  & $0$  & $0$  & $0$  & $0$  & -$1$  & -$1$  & -$1$  & $0$  & $0$  & $0$  & $1$  & $1$  & $0$  & $0$  & $0$  & -$1$  & -$1$ \\
		 $\chi_{39}$ & $0$  & $0$  & -$1$  & -$1$  & $0$  & $0$  & -$1$  & $0$  & $0$  & -$1$  & -$1$  & $0$  & $0$  & $0$  & $1$  & $0$  & $0$  & $0$  & $0$  & $1$  & $1$  & $1$  & $0$  & $0$ \\
		 $\chi_{40}$ & -$1$  & -$1$  & $2$  & -$1$  & $0$  & $0$  & $0$  & $0$  & $0$  & $0$  & $0$  & $0$  & $0$  & $0$  & $0$  & $1$  & $1$  & $0$  & $0$  & $0$  & $0$  & $0$  & -$1$  & -$1$ \\
		 $\chi_{41}$ & $0$  & $0$  & $0$  & $0$  & $0$  & $0$  & $0$  & $0$  & $0$  & $1$  & $1$  & $0$  & $0$  & $0$  & -$1$  & $0$  & $0$  & $0$  & $0$  & $0$  & $0$  & $0$  & $0$  & $0$ \\
		 $\chi_{42}$ & $0$  & $0$  & $0$  & $0$  & $0$  & $1$  & $0$  & $0$  & $0$  & $0$  & $0$  & $0$  & $0$  & $0$  & -$1$  & $0$  & $0$  & -$1$  & -$1$  & $0$  & $0$  & $0$  & $0$  & $0$ \\
		 $\chi_{43}$ & $0$  & $0$  & $0$  & $0$  & $1$  & $0$  & $1$  & -$1$  & -$1$  & $0$  & $0$  & $0$  & $0$  & $0$  & -$1$  & $0$  & $0$  & $0$  & $0$  & $0$  & $0$  & $0$  & $1$  & $1$ \\
		 $\chi_{44}$ & $0$  & $0$  & -$2$  & $1$  & $1$  & $0$  & -$1$  & $0$  & $0$  & $0$  & $0$  & $0$  & $0$  & $0$  & $0$  & $0$  & $0$  & $0$  & $0$  & $0$  & $0$  & $0$  & $0$  & $0$ \\
		 $\chi_{45}$ & $0$  & $0$  & $0$  & $0$  & $0$  & -$1$  & -$1$  & $1$  & $1$  & $0$  & $0$  & $0$  & $0$  & $0$  & $1$  & $0$  & $0$  & $0$  & $0$  & $0$  & $0$  & $0$  & -$1$  & -$1$ \\
		 $\chi_{46}$ & $0$  & $0$  & $0$  & $0$  & $0$  & $0$  & $0$  & $0$  & $0$  & $0$  & $0$  & $0$  & $0$  & $0$  & $0$  & $0$  & $0$  & $0$  & $0$  & $0$  & $0$  & $0$  & $0$  & $0$ \\
		 $\chi_{47}$ & $0$  & $0$  & $0$  & $0$  & $0$  & $0$  & $1$  & $0$  & $0$  & $0$  & $0$  & $0$  & $0$  & $0$  & -$1$  & $0$  & $0$  & $0$  & $0$  & $0$  & $0$  & $0$  & $1$  & $1$ \\
		 $\chi_{48}$ & $0$  & $0$  & $0$  & $0$  & -$1$  & $0$  & $0$  & $0$  & $0$  & $0$  & $0$  & $0$  & $0$  & $0$  & $1$  & $0$  & $0$  & $0$  & $0$  & $0$  & $0$  & $0$  & $0$  & $0$ \\
		 \bottomrule
	\end{tabular}
	\end{small}
	\end{center}
    \caption{The character table of the Thompson group, part IV.}\label{thchartabfour}
	\end{table}
	\end{landscape}

\begin{table}[h!]
\centering
\begin{small}
\begin{tabular}{c | c c c c c c c c c}
\toprule
$n$X & $\mathrm{1A}$ & $\mathrm{2A}$ & $\mathrm{2B}$ & $\mathrm{3A}$ & $\mathrm{3B}$ & $\mathrm{4A}$ & $\mathrm{4B}$ & $\mathrm{4C}$ & $\mathrm{5A}$\\
$m$Y & $\mathrm{1A}$ & $\mathrm{2A}$ & $\mathrm{2B}$ & $\mathrm{3A}$ & $\mathrm{3B}$ & $\mathrm{4A}$ & $\mathrm{4B}$ & $\mathrm{4D}$ & $\mathrm{5A}$\\\midrule
$n$X & $\mathrm{5B}$ & $\mathrm{5C}$ & $\mathrm{5D}$ & $\mathrm{5E}$ & $\mathrm{6A}$ & $\mathrm{6B}$ & $\mathrm{6C}$ & $\mathrm{7A}$ & $\mathrm{8A}$\\
$m$Y & $\mathrm{5B}$ & $\mathrm{5B}$ & $\mathrm{5B}$ & $\mathrm{5A}$ & $\mathrm{6A}$ & $\mathrm{6C}$ & $\mathrm{6B}$ & $\mathrm{7A}$ & $\mathrm{8C}$\\\midrule
$n$X & $\mathrm{8B}$ & $\mathrm{9A}$ & $\mathrm{10A}$ & $\mathrm{10B}$ & $\mathrm{10C}$ & $\mathrm{10D}$ & $\mathrm{10E}$ & $\mathrm{10F}$ & $\mathrm{10G}$\\
$m$Y & $\mathrm{8B}$ & $\mathrm{9A}$ & $\mathrm{10C}$ & $\mathrm{10A}$ & $\mathrm{10E}$ & $\mathrm{10D}$ & $\mathrm{10D}$ & $\mathrm{10A}$ & $\mathrm{10B}$\\\midrule
$n$X & $\mathrm{10H}$ & $\mathrm{11A}$ & $\mathrm{12A}$ & $\mathrm{12B}$ & $\mathrm{12C}$ & $\mathrm{14A}$ & $\mathrm{15A}$ & $\mathrm{15B}$ & $\mathrm{15C}$\\
$m$Y & $\mathrm{10B}$ & $\mathrm{11A}$ & $\mathrm{12C}$ & $\mathrm{12A}$ & $\mathrm{12F}$ & $\mathrm{14A}$ & $\mathrm{15A}$ & $\mathrm{15C}$ & $\mathrm{15C}$\\\midrule
$n$X & $\mathrm{19A}$ & $\mathrm{19B}$ & $\mathrm{20A}$ & $\mathrm{20B}$ & $\mathrm{20C}$ & $\mathrm{20D}$ & $\mathrm{20E}$ & $\mathrm{21A}$ & $\mathrm{22A}$\\
$m$Y & $\mathrm{19A}$ & $\mathrm{19A}$ & $\mathrm{20C}$ & $\mathrm{20C}$ & $\mathrm{20B}$ & $\mathrm{20E}$ & $\mathrm{20E}$ & $\mathrm{21A}$ & $\mathrm{22A}$\\\midrule
$n$X & $\mathrm{25A}$ & $\mathrm{25B}$ & $\mathrm{30A}$ & $\mathrm{30B}$ & $\mathrm{30C}$ & $\mathrm{35A}$ & $\mathrm{35B}$ & $\mathrm{40A}$ & $\mathrm{40B}$\\
$m$Y & $\mathrm{25A}$ & $\mathrm{25A}$ & $\mathrm{30B}$ & $\mathrm{30A}$ & $\mathrm{30A}$ & $\mathrm{35A}$ & $\mathrm{35A}$ & $\mathrm{40A}$ & $\mathrm{40A}$\\
\bottomrule
\end{tabular}
\caption{The fusion of conjugacy classes in $\tsl{HN}$ into conjugacy classes of $\mathbb{M}$. The notation $n$X indicates a conjugacy class of $\tsl{HN}$, and $m$Y indicates a conjugacy class of $\mathbb{M}$, with both following the labeling conventions of the Atlas of Finite Groups \cite{atlas}. This data was computed using Gap \cite{GAP}. \label{HN Fusion table}}
\end{small}
\end{table}

\begin{sidewaystable}[t!]
\begin{center}
\begin{footnotesize}
\begin{tabular}{c | c c c c c c c c c c}
\toprule
$(\omega^k,n\mathrm{X})$ & $(1,\mathrm{1A})$ & $(1,\mathrm{2A})$ & $(1,\mathrm{3A})$ & $(1,\mathrm{3B})$ & $(1,\mathrm{3C})$ & $(1,\mathrm{4A})$ & $(1,\mathrm{4B})$ & $(1,\mathrm{5A})$ & $(1,\mathrm{6A})$ & $(1,\mathrm{6B})$\\
$m$Y & $\mathrm{1A}$ & $\mathrm{2B}$ & $\mathrm{3A}$ & $\mathrm{3B}$ & $\mathrm{3B}$ & $\mathrm{4A}$ & $\mathrm{4D}$ & $\mathrm{5B}$ & $\mathrm{6B}$ & $\mathrm{6C}$\\
$(\omega^k,n\mathrm{X})$ & $(1,\mathrm{6C})$ & $(1,\mathrm{7A})$ & $(1,\mathrm{8A})$ & $(1,\mathrm{8B})$ & $(1,\mathrm{9A})$ & $(1,\mathrm{9B})$ & $(1,\mathrm{9C})$ & $(1,\mathrm{10A})$ & $(1,\mathrm{12A})$\\
$m$Y & $\mathrm{6E}$ & $\mathrm{7A}$ & $\mathrm{8B}$ & $\mathrm{8F}$ & $\mathrm{9B}$ & $\mathrm{9B}$ & $\mathrm{9A}$ & $\mathrm{10D}$ & $\mathrm{12A}$\\
$(\omega^k,n\mathrm{X})$ & $(1,\mathrm{12B})$ & $(1,\mathrm{12C})$ & $(1,\mathrm{12D})$ & $(1,\mathrm{13A})$ & $(1,\mathrm{14A})$ & $(1,\mathrm{15A})$ & $(1,\mathrm{15B})$ & $(1,\mathrm{18A})$ & $(1,\mathrm{18B})$ & $(1,\mathrm{19A})$\\
$m$Y & $\mathrm{12A}$ & $\mathrm{12B}$ & $\mathrm{12F}$ & $\mathrm{13A}$ & $\mathrm{14B}$ & $\mathrm{15C}$ & $\mathrm{15C}$ & $\mathrm{18D}$ & $\mathrm{18C}$ & $\mathrm{19A}$\\
$(\omega^k,n\mathrm{X})$ & $(1,\mathrm{20A})$ & $(1,\mathrm{21A})$ & $(1,\mathrm{24A})$ & $(1,\mathrm{24B})$ & $(1,\mathrm{24C})$ & $(1,\mathrm{24D})$ & $(1,\mathrm{27A})$ & $(1,\mathrm{27B})$ & $(1,\mathrm{27C})$\\
$m$Y & $\mathrm{20E}$ & $\mathrm{21A}$ & $\mathrm{24A}$ & $\mathrm{24A}$ & $\mathrm{24F}$ & $\mathrm{24F}$ & $\mathrm{27A}$ & $\mathrm{27A}$ & $\mathrm{27A}$\\
$(\omega^k,n\mathrm{X})$ & $(1,\mathrm{28A})$ & $(1,\mathrm{30A})$ & $(1,\mathrm{30B})$ & $(1,\mathrm{31A})$ & $(1,\mathrm{31B})$ & $(1,\mathrm{36A})$ & $(1,\mathrm{36B})$ & $(1,\mathrm{36C})$ & $(1,\mathrm{39A})$ & $(1,\mathrm{39B})$\\
$m$Y & $\mathrm{28B}$ & $\mathrm{30A}$ & $\mathrm{30A}$ & $\mathrm{31A/B}$ & $\mathrm{31B/A}$ & $\mathrm{36B}$ & $\mathrm{36B}$ & $\mathrm{36B}$ & $\mathrm{39A}$ & $\mathrm{39A}$\\
\midrule
$(\omega^k,n\mathrm{X})$ & $(\omega,\mathrm{1A})$ & $(\omega,\mathrm{2A})$ & $(\omega,\mathrm{3A})$ & $(\omega,\mathrm{3B})$ & $(\omega,\mathrm{3C})$ & $(\omega,\mathrm{4A})$ & $(\omega,\mathrm{4B})$ & $(\omega,\mathrm{5A})$ & $(\omega,\mathrm{6A})$ & $(\omega,\mathrm{6B})$\\
$m$Y & $\mathrm{3C}$ & $\mathrm{6F}$ & $\mathrm{3C}$ & $\mathrm{3C}$ & $\mathrm{3C}$ & $\mathrm{12D}$ & $\mathrm{12J}$ & $\mathrm{15D}$ & $\mathrm{6F}$ & $\mathrm{6F}$\\
$(\omega^k,n\mathrm{X})$ & $(\omega,\mathrm{6C})$ & $(\omega,\mathrm{7A})$ & $(\omega,\mathrm{8A})$ & $(\omega,\mathrm{8B})$ & $(\omega,\mathrm{9A})$ & $(\omega,\mathrm{9B})$ & $(\omega,\mathrm{9C})$ & $(\omega,\mathrm{10A})$ & $(\omega,\mathrm{12A})$\\
$m$Y & $\mathrm{6F}$ & $\mathrm{21C}$ & $\mathrm{24E}$ & $\mathrm{24J}$ & $\mathrm{9A}$ & $\mathrm{9A}$ & $\mathrm{9B}$ & $\mathrm{30E}$ & $\mathrm{12D}$\\
$(\omega^k,n\mathrm{X})$ & $(\omega,\mathrm{12B})$ & $(\omega,\mathrm{12C})$ & $(\omega,\mathrm{12D})$ & $(\omega,\mathrm{13A})$ & $(\omega,\mathrm{14A})$ & $(\omega,\mathrm{15A})$ & $(\omega,\mathrm{15B})$ & $(\omega,\mathrm{18A})$ & $(\omega,\mathrm{18B})$ & $(\omega,\mathrm{19A})$\\
$m$Y & $\mathrm{12D}$ & $\mathrm{12D}$ & $\mathrm{12J}$ & $\mathrm{39B}$ & $\mathrm{42C}$ & $\mathrm{15D}$ & $\mathrm{15D}$ & $\mathrm{18C}$ & $\mathrm{18E}$ & $\mathrm{57A}$\\
$(\omega^k,n\mathrm{X})$ & $(\omega,\mathrm{20A})$ & $(\omega,\mathrm{21A})$ & $(\omega,\mathrm{24A})$ & $(\omega,\mathrm{24B})$ & $(\omega,\mathrm{24C})$ & $(\omega,\mathrm{24D})$ & $(\omega,\mathrm{27A})$ & $(\omega,\mathrm{27B})$ & $(\omega,\mathrm{27C})$\\
$m$Y & $\mathrm{60F}$ & $\mathrm{21C}$ & $\mathrm{24E}$ & $\mathrm{24E}$ & $\mathrm{24J}$ & $\mathrm{24J}$ & $\mathrm{27B}$ & $\mathrm{27B}$ & $\mathrm{27B}$\\
$(\omega^k,n\mathrm{X})$ & $(\omega,\mathrm{28A})$ & $(\omega,\mathrm{30A})$ & $(\omega,\mathrm{30B})$ & $(\omega,\mathrm{31A})$ & $(\omega,\mathrm{31B})$ & $(\omega,\mathrm{36A})$ & $(\omega,\mathrm{36B})$ & $(\omega,\mathrm{36C})$ & $(\omega,\mathrm{39A})$ & $(\omega,\mathrm{39B})$\\
$m$Y & $\mathrm{84C}$ & $\mathrm{30E}$ & $\mathrm{30E}$ & $\mathrm{93A/B}$ & $\mathrm{93B/A}$ & $\mathrm{36A}$ & $\mathrm{36A}$ & $\mathrm{36A}$ & $\mathrm{39B}$ & $\mathrm{39B}$\\\midrule
$(\omega^k,n\mathrm{X})$ & $(\omega^2,\mathrm{1A})$ & $(\omega^2,\mathrm{2A})$ & $(\omega^2,\mathrm{3A})$ & $(\omega^2,\mathrm{3B})$ & $(\omega^2,\mathrm{3C})$ & $(\omega^2,\mathrm{4A})$ & $(\omega^2,\mathrm{4B})$ & $(\omega^2,\mathrm{5A})$ & $(\omega^2,\mathrm{6A})$ & $(\omega^2,\mathrm{6B})$\\
$m$Y & $\mathrm{3C}$ & $\mathrm{6F}$ & $\mathrm{3C}$ & $\mathrm{3C}$ & $\mathrm{3C}$ & $\mathrm{12D}$ & $\mathrm{12J}$ & $\mathrm{15D}$ & $\mathrm{6F}$ & $\mathrm{6F}$\\
$(\omega^k,n\mathrm{X})$ & $(\omega^2,\mathrm{6C})$ & $(\omega^2,\mathrm{7A})$ & $(\omega^2,\mathrm{8A})$ & $(\omega^2,\mathrm{8B})$ & $(\omega^2,\mathrm{9A})$ & $(\omega^2,\mathrm{9B})$ & $(\omega^2,\mathrm{9C})$ & $(\omega^2,\mathrm{10A})$ & $(\omega^2,\mathrm{12A})$\\
$m$Y & $\mathrm{6F}$ & $\mathrm{21C}$ & $\mathrm{24E}$ & $\mathrm{24J}$ & $\mathrm{9A}$ & $\mathrm{9A}$ & $\mathrm{9B}$ & $\mathrm{30E}$ & $\mathrm{12D}$\\
$(\omega^k,n\mathrm{X})$ & $(\omega^2,\mathrm{12B})$ & $(\omega^2,\mathrm{12C})$ & $(\omega^2,\mathrm{12D})$ & $(\omega^2,\mathrm{13A})$ & $(\omega^2,\mathrm{14A})$ & $(\omega^2,\mathrm{15A})$ & $(\omega^2,\mathrm{15B})$ & $(\omega^2,\mathrm{18A})$ & $(\omega^2,\mathrm{18B})$ & $(\omega^2,\mathrm{19A})$\\
$m$Y & $\mathrm{12D}$ & $\mathrm{12D}$ & $\mathrm{12J}$ & $\mathrm{39B}$ & $\mathrm{42C}$ & $\mathrm{15D}$ & $\mathrm{15D}$ & $\mathrm{18C}$ & $\mathrm{18E}$ & $\mathrm{57A}$\\
$(\omega^k,n\mathrm{X})$ & $(\omega^2,\mathrm{20A})$ & $(\omega^2,\mathrm{21A})$ & $(\omega^2,\mathrm{24A})$ & $(\omega^2,\mathrm{24B})$ & $(\omega^2,\mathrm{24C})$ & $(\omega^2,\mathrm{24D})$ & $(\omega^2,\mathrm{27A})$ & $(\omega^2,\mathrm{27B})$ & $(\omega^2,\mathrm{27C})$\\
$m$Y & $\mathrm{60F}$ & $\mathrm{21C}$ & $\mathrm{24E}$ & $\mathrm{24E}$ & $\mathrm{24J}$ & $\mathrm{24J}$ & $\mathrm{27B}$ & $\mathrm{27B}$ & $\mathrm{27B}$\\
$(\omega^k,n\mathrm{X})$ & $(\omega^2,\mathrm{28A})$ & $(\omega^2,\mathrm{30A})$ & $(\omega^2,\mathrm{30B})$ & $(\omega^2,\mathrm{31A})$ & $(\omega^2,\mathrm{31B})$ & $(\omega^2,\mathrm{36A})$ & $(\omega^2,\mathrm{36B})$ & $(\omega^2,\mathrm{36C})$ & $(\omega^2,\mathrm{39A})$ & $(\omega^2,\mathrm{39B})$\\
$m$Y & $\mathrm{84C}$ & $\mathrm{30E}$ & $\mathrm{30E}$ & $\mathrm{93A/B}$ & $\mathrm{93B/A}$ & $\mathrm{36A}$ & $\mathrm{36A}$ & $\mathrm{36A}$ & $\mathrm{39B}$ & $\mathrm{39B}$\\
\bottomrule
\end{tabular}
\captionsetup{width=0.83\textwidth}
\caption{\footnotesize The fusion of conjugacy classes in $\mathbb{Z}_3\times\tsl{Th}$ into conjugacy classes of $\mathbb{M}$. A conjugacy class in $\mathbb{Z}_3\times\Th$ is denoted as $(\omega^k,n\mathrm{X})$ where $\omega^k$ for $k=0,1,2$ are the three conjugacy classes of $\mathbb{Z}_3$, and $n$X is a conjugacy class of $\Th$, following the labeling in the Atlas of Finite Groups \cite{atlas}. Monster conjugacy classes are labeled according to the Atlas as well. This table was computed using Gap \cite{GAP}, which provides two possibilities for the fusions of the classes $(\omega^k,31\mathrm{AB})$; the ambiguity is of no consequence for the bilinears in equations \eqref{Thompson Gen bil} and \eqref{Thompson Gen bilinear}, since $J_{\mathrm{31A}} = J_{31B}$ and $J_{\mathrm{93A}} = J_{\mathrm{93B}}$. \label{Th Fusion table}}
\end{footnotesize}
\end{center}
\end{sidewaystable}

\clearpage

\section{Alternative derivation of the characters of $\normalfont{\tsl{VF}}^\natural_{22}$}\label{app:C}
In this appendix, we give an alternative derivation of the characters of $\tsl{VF}^\natural_{22}$. The basic idea is that, although the Hecke method does not work out of the box, one can perform intermediate deconstructions for which the Hecke method \emph{is} effective. Although we work purely at the level of modular forms, our steps are motivated by the following algebraic manipulations.
\begin{enumerate}
\item We first decompose the moonshine module into (an extension of) one of its $\mc{L}(\tfrac12,0)\otimes \mc{L}(\tfrac45,0)\otimes\mc{L}(\tfrac{7}{10},0)$ subalgebras\footnote{Such a subalgebra exists, as proven in \cite{dong1996associative}.} and its commutant. Here, the Hecke method is effective in producing the dual characters.
\item It is straightforward to infer from the previous step how the moonshine module decomposes into just $\mc{L}(\tfrac12,0)\otimes\mc{L}(\tfrac45,0)$ and its commutant. From the fact that $\mc{P}(2)\cong \mc{L}(\tfrac12,0)$ and $\mc{P}(3) \cong \mc{L}(\tfrac45,0)\oplus\mc{L}(\tfrac45,3)$, we will be able to re-interpret this as a decomposition of $V^\natural$ into a $\mc{P}(2)\otimes \mc{P}(3)$ subalgebra and its commutant. 
\item The previous step will give us a bilinear of the form $J(\tau) = \sum_i g_i(\tau) \widetilde{g}_{i}(\tau)$ which we can set equal to the bilinear in equation \eqref{Bilinear k236} to extract expressions for the dual characters $\chi_{\tsl{VF}^\natural_{22}(\alpha)}(\tau)$.
\end{enumerate}
We start by constructing the characters of an extension of $\mc{P}(2)\otimes\mc{P}(3)\otimes \mc{P}(6)$, using the block-diagonalization method outlined in \S\ref{subsubsec:modules}. We label the S-matrices of the VOAs $\mc{P}(2)$, $\mc{P}(3)$, and $\mc{P}(6)$ as $\mathcal{S}^{(2)}, \mathcal{S}^{(3)}$, and $\mathcal{S}^{(6)}$. One can show that the matrix of the tensor product theory, $\mathcal{S}^{(2)} \otimes \mathcal{S}^{(3)} \otimes \mathcal{S}^{(6)}$, can be block-diagonalized into a $14 \times 14$ block and its complement. This suggests the existence of a unitary RCFT described by 14 characters which can be expressed in terms of parafermion characters as
\begin{align}
\label{m236 characters}
\begin{split}
\chi_{0} &= \psi^{(2)}_{2,2} \psi^{(3)}_{3,3} \psi^{(6)}_{6,6} + \psi^{(2)}_{2,2} \psi^{(3)}_{3,1} \psi^{(6)}_{6,-2} + \psi^{(2)}_{2,2} \psi^{(3)}_{3,-1}\psi^{(6)}_{6,2} \\
&\hspace{.8in} +\psi^{(2)}_{2,0}\psi^{(3)}_{3,3}\psi^{(6)}_{6,0} +\psi^{(2)}_{2,0}\psi^{(3)}_{3,-1}\psi^{(6)}_{6,-4}+ \psi^{(2)}_{2,0}\psi^{(3)}_{3,1}\psi^{(6)}_{6,4},  \\
\chi_{1} &= \psi^{(2)}_{2,2} \psi^{(3)}_{3,3} \psi^{(6)}_{2,0} + \psi^{(2)}_{2,2} \psi^{(3)}_{3,1} \psi^{(6)}_{4,-2} + \psi^{(2)}_{2,2} \psi^{(3)}_{3,-1} \psi^{(6)}_{4,2} \\
&\hspace{.8in} + \psi^{(2)}_{2,0} \psi^{(3)}_{3,3} \psi^{(6)}_{4,0}  + \psi^{(2)}_{2,0} \psi^{(3)}_{3,-1} \psi^{(6)}_{2,2} + \psi^{(2)}_{2,0} \psi^{(3)}_{3,1} \psi^{(6)}_{4,4}  , \\
\chi_{2} &=  \psi^{(2)}_{2,2} \psi^{(3)}_{3,3} \psi^{(6)}_{4,0} + \psi^{(2)}_{2,2} \psi^{(3)}_{3,1} \psi^{(6)}_{4,4} + \psi^{(2)}_{2,2} \psi^{(3)}_{3,-1} \psi^{(6)}_{2,2} \\
&\hspace{.8in}+ \psi^{(2)}_{2,0} \psi^{(3)}_{3,3} \psi^{(6)}_{2,0}   + \psi^{(2)}_{2,0} \psi^{(3)}_{3,-1} \psi^{(6)}_{4,2} + \psi^{(2)}_{2,0} \psi^{(3)}_{3,1} \psi^{(6)}_{4,-2}  , \\
\chi_{3} &=  \psi^{(2)}_{2,2} \psi^{(3)}_{3,3} \psi^{(6)}_{6,0} + \psi^{(2)}_{2,2} \psi^{(3)}_{3,1} \psi^{(6)}_{6,4} + \psi^{(2)}_{2,2} \psi^{(3)}_{3,-1} \psi^{(6)}_{6,-4} \\
&\hspace{.8in}+ \psi^{(2)}_{2,0} \psi^{(3)}_{3,3} \psi^{(6)}_{6,6}    + \psi^{(2)}_{2,0} \psi^{(3)}_{3,-1} \psi^{(6)}_{6,2} + \psi^{(2)}_{2,0} \psi^{(3)}_{3,1} \psi^{(6)}_{6,-2}  , \\
\chi_{4} &= \psi^{(2)}_{2,2} \psi^{(3)}_{2,0} \psi^{(6)}_{4,0} + \psi^{(2)}_{2,2} \psi^{(3)}_{2,2} \psi^{(6)}_{2,2} + \psi^{(2)}_{2,2} \psi^{(3)}_{1,1} \psi^{(6)}_{4,4} \\
&\hspace{.8in} + \psi^{(2)}_{2,0} \psi^{(3)}_{2,0} \psi^{(6)}_{2,0}    + \psi^{(2)}_{2,0} \psi^{(3)}_{2,2} \psi^{(6)}_{4,2} + \psi^{(2)}_{2,0} \psi^{(3)}_{1,1} \psi^{(6)}_{4,-2}  , \\
\chi_{5} &=  \psi^{(2)}_{2,2} \psi^{(3)}_{2,0} \psi^{(6)}_{2,0} + \psi^{(2)}_{2,2} \psi^{(3)}_{2,2} \psi^{(6)}_{4,2} + \psi^{(2)}_{2,2} \psi^{(3)}_{1,1} \psi^{(6)}_{4,-2} \\
&\hspace{.8in}+ \psi^{(2)}_{2,0} \psi^{(3)}_{1,1} \psi^{(6)}_{4,4}   + \psi^{(2)}_{2,0} \psi^{(3)}_{2,0} \psi^{(6)}_{4,0} + \psi^{(2)}_{2,0} \psi^{(3)}_{2,2} \psi^{(6)}_{2,2}  , \\
\chi_{6} &=  \psi^{(2)}_{2,2} \psi^{(3)}_{2,0} \psi^{(6)}_{6,0} + \psi^{(2)}_{2,0} \psi^{(3)}_{2,0} \psi^{(6)}_{6,6} + \psi^{(2)}_{2,2} \psi^{(3)}_{2,2} \psi^{(6)}_{6,-4} \\
&\hspace{.8in}+ \psi^{(2)}_{2,2} \psi^{(3)}_{1,1} \psi^{(6)}_{6,4}   + \psi^{(2)}_{2,0} \psi^{(3)}_{1,1} \psi^{(6)}_{6,-2} + \psi^{(2)}_{2,0} \psi^{(3)}_{2,2} \psi^{(6)}_{6,2}, \\
\chi_{7} &=  \psi^{(2)}_{2,2} \psi^{(3)}_{2,0} \psi^{(6)}_{6,6} + \psi^{(2)}_{2,2} \psi^{(3)}_{2,2} \psi^{(6)}_{6,2} + \psi^{(2)}_{2,2} \psi^{(3)}_{1,1} \psi^{(6)}_{6,-2} \\
&\hspace{.8in}+ \psi^{(2)}_{2,0} \psi^{(3)}_{2,0} \psi^{(6)}_{6,0}    + \psi^{(2)}_{2,0} \psi^{(3)}_{1,1} \psi^{(6)}_{6,4} + \psi^{(2)}_{2,0} \psi^{(3)}_{2,2} \psi^{(6)}_{6,-4}  , \\
\chi_{8} &=  \psi^{(2)}_{1,1} \psi^{(3)}_{3,3} \psi^{(6)}_{3,3} + \psi^{(2)}_{1,1} \psi^{(3)}_{3,1} \psi^{(6)}_{3,1} +  \psi^{(2)}_{1,1} \psi^{(3)}_{3,-1} \psi^{(6)}_{3,-1}, \\
\chi_{9} &=  \psi^{(2)}_{1,1} \psi^{(3)}_{3,3} \psi^{(6)}_{3,3} + \psi^{(2)}_{1,1} \psi^{(3)}_{3,1} \psi^{(6)}_{3,1} +  \psi^{(2)}_{1,1} \psi^{(3)}_{3,-1} \psi^{(6)}_{3,-1}, \\
\chi_{10} &=  \psi^{(2)}_{1,1} \psi^{(3)}_{3,3} \psi^{(6)}_{5,-3} + \psi^{(2)}_{1,1} \psi^{(3)}_{3,3} \psi^{(6)}_{5,3} + \psi^{(2)}_{1,1} \psi^{(3)}_{3,1} \psi^{(6)}_{1,1} \\
&\hspace{.8in} + \psi^{(2)}_{1,1} \psi^{(3)}_{3,-1} \psi^{(6)}_{5,5}  + \psi^{(2)}_{1,1} \psi^{(3)}_{3,1} \psi^{(6)}_{5,1} + \psi^{(2)}_{1,1} \psi^{(3)}_{3,-1} \psi^{(6)}_{5,-1}  , \\
\chi_{11}  &=  \psi^{(2)}_{1,1} \psi^{(3)}_{2,0} \psi^{(6)}_{5,-3} + \psi^{(2)}_{1,1} \psi^{(3)}_{2,0} \psi^{(6)}_{5,3} + \psi^{(2)}_{1,1} \psi^{(3)}_{1,1} \psi^{(6)}_{1,1} \\
&\hspace{.8in} + \psi^{(2)}_{1,1} \psi^{(3)}_{2,2} \psi^{(6)}_{5,5}  + \psi^{(2)}_{1,1} \psi^{(3)}_{1,1} \psi^{(6)}_{5,1} + \psi^{(2)}_{1,1} \psi^{(3)}_{2,2} \psi^{(6)}_{5,-1}  , \\
\chi_{12} &=  \psi^{(2)}_{1,1} \psi^{(3)}_{2,0} \psi^{(6)}_{3,3} + \psi^{(2)}_{1,1} \psi^{(3)}_{1,1} \psi^{(6)}_{3,1} +  \psi^{(2)}_{1,1} \psi^{(3)}_{2,2} \psi^{(6)}_{3,-1}, \\
\chi_{13} &=  \psi^{(2)}_{1,1} \psi^{(3)}_{2,0} \psi^{(6)}_{3,3} + \psi^{(2)}_{1,1} \psi^{(3)}_{1,1} \psi^{(6)}_{3,1} +  \psi^{(2)}_{1,1} \psi^{(3)}_{2,2} \psi^{(6)}_{3,-1}.
\end{split}
\end{align}
The modular properties of the characters in \eqref{m236 characters} are governed by the S-matrix $\mathcal{S}$
\begin{align}
\begin{tiny}
    \frac{1}{4}
    \begin{pmatrix}
    \sqrt{2} s_1 \alpha & \frac{\alpha}{2 s_1}  & \frac{\alpha}{2 s_1}  & \sqrt{2} s_1 \alpha & \frac{\beta}{2 s_1}  & \frac{\beta}{2 s_1} & \sqrt{2} s_1 \beta & \sqrt{2} s_1 \beta & \alpha & \alpha & \sqrt{2} \alpha & \sqrt{2} \beta & \beta & \beta \\
    \frac{\alpha}{2s_1} & -\sqrt{2} s_1 \alpha & -\sqrt{2} s_1 \alpha & \frac{\alpha}{2s_1} & -\sqrt{2} s_1\beta & -\sqrt{2} s_1 \beta & \frac{\beta}{2s_1} & \frac{\beta}{2s_1} & -\alpha & -\alpha & \sqrt{2} \alpha & \sqrt{2} \beta & -\beta & -\beta \\
    \frac{\alpha}{2s_1} & -\sqrt{2} s_1 \alpha & -\sqrt{2} s_1 \alpha & \frac{\alpha}{2 s_1} & -\sqrt{2} s_1 \beta & -\sqrt{2} s_1 \beta & \frac{\beta}{2s_1} & \frac{\beta}{2s_1} & \alpha & \alpha & -\sqrt{2} \alpha & -\sqrt{2} \beta & \beta & \beta \\
    \sqrt{2} s_1 \alpha & \frac{\alpha}{2s_1} & \frac{\alpha}{2s_1} & \sqrt{2} s_1 \alpha & \frac{\beta}{2s_1} & \frac{\beta}{2s_1} & \sqrt{2} s_1 \beta & \sqrt{2} s_1 \beta & -\alpha & -\alpha & -\sqrt{2} \alpha & -\sqrt{2} \beta & -\beta & -\beta \\
    \frac{\beta}{2s_1} & -\sqrt{2} s_1 \beta & -\sqrt{2} s_1 \beta & \frac{\beta}{2s_1} & \sqrt{2} s_1 \alpha & \sqrt{2} s_1 \alpha & -\frac{\alpha}{2s_1} & -\frac{\alpha}{2s_1} & \beta & \beta & -\sqrt{2}\beta & \sqrt{2} \alpha & -\alpha & -\alpha \\
    \frac{\beta}{2s_1} & -\sqrt{2} s_1 \beta & -\sqrt{2} s_1 \beta & \frac{\beta}{2s_1} & \sqrt{2} s_1 \alpha & \sqrt{2} s_1 \alpha & -\frac{\alpha}{2s_1} & -\frac{\alpha}{2s_1} & -\beta & -\beta & \sqrt{2} \beta & -\sqrt{2} \alpha & \alpha & \alpha \\
    \sqrt{2} s_1 \beta & \frac{\beta}{2 s_1} & \frac{\beta}{ 2s_1} & \sqrt{2} s_1 \beta & -\frac{\alpha}{2 s_1} & -\frac{\alpha}{2 s_1} & -\sqrt{2} s_1 \alpha & -\sqrt{2} s_1 \alpha & -\beta & -\beta & -\sqrt{2} \beta & \sqrt{2} \alpha & \alpha & \alpha \\
    \sqrt{2} s_1 \beta & \frac{\beta}{2 s_1} & \frac{\beta}{2 s_1} & \sqrt{2} s_1 \beta & -\frac{\alpha}{2s_1} & -\frac{\alpha}{2s_1} & -\sqrt{2} s_1 \alpha & -\sqrt{2} s_1 \alpha & \beta & \beta & \sqrt{2} \beta & -\sqrt{2} \alpha & -\alpha & -\alpha \\
    \alpha & -\alpha & \alpha & -\alpha & \beta & -\beta & -\beta & \beta & -\sqrt{2} \alpha &\sqrt{2} \alpha & 0 & 0 & \sqrt{2} \beta & -\sqrt{2} \beta \\
    \alpha & -\alpha & \alpha & -\alpha & \beta & -\beta & -\beta & \beta & \sqrt{2} \alpha & -\sqrt{2} \alpha & 0 & 0 & -\sqrt{2} \beta & \sqrt{2} \beta \\
    \sqrt{2} \alpha & \sqrt{2} \alpha & -\sqrt{2} \alpha & -\sqrt{2} \alpha & -\sqrt{2} \beta & \sqrt{2} \beta & -\sqrt{2} \beta & \sqrt{2} \beta & 0 & 0 & 0 & 0& 0 & 0 \\
    \sqrt{2} \beta & \sqrt{2} \beta & - \sqrt{2} \beta & -\sqrt{2} \beta & \sqrt{2} \alpha & -\sqrt{2} \alpha & \sqrt{2} \alpha & -\sqrt{2} \alpha & 0 & 0 & 0 & 0 & 0 & 0 \\
    \beta & -\beta & \beta & -\beta & -\alpha & \alpha & \alpha & -\alpha & \sqrt{2} \beta & -\sqrt{2} \beta & 0 & 0 & \sqrt{2} \alpha & -\sqrt{2} \alpha \\
    \beta & -\beta & \beta & -\beta & -\alpha & \alpha & \alpha & -\alpha & -\sqrt{2} \beta & \sqrt{2} \beta & 0 & 0 & -\sqrt{2} \alpha & \sqrt{2} \alpha
    \end{pmatrix} \nonumber
\end{tiny}
\end{align}
where $\alpha = \sqrt{1-\frac{1}{\sqrt{5}}}, \ \beta = \sqrt{1+\frac{1}{\sqrt{5}}}$, and $s_1 = \text{sin}\left( \frac{\pi}{8}\right)$. The T-matrix  $\mathcal{T}$ reads
\begin{align}
\begin{split}
\text{diag} \Big( e^{-\frac{17 i \pi}{80}},  e^{\frac{23 i \pi }{80}},  e^{-\frac{57 i \pi}{80}}, e^{\frac{63 i \pi }{80}}, e^{\frac{7 i \pi }{80}}, e^{-\frac{73 i \pi }{80} },  e^{-\frac{33 i \pi}{80}},  e^{\frac{47 i \pi }{80}}, e^{\frac{i \pi }{10}},  e^{\frac{i \pi }{10}},  e^{-\frac{13 i \pi}{20}},  e^{\frac{3 i \pi }{20}},  e^{\frac{9 i \pi }{10}},  e^{\frac{9 i \pi }{10}} \Big). \nonumber
\end{split}
\end{align}
One can check that these matrices furnish a representation of $\tsl{PSL}_2(\mathbb{Z})$,  namely  $\mathcal{S}^2 = 1$ and $(\mathcal{S} \cdot \mathcal{T})^3 = 1$.

Our goal is to find the characters dual to those in equation \eqref{m236 characters}. The solution we present, although inspired by the idea of performing a series of intermediate deconstructions, unfortunately features steps which do not quite have consistent VOA interpretations. The contents of the remainder of this appendix should therefore be thought of strictly as manipulations at the level of modular forms which produce the right answer for the characters $\chi_{\tsl{VF}_{22}^\natural}(\tau)$, although we do believe it should be possible to improve upon our results.

\paragraph{Step 1}
We start off by constructing the fictitious characters\footnote{We use the term fictitious because these characters e.g.\ do not lead to consistent fusion rules, and the bilinear they participate in, equation \eqref{Bilinear m345}, has fractional coefficients $\alpha_i$.} of an extension $\U$ of $\mc{L}(\tfrac12,0)\otimes \mc{L}(\tfrac45,0)\otimes \mc{L}(\tfrac{7}{10},0)$. This extension has 16 states of conformal weight
\begin{align}
\label{m345 states}
h = \left( 0, 1, \frac{1}{2}, \frac{1}{2}, \frac{1}{6}, \frac{7}{6}, \frac{2}{3}, \frac{2}{3}, \frac{2}{5}, \frac{1}{15}, \frac{1}{10}, \frac{23}{30}, \frac{3}{5}, \frac{19}{15}, \frac{9}{10}, \frac{17}{30} \right),
\end{align}
whose characters $f_j$ with conductor $N=60$ can be written in terms of minimal model characters as 
\begin{align}\label{345 subalgebra}
\begin{split}
f_{0} &= \chi^{(3)}_{1,1} \chi^{(4)}_{1,1} \chi^{(5)}_{1,+} + \chi^{(3)}_{2,1} \chi^{(4)}_{3,1} \chi^{(5)}_{1,+}, \quad f_{1} = \chi^{(3)}_{2,1} \chi^{(4)}_{1,2} \chi^{(5)}_{2,+} + \chi^{(3)}_{1,1} \chi^{(4)}_{1,3} \chi^{(5)}_{2,+}, \\
f_{2} &=
\chi^{(3)}_{1,1} \chi^{(4)}_{1,2} \chi^{(5)}_{2,+} +\chi^{(3)}_{2,1} \chi^{(4)}_{1,3} \chi^{(5)}_{2,+} + 2 \chi^{(3)}_{1,2} \chi^{(4)}_{2,2} \chi^{(5)}_{2,+},  \\
f_{3} &=
\chi^{(3)}_{1,1} \chi^{(4)}_{1,4} \chi^{(5)}_{1,+} +\chi^{(3)}_{2,1} \chi^{(4)}_{1,1} \chi^{(5)}_{1,+} + 2 \chi^{(3)}_{1,2} \chi^{(4)}_{2,1} \chi^{(5)}_{1,+},  \\
f_{4} &=  \chi^{(3)}_{1,1} \chi^{(4)}_{1,2} \chi^{(5)}_{2,3} + \chi^{(3)}_{2,1} \chi^{(4)}_{1,3} \chi^{(5)}_{2,3} + 2 \chi^{(3)}_{1,2} \chi^{(4)}_{2,2} \chi^{(5)}_{2,3} , \\
f_{5} &=\chi^{(3)}_{2,1} \chi^{(4)}_{1,1} \chi^{(5)}_{1,3} + \chi^{(3)}_{1,1} \chi^{(4)}_{3,1} \chi^{(5)}_{1,3} + 2 \chi^{(3)}_{1,2} \chi^{(4)}_{2,1} \chi^{(5)}_{1,3} , \\
f_{6} &= \chi^{(3)}_{1,1} \chi^{(4)}_{1,1} \chi^{(5)}_{1,3} + \chi^{(3)}_{2,1} \chi^{(4)}_{3,1} \chi^{(5)}_{1,3}, \quad f_{7} = \chi^{(3)}_{1,1} \chi^{(4)}_{1,3} \chi^{(5)}_{2,3} + \chi^{(3)}_{2,1} \chi^{(4)}_{1,2} \chi^{(5)}_{2,3}, \\
f_{8} &= \chi^{(3)}_{1,1} \chi^{(4)}_{1,1} \chi^{(5)}_{2,+} + \chi^{(3)}_{2,1} \chi^{(4)}_{3,1} \chi^{(5)}_{2,+}, \quad f_{9} = \chi^{(3)}_{1,1} \chi^{(4)}_{1,1} \chi^{(5)}_{2,3}  + \chi^{(3)}_{2,1} \chi^{(4)}_{3,1} \chi^{(5)}_{2,3}, \\
f_{10} &=  \chi^{(3)}_{1,1} \chi^{(4)}_{1,2} \chi^{(5)}_{1,+} +\chi^{(3)}_{2,1} \chi^{(4)}_{1,3} \chi^{(5)}_{1,+} +2\chi^{(3)}_{1,2} \chi^{(4)}_{2,2} \chi^{(5)}_{1,+} , \\
f_{11} &=   \chi^{(3)}_{1,1} \chi^{(4)}_{1,2} \chi^{(5)}_{1,3}+\chi^{(3)}_{2,1} \chi^{(4)}_{1,3} \chi^{(5)}_{1,3} +2\chi^{(3)}_{1,2} \chi^{(4)}_{2,2} \chi^{(5)}_{1,3}, \\
f_{12}&= \chi^{(3)}_{1,1} \chi^{(4)}_{1,3} \chi^{(5)}_{1,+}+ \chi^{(3)}_{2,1} \chi^{(4)}_{(1,2)} \chi^{(5)}_{1,+}, \quad f_{13} = \chi^{(3)}_{1,1} \chi^{(4)}_{1,3} \chi^{(5)}_{1,3}+ \chi^{(3)}_{2,1} \chi^{(4)}_{1,2} \chi^{(5)}_{1,3}, \\
f_{14} &=  \chi^{(3)}_{1,1} \chi^{(4)}_{3,1} \chi^{(5)}_{2,+} + \chi^{(3)}_{2,1} \chi^{(4)}_{1,1} \chi^{(5)}_{2,+} + 2 \chi^{(3)}_{1,2} \chi^{(4)}_{2,1} \chi^{(5)}_{2,+} , \\
f_{15} &=  \chi^{(3)}_{1,1} \chi^{(4)}_{3,1} \chi^{(5)}_{2,3} + \chi^{(3)}_{2,1} \chi^{(4)}_{1,1} \chi^{(5)}_{2,3} + 2 \chi^{(3)}_{1,2} \chi^{(4)}_{2,1} \chi^{(5)}_{2,3},
\end{split}
\end{align}
where $\chi^{(5)}_{1,+} \equiv \chi_{1,1}^{(5)} + \chi_{1,5}^{(5)}$ and $\chi^{(5)}_{2,+} \equiv \chi_{2,1}^{(5)} + \chi_{2,5}^{(5)}$.

The central charge of $\mc{L}(\tfrac12,0)\otimes\mc{L}(\tfrac{7}{10},0)\otimes\mc{L}(\tfrac45,0)$ is $2$, and its commutant in $V^\natural$ has central charge $22=2\cdot 11$, so we have a chance at finding the dual characters as the Hecke image of $f$ under $\mathsf{T}_{11}$. We provide the $q$-expansions of the components of ${\mathsf T}_{11}f$ below,
\begin{align}
\label{Hecke image m345}
\begin{split}
\widetilde{f}_0(\tau) &\equiv {\mathsf T}_{11} f_{0}(\tau) = q^{-\frac{11}{12}}(1+ 13959 q^2 + 1083742 q^3+ 34869263 q^4 + \cdots), \\
\widetilde{f}_1(\tau) &\equiv {\mathsf T}_{11} f_{1}(\tau) = q^{\frac{1}{12}}(22 + 36212 q + 2838132 q^2 + 91279606 q^3 + \cdots), \\
\widetilde{f}_2(\tau) &\equiv {\mathsf T}_{11} f_{2}(\tau) = q^{\frac{7}{12}}(6072  + 1124640 q + 52185936 q^2 + 1273841712 q^3 + \cdots), \\
\widetilde{f}_3(\tau) &\equiv {\mathsf T}_{11} f_{3}(\tau) = q^{\frac{7}{12}}(2376  + 429792 q + 19934640 q^2 + 486569424 q^3 + \cdots), \\
\widetilde{f}_4(\tau) &\equiv {\mathsf T}_{11} f_{4}(\tau) = q^{\frac{11}{12}}(45048 + 4456584 q + 159935952 q^2  + \cdots), \\
\widetilde{f}_5(\tau) &\equiv {\mathsf T}_{11} f_{5}(\tau) =  q^{\frac{11}{12}}(17160 + 1702008 q + 61089072 q^2 + \cdots), \\
\widetilde{f}_6(\tau) &\equiv {\mathsf T}_{11} f_{6}(\tau) =  q^{\frac{5}{12}}(253 + 68321 q+ 3703205 q^2 +  98302325 q^3 + \cdots), \\
\widetilde{f}_7(\tau) &\equiv {\mathsf T}_{11} f_{7}(\tau) =  q^{\frac{5}{12}}(638 + 179377 q+ 9692980 q^2 +  257372401 q^3 + \cdots), \\
\widetilde{f}_{8}(\tau) &\equiv {\mathsf T}_{11} f_{12}(\tau) =  q^{\frac{41}{60}}(2387 + 355014 q + 15143865 q^2 + \cdots )\\
\widetilde{f}_{9}(\tau) &\equiv {\mathsf T}_{11} f_{13}(\tau)  = q^{\frac{61}{60}}(15884  + 1357477 q + 45571669 q^2  + \cdots) \\
\widetilde{f}_{10}(\tau) &\equiv {\mathsf T}_{11} f_{14}(\tau)  =  q^{\frac{59}{60}}(39864 + 3578784 q + 122770296 q^2  + \cdots )\\
\widetilde{f}_{11}(\tau) &\equiv {\mathsf T}_{11} f_{15}(\tau)  =  q^{\frac{19}{60}}(528 + 209880 q + 12540264 q^2 + 351454488 q^3 + \cdots) \\
\widetilde{f}_{12}(\tau) &\equiv {\mathsf T}_{11} f_{8}(\tau)  =   q^{\frac{29}{60}}(638 + 149402 q + 7586128 q^2 + 194589330 q^3 + \cdots )\\
\widetilde{f}_{13}(\tau) &\equiv  {\mathsf T}_{11} f_{9}(\tau)  = q^{-\frac{11}{60}}(1 + 5258 q + 615197 q^2 + 23698356 q^3 + \cdots) \\
\widetilde{f}_{14}(\tau) &\equiv {\mathsf T}_{11} f_{10}(\tau)  =  q^{\frac{11}{60}}(168 + 110880 q + 7675800 q^2 + 232188528 q^3 + \cdots) \\
\widetilde{f}_{15}(\tau) &\equiv {\mathsf T}_{11} f_{11}(\tau)  =  q^{\frac{31}{60}}(2376 + 519816 q + 25579224 q^2 +  645255336 q^3 + \cdots) .
\end{split}
\end{align}
One can check that the Hecke images \eqref{Hecke image m345} satisfy the following bilinear,
\begin{align}
\label{Bilinear m345}
\begin{split}
J(\tau) &= \sum_{i=0}^{15} \alpha_{i} f_i(\tau) \widetilde{f}_{i}(\tau).
\end{split}
\end{align}
where the $\alpha_{i}$ take the values
\begin{align}
\begin{split}
\alpha_{i} &= 1, 1, \frac{1}{3}, \frac{1}{3}, \frac{2}{3}, \frac{2}{3}, 2, 2, 1, 2, \frac{1}{3}, \frac{2}{3}, 1, 2, \frac{1}{3}, \frac{2}{3},
\end{split}
\end{align}
for $i=0,1, \dots, 15$.

\paragraph{Step 2}
Next, let us construct the fictitious characters\footnote{C.f.\ the previous footnote.} of a particular extension of $\mc{L}(\tfrac12,0)\otimes \mc{L}(\tfrac45,0)$; its irreducible modules have highest weights
\begin{align}
h = \left( 0, \frac{1}{16}, \frac{1}{2}, \frac{2}{3}, \frac{35}{48}, \frac{7}{6}, \frac{2}{5}, \frac{37}{80}, \frac{9}{10}, \frac{1}{15}, \frac{31}{240}, \frac{17}{30} \right),
\end{align}
and its characters can be written as
\begin{align}
\begin{split}
g_{0}(\tau) &= \chi_{1,1}^{(3)}(\tau) \chi_{1,+}^{(5)}(\tau), \quad g_{1}(\tau) = \chi_{1,2}^{(3)}(\tau) \chi_{1,+}^{(5)}(\tau), \quad g_{2}(\tau) = \chi_{1,3}^{(3)}(\tau) \chi_{1,+}^{(5)}(\tau), \\
g_{3}(\tau) &= \chi_{1,1}^{(3)}(\tau) \chi_{1,3}^{(5)}(\tau), \quad g_{4}(\tau) = \chi_{1,2}^{(3)}(\tau) \chi_{1,3}^{(5)}(\tau), \quad g_{5}(\tau) = \chi_{1,3}^{(3)}(\tau) \chi_{1,3}^{(5)}(\tau), \\
g_{6}(\tau) &= \chi_{1,1}^{(3)}(\tau) \chi_{2,+}^{(5)}(\tau), \quad g_{7}(\tau) = \chi_{1,2}^{(3)}(\tau) \chi_{2,+}^{(5)}(\tau), \quad g_{8}(\tau) = \chi_{1,3}^{(3)}(\tau) \chi_{2,+}^{(5)}(\tau), \\
g_{9}(\tau) &= \chi_{1,1}^{(3)}(\tau) \chi_{2,3}^{(5)}(\tau), \quad g_{10}(\tau) = \chi_{1,2}^{(3)}(\tau) \chi_{2,3}^{(5)}(\tau), \quad g_{11}(\tau) = \chi_{1,3}^{(3)}(\tau) \chi_{2,3}^{(5)}(\tau).
\end{split}
\end{align}
We would like to find the dual characters, which by assumption should satisfy a bilinear of the form
\begin{align}
\label{Bilinear m35}
\begin{split}
J(\tau) &= \sum_{i} g_i(\tau) \widetilde{g}_{i}(\tau).
\end{split}
\end{align}
By comparing \eqref{Bilinear m345} and \eqref{Bilinear m35}, one can express $\widetilde{g}_{i}(\tau)$ in terms of \eqref{Hecke image m345} and the characters of $\mc{L}(\tfrac{7}{10},0)$. We find
\begin{align}
\label{dual ch m35}
\begin{split}
\widetilde{g}_{0}(\tau) &= \frac{1}{3} \chi_{1,4}^{(4)} \widetilde{f}_3(\tau) + \frac{1}{3} \chi_{1,2}^{(4)} \widetilde{f}_{10}(\tau) + \chi_{1,1}^{(4)} \widetilde{f}_0(\tau) +  \chi_{1,3}^{(4)} \widetilde{f}_{12}(\tau), \\
\widetilde{g}_{1}(\tau) &= \frac{2}{3} \chi_{2,1}^{(4)} \widetilde{f}_{3}(\tau) + \frac{2}{3} \chi_{2,2}^{(4)} \widetilde{f}_{10}(\tau), \quad \widetilde{g}_{4}(\tau) = \frac{4}{3} \chi_{2,1}^{(4)} \widetilde{f}_{5}(\tau) + \frac{4}{3} \chi_{2,2}^{(4)} \widetilde{f}_{11}(\tau), \\
\widetilde{g}_{2}(\tau) &= \frac{1}{3} \chi_{1,1}^{(4)} \widetilde{f}_{3}(\tau) + \frac{1}{3} \chi_{1,3}^{(4)} \widetilde{f}_{10}(\tau) + \chi_{3,1}^{(4)} \widetilde{f}_{0}(\tau) + \chi_{1,2}^{(4)} \widetilde{f}_{12}(\tau), \\
\widetilde{g}_{3}(\tau) &= \frac{1}{3} \chi_{3,1}^{(4)} \widetilde{f}_{5}(\tau) + \frac{2}{3} \chi_{1,2}^{(4)} \widetilde{f}_{11}(\tau) + 2 \chi_{1,1}^{(4)} \widetilde{f}_{6}(\tau) + 2 \chi_{1,3}^{(4)} \widetilde{f}_{13}(\tau), \\
\widetilde{g}_{5}(\tau) &= \frac{2}{3} \chi_{1,1}^{(4)} \widetilde{f}_{5}(\tau) + \frac{2}{3} \chi_{1,3}^{(4)} \widetilde{f}_{11}(\tau) + 2\chi_{3,1}^{(4)} \widetilde{f}_{6}(\tau) + 2 \chi_{1,2}^{(4)} \widetilde{f}_{13}(\tau), \\
\widetilde{g}_{6}(\tau) &= \frac{1}{3} \chi_{1,2}^{(4)} \widetilde{f}_{2}(\tau) + \frac{1}{3} \chi_{3,1}^{(4)} \widetilde{f}_{14}(\tau) + \chi_{1,3}^{(4)} \widetilde{f}_1(\tau) + \chi_{1,1}^{(4)} \widetilde{f}_{8}(\tau), \\
\widetilde{g}_{7}(\tau) &= \frac{2}{3} \chi_{2,2}^{(4)} \widetilde{f}_{2}(\tau) + \frac{2}{3} \chi_{2,1}^{(4)} \widetilde{f}_{14}(\tau), \quad
\widetilde{g}_{10}(\tau) = \frac{4}{3} \chi_{2,2}^{(4)} \widetilde{f}_{4}(\tau) + \frac{4}{3} \chi_{2,1}^{(4)} \widetilde{f}_{15}(\tau), \\
\widetilde{g}_{8}(\tau) &= \frac{1}{3} \chi_{1,3}^{(4)} \widetilde{f}_{2}(\tau) + \frac{1}{3} \chi_{1,1}^{(4)} \widetilde{f}_{14}(\tau) + \chi_{1,2}^{(4)} \widetilde{f}_1(\tau) + \chi_{3,1}^{(4)} \widetilde{f}_{8}(\tau), \\
\widetilde{g}_{9}(\tau) &= \frac{2}{3} \chi_{1,2}^{(4)} \widetilde{f}_{4}(\tau) + \frac{2}{3} \chi_{3,1}^{(4)} \widetilde{f}_{15}(\tau) + 2 \chi_{1,3}^{(4)} \widetilde{f}_{7}(\tau) + 2 \chi_{1,1}^{(4)} \widetilde{f}_{9}(\tau), \\
\widetilde{g}_{11}(\tau) &= \frac{2}{3} \chi_{1,3}^{(4)} \widetilde{f}_{4}(\tau) + \frac{2}{3} \chi_{1,1}^{(4)} \widetilde{f}_{15}(\tau) + 2 \chi_{1,2}^{(4)} \widetilde{f}_{7}(\tau) + 2 \chi_{3,1}^{(4)} \widetilde{f}_9(\tau).
\end{split}
\end{align}

\paragraph{Step 3}
Note that the parafermion theories $\mc{P}(2)$ and $\mc{P}(3)$ are the same as $\mc{L}(\tfrac12,0)$ and $\mc{L}(\tfrac45,0)\oplus\mc{L}(\tfrac45,3)$, respectively. Thus, we can replace the characters of the $\mathbb{Z}_2$ parafermion theory $\psi^{(2)}_{\ell,m}$ with the characters $\chi^{(3)}_{r,s}$ in equation \eqref{m236 characters}. The relation between Ising and $\mc{P}(2)$ characters is
\begin{align}
\psi^{(2)}_{2,2} = \chi^{(3)}_{1,1}, \quad \psi^{(2)}_{2,0} = \chi^{(3)}_{2,1}, \quad \psi^{(2)}_{1,1} = \chi^{(3)}_{2,2}.
\end{align}
Similarly, we can substitute $\chi^{(5)}_{r,s}$ for the  characters of the $\mathbb{Z}_3$ parafermion theory $\psi^{(3)}_{\ell,m}$ as
\begin{align}
\begin{split}
\psi^{(3)}_{3,3} &= \chi^{(5)}_{1,1} + \chi^{(5)}_{1,5}, \quad \psi^{(3)}_{1,1} + \psi^{(3)}_{2,2} = 2\chi^{(5)}_{2,3}, \\
\psi^{(3)}_{2,0} &= \chi^{(5)}_{2,1} + \chi^{(5)}_{2,5}, \quad \psi^{(3)}_{3,1} + \psi^{(3)}_{3,-1} = 2\chi^{(5)}_{1,3}.
\end{split}
\end{align}
The next step is to find the relations among the characters $\widetilde{g}_{i}(\tau)$ and $\chi_{\tsl{VF}^\natural_{22}(\alpha)}$ by comparing equations \eqref{Bilinear m35} and \eqref{Bilinear k236}. Setting them to be equal, we get a relation of the form 
\begin{align}
\label{gtoFi22}
\begin{split}
\widetilde{g}_{0} &= \psi^{(6)}_{2,0}  \widetilde{\chi}_{1} + \psi^{(6)}_{4,0}  \widetilde{\chi}_{2}  + \psi^{(6)}_{6,0} {\widetilde{\chi}}_{3} + \psi^{(6)}_{6,6} {\widetilde{\chi}}_{0},  \\
\widetilde{g}_{1} &=  \psi^{(6)}_{3,3} \widetilde{\chi}_{8} +  \psi^{(6)}_{3,3} \widetilde{\chi}_{9}+ \psi^{(6)}_{5,-3} \widetilde{\chi}_{10} +\psi^{(6)}_{5,3} \widetilde{\chi}_{10},  \\
\widetilde{g}_{2} &= \psi^{(6)}_{2,0} \widetilde{\chi}_{2} + \psi^{(6)}_{4,0} \widetilde{\chi}_{1} + \psi^{(6)}_{6,0} \widetilde{\chi}_{0} + \psi^{(6)}_{6,6} \widetilde{\chi}_{3},  \\
\widetilde{g}_{3} &= \psi^{(6)}_{2,2} \widetilde{\chi}_{2} + \psi^{(6)}_{4,-2} \widetilde{\chi}_{1} + \psi^{(6)}_{4,2} \widetilde{\chi}_{1} + \psi^{(6)}_{4,4} \widetilde{\chi}_{2}  + \psi^{(6)}_{6,-4} \widetilde{\chi}_{3} + \psi^{(6)}_{6,-2} \widetilde{\chi}_{0} + \psi^{(6)}_{6,2} \widetilde{\chi}_{0} + \psi^{(6)}_{6,4} \widetilde{\chi}_{3},  \\
\widetilde{g}_{4} &= \psi^{(6)}_{1,1} \widetilde{\chi}_{10} + \psi^{(6)}_{3,-1} \widetilde{\chi}_{8} + \psi^{(6)}_{3,-1} \widetilde{\chi}_{9} + \psi^{(6)}_{3,1} \widetilde{\chi}_{8}  + \psi^{(6)}_{3,1} \widetilde{\chi}_{9} + \psi^{(6)}_{5,-1} \widetilde{\chi}_{10} + \psi^{(6)}_{5,1} \widetilde{\chi}_{10} + \psi^{(6)}_{5,5} \widetilde{\chi}_{10},  \\
\widetilde{g}_{5} &=  \psi^{(6)}_{2,2} \widetilde{\chi}_{1} + \psi^{(6)}_{4,-2} \widetilde{\chi}_{2} + \psi^{(6)}_{4,2} \widetilde{\chi}_{2} + \psi^{(6)}_{4,4} \widetilde{\chi}_{1}  + \psi^{(6)}_{6,-4} \widetilde{\chi}_{0} + \psi^{(6)}_{6,-2} \widetilde{\chi}_{3} + \psi^{(6)}_{6,2} \widetilde{\chi}_{3} + \psi^{(6)}_{6,4} \widetilde{\chi}_{0} ,  \\
\widetilde{g}_{6} &= \psi^{(6)}_{2,0} \widetilde{\chi}_{5} + \psi^{(6)}_{4,0} \widetilde{\chi}_{4} + \psi^{(6)}_{6,0} \widetilde{\chi}_{6} + \psi^{(6)}_{6,6} \widetilde{\chi}_{7},  \\
\widetilde{g}_{7} &= \psi^{(6)}_{3,3} \widetilde{\chi}_{12} + \psi^{(6)}_{3,3} \widetilde{\chi}_{13} + \psi^{(6)}_{5,-3} \widetilde{\chi}_{11} + \psi^{(6)}_{5,3} \widetilde{\chi}_{11},  \\
\widetilde{g}_{8} &= \psi^{(6)}_{2,0} \widetilde{\chi}_{4} + \psi^{(6)}_{4,0} \widetilde{\chi}_{5} + \psi^{(6)}_{6,0} \widetilde{\chi}_{7} + \psi^{(6)}_{6,6} \widetilde{\chi}_{6}, \\
\widetilde{g}_{9} &= \psi^{(6)}_{2,2}\widetilde{\chi}_{4} + \psi^{(6)}_{4,-2} \widetilde{\chi}_{5}+ \psi^{(6)}_{4,2} \widetilde{\chi}_{5} + \psi^{(6)}_{4,4} \widetilde{\chi}_{4}  + \psi^{(6)}_{6,-4}\widetilde{\chi}_{6} + \psi^{(6)}_{6,-2} \widetilde{\chi}_{7}+ \psi^{(6)}_{6,2} \widetilde{\chi}_{7} + \psi^{(6)}_{6,4} \widetilde{\chi}_{6} , \\
\widetilde{g}_{10} &=  \psi^{(6)}_{1,1}\widetilde{\chi}_{11} + \psi^{(6)}_{3,-1} \widetilde{\chi}_{12}+ \psi^{(6)}_{3,-1} \widetilde{\chi}_{13} + \psi^{(6)}_{3,1} \widetilde{\chi}_{12}  + \psi^{(6)}_{3,1}\widetilde{\chi}_{13} + \psi^{(6)}_{5,-1} \widetilde{\chi}_{11}+ \psi^{(6)}_{5,1} \widetilde{\chi}_{11} + \psi^{(6)}_{5,5} \widetilde{\chi}_{11}  ,  \\
\widetilde{g}_{11} &= \psi^{(6)}_{2,2}\widetilde{\chi}_{5} + \psi^{(6)}_{4,-2} \widetilde{\chi}_{4}+ \psi^{(6)}_{4,2} \widetilde{\chi}_{4} + \psi^{(6)}_{4,4} \widetilde{\chi}_{5} + \psi^{(6)}_{6,-4}\widetilde{\chi}_{7} + \psi^{(6)}_{6,-2} \widetilde{\chi}_{6}+ \psi^{(6)}_{6,2} \widetilde{\chi}_{6} + \psi^{(6)}_{6,4} \widetilde{\chi}_{7}.
\end{split}
\end{align}
Here, $\widetilde{\chi}_{\alpha}(\tau)$ is short notation for $\chi_{\tsl{VF}^\natural_{22}(\alpha)}(\tau)$.
Using the expressions for the $\widetilde{g}_{i}(\tau)$ in equation \eqref{dual ch m35}, one can find the $q$-expansions of $\chi_{\tsl{VF}^\natural_{22}(\alpha)}(\tau)$. Here, we assumed $\widetilde{\chi}_{8}(\tau)=\widetilde{\chi}_{9}(\tau)$ and $\widetilde{\chi}_{12}(\tau)=\widetilde{\chi}_{13}(\tau)$, because ${\chi}_{8}(\tau)={\chi}_{9}(\tau)$ and ${\chi}_{12}(\tau)={\chi}_{13}(\tau)$. With these extra conditions, \eqref{gtoFi22} recovers the $q$-expansions \eqref{Fi22 dual characters}.


\clearpage

\begin{thebibliography}{99}

\bibitem{atlas}
 J.~H.~Conway, R.~T.~Curtis, S.~P.~Norton, R.~A.~Parker and R.~A.~Wilson,
``The ATLAS of Finite Groups,''
{\em Oxford University Press}, (1985).

\bibitem{griess1982friendly}
R.~L.~Griess~Jr.,
\newblock ``The friendly giant,''
\newblock {\em Inventiones mathematicae}, 69(1):1--102, (1982).

\bibitem{witt19375}
E.~Witt,
\newblock ``Die 5-fach transitiven gruppen von mathieu,''
\newblock {\em Abhandlungen aus dem mathematischen Seminar der
  Universit{\"a}t Hamburg}, volume~12, pages 256--264. Springer, (1937).
  
\bibitem{thompson1976conjugacy}
J.~G.~Thompson,
\newblock ``A conjugacy theorem for E$_8$,''
\newblock {\em Journal of Algebra}, 38(2):525--530, (1976).
  
\bibitem{smith1976simple}
P.~E.~Smith,
\newblock ``A simple subgroup of M? and E$_8$(3),''
\newblock {\em Bulletin of the London Mathematical Society}, 8(2):161--165, (1976).
  
\bibitem{flm}
I.~B.~Frenkel, J.~Lepowsky and A.~Meurman,
``A natural representation of the Fischer-Griess Monster with the modular function $J$ as character,''
{\em Proc. Nat. Acad. Sci. U.S.A.} {\bf 81} no. 10, pp 3256--3260 (1984) .

\bibitem{Borcherds}
R.~Borcherds,
``Monstrous moonshine and monstrous Lie superalgebras,"
{\em Invent. Math.} {\bf 109, No.2} pp. 405-444  (1992).

\bibitem{flma}
I.~B.~Frenkel, J.~Lepowsky and A.~Meurman,
``A moonshine module for the Monster," in {\it Vertex Operators in Mathematics and Physics}, Math. Sci. Res. Inst. Publ. 3, Springer, New York, 1985, 231-273.

\bibitem{Duncan-super}
J.~F.~R.~Duncan,
``Super-moonshine for Conway's largest sporadic group,"
{\em Duke Math. J.} 139, no. 2, 255-315, (2007).

\bibitem{Duncan-Mack}
  J.~F.~R.~Duncan and S.~Mack-Crane,
  ``The Moonshine Module for Conway's Group,''
  {\em SIGMA} {\bf 3}, e10 (2015).
  

\bibitem{Eguchi:2010ej}
  T.~Eguchi, H.~Ooguri and Y.~Tachikawa,
   ``Notes on the K3 Surface and the Mathieu group $M_{24}$,''
  {\em Exper.\ Math.\ }  {\bf 20}, 91 (2011)
[\href{https://arxiv.org/abs/1004.0956}{arXiv:1004.0956}].
 
 \bibitem{Cheng} 
  M.~C.~N.~Cheng,
  ``K3 Surfaces, N=4 Dyons, and the Mathieu Group M24,''
  Commun.\ Num.\ Theor.\ Phys.\  {\bf 4}, 623 (2010)
  doi:10.4310/CNTP.2010.v4.n4.a2
  [arXiv:1005.5415 [hep-th]].
\bibitem{Gaberdiel} 
  M.~R.~Gaberdiel, S.~Hohenegger and R.~Volpato,
  ``Mathieu Moonshine in the elliptic genus of K3,''
  JHEP {\bf 1010}, 062 (2010)
  doi:10.1007/JHEP10(2010)062
  [arXiv:1008.3778 [hep-th]].
\bibitem{Eguchi} 
  T.~Eguchi and K.~Hikami,
  ``Note on twisted elliptic genus of $K3$ surface,''
  Phys.\ Lett.\ B {\bf 694}, 446 (2011)
  doi:10.1016/j.physletb.2010.10.017
  [arXiv:1008.4924 [hep-th]].
\bibitem{gannon}
  T.~Gannon,
  ``Much ado about Mathieu,''
  {\em Adv. Math.} 301, 322 (2016)
[\href{https://arxiv.org/abs/1211.5531}{arXiv:1211.5531}]. 
\bibitem{UM}
  M.~C.~N.~Cheng, J.~F.~R.~Duncan and J.~A.~Harvey,
  ``Umbral Moonshine,''
  {\em Commun.\ Num.\ Theor.\ Phys.\ }  {\bf 08}, 101 (2014)
  doi:10.4310/CNTP.2014.v8.n2.a1
[\href{https://arxiv.org/abs/1204.2779}{arXiv:1204.2779}].

\bibitem{UMNL}
  M.~C.~N.~Cheng, J.~F.~R.~Duncan and J.~A.~Harvey,
  ``Umbral Moonshine and the Niemeier Lattices,''
  {\em Research in the Mathematical Sciences}, 2014, vol. 1.
[\href{https://arxiv.org/abs/1307.5793}{arXiv:1307.5793}]. 
 \bibitem{DGO}
  J.~F.~R.~Duncan, M.~J.~Griffin and K.~Ono,
  ``Proof of the Umbral Moonshine Conjecture,''
 {\em Res. Math. Sci.} 2, Art. 26 (2015).
[\href{https://arxiv.org/abs/1503.01472}{arXiv:1503.01472}].
\bibitem{Harvey:2015mca}
  J.~A.~Harvey and B.~C.~Rayhaun,
  ``Traces of Singular Moduli and Moonshine for the Thompson Group,''
  {\em Commun.\ Num.\ Theor.\ Phys.\ }{\bf 10}, 23 (2016)
[\href{https://arxiv.org/abs/1504.08179}{arXiv:1504.08179}].
   
\bibitem{griffin}
M.~J.~Griffin and M.~Mertens,
``A proof of the Thompson Moonshine Conjecture,"
{\em Res. Math. Sci.} {\bf 3}, no. One 36, 2016.
[\href{https://arxiv.org/abs/1607.03078}{arXiv:1607.03078}].  

\bibitem{onan}
J.~F.~R.~Duncan, M.~H.~Mertens and K.~Ono,
``Pariah moonshine,"
{\em Nature Communications}, {\bf 8} no. 670 (2017). 

\bibitem{anatomy}
S. Norton,
``Anatomy of the Monster I,"
in {\it The Atlas of Finite Groups--Ten Years on} (1998).

\bibitem{howe1995perspectives}
R.~Howe.
\newblock ``Perspectives on invariant theory: Schur duality, multiplicity-free
  actions and beyond,''
\newblock {\em The Schur lectures (1992)(Tel Aviv)}, pages 1--182, 1995.

\bibitem{howe1979fi}
R.~Howe,
\newblock ``$\theta$-series and invariant theory,''
\newblock {\em Automorphic Forms, Representations and $L$-Functions:
  Automorphic Forms, Representations and L-functions}, 1(Part 1):275--285,
  1979.

\bibitem{howe1989remarks}
R.~Howe,
\newblock ``Remarks on classical invariant theory,''
\newblock {\em Transactions of the American Mathematical Society},
  313(2):539--570, 1989.

\bibitem{dmz}
C.~Dong, G.~Mason and Y.~Zhu,
``Discrete Series of the Virasoro Algebra and the Moonshine Module,"
{\em Proc. of Symp.} in {\em Pure Mathematics}, Vol. 56, Part 2 (1994).

\bibitem{hoehnbaby}
G. H\"ohn,
``Selbstdual Vertesoperatorsuperalgebren und das Babymonster,"
{\it Ph.D. thesis, Bonn University} (1995)
[\href{https://arxiv.org/abs/0706.0236}{arXiv:0706.0236}].

\bibitem{hohn2012mckaye6}
G.~H{\"o}hn, C.~H.~Lam, and H.~Yamauchi.
``Mckay’s E6 observation on the largest Fischer group,''
\newblock {\em Communications in Mathematical Physics}, 310(2):329--365, (2012).


\bibitem{Goddard:1984vk} 
  P.~Goddard, A.~Kent and D.~I.~Olive,
  ``Virasoro Algebras and Coset Space Models,''
  {\it Phys.\ Lett.\ }{\bf 152B}, 88 (1985).
  

\bibitem{Goddard:1986ee} 
  P.~Goddard, A.~Kent and D.~I.~Olive,
  ``Unitary Representations of the Virasoro and Supervirasoro Algebras,''
  {\it Commun.\ Math.\ Phys.\ }{\bf 103}, 105 (1986).
 
\bibitem{fz}
V.~A.~Fateev and A.~B.~Zamolodchikov,
``Conformal quantum field theory models in two dimensions having $\mathbb{Z}_3$ symmetry,"
{\it Nucl. Phys. B} {\bf 280}, 644-660 (1987).

\bibitem{flone}
 V.~A.~Fateev and S.~L.~Lukyanov,
 ``The Models of Two-Dimensional Conformal Quantum Field Theory with $\mathbb{Z}_n$ Symmetry,''
 {\it Int.\ J.\ Mod.\ Phys.\ A} {\bf 3}, 507 (1988).

\bibitem{conway1985simple}
J.~H.~Conway.
``A simple construction for the Fischer-Griess monster group,''
\newblock {\em Inventiones mathematicae}, 79(3):513--540, (1985).

\bibitem{Sakuma}
S.~Sakuma,
``6-transposition property of $\tau$-involutions of vertex operator algebras,"
    [\href{https://arxiv.org/abs/math/0608709}{arXiv:math/0608709}].
    
\bibitem{griess2008ee_8}
R.~L.~Griess~Jr., and C.~H.~Lam,
\newblock ``$ EE_8$-lattices and dihedral groups,''
\newblock {\em Pure Appl. Math. Q.}7, no. 3, 621-743. (2011)

\bibitem{miyaconf}
M.~Miyamoto,
``VOAs generated by two conformal vectors whose $\tau$-involutions generate $S_3$,''
{\em J. Algebra} {\bf{268}}, no. 2 (2003) 653-671. 

\bibitem{sakuma2003vertex}
S.~Sakuma and H.~Yamauchi.
``Vertex operator algebra with two Miyamoto involutions generating $S_3$,''
\newblock {\em Journal of Algebra}, 267(1):272--297, (2003).


\bibitem{lametal}
C.~H.~Lam, H.~Yamada and H.~Yamauchi,
``McKay's observation and vertex operator algebras generated by two conformal vectors of central charge $1/2$,''
{\em Int Math Res Papers} Vol. 2005, 117 (2005) 
[\href{https://arxiv.org/abs/math/0503239}{arXiv:math/0503239}].

\bibitem{miyamotogriess}
M.~Miyamoto,
``Griess Algebras and Conformal Vectors in Vertex Operator Algebras,''
{\em Journal of Algebra}, {\bf 179}, 523 (1996).
 
\bibitem{creutzig2019schur}
T.~Creutzig, S.~Kanade, A.~R.~Linshaw, and D.~Ridout,
\newblock ``Schur--Weyl duality for Heisenberg cosets,''
\newblock {\em Transformation Groups}, 24(2):301--354, 2019.

\bibitem{lin2017mirror}
X.~Lin,
\newblock ``Mirror extensions of rational vertex operator algebras,''
\newblock {\em Transactions of the American Mathematical Society},
  369(6):3821--3840, 2017.

\bibitem{HarveyWu}
  J.~A.~Harvey and Y.~Wu,
  ``Hecke Relations in Rational Conformal Field Theory,''
  {\it JHEP} {\bf 1809}, 032 (2018)
    [\href{https://arxiv.org/abs/1804.06860}{arXiv:1804.06860}].


\bibitem{frenkel2004vertex}
E.~Frenkel and D.~Ben-Zvi,
``Vertex algebras and algebraic curves,''
Number~88, {\em American Mathematical Soc.}, (2004).

\bibitem{francesco2012conformal}
P.~Francesco, P.~Mathieu, and D.~ S{\'e}n{\'e}chal,
``Conformal field theory,''
\newblock {\em Springer Science \& Business Media}, (2012).

\bibitem{ginsparg1988applied}
P.~Ginsparg,
``Applied conformal field theory,''  
[\href{https://arxiv.org/abs/hep-th/9108028}{arXiv:hep-th/9108028}].


\bibitem{Dong:1994wn} 
  C.~Y.~Dong and G.~Mason,
  ``Nonabelian orbifolds and the boson - fermion correspondence,''
  {\em Commun.\ Math.\ Phys.\ }{\bf 163}, 523 (1994).
  

\bibitem{zhu1996modular}
Y. Zhu,
\newblock Modular invariance of characters of vertex operator algebras.
\newblock {\em Journal of the American Mathematical Society}, 9(1):237--302,
  (1996).
  
\bibitem{moore1989classical}
G.~Moore and N.~Seiberg,
``Classical and quantum conformal field theory,''
\newblock {\em Communications in Mathematical Physics}, 123(2):177--254, (1989).

\bibitem{verlinde1988fusion}
E.~Verlinde,
``Fusion rules and modular transformations in 2d conformal field theory,''
\newblock {\em Nucl. Phys. B}, 300:360--376, (1988).
  
\bibitem{Dong:1997ea} 
  C.~Y.~Dong, H.~S.~Li and G.~Mason,
  ``Modular invariance of trace functions in orbifold theory,''
  {\em Commun.\ Math.\ Phys.\ } {\bf 214}, 1 (2000)
[\href{https://arxiv.org/abs/q-alg/9703016}{arXiv:q-alg/9703016}].
    

\bibitem{Dong:2005xj} 
  C.~Y.~Dong and Z.~P.~Zhao,
  ``Modularity in orbifold theory for vertex operator superalgebras,''
  {\em Commun.\ Math.\ Phys.\ }{\bf 260}, 227 (2005).


\bibitem{beauty}
 L.~J.~Dixon, P.~H.~Ginsparg and J.~A.~Harvey,
  ``Beauty and the Beast: Superconformal Symmetry in a Monster Module,''
  {\it Commun.\ Math.\ Phys.\ }{\bf 119}, 221 (1988).

\bibitem{flmbook}
I.~Frenkel, J.~Lepowsky and A.~Meurman,
``Vertex Operator Algebras and the Monster,''
Volume 134, {\em Academic Press}, (1989). 
    
\bibitem{lam2003extension}
C.~H.~Lam, N.~Lam, and H.~Yamauchi.
`` Extension of unitary virasoro vertex operator algebra by a simple module,''
\newblock {\em International Mathematics Research Notices}, 2003(11):577--611, (2003).

\bibitem{zam}
A.~B.~Zamolodchikov,
``Infinite additional symmetries in two-dimensional conformal quantum field theory,''
{\it Teoreticheskaya i Matematicheskaya Fizika}, Vol. 65, No. 3, 347-359 (1985).

\bibitem{Mercat:2001pu}
  C.~Mercat and P.~A.~Pearce,
  ``Integrable and conformal boundary conditions for $\mathbb{Z}_k$ parafermions on a cylinder,''
  {\it J.\ Phys.\ A} {\bf 34}, 5751 (2001)
    [\href{https://arxiv.org/abs/hep-th/0103232}{arXiv:hep-th/0103232}].

\bibitem{dgm} 
  L.~Dolan, P.~Goddard and P.~Montague,
  ``Conformal field theories, representations and lattice constructions,''
  {\em Commun.\ Math.\ Phys.\ }  {\bf 179}, 61 (1996).

  
\bibitem{frenkel1992vertex}
I.~B.~Frenkel and Y.~Zhu.
``Vertex operator algebras associated to representations of affine and
  Virasoro algebras,''
\newblock {\em Duke Mathematical Journal}, 66(1):123--168, (1992).


\bibitem{Dijkgraaf:1989hb} 
  R.~Dijkgraaf, C.~Vafa, E.~P.~Verlinde and H.~L.~Verlinde,
  ``The Operator Algebra of Orbifold Models,''
  {\em Commun.\ Math.\ Phys.\ }{\bf 123}, 485 (1989).
 

\bibitem{dong1999representations}
C.~Dong and K.~Nagatomo,
``Representations of vertex operator algebra $V_L^+$ for rank one lattice $L$,''
\newblock {\em Communications in mathematical physics}, 202(1):169--195, (1999).

\bibitem{abe2004classification}
T.~Abe and C.~Dong,
``Classification of irreducible modules for the vertex operator algebra
 $V_L^+$: general case,''
\newblock {\em Journal of Algebra}, 273(2):657--685, (2004).

\bibitem{Gepner:1987vz} 
  D.~Gepner,
  ``Exactly Solvable String Compactifications on Manifolds of SU(N) Holonomy,''
  {\em Phys.\ Lett.\ B} {\bf 199}, 380 (1987).


\bibitem{Dong:1997xy} 
  C.~Dong, R.~L.~Griess Jr. and G.~Hoehn,
  ``Framed vertex operator algebras, codes and the moonshine module,''
  {\em Commun.\ Math.\ Phys.\ }{\bf 193}, 407 (1998)
[\href{https://arxiv.org/abs/q-alg/9707008}{arXiv:q-alg/9707008}].
 

\bibitem{DH}
L.~Dixon and J.~A.~Harvey, unpublished.

\bibitem{dong1996associative}
C.~Dong, H.~Li, G.~Mason, and S.~P.~Norton,
``Associative subalgebras of the griess algebra and related topics,''
\newblock In {\em Proc. of the Conference on the Monster and Lie algebras at
  The Ohio State University}, pp 27--42, (1996).

\bibitem{creutzig2019glueing}
T.~Creutzig, S.~Kanade, and R.~McRae,
\newblock ``Glueing vertex algebras,''
[\href{https://arxiv.org/abs/1906.00119}{arXiv:1906.00119}].

\bibitem{Mathur:1988na} 
  S.~Mathur, and S.~Mukhi, and A.~Sen,
  ``On the Classification of Rational Conformal Field Theories,''
  {\em Phys. Lett. B} 213, 303 (1988).

\bibitem{Mukhi:2019xjy} 
  S.~Mukhi,
  ``Classification of RCFT from Holomorphic Modular Bootstrap: A Status Report,''
[\href{https://arxiv.org/abs/1910.02973}{arXiv:1910.02973}].

\bibitem{arakawa2018quasi}
T.~Arakawa and K.~Kawasetsu,
\newblock ``Quasi-lisse vertex algebras and modular linear differential
  equations,''
\newblock In {\em Lie Groups, Geometry, and Representation Theory}, pages
  41--57. Springer, 2018.

\bibitem{gaberdiel2008modular}
M.~R.~Gaberdiel and C.~A.~Keller,
\newblock ``Modular differential equations and null vectors,''
\newblock {\em Journal of High Energy Physics}, 2008(09):079, 2008.

\bibitem{beem2018vertex}
C.~Beem and L.~Rastelli,
\newblock ``Vertex operator algebras, Higgs branches, and modular differential
  equations,''
\newblock {\em Journal of High Energy Physics}, 2018(8):114, 2018.

\bibitem{francmason}
C.~Franc and G.~Mason,
``Hypergeometric series, modular linear differential equations, and vector-valued modular forms,"
{\it Ramanujan J.} 41, No. 1-3, 233-267 (2016).

\bibitem{bantay}
P.~Bantay,
``The kernel of the modular representation and the Galois action in RCFT,"
{\it Commun. Math. Phys.} {\bf 233},423 (2003).



\bibitem{poincare1911fonctions}
H.~Poincar{\'e},
\newblock ``Fonctions modulaires et fonctions fuchsiennes,''
\newblock In {\em Annales de la Facult{\'e} des sciences de Toulouse:
  Math{\'e}matiques}, volume~3, pages 125--149, 1911.

\bibitem{rademacher1938fourier}
H.~Rademacher,
\newblock ``The Fourier coefficients of the modular invariant $J(\tau)$,''
\newblock {\em American Journal of Mathematics}, 60(2):501--512, 1938.

\bibitem{rademacher1938partition}
H.~Rademacher,
\newblock ``On the partition function $p(n)$,''
\newblock {\em Proceedings of the London Mathematical Society}, 2(1):241--254,
  1938.

\bibitem{niebur1974construction}
D.~Niebur,
\newblock ``Construction of automorphic forms and integrals,''
\newblock {\em Transactions of the American Mathematical Society},
  191:373--385, 1974.

\bibitem{whalen2014vector}
D.~Whalen,
\newblock ``Vector-valued Rademacher sums and automorphic integrals,''
\newblock {\em arXiv preprint arXiv:1406.0571}, 2014.


\bibitem{Duncan:2009sq} 
  J.~F.~Duncan and I.~B.~Frenkel,
  ``Rademacher sums, Moonshine and Gravity,''
  Commun.\ Num.\ Theor.\ Phys.\  {\bf 5}, 849 (2011)
  doi:10.4310/CNTP.2011.v5.n4.a4
  [arXiv:0907.4529 [math.RT]].

\bibitem{Cheng:2011ay} 
  M.~C.~N.~Cheng and J.~F.~R.~Duncan,
  ``On Rademacher Sums, the Largest Mathieu Group, and the Holographic Modularity of Moonshine,''
  Commun.\ Num.\ Theor.\ Phys.\  {\bf 6}, 697 (2012)
  doi:10.4310/CNTP.2012.v6.n3.a4
  [arXiv:1110.3859 [math.RT]].


\bibitem{dijkgraaf2000black}
R.~Dijkgraaf, J.~Maldacena, G.~Moore, and E.~Verlinde,
\newblock ``A black hole farey tail,''
\newblock {\em arXiv preprint hep-th/0005003}, 2000.

\bibitem{deBoer:2006vg} 
  J.~de Boer, M.~C.~N.~Cheng, R.~Dijkgraaf, J.~Manschot and E.~Verlinde,
  ``A Farey Tail for Attractor Black Holes,''
  JHEP {\bf 0611}, 024 (2006)
  doi:10.1088/1126-6708/2006/11/024
  [hep-th/0608059].

\bibitem{Manschot:2007ha} 
  J.~Manschot and G.~W.~Moore,
  ``A Modern Farey Tail,''
  Commun.\ Num.\ Theor.\ Phys.\  {\bf 4}, 103 (2010)
  doi:10.4310/CNTP.2010.v4.n1.a3
  [arXiv:0712.0573 [hep-th]].

\bibitem{Maloney:2016kee} 
  A.~Maloney, H.~Maxfield and G.~S.~Ng,
  ``A conformal block Farey tail,''
  JHEP {\bf 1706}, 117 (2017)
  doi:10.1007/JHEP06(2017)117
  [arXiv:1609.02165 [hep-th]].

\bibitem{Alday:2019vdr} 
  L.~F.~Alday and J.~Bae,
  ``Rademacher Expansions and the Spectrum of 2d CFT,''
  arXiv:2001.00022 [hep-th].

\bibitem{nally2019exact}
R.~Nally,
\newblock ``Exact half-BPS black hole entropies in CHL models from Rademacher
  series,''
\newblock {\em Journal of High Energy Physics}, 2019(1):60, 2019.

\bibitem{Ferrari:2017kbp} 
  F.~Ferrari and S.~M.~Harrison,
  ``Properties of extremal CFTs with small central charge,''
  arXiv:1710.10563 [hep-th].



\bibitem{cheng2014rademacher}
M.~C.~N.~Cheng and J.~F.~R.~Duncan.
\newblock Rademacher sums and Rademacher series.
\newblock In {\em Conformal field theory, automorphic forms and related
  topics}, pages 143--182. Springer, 2014.

\bibitem{Cardy:1986ie} 
  J.~L.~Cardy,
  ``Operator Content of Two-Dimensional Conformally Invariant Theories,''
  Nucl.\ Phys.\ B {\bf 270}, 186 (1986).
  doi:10.1016/0550-3213(86)90552-3

\bibitem{pmo}
	J.~F.~R.~Duncan, J.~A.~Harvey, and B.~C.~Rayhaun, 
	``An overview of penumbral moonshine,''
	\emph{in preparation}.


\bibitem{conwaynorton}
J.~H.~Conway and S.~P.~Norton,
``Monstrous Moonshine,''
{\em Bull. London Math. Soc.} {\bf 11}, 308 (1979).

\bibitem{mckaye8}
J.~McKay,
``Graphs, singularities and finite groups," The Santa Cruz Conference on Finite Groups (Santa Cruz, 1979), Proc. Symp. Pure Math. vol. 37, Amer. Math. Soc., Providence RI, 1980, pp 183-186. 

\bibitem{glaubnorton}
G.~Glauberman and S.~P.~Norton,
``On McKay's connection between the affine $E_8$ diagram and the Monster,"
CRM Proceedings and Lecture Notes, {\bf 30}, 2001. 

\bibitem{Griess_topics} 
R.~Griess~Jr.,
``Research topics in finite groups and vertex algebras,"
    [\href{https://arxiv.org/abs/1903.08805}{arXiv:1903.08805}].

\bibitem{hohn2010group}
G. H{\"o}hn.
``The group of symmetries of the shorter moonshine module,''
\newblock in {\em Abhandlungen aus dem Mathematischen Seminar der
  Universit{\"a}t Hamburg}, volume~80, pp 275--283. Springer, (2010).
  
\bibitem{yamauchi20052a}
H. Yamauchi.
`` 2A-orbifold construction and the baby-monster vertex operator
  superalgebra,''
\newblock {\em Journal of Algebra}, 284(2):645--668, (2005).
    
\bibitem{hohn2012mckaye7}
G.~H{\"o}hn, C.~H.~Lam, and H.~Yamauchi.
``Mckay’s E7 observation on the baby monster,''
\newblock {\em International Mathematics Research Notices}, 2012(1):166--212,
  (2012).

\bibitem{Bae:2018qfh}
  J.~Bae, K.~Lee and S.~Lee,
  ``Monster Anatomy,''
  {\it JHEP} {\bf 1907}, 026 (2019)
    [\href{https://arxiv.org/abs/1811.12263}{arXiv:1811.12263}].

\bibitem{Hampapura:2016mmz}
H.~Hampapura, and S.~Mukhi,,
``Two-dimensional RCFT’s without Kac-Moody symmetry,"
  {\em JHEP} {\bf 1607}, 138 (2016)
[\href{https://arxiv.org/abs/1605.03314}{arXiv:1605.03314}].

\bibitem{shimakura2002decompositions}
H.~Shimakura.
``Decompositions of the moonshine module with respect to subVOAs
  associated to codes over $\mathbb{Z}_{2k}$,"
\newblock {\em Journal of Algebra}, 251(1):308--322, (2002).

\bibitem{hohnmason}
G.~H{\"o}hn and G.~Mason,
``The 290 fixed-point sublattices of the Leech lattice,''
{\em J. Algebra} {\bf 448} (2016) 628-637.
  
\bibitem{MR1484478}
W.~Bosma, J.~Cannon, and C.~Playoust,
``The {M}agma algebra system {I} : {T}he user language,''
\newblock {\em J. Symbolic Comput.}, 24(3-4):235--265, 1997,
\newblock {\em Computational algebra and number theory} (London, 1993).

\bibitem{splag}
J.~H.~Conway and N.~J.~A.~Sloane,
``Sphere Packings, Lattices and Groups,''
Third edition, Springer 1993.

\bibitem{MN}
W.~Meyer and W.~Neutsch,
``Associative Subalgebras of the Griess Algebra,"
{\it Journal of Algebra} {\bf 158}, 1-17 (1993).

\bibitem{lam2000z2}
C.~H.~Lam and H.~Yamada.
``$\mathbb{Z}_2\times\mathbb{Z}_2$ codes and vertex operator algebras,''
\newblock {\em Journal of Algebra}, 224(2):268--291, (2000).

\bibitem{wilson1999maximal}
R.~A.~Wilson.
``The maximal subgroups of the baby monster, I,''
\newblock {\em Journal of Algebra}, 211(1):1--14, (1999).

\bibitem{GAP} The GAP Group, GAP - Groups, Algorithms, and Programming,
      Version 4.8.8; 2017 (https://www.gap-system.org).

\bibitem{linton1991maximal}
S.~A.~Linton and R.~A.~Wilson.
``The maximal subgroups of the Fischer groups $\tsl{Fi}_{24}$ and $\Fi'_{24}$,''
\newblock {\em Proceedings of the London Mathematical Society}, 3(1):113--164,
  (1991).
  
\bibitem{abe2001fusion}
T.~Abe,
\newblock ``Fusion rules for the charge conjugation orbifold,''
\newblock {\em Journal of Algebra}, 2(242):624--655, 2001.

\bibitem{abe2005fusion}
T.~Abe, C.~Dong, and H.~Li,
\newblock ``Fusion rules for the vertex operator algebras $M(1)^+$ and $V_L^+$,''
\newblock {\em Communications In Mathematical Physics}, 253(1):171--219, 2005.

\bibitem{Dong:2019ihr} 
  C.~Dong, X.~Jiao and N.~Yu,
  ``6$A$-Algebra and its representations,''
  J.\ Algebra {\bf 533}, 174 (2019)
  doi:10.1016/j.jalgebra.2019.06.003
  [arXiv:1902.06951 [math.QA]].
  
\bibitem{pahlings2007character}
H.~Pahlings.
``The character table of $2_+^{1+ 22}.\tsl{Co}_2$,''
\newblock {\em Journal of Algebra}, 315(1):301--325, (2007).

\bibitem{Ginsparg:1987eb} 
  P.~H.~Ginsparg,
  ``Curiosities at $c = 1$,''
  Nucl.\ Phys.\ B {\bf 295}, 153 (1988).
 
 
\bibitem{Lin:2019hks} 
  Y.~H.~Lin and S.~H.~Shao,
  ``Duality Defect of the Monster CFT,''
[\href{https://arxiv.org/abs/1911.00042}{arXiv:1911.00042}].
  

\bibitem{Johnson-Freyd:2019wgb} 
  T.~Johnson-Freyd,
  ``Supersymmetry and the Suzuki chain,''
[\href{https://arxiv.org/abs/1908.11012}{arXiv:1908.11012}].
  
\bibitem{sage}
\emph{{S}ageMath, the {S}age {M}athematics {S}oftware {S}ystem ({V}ersion
  8.8)}, The Sage Developers, 2019, {\tt https://www.sagemath.org}.



\end{thebibliography}
\end{document}